\documentclass{article}
\usepackage{epubtk}
\usepackage{graphicx}
\usepackage{amssymb}
\usepackage{amsthm}

\newcommand{\re}{\mbox{Re}}
\newcommand{\im}{\mbox{Im}}

\newcommand{\diag}{\mbox{diag}}

\newcommand{\Integer}{\mathbb{Z}}
\newcommand{\Real}{\mathbb{R}}
\newcommand{\Complex}{\mathbb{C}}

\newcommand{\gz}{\mbox{\em \r{g}\hspace{0.3mm}}}  % Zero order approximation.
\newcommand{\Rz}{\mbox{\em \r{R}}}
\newcommand{\Dz}{\mbox{\em \r{D}\hspace{0.3mm}}}
\newcommand{\nablaz}{\nabla\hspace{-0.27cm}{}^{\mbox{\r{~}}}{}\hspace{-0.12cm}}
\newcommand{\alphaz}{\alpha\hspace{-0.23cm}{}^{\mbox{\r{~}}}{}\hspace{-0.12cm}}
\newcommand{\betaz}{\beta\hspace{-0.23cm}{}^{\mbox{\r{~}}}{}\hspace{-0.12cm}}
\newcommand{\gammaz}{\gamma\hspace{-0.23cm}{}^{\mbox{\r{~}}}{}\hspace{-0.12cm}}
\newcommand{\Gammaz}{\Gamma\hspace{-0.25cm}{}^{\mbox{\r{~}}}{}\hspace{-0.12cm}}

\newcommand{\ds}{\displaystyle}
\newcommand{\la}{\langle}
\newcommand{\ra}{\rangle}

\newtheorem{definition}{Definition}
\newtheorem{lemma}{Lemma}

\newtheorem{theorem}{Theorem}

\newcounter{example}
\newenvironment{example}[1][]{\refstepcounter{example}\par\medskip\noindent%
   \textbf{Example~\theexample. #1} \rmfamily}{\medskip}
   
\newenvironment{dedication}
{
   \cleardoublepage
   \thispagestyle{empty}
   \vspace*{\stretch{1}}
   \hfill\begin{minipage}[t]{0.66\textwidth}
   \raggedright
}%
{
   \end{minipage}
   \vspace*{\stretch{3}}
   \clearpage
}

\newcommand{\be}{\begin{equation}}
\newcommand{\ee}{\end{equation}}
\newcommand{\bea}{\begin{eqnarray}}
\newcommand{\eea}{\end{eqnarray}}

\begin{document}

\title{Continuum and Discrete Initial-Boundary-Value Problems and Einstein's Field Equations}

\author{%
\epubtkAuthorData{Olivier Sarbach}{%
Instituto de F\'{\i}sica y Matem\'aticas \\
Universidad Michoacana de San Nicol\'as de Hidalgo  \\
Edificio C-3, Ciudad Universitaria \\
58040 Morelia, Michoac\'an, Mexico}{%
sarbach@ifm.umich.mx}{%
http://www.ifm.umich.mx/~sarbach}
\and
\epubtkAuthorData{Manuel Tiglio}{%
Center for Scientific Computation and Mathematical Modeling, \\
Department of Physics, \\
Joint Space Sciences Institute. \\
Maryland Center for Fundamental Physics,  \\
University of Maryland \\
College Park, MD 20742, USA}{%
tiglio@umd.edu}{%
http://www.cscamm.umd.edu/people/faculty/tiglio}
}

\date{}
\maketitle

\begin{abstract}
Many evolution problems in physics are described by partial differential equations on an infinite domain; therefore, one is interested in the solutions to such problems for a given initial dataset.  A prominent example is the binary black hole problem within Einstein's theory of gravitation, in which one computes the gravitational radiation emitted  from the inspiral of the two black holes, merger and ringdown. Powerful mathematical tools can be used to establish qualitative statements about the solutions, such as their existence, uniqueness, continuous dependence on the initial data, or their asymptotic behavior over large time scales. However, one is often interested in computing the solution itself, and unless the partial differential equation is very simple, or the initial data possesses a high degree of symmetry, this computation requires approximation by numerical discretization. When solving such discrete problems on a machine, one is faced with a finite limit to computational resources, which leads to the replacement of the infinite continuum domain with a finite computer grid. This, in turn, leads to a discrete initial-boundary value problem.  The hope is to recover, with high accuracy, the exact solution in the limit where the grid spacing converges to zero with the boundary being pushed to infinity.

The goal of this article is to review some of the theory necessary to understand the continuum and discrete initial-boundary value problems arising from hyperbolic partial differential equations and to discuss its applications to numerical relativity; in particular, we present well-posed initial and initial-boundary value formulations of Einstein's equations, and we discuss multi-domain high-order finite difference and spectral methods to solve them.
\end{abstract}

\epubtkKeywords{}

%===================================================================
\newpage
\thispagestyle{empty}
\begin{dedication}

To:
\vspace{1.0 cm}

\hspace{1.0cm} Susana and Eliana

\vspace{1.0 cm}

\hspace{1.0cm} The memory of Marta Pe\~na

\vspace{1.0 cm}

\hspace{1.0cm} Romina

\end{dedication}
%===================================================================

\newpage
\tableofcontents

\newpage 
%===================================================================
\section{Introduction}
\label{section:introduction}
%===================================================================

This review discusses fundamental tools from the analytical and numerical theory underlying the Einstein field equations as an evolution problem on a finite computational domain. The process of reaching the current status of numerical relativity after decades of effort not only has driven the community to use state of the art techniques but also to extend and work out new approaches and methodologies of its own. This review discusses some of the theory involved in setting up the problem and numerical approaches for solving it. Its scope is rather broad: it ranges from analytical aspects related to the well posedness of the Cauchy problem to numerical discretization schemes guaranteeing stability and convergence to the exact solution.

At the continuum, emphasis is placed on setting up the initial-boundary value problem (IBVP) for Einstein's equations properly, by which we mean obtaining a well posed formulation which is flexible enough to incorporate coordinate conditions which allow for long-term and accurate stable numerical evolutions. Here, the well posedness property is essential in that it guarantees existence of a unique solution which depends continuously on the initial and boundary data. In particular, this assures that small perturbations in the data do not get arbitrarily amplified. Since such small perturbations do appear in numerical simulations because of discretization errors or finite machine precision, if such unbounded growth were allowed, the numerical solution would not converge to the exact one as resolution is increased. This picture is at the core of Lax' historical theorem, which implies that the consistency of the numerical scheme is not sufficient for its solution to converge to the exact one. Instead, the scheme also needs to be numerically stable, a property which is the discrete counterpart of well posedness of the continuum problem.

While the well posedness of the Cauchy problem in General Relativity in the absence of boundaries was established a long time ago, only relatively recently has the IBVP been addressed and well posed problems formulated. This is mainly due to the fact that the IBVP presents several new challenges, related to constraint preservation, the minimization of spurious reflections, and well posedness. In fact, it is only very recently that such a well posed problem has been found for a metric based formulation used in numerical relativity, and there are still open issues that need to be sorted out. It is interesting to point out that the IBVP in General Relativity has driven research which has led to well posedness results for second order systems with a new large class of boundary conditions which, in addition to Einstein's equations, are also applicable to Maxwell's equations in their potential formulation.

At the discrete level, the focus of this review is mainly on designing numerical schemes for which fast convergence to the exact solution is guaranteed. Unfortunately, no or very few general results are known for nonlinear equations and, therefore, we concentrate on schemes for which stability and convergence can be shown at the linear level, at least. If the exact solution is smooth, as expected for vacuum solutions of Einstein's field equations with smooth initial data and appropriate gauge conditions, at least as long a no curvature singularities form, it is not unreasonable to expect that  schemes guaranteeing stability at the linearized level, perhaps with some additional filtering, are also stable for the nonlinear problem. Furthermore, since the solutions are expected to be smooth, emphasis is here placed on using fast converging space discretizations, such as high order finite difference or spectral methods, especially those which can be applied to multi-domain implementations.

The organization of this review is at follows. Section~\ref{section:ivp} starts with a discussion of well posedness for initial value problems for evolution problems in general, with special emphasis on hyperbolic ones, including their algebraic characterization. Next, in Section~\ref{section:IVFEinstein} we review some formulations of Einstein's equations which yield a well posed initial value problem. Here, we mainly focus on the harmonic and BSSN formulations which are the two most widely used ones in numerical relativity, as well as the ADM formulation with different gauge conditions. Actual numerical simulations always involve the presence of computational boundaries, which raises the need of analyzing well posedness of the IBVP. For this reason, the theory of IBVP for hyperbolic problems is reviewed in Section~\ref{section:ibvp}, followed by a presentation of the state of the art of boundary conditions for the harmonic and BSSN formulations of Einstein's equations in Section~\ref{section:ibvpEinstein}, where open problems related with gauge uniqueness are also described.

Section~\ref{sec:num_stability} reviews some of the numerical stability theory, including necessary eigenvalue conditions. These are quite useful in practice for analyzing complicated systems or discretizations. We also discuss necessary and sufficient conditions for stability within the method of lines, and Runge--Kutta methods. Sections~\ref{sec:fd} and \ref{sec:spec} are devoted to two classes of spatial approximations: finite differences and spectral methods. Finite differences are rather standard and widespread, so in Section~\ref{sec:fd} we mostly focus on the construction of optimized operators of arbitrary high order satisfying the Summation By Parts property, which is useful in stability analyses. We also briefly mention classical polynomial interpolation and how to systematically construct finite difference operators from it. In Section~\ref{sec:spec} we present the main elements and theory of spectral methods, including spectral convergence from solutions to Sturm--Liouville problems, expansions in orthogonal polynomials, Gauss quadratures, spectral differentiation, and spectral viscosity. We present several explicit formulae for the families of polynomials most widely used: Legendre and Chebyshev. Section~\ref{sec:num_boundary} describes  boundary closures. In the present context they refer to procedures for imposing boundary conditions leading to stability results. We emphasize the penalty technique, which applies to both finite difference methods of arbitrary high order and spectral ones, as well as outer and interface boundaries, such as those appearing when there are multiple grids as in complex geometries domain decompositions. We also discuss absorbing boundary conditions for Einstein's equations. Finally, Section~\ref{sec:complex} presents a random sample of approaches in numerical relativity using multiple, semi-structured grids, and/or curvilinear coordinates. In particular, some of these examples illustrate many of the methods discussed in this review in realistic simulations. 

There are many topics related to numerical relativity which are not covered by this review. It does not include discussions of physical results in General Relativity obtained through numerical simulations, such as critical phenomena or gravitational waveforms computed from binary black hole mergers. For reviews on these topics we refer the reader to~\cite{cGjG07} and~\cite{fP09,jCjBbKjvM10}, respectively. See also Refs.~\cite{Alcubierre-Book,BaumgarteShapiro-Book} for recent books on numerical relativity. Next, we do not discuss setting up initial data and solving the Einstein constraints, and refer to~\cite{gC00}. For reviews on the characteristic and conformal approach, which are only briefly mentioned in Section~\ref{SubSec:OutToInfinity}, we refer the reader to~\cite{Winicour:2008tr} and \cite{jF04}, respectively. Most of the results specific to Einstein's field equations in Sections~\ref{section:IVFEinstein} and \ref{section:ibvpEinstein} apply to four-dimensional gravity only, though it should be possible to generalize some of them to higher-dimensional theories. Also, as we have already mentioned, the results described here mostly apply to the vacuum field equations, in which case the solutions are expected to be smooth. For aspects involving the presence of shocks, such as those present in relativistic hydrodynamics we refer the reader to Refs.~\cite{Font:2008zz, Marti:2003wi}. Finally, see~\cite{oR98} for a more detailed review on hyperbolic formulations of Einstein's equations, and \cite{aR05} for one on global existence theorems in General Relativity. Spectral methods in numerical relativity are discussed in detail in \cite{GrandclementNovakLR}. The $3+1$ approach to General Relativity is thoroughly reviewed in \cite{Gourgoulhon}. Finally, we refer the reader to~\cite{ChoquetBruhat-Book} for a recent book on General Relativity and the Einstein equations which, among many other topics, discusses local and global aspects of the Cauchy problem, the constraint equations, and self-gravitating matter fields such as relativistic fluids and the relativistic kinetic theory of gases.

Except for a few historical remarks, this review does not discuss much of the historical path to the techniques and tools presented, but rather describes the state of the art of a subset of those which appear to be useful. Our choice of topics is mostly influenced by those for which some analysis is available or possible.  

We have tried to make each section as self-consistent as possible within the scope of a manageable review, so that they can be read separately, though each of them builds from the previous ones. Numerous examples are included. 

\newpage

%===================================================================
\section{Notation and Conventions}
\label{section:notation}
%===================================================================

Throughout this article, we use the following notation and conventions. For a complex vector $u\in \Complex^m$, we denote by $u^*$ its transposed, complex conjugate, such that $u\cdot v := u^* v$ is the standard scalar product for two vectors $u,v\in\Complex^m$. The corresponding norm is defined by $|u|:=\sqrt{u^* u}$. The norm of a complex, $m\times k$ matrix $A$ is
\begin{displaymath}
|A| := \sup\limits_{u\in\Complex^k\setminus\{ 0 \}} \frac{|A u|}{|u|}.
\end{displaymath}
The transposed, complex conjugate of $A$ is denoted by $A^*$, such that $v\cdot (Au) = (A^* v)\cdot u$ for all $u\in\Complex^k$ and $v\in\Complex^m$. For two Hermitian $m\times m$ matrices $A=A^*$ and $B=B^*$, the inequality $A\leq B$ means $u\cdot A u \leq u\cdot Bu$ for all $u\in\Complex^m$. The identity matrix is denoted by $I$.

The spectrum of a complex, $m\times m$ matrix $A$ is the set of all eigenvalues of $A$,
\begin{displaymath}
\sigma(A) := \{ \lambda\in\Complex : \lambda I - A \hbox{ is not invertible } \},
\end{displaymath}
which is real for Hermitian matrices. The spectral radius of $A$ is defined as 
\begin{displaymath}
\rho(A):= \max \{ |\lambda |: \lambda \in \sigma(A) \}. 
\end{displaymath}
Then, the matrix norm $|B|$ of a complex $m\times k$ matrix $B$ can also be computed as $|B| = \sqrt{\rho(B^* B)}$.

Next, we denote by $L^2(U)$ the class of measurable functions $f: U\subset\Real^n \to \Complex^m$ on the open subset $U$ of $\Real^n$ which are square-integrable. Two functions $f,g\in L^2(U)$ which differ from each other only by a set of measure zero are identified. The scalar product on $L^2(U)$ is defined as
\begin{displaymath}
\la f,g \ra := \int\limits_U f(x)^* g(x) d^n x,\qquad f,g\in L^2(U),
\end{displaymath}
and the corresponding norm is $\| f \| := \sqrt{\la f,f \ra}$. According to the Cauchy-Schwarz inequality we have
\begin{equation}
\la f,g \ra \leq \| f \| \| g \|,\qquad f,g\in L^2(U).
\end{equation}
The Fourier transform of a function $f$ belonging to the class $C_0^\infty(\Real^n)$ of infinitely differentiable functions with compact support is defined as
\begin{equation}
\hat{f}(k) := \frac{1}{(2\pi)^{n/2}} \int e^{-i k\cdot x} f(x) d^n x,
\qquad k\in\Real^n.
\end{equation}
According to Parseval's identities, $\la \hat{f},\hat{g} \ra = \la f,g \ra$ for all $f,g\in C_0^\infty(\Real^n)$, and the map $C_0^\infty(\Real^n) \to L^2(\Real^n)$, $f\mapsto \hat{f}$ can be extended to a linear, unitary map ${\cal F}: L^2(\Real^n) \to L^2(\Real^n)$ called the Fourier-Plancharel operator, see for example Ref.~\cite{ReedSimon80II}. Its inverse is given by ${\cal F}^{-1}(f)(x) = \hat{f}(-x)$ for $f\in L^2(\Real^n)$ and $x\in\Real^n$.

For a differentiable function $u$, we denote by $u_t$, $u_x$, $u_y$, $u_z$. its partial derivatives with respect to $t$, $x$, $y$, $z$.

Indices labeling gridpoints and number of basis functions range from $0$ to $N$. Superscripts and subscripts are used to denote the numerical solution at some discrete timestep and gridpoint, as in 
$$
v^k_j:= v(t_k, x_j) \,  . 
$$
We use boldface fonts for gridfunctions, as in 
$$
{\bf v}^k:= \{ v(t_k, x_j) \}_{j=0}^N \, . 
$$

\newpage
%===================================================================
%===================================================================
\section{The Initial Value Problem}
\label{section:ivp}
%===================================================================
%===================================================================

We start here with a discussion of hyperbolic evolution problems on
the infinite domain $\Real^n$. This is usually the situation one
encounters in the mathematical description of isolated systems, where
some strong field phenomena takes place ``near the origin'' and
generates waves which are emitted toward ``infinity''. Therefore, the goal
of this section is to analyze the well-posedness of the Cauchy problem
for quasilinear hyperbolic evolution equations without
boundaries. The case with boundaries is the subject of the
Section~\ref{section:ibvp}. As mentioned in the introduction, the
well-posedness results are fundamental in the sense that they give
existence (at least local in time if the problem is nonlinear) and
uniqueness of solutions and show that these depend continuously on the
initial data. Of course, how the solution actually looks like in
detail needs to be established by more sophisticated mathematical
tools or by numerical experiments, but it is clear that it does not
make sense to speak about ``the solution'' if the problem is not well
posed.

Our presentation starts with the simplest case of linear constant coefficient problems in Section~\ref{SubSec:LPCC}, where solutions can be constructed explicitly using Fourier transform. Then, we consider in Section~\ref{SubSec:LPVC} linear problems with variable coefficients which we reduce to the constant coefficient case using the localization principle. Next, in Section~\ref{SubSec:QLP} we treat first order quasilinear equations which we reduce to the previous case by the principle of linearization. Finally, in Section~\ref{SubSec:SemiGroups} we summarize some basic results about abstract evolution operators, which give the general framework for treating evolution problems including not only those described by local partial differential operators, but also more general ones.

Much of the material from the first three subsections is taken from
the book by Kreiss and Lorenz \cite{KL89}. However, our summary also
includes recent results concerning second order equations, examples of
wave systems on curved spacetimes, and a very brief review of semigroup
theory.

%===================================================================
\subsection{Linear, constant coefficient problems}
\label{SubSec:LPCC}
%===================================================================

We consider an evolution equation on $n$-dimensional space of the following form:  
\begin{equation}
u_t = P(\partial/\partial x) u
  \equiv \sum\limits_{|\nu| \leq p} A_\nu D_\nu u,
\qquad x\in \Real^n, \quad t \geq 0.
\label{Eq:LinearCCPDE}
\end{equation}
Here, $u = u(t,x)\in \Complex^m$ is the state vector, and $u_t$ its partial derivative with respect to $t$. Next, the $A_\nu$'s denote complex, $m\times m$ matrices where $\nu = (\nu_1,\nu_2,\ldots,\nu_n)$ denotes a multi-index with components $\nu_j\in \{ 0,1,2,3,\ldots \}$ and $|\nu| := \nu_1 + \ldots + \nu_n$. Finally, $D_\nu$ denotes the partial derivative operator
\begin{displaymath}
D_\nu := \frac{\partial^{|\nu|}}{\partial x_1^{\nu_1}\cdot\cdot\cdot \partial x_n^{\nu_n}}
\end{displaymath}
of order $|\nu|$, where $D_0 := I$. Here are a few representative examples:

\begin{example}
The advection equation $u_t(t,x) = \lambda u_x(t,x)$ with speed $\lambda\in\Real$ in the negative $x$ direction.
\end{example}

\begin{example}
The heat equation $u_t(t,x) = \Delta u(t,x)$, where
\begin{displaymath}
\Delta := \frac{\partial^2}{\partial x_1^2} + \frac{\partial^2}{\partial x_2^2} + \ldots
 + \frac{\partial^2}{\partial x_n^2}  
\end{displaymath}
denotes the Laplace operator.
\end{example}

\begin{example}
The Schr\"odinger equation $u_t(t,x) = i \Delta u(t,x)$.
\end{example}

\begin{example}
The wave equation $U_{tt} = \Delta U$, which can be cast into the form of Equation~(\ref{Eq:LinearCCPDE}),
\begin{displaymath}
u_t = \left( \begin{array}{cc} 0 & 1 \\ \Delta & 0 \end{array} \right) u,\qquad
u = \left( \begin{array}{l} U \\ V \end{array} \right).
\end{displaymath}
\end{example}

We can find solutions of Equation~(\ref{Eq:LinearCCPDE}) by Fourier transformation in space,
\begin{displaymath}
\hat{u}(t,k) = \frac{1}{(2\pi)^{n/2}}\int e^{-i k\cdot x} u(t,x) d^n x,
\qquad k\in\Real^n,\quad t \geq 0.
\end{displaymath}
Applied to Equation~(\ref{Eq:LinearCCPDE}) this yields the system of linear ordinary differential equations
\begin{equation}
\hat{u}_t = P(i k)\hat{u},\qquad t \geq 0,
\label{Eq:LinearCCPDEFourier}
\end{equation}
for each wave vector $k\in \Real^n$ where $P(i k)$, called the \textbf{symbol} of the differential operator $P(\partial/\partial x)$, is defined as
\begin{equation}
P(i k) := \sum\limits_{|\nu| \leq p} A_\nu(i k_1)^{\nu_1} \cdot\cdot\cdot (i k_n)^{\nu_n},
\qquad k\in\Real^n.
\label{Eq:Symbol}
\end{equation}
The solution of Equation~(\ref{Eq:LinearCCPDEFourier}) is given by
\begin{displaymath}
\hat{u}(t,k) = e^{P(ik) t} \hat{u}(0,k),\qquad t \geq 0,
\end{displaymath}
where $\hat{u}(0,k)$ is determined by the initial data for $u$ at $t=0$. Therefore, the formal solution of the Cauchy problem
\begin{eqnarray}
u_t(t,x) = P(\partial/\partial x) u(t,x),
&& x\in \Real^n, \quad t \geq 0,
\label{Eq:LinearCCPDECauchy1}\\
u(0,x) = f(x),
&& x\in\Real^n,
\label{Eq:LinearCCPDECauchy2}
\end{eqnarray}
with given initial data $f$ for $u$ at $t=0$ is
\begin{equation}
u(t,x) = \frac{1}{(2\pi)^{n/2}}\int e^{i k\cdot x} e^{P(ik)t}\hat{f}(k) d^n k,
\qquad x\in \Real^n,\quad t \geq 0,
\label{Eq:FormalIntegralRepresentation}
\end{equation}
where $\hat{f}(k) = \frac{1}{(2\pi)^{n/2}}\int e^{-i k\cdot x} f(x) d^n x$.

\subsubsection{Well-posedness}

At this point, however, we have to ask ourselves if the expression~(\ref{Eq:FormalIntegralRepresentation}) makes sense. In fact, we do not expect the integral to converge in general. Even if $\hat{f}$ is smooth and decays to zero as $|k|\to\infty$ we could still have problems if $|e^{P(ik)t}|$ diverges as $|k| \to \infty$. One simple, but very restrictive, possibility to control this problem is to limit ourselves to initial data $f$ in the class ${\cal S}^\omega$ of functions which are the Fourier transform of a $C^\infty$-function with compact support, i.e. $f\in {\cal S}^\omega$, where
\begin{displaymath}
{\cal S}^\omega := \left\{ v(\cdot)
 = \frac{1}{(2\pi)^{n/2}} \int e^{i k\cdot (\cdot)} \hat{v}(k) d^n k : 
\hat{v}\in C^\infty_0(\Real^n) \right\}.
\end{displaymath}
A function in this space is real analytic and decays faster than any polynomial as $|x|\to \infty$.\epubtkFootnote{More precisely, it follows from the Paley-Wiener theorem (see Theorem IX.11 in Ref.~\cite{ReedSimon80II}) that $f\in  {\cal S}^\omega$ with $\hat{f}(k) = 0$ for $|k|\geq R$ if and only if $f$ possesses an analytic extension $\bar{f}:\Complex^n\to\Complex$ such that for each $N=0,1,2,\ldots$ there exists a constant $C_N$ with
\begin{displaymath}
|\bar{f}(\zeta)| \leq C_N\frac{e^{R|\im(\zeta)|}}{(1+|\zeta|)^N},
\qquad \zeta\in\Complex^n.
\end{displaymath}} If $f\in {\cal S}^\omega$ the integral in Equation~(\ref{Eq:FormalIntegralRepresentation}) is well-defined and we obtain a solution of the Cauchy problem~(\ref{Eq:LinearCCPDECauchy1}, \ref{Eq:LinearCCPDECauchy2}) which, for each $t\geq 0$ lies in this space. However, this possibility suffers from several unwanted features:
\begin{itemize}
\item The space of admissible initial data is very restrictive. Indeed, since $f\in {\cal S}^\omega$ is necessarily analytic it is not possible to consider nontrivial data with, say, compact support and study the propagation of the support for such data.
\item For fixed $t > 0$, the solution may grow without bound when perturbations with arbitrarily small amplitude but higher and higher frequency components are considered. Such an effect is illustrated in Example~\ref{Example:BackwardsHeat} below.
\item The function space ${\cal S}^\omega$ does not seem to be useful as a solution space when considering linear variable coefficient or quasilinear problems, since for such problems the different $k$ modes do not decouple from each other. Hence, mode coupling can lead to components with arbitrarily high frequencies.\epubtkFootnote{In this regard we should mention the Cauchy-Kovalevskaya theorem (see, for example Ref.~\cite{Evans-Book}) which always provides unique local analytic solutions to the Cauchy problem for quasilinear partial differential equations with analytic coefficients and data on a non-characteristic surface. However, this theorem does not say anything about causal propagation and stability with respect to high-frequency perturbations.}
\end{itemize}
For these reasons, it is desirable to consider initial data of a more general class than ${\cal S}^\omega$. For this, we need to control the growth of $e^{P(ik)t}$. This is captured in the following

\begin{definition}
\label{Def:WP}
The Cauchy problem~(\ref{Eq:LinearCCPDECauchy1}, \ref{Eq:LinearCCPDECauchy2}) is called \textbf{well posed} if there are constants $K\geq 1$ and $ \alpha\in\Real$ such that
\begin{equation}
|e^{P(ik) t}| \leq K e^{\alpha t} \qquad
\hbox{for all $ t\geq 0 $ and all $ k\in\Real^n $}.
\label{Eq:WPEstimate}
\end{equation}
\end{definition}

The importance of this definition relies in the property that for each fixed time $t > 0$ the norm $|e^{P(ik)t}|$ of the propagator is bounded by the constant $C(t):=K e^{\alpha t}$ which is \emph{independent of the wave vector $k$}. The definition does not state anything about the growth of the solution with time other that this growth is bounded by an exponential. In this sense, unless one can choose $\alpha\leq 0$ or $\alpha > 0$ arbitrarily small, well posedness is not a statement about the stability in time, but rather about stability with respect to mode fluctuations.

Let us illustrate the meaning of Definition~\ref{Def:WP} with a few examples:

\begin{example}
The heat equation $u_t(t,x) = \Delta u(t,x)$.\\
Fourier transformation converts this equation into $\hat{u}_t(t,k) = -|k|^2\hat{u}(t,k)$. Hence, the symbol is $P(ik) = -|k|^2$ and $|e^{P(ik) t}| = e^{-|k|^2 t} \leq 1$. The problem is well posed.
\end{example}

\begin{example}
\label{Example:BackwardsHeat}
The backwards heat equation $u_t(t,x) = -\Delta u(t,x)$.\\
In this case the symbol is $P(ik) = +|k|^2$, and $|e^{P(ik) t}| = e^{|k|^2 t}$. In contrast to the previous case, $e^{P(ik) t}$ exhibits exponential frequency-dependent growth for each fixed $t > 0$ and the problem is not well posed. Notice that small initial perturbations with large $|k|$ are amplified by a factor that becomes larger and larger as $|k|$ increases. Therefore, after an arbitrarily small time, the solution is contaminated by high-frequency modes.
\end{example}

\begin{example}
The Schr\"odinger equation $u_t(t,x) = i \Delta u(t,x)$.\\
In this case we have $P(ik) = i |k|^2$ and $|e^{P(ik) t}| = 1$. The problem is well posed. Furthermore, the evolution is unitary, and we can evolve forward and backwards in time. When compared to the previous example, it is the factor $i$ in front of the Laplace operator that saves the situation and allows the evolution backwards in time.
\end{example}

\begin{example}
The one-dimensional wave equation written in first order form,
\begin{displaymath}
u_t(t,x) = A u_x(t,x), \qquad
A = \left( \begin{array}{ll} 0 & 1 \\ 1 & 0 \end{array} \right). 
\end{displaymath}
The symbol is $P(ik) = ik A$. Since the matrix $A$ is symmetric and has eigenvalues $\pm 1$, there exists an orthogonal transformation $U$ such that
\begin{displaymath}
A = U\left( \begin{array}{ll} 1 & 0 \\ 0 & -1 \end{array} \right) 
U^{-1}, \qquad
e^{ik A t} = U
\left( \begin{array}{ll} e^{ikt} & 0 \\ 0 & e^{-ikt} \end{array} \right) 
U^{-1}.
\end{displaymath}
Therefore, $|e^{P(ik) t}| = 1$, and the problem is well posed.
\end{example}

\begin{example}
Perturb the previous problem by a lower order term,
\begin{displaymath}
u_t(t,x) = A u_x(t,x) + \lambda u(t,x), \qquad
A = \left( \begin{array}{ll} 0 & 1 \\ 1 & 0 \end{array} \right),\qquad \lambda\in\Real.
\end{displaymath}
The symbol is $P(ik) = ik A + \lambda I$, and $|e^{P(ik) t}| = e^{\lambda t}$. The problem is well posed, even though the solution grows exponentially in time if $\lambda > 0$.
\end{example}

More generally one can show, see Theorem~2.1.2 in Ref.~\cite{KL89}:
\begin{lemma}
\label{Lem:1DWP}
The Cauchy problem for the first order equation $u_t = A u_x + B$ with complex $m\times m$ matrices $A$ and $B$ is well posed if and only if $A$ is diagonalizable and has only real eigenvalues.
\end{lemma}

By considering the eigenvalues of the symbol $P(ik)$ we obtain the following simple necessary condition for well posedness:

\begin{lemma}[Petrovskii condition]
\label{Lem:Petrovskii}
Suppose the Cauchy problem~(\ref{Eq:LinearCCPDECauchy1}, \ref{Eq:LinearCCPDECauchy2}) is well posed. Then, there is a constant $\alpha \in \Real$ such that
\begin{equation}
\re(\lambda) \leq \alpha
\label{Eq:Petrovskii}
\end{equation}
for all eigenvalues $\lambda$ of $P(ik)$.
\end{lemma}

\proof
Suppose $\lambda$ is an eigenvalue of $P(ik)$ with corresponding eigenvector $v$, $P(ik) v = \lambda v$. Then, if the problem is well posed,
\begin{displaymath}
K e^{\alpha t}|v| \geq |e^{P(ik)t} v| = |e^{\lambda t} v | = e^{\re(\lambda) t} |v|,
\end{displaymath}
for all $t\geq 0$ and all $k\in\Real^n$ which implies that $e^{\re(\lambda) t} \leq K e^{\alpha t}$ for all $t\geq 0$, and hence $\re(\lambda)\leq \alpha$.
\qed

Although the Petrovskii condition is a very simple necessary condition, we stress that it is not sufficient in general. Counterexamples are first order systems which are weakly but not strongly hyperbolic, see Example~\ref{Example:WH} below.

\subsubsection{Extension of solutions}
\label{SubSubSec:ExtensionOfSolutions}

Now that we have defined and illustrated the notion of well-posedness, let us see how it can be used to solve the Cauchy problem~(\ref{Eq:LinearCCPDECauchy1}, \ref{Eq:LinearCCPDECauchy2}) for initial data more general than in ${\cal S}^\omega$. Suppose first that $f\in {\cal S}^\omega$, as before. Then, if the problem is well posed, Parseval's identities imply that the solution~(\ref{Eq:FormalIntegralRepresentation}) must satisfy
\begin{displaymath}
\| u(t,.) \| = \| \hat{u}(t,.) \| = \| e^{P(i\cdot) t}\hat{f} \|
 \leq K e^{\alpha t} \| \hat{f} \| = K e^{\alpha t} \| f \|,\qquad t\geq 0.
\end{displaymath}
Therefore, the ${\cal S}^\omega$-solution satisfies the following estimate
\begin{equation}
\| u(t,.) \| \leq K e^{\alpha t} \| f \|,\qquad t \geq 0,
\label{Eq:WPL2Norms}
\end{equation}
for all $f\in {\cal S}^\omega$. This estimate is important because it allows us to extend the solution to the much larger space $L^2(\Real^n)$. This extension is defined in the following way: let $f\in L^2(\Real^n)$. Since ${\cal S}^\omega$ is dense in $L^2(\Real^n)$ there exists a sequence $\{ f_j \}$ in ${\cal S}^\omega$ such that $\| f_j - f \| \to 0$. Therefore, if the problem is well posed, it follows from the estimate~(\ref{Eq:WPL2Norms}) that the corresponding solutions $u_j$ defined by Equation~(\ref{Eq:FormalIntegralRepresentation}) form a Cauchy-sequence in $L^2(\Real^n)$, and we can define
\begin{equation}
U(t) f(x) := \lim\limits_{j\to\infty}
\frac{1}{(2\pi)^{n/2}}\int e^{i k\cdot x} e^{P(ik)t}\hat{f}_j(k) d^n k,
\qquad x\in\Real^n,\quad t\geq 0,
\end{equation}
where the limit exists in the $L^2(\Real^n)$ sense. The linear map $U(t): L^2(\Real^n) \to L^2(\Real^n)$ satisfies the following properties:
\begin{enumerate}
\item[(i)] $U(0) = I$ is the identity map.
\item[(ii)] $U(t + s) = U(t) U(s)$ for all $t,s\geq 0$.
\item[(iii)] For $f\in {\cal S}^\omega$, $u(t,.) = U(t) f$ is the unique solution to the Cauchy problem~(\ref{Eq:LinearCCPDECauchy1}, \ref{Eq:LinearCCPDECauchy2}).
\item[(iv)] $ \| U(t) f \| \leq K e^{\alpha t} \| f \|$ for all $f\in L^2(\Real^n)$ and all $t\geq 0$.
\end{enumerate}
The family $\{ U(t) : t\geq 0 \}$ is called a semi-group on $L^2(\Real^n)$. In general, $U(t)$ cannot be extended to negative $t$ as the example of the backwards heat equation, Example~\ref{Example:BackwardsHeat}, shows.

For $f\in L^2(\Real^n)$ the function $u(t,x):= U(t) f(x)$ is called a \textbf{weak solution} of the Cauchy problem~(\ref{Eq:LinearCCPDECauchy1}, \ref{Eq:LinearCCPDECauchy2}). It can also be constructed in an abstract way by using the Fourier--Plancharel operator ${\cal F}: L^2(\Real^n) \to L^2(\Real^n)$. If the problem is well posed, then for each $f\in L^2(\Real^n)$ and $t\geq 0$ the map $k\mapsto e^{P(ik) t}{\cal F}(f)(k)$ defines an $L^2(\Real^n)$-function, and hence we can define
\begin{displaymath}
u(t,\cdot) := {\cal F}^{-1}\left( e^{P(i\cdot) t}{\cal F} f \right), \qquad t\geq 0.
\end{displaymath}

According to \textbf{Duhamel's principle}, the semi-group $U(t)$ can also be used to construct weak solutions of the inhomogeneous problem,
\begin{eqnarray}
u_t(t,x) = P(\partial/\partial x) u(t,x) + F(t,x),
&& x\in \Real^n, \quad t \geq 0,
\label{Eq:LinearCCPDECauchy1Inhomo}\\
u(0,x) = f(x),
&& x\in\Real^n,
\label{Eq:LinearCCPDECauchy2Inhomo}
\end{eqnarray}
where $F: [0,\infty) \to L^2(\Real^n)$, $t\mapsto F(t,\cdot)$ is continuous:
\begin{equation}
u(t,\cdot) = U(t) f + \int\limits_0^t U(t-s) F(s,\cdot) ds.
\label{Eq:DuhamelFormula}
\end{equation}
For a discussion on semi-groups in a more general context see Section~\ref{SubSec:SemiGroups} below.

\subsubsection{Algebraic characterization}

In order to extend the solution concept to initial data more general than analytic one, we have introduced the concept of well-posedness in Definition~\ref{Def:WP}. However, given a symbol $P(ik)$, it is not always a simple task to determine whether or not constants $K \geq 0$ and $\alpha\in\Real$ exist such that $|e^{P(ik)t}| \leq K e^{\alpha t}$ for all $t\geq 0$ and $k\in\Real^n$. Fortunately, the matrix theorem by Kreiss~\cite{hK59} provides necessary and sufficient conditions on the symbol $P(ik)$ for well-posedness.

\begin{theorem}
\label{Thm:MatrixTheorem}
Let $P(ik)$, $k\in\Real^n$, be the symbol of a constant coefficient linear problem, see Equation~(\ref{Eq:Symbol}), and let $\alpha\in\Real$. Then, the following conditions are equivalent:
\begin{enumerate}
\item[(i)] There exists a constant $K \geq 0$ such that
\begin{equation}
|e^{P(ik) t}| \leq K e^{\alpha t}
\label{Eq:WPBis}
\end{equation}
for all $t \geq 0$ and $k\in\Real^n$.
\item[(ii)] There exists a constant $M > 0$ and a family $H(k)$ of $m\times m$ Hermitian matrices such that
\begin{equation}
M^{-1} I \leq H(k) \leq M I ,\qquad
H(k)P(ik) + P(ik)^* H(k) \leq 2\alpha H(k)
\label{Eq:Symmetrizer}
\end{equation}
for all $ k\in\Real^n$.
\end{enumerate}
\end{theorem}

A generalization and complete proof of this theorem can be found in Ref.~\cite{KL89}. 
However, let us show here the implication (ii) $\Rightarrow$ (i) since it illustrates the concept of energy estimates which will be used quite often throughout this review (see Section~\ref{sec:EnergyEstimate} below for a more general discussion of these estimates). Hence, let $H(k)$ be a family of $m\times m$ Hermitian matrices satisfying the condition~(\ref{Eq:Symmetrizer}). Let $k\in \Real^n$ and $v_0\in\Complex^m$ be fixed, and define $v(t) := e^{P(ik)t} v_0$ for $t\geq 0$. Then, we have the following estimate for the ``energy'' density $v(t)^* H(k) v(t)$,
\begin{eqnarray*}
\frac{d}{dt} v(t)^* H(k) v(t) 
 &=& [P(ik) v(t)]^* H(k) v(t) + v(t)^* H(k) P(ik) v(t)
\nonumber\\
 &=& v(t)^*\left[ P(ik)^* H(k) + H(k) P(ik) \right]v(t)
\nonumber\\
 &\leq& 2\alpha\,v(t)^* H(k) v(t),
\end{eqnarray*}
which implies the differential inequality
\begin{displaymath}
\frac{d}{dt} \left[ e^{-2\alpha t}\, v(t)^* H(k) v(t)Ê\right] \leq 0,
\qquad t\geq 0,\quad k\in \Real^n.
\end{displaymath}
Integrating, we find
\begin{displaymath}
M^{-1} |v(t)|^2 \leq v(t)^* H(k) v(t) \leq
e^{2\alpha t} v_0^* H(k) v_0  \leq M e^{2\alpha t} |v_0|^2,
\end{displaymath}
which implies the inequality~(\ref{Eq:WPBis}) with $K = M$.

\subsubsection{First order systems}

Many systems in physics, like Maxwell's equations, the Dirac equation, and certain formulation of Einstein's equations, are described by first order partial differential equations (PDEs). Even if the system is described by a higher order PDE, it can be reduced to a first order one by introducing new variables. Therefore, let us specialize the above results to a first order linear problem of the form 
\begin{equation}
u_t = P(\partial/\partial x) u
  \equiv \sum\limits_{j=1}^n A^j \frac{\partial}{\partial x^j} u + B u,
\qquad x\in \Real^n, \quad t \geq 0,
\label{Eq:LinearFOS}
\end{equation}
where $A^1, \ldots ,A^n,B$ are complex $m\times m$ matrices. We split $P(ik) = P_0(ik) + B$ into its \textbf{principal symbol}, $P_0(ik) = i\sum\limits_{j=1}^n k_j A^j$, and the lower order term $B$. The principal part is the one that dominates for large $|k|$ and hence the one that turns out to be important for well-posedness. Notice that $P_0(ik)$ depends linearly on $k$. With these observations in mind we note:
\begin{itemize}
\item A necessary condition for the problem to be well posed is that for each $k\in\Real^n$ with $|k| = 1$ the symbol $P_0(ik)$ is diagonalizable and has only purely imaginary eigenvalues. To see this, we require the inequality
\begin{displaymath}
| e^{|k| P_0(i k') t + B t} | \leq K e^{\alpha t},\qquad k' := \frac{k}{|k|},
\end{displaymath}
for all $t\geq 0$ and $k\in\Real^n$, $k\neq 0$, replace $t$ by $t/|k|$, and take the limit $|k|\to\infty$, which yields $| e^{P_0(ik')t}| \leq K$ for all $k'\in\Real^n$ with $|k'|=1$. Therefore, there must exist for each such $k'$ a complex $m\times m$ matrix $S(k')$ such that $S(k')^{-1} P_0(ik') S(k') = i\Lambda(k')$, where $\Lambda(k')$ is a diagonal real matrix (cf. Lemma~\ref{Lem:1DWP}).
\item In this case the family of Hermitian $m\times m$ matrices $H(k') := (S(k')^{-1})^* S(k')^{-1}$ satisfies
\begin{displaymath}
H(k') P_0(ik') + P_0(ik')^* H(k') = 0
\end{displaymath}
for all $k'\in \Real^n$ with $|k'|=1$.
\item However, in order to obtain the energy estimate, one also needs the condition
$M^{-1} I \leq H(k') \leq M I$, that is, $H(k')$ must be uniformly bounded and positive.  This follows automatically if $H(k')$ depends continuously on $k'$, since $k'$ varies over the $(n-1)$-dimensional unit sphere which is compact.\epubtkFootnote{In fact, we will see in Section~\ref{SubSec:LPVC} that in the variable coefficient case, smoothness of the symmetrizer in $k'$ is required.} In turn, it follows that $H(k')$ depends continuously on $k'$ if $S(k')$ does. However, although this may hold in many situations, continuous dependence of $S(k')$ on $k'$ cannot always be established, see Example~\ref{Example:EigenVectors} below for a counterexample.
\end{itemize}

These observations motivate the following three notions of hyperbolicity, each of them being a stronger condition than the previous one:
\begin{definition}
\label{Def:Hyperbolicity}
The first order system~(\ref{Eq:LinearFOS}) is called
\begin{enumerate}
\item[(i)] \textbf{weakly hyperbolic} if all the eigenvalues of its principal symbol $P_0(ik)$ are purely imaginary.
\item[(ii)] \textbf{strongly hyperbolic} if there exists a constant $M > 0$ and a family of Hermitian $m\times m$ matrices $H(k)$, $k\in S^{n-1}$, satisfying
\begin{equation}
M^{-1} I \leq H(k) \leq M I,\qquad
H(k)P_0(ik) + P_0(ik)^* H(k) = 0,
\label{Eq:SymmetrizerConditions}
\end{equation}
for all $k\in S^{n-1}$, where $S^{n-1} := \{ k\in\Real^n : |k| = 1 \}$ denotes the unit sphere.
\item[(iii)] \textbf{symmetric hyperbolic} if there exists a Hermitian, positive definite $m\times m$ matrix $H$ (which is independent of $k$) such that
\begin{displaymath}
H P_0(ik) + P_0(ik)^* H = 0,
\end{displaymath}
for all $k\in S^{n-1}$.
\end{enumerate}
\end{definition}
The matrix theorem implies the following statements:
\begin{itemize}
\item Strongly and symmetric hyperbolic systems give rise to a well posed Cauchy problem. According to Theorem~\ref{Thm:MatrixTheorem} their principal symbol satisfies 
\begin{displaymath}
|e^{P_0(ik) t}| \leq K, \qquad
k\in\Real^n ,\quad t\in\Real,
\end{displaymath}
and this property is stable with respect to lower order perturbations,
\begin{displaymath}
|e^{P(ik) t}| = |e^{P_0(ik) t + B t} |\leq K e^{K |B| t}, \qquad
k\in\Real^n ,\quad t\in\Real.
\end{displaymath}
The last inequality can be proven by applying Duhamel's formula~(\ref{Eq:DuhamelFormula}) to the function $\hat{u}(t) := e^{P(ik) t}\hat{f}$ which satisfies
$\hat{u}_t(t) = P_0(ik)\hat{u}(t) + F(t)$ with $F(t) = B\hat{u}(t)$. The solution formula~(\ref{Eq:DuhamelFormula}) then gives $|\hat{u}(t)| \leq K( |\hat{f}| + |B|\int_0^t |\hat{u}(s)| ds)$ which yields $|\hat{u}(t)| \leq K e^{K|B|t} |\hat{f}|$ upon integration.
\item As we have anticipated above, a necessary condition for well-posedness is the existence of a complex $m\times m$ matrix $S(k)$ for each $k\in S^{n-1}$ on the unit sphere which brings the principal symbol $P_0(ik)$ into diagonal, purely imaginary form. If, furthermore, $S(k)$ can be chosen such that $|S(k)|$ and $|S(k)^{-1}|$ are uniformly bounded for all $k\in S^{n-1}$, then $H(k) := (S(k)^{-1})^* S(k)^{-1}$ satisfies the conditions~(\ref{Eq:SymmetrizerConditions}) for strong hyperbolicity. If the system is well posed, Theorem 2.4.1 in Ref.~\cite{KL89} shows that it is always possible to construct a symmetrizer $H(k)$ satisfying the conditions~(\ref{Eq:SymmetrizerConditions}) in this manner, and hence, strong hyperbolicity is also a  necessary condition for well posedness. The symmetrizer construction $H(k) := (S(k)^{-1})^* S(k)^{-1}$ is useful for applications, since $S(k)$ is easily constructed from the eigenvectors and $S(k)^{-1}$ from the eigenfields of the principal symbol, see Example~\ref{Example:FatMaxwell} below.
\item Weakly hyperbolic systems are not well posed in general because $|e^{P_0(ik)t}|$ might exhibit polynomial growth in $|k| t$. Although one might consider such polynomial growth as acceptable, such systems are unstable with respect to lower order perturbations. As the next example shows, it is possible that $|e^{P(ik) t}|$ grows exponentially in $|k|$ if the system is weakly hyperbolic.
\end{itemize}

\begin{example} 
\label{Example:WH}
Consider the weakly hyperbolic system \cite{KL89}
\begin{equation}
u_t = \left( \begin{array}{ll} 1 & 1 \\ 0 & 1 \end{array} \right)u_x
 + a\left( \begin{array}{ll} -1 & +1 \\ -1 & -1 \end{array} \right)u,
\label{Eq:WeaklyHyperbolicEq}
\end{equation}
with $a\in\Real$ a parameter. The principal symbol is $P_0(ik) = ik\left( \begin{array}{ll} 1 & 1 \\ 0 & 1 \end{array} \right)$ and
\begin{displaymath}
e^{P_0(ik) t} = e^{ikt}\left( \begin{array}{ll} 1 & ik t \\ 0 & 1 \end{array} \right).
\end{displaymath}
Using the tools described in Section~\ref{section:notation} we find for the norm
\begin{displaymath}
|e^{P_0(ik) t}| 
 = \sqrt{ 1 + \frac{k^2 t^2}{2} + \sqrt{ \left( 1 + \frac{k^2 t^2}{2} \right)^2 - 1 }},
\end{displaymath}
which is approximately equal to $|k| t$ for large $|k| t$. Hence, the solutions to Equation~(\ref{Eq:WeaklyHyperbolicEq}) contain modes which grow linearly in $|k| t$ for large $|k| t$ when $a=0$, i.e., when there are no lower order terms.

However, when $a\neq 0$, the eigenvalues of $P(ik)$ are
\begin{displaymath}
\lambda_{\pm} =  ik - a \pm i\sqrt{a(a+ik)},
\end{displaymath}
which, for large $k$ has real part $\re(\lambda_{\pm}) \approx \pm \sqrt{|a\|k|/2}$. The eigenvalue with positive real part gives rise to solutions which, for fixed $t$, grow exponentially in $|k|$.
\end{example}

\begin{example}
For the system \cite{oR04},
\begin{displaymath}
u_ t = A^1 u_x + A^2 u_y,\qquad
A^1 = \left( \begin{array}{ll} 1 & 1 \\ 0 & 2 \end{array} \right),\quad
A^2 = \left( \begin{array}{ll} 1 & 0 \\ 0 & 2 \end{array} \right),
\end{displaymath}
the principal symbol, $P_0(ik) = i\left( \begin{array}{cc} k_1 + k_2 & k_1 \\ 0 & 2(k_1 + k_2) \end{array} \right)$, is diagonalizable for all vectors $k = (k_1,k_2)\in S^1$ except for those with $k_1 + k_2 = 0$. In particular, $P_0(ik)$ is diagonalizable for $k = (1,0)$ and $k = (0,1)$. This shows that in general, it is not sufficient to check that the $n$ matrices $A^1$, $A^2$,\ldots, $A^n$ alone are diagonalizable and have real eigenvalues, one has to consider all possible linear combinations $\sum\limits_{j=1}^n A^j k_j$ with $k\in S^{n-1}$.
\end{example}

\begin{example}
\label{Example:EigenVectors}
Next, we present a system for which the eigenvectors of the principal symbol cannot be chosen to be continuous functions of $k$:
\begin{displaymath}
u_ t = A^1 u_x + A^2 u_y + A^3 u_z,\qquad
A^1 = \left( \begin{array}{ll} 1 & 0 \\ 0 & -1 \end{array} \right),\quad
A^2 = \left( \begin{array}{ll} 0 & 1 \\ 1 & 0 \end{array} \right),\quad
A^3 = \left( \begin{array}{ll} 0 & 0 \\ 0 & 0 \end{array} \right).
\end{displaymath}
The principal symbol $P_0(ik) = i\left( \begin{array}{cc} k_1 & k_2 \\ k_2 & -k_1 \end{array} \right)$ has eigenvalues $\lambda_\pm(k) = \pm i\sqrt{k_1^2 + k_2^2}$ and for $(k_1,k_2)\neq (0,0)$ the corresponding eigenprojectors are
\begin{displaymath}
P_\pm(k_1,k_2) = \frac{1}{2\lambda_\pm(k)}
\left( \begin{array}{cc} \lambda_\pm(k) + i k_1 & i k_2 \\ 
 i k_2 & \lambda_\pm(k) - i k_1 \end{array} \right).
\end{displaymath}
When $(k_1,k_2)\to (0,0)$ the two eigenvalues fall together, and $A(k)$ converges to the zero matrix. However, it is not possible to continuously extend $P_\pm(k_1,k_2)$ to $(k_1,k_2) = (0,0)$. For example,
$$
P_+(h,0) = \left( \begin{array}{cc} 1 & 0 \\ 0 & 0 \end{array} \right),\quad
P_+(-h,0) = \left( \begin{array}{cc} 0 & 0 \\ 0 & 1 \end{array} \right),
$$
for positive $h > 0$. Therefore, any choice for the matrix $S(k)$ which diagonalizes $A(k)$ must be discontinuous at $k = (0,0,\pm 1)$ since the columns of $S(k)$ are the eigenvectors of $A(k)$.

Of course, $A(k)$ is symmetric and so $S(k)$ can be chosen to be unitary which yields the trivial symmetrizer $H(k) = I$. Therefore, the system is symmetric hyperbolic and yields a well posed Cauchy problem; however, this example shows that it is not always possible to choose $S(k)$ as a continuous function of $k$.
\end{example}

\begin{example}
\label{Example:KleinGordonFlat}
Consider the Klein--Gordon equation
\begin{equation}
\Phi_{tt} = \Delta\Phi - m^2\Phi ,
\label{Eq:KG}
\end{equation}
in two spatial dimensions, where $m\in\Real$ is a parameter which is proportional to the mass of the field $\Phi$. Introducing the variables $u = (\Phi,\Phi_t,\Phi_x,\Phi_y)$ we obtain the first order system
\begin{displaymath}
u_t = \left( \begin{array}{llll}
 0 & 0 & 0 & 0 \\
 0 & 0 & 1 & 0 \\
 0 & 1 & 0 & 0 \\
 0 & 0 & 0 & 0 \end{array} \right) u_x
 + \left( \begin{array}{llll}
 0 & 0 & 0 & 0 \\
 0 & 0 & 0 & 1 \\
 0 & 0 & 0 & 0 \\
 0 & 1 & 0 & 0 \end{array} \right) u_y
 +  \left( \begin{array}{rrrr}
 0   & 1 & 0 & 0 \\
-m^2 & 0 & 0 & 0 \\
 0   & 0 & 0 & 0 \\
 0   & 0 & 0 & 0 \end{array} \right) u.
\end{displaymath}
The matrix coefficients in front of $u_x$ and $u_y$ are symmetric; hence the system is symmetric hyperbolic with trivial symmetrizer $H = \diag(m^2,1,1,1)$.\epubtkFootnote{Here, the factor $m^2$ could in principle be replaced by any positive number, which shows that the symmetrizer is not always unique. The choice here is such that the expression $u^* H u$ is proportional to the physical energy density of the system. Notice, however, that for the massless case $m=0$ one must replace $m^2=0$ by a positive constant in order for the symmetrizer to be positive definite.}
The corresponding Cauchy problem is well posed. However, a problem with this first order system is that it is only equivalent to the original, second order equation~(\ref{Eq:KG}) if the constraints $(u_1)_x = u_3$ and $(u_1)_y = u_4$ are satisfied.

An alternative symmetric hyperbolic first order reduction of the Klein--Gordon equation which does not require the introduction of constraints is the Dirac equation in two spatial dimensions,
\begin{equation}
v_t = \left( \begin{array}{rr}
 1 & 0 \\
 0 & -1 \end{array} \right) v_x
 + \left( \begin{array}{rr}
 0 & 1 \\
 1 & 0 \end{array} \right) v_y
 + m\left( \begin{array}{rr}
 0 & 1 \\
 -1 & 0 \end{array} \right) v,
 \qquad
v = \left( \begin{array}{r} v_1 \\ v_2 \end{array} \right).
\label{Eq:2DDirac}
\end{equation}
This system implies the Klein--Gordon equation~(\ref{Eq:KG}) for either of the two components of $v$.

Yet another way of reducing second order equations to first order ones without introducing constraints will be discussed in Section~\ref{SubSubSec:SecondOrderCC} below.
\end{example}

\begin{example}
\label{Example:Maxwell}
In terms of the electric and magnetic fields $u = (E,B)$, Maxwell's evolution equations,
\begin{eqnarray}
E_t &=& +\nabla\wedge B - J,
\label{Eq:Maxwell1}\\
B_t &=& -\nabla\wedge E,
\label{Eq:Maxwell2}
\end{eqnarray}
constitute a symmetric hyperbolic system. Here, $J$ is the current density and $\nabla$ and $\wedge$ denote the nabla operator and the vector product, respectively. The principal symbol is
\begin{displaymath}
P_0(ik)\left( \begin{array}{c} E \\ B \end{array} \right)
 = i\left( \begin{array}{c} +k\wedge B \\ -k\wedge E \end{array} \right)
\end{displaymath}
and a symmetrizer is given by the physical energy density,
\begin{displaymath}
u^* H u = \frac{1}{2}\left( |E|^2 + |B|^2 \right),
\end{displaymath}
in other words, $H = 2^{-1} I$ is trivial. The constraints $\nabla\cdot E = \rho$ and $\nabla\cdot B = 0$ propagate as a consequence of Equations~(\ref{Eq:Maxwell1}, \ref{Eq:Maxwell2}), provided that the continuity equation holds:
$(\nabla\cdot E - \rho)_t = -\nabla\cdot J - \rho_t = 0$, $(\nabla\cdot B)_t = 0$.
\end{example}

\begin{example}
\label{Example:FatMaxwell}
There are many alternative ways for writing Maxwell's equations. The following system~\cite{oR04,lLmSlKhPdSsT04} was originally motivated by an analogy with certain parametrized first order hyperbolic formulations of the Einstein equations, and provides an example of a system that can be symmetric, strongly, weakly or not hyperbolic at all, depending on the parameter values. Using the Einstein summation convention, the evolution system in vacuum has the form
\begin{eqnarray}
\partial_t E_i &=& \partial^j (W_{ij} - W_{ji}) - \alpha(\partial_i W^j{}_j - \partial^j W_{ij}),
\label{Eq:FatMaxwell1}\\
\partial_t W_{ij} &=& -\partial_i E_j - \frac{\beta}{2}\delta_{ij}\partial^k E_k,
\label{Eq:FatMaxwell2}
\end{eqnarray}
where $E_i$ and $W_{ij} = \partial_i A_j$, $i=1,2,3$, represent the Cartesian components of the electric field and the gradient of the magnetic potential $A_j$, respectively, and where the real parameters $\alpha$ and $\beta$ determine the dynamics off the constraint hypersurface defined by $\partial^k E_k = 0$ and $\partial_k W_{ij} - \partial_i W_{kj} = 0$.

In order to analyze under which conditions on $\alpha$ and $\beta$ the system~(\ref{Eq:FatMaxwell1}, \ref{Eq:FatMaxwell2}) is strongly hyperbolic we consider the corresponding symbol,
\begin{displaymath}
P_0(ik)u
 = i\left( \begin{array}{c} (1+\alpha) k^j W_{ij} - k^j W_{ji} - \alpha k_i W^j{}_j \\
 -k_i E_j - \frac{\beta}{2}\delta_{ij} k^l E_l \end{array} \right),\qquad
u = \left( \begin{array}{c} E_i \\ W_{ij} \end{array} \right),\quad
k\in S^2.
\end{displaymath}
Decomposing $E_i$ and $W_{ij}$ into components parallel and orthogonal to $k_i$,
\begin{displaymath}
E_i = \bar{E} k_i + \bar{E}_i,\qquad
W_{ij} = \bar{W} k_i k_j + \bar{W}_i k_j + k_i\bar{V}_j + \bar{W}_{ij}
 + \frac{1}{2}\gamma_{ij}\bar{U},
\end{displaymath}
where in terms of the projector $\gamma_i{}^j := \delta_i{}^j - k_i k^j$ orthogonal to $k$ we have have defined $\bar{E}:=k^l E_l$, $\bar{E}_i := \gamma_i{}^j E_j$ and $\bar{W} := k^i k^j W_{ij}$, $\bar{W}_i := \gamma_i{}^k W_{kj} k^j$, $\bar{V}_j := k^i W_{ik}\gamma^k{}_j$, $\bar{U} := \gamma^{ij} W_{ij}$, and $\bar{W}_{ij} := (\gamma_i{}^k\gamma_j{}^l - 2^{-1}\gamma_{ij}\gamma^{kl})W_{kl}$,\epubtkFootnote{Notice that $\bar{E}_i$ has only two degrees of freedom since it is orthogonal to $k$. Likewise, the quantities $\bar{W}_i$, $\bar{V}_i$ have two degrees of freedom and $\bar{W}_{ij}$ has three since it its orthogonal to $k$ and trace-free.} we can write the eigenvalue problem $P_0(ik) u = i\lambda u$ as
\begin{eqnarray*}
\lambda \bar{E} &=& -\alpha \bar{U},\\
\lambda \bar{U} &=& -\beta \bar{E},\\
\lambda \bar{W} &=& -\left( 1 + \frac{\beta}{2} \right)\bar{E},\\
\lambda \bar{E}_i &=& (1+\alpha)\bar{W}_i - \bar{V}_i,\\
\lambda \bar{V}_i &=& -\bar{E}_i,\\
\lambda \bar{W}_{i} &=& 0,\\
\lambda \bar{W}_{ij} &=& 0.
\end{eqnarray*}
It follows that $P_0(ik)$ is diagonalizable with purely complex eigenvalues if and only if $\alpha\beta > 0$. However, in order to show that in this case the system is strongly hyperbolic one still needs to construct a bounded symmetrizer $H(k)$. For this, we set $\mu := \sqrt{\alpha\beta}$ and diagonalize $P_0(ik) = i S(k)\Lambda(k) S(k)^{-1}$ with $\Lambda(k) = \diag(\mu,-\mu,0,1,-1,0,0)$ and
\begin{equation}
S(k)^{-1} u
 = \left( \begin{array}{c} 
 \bar{E} - \frac{\mu}{\beta}\bar{U} \\ 
 \bar{E} + \frac{\mu}{\beta}\bar{U} \\ 
  \beta\bar{W} - \left( 1 + \frac{\beta}{2} \right)\bar{U} \\
  \bar{E}_i - \bar{V}_i + (1+\alpha)\bar{W}_i \\
  \bar{E}_i + \bar{V}_i - (1+\alpha)\bar{W}_i \\
  \bar{W}_i \\
  \bar{W}_{ij}  \end{array} \right).
\label{Eq:FatMaxwellCharVars}
\end{equation}
Then, the quadratic form associated with the symmetrizer is
\begin{eqnarray}
u^* H(k) u &=& u^* (S(k)^{-1})^* S(k)^{-1} u
\nonumber\\
 &=& 2 |\bar{E}|^2 + 2\frac{\alpha}{\beta} |\bar{U}|^2 
 + \left| \beta\bar{W} - \left( 1 + \frac{\beta}{2} \right)\bar{U} \right|^2+ 2\bar{E}^i\bar{E}_i 
\nonumber\\
 &+& 2\left[ \bar{V}^i - (1+\alpha)\bar{W}^i \right]
 \left[ \bar{V}_i - (1+\alpha)\bar{W}_i \right] + \bar{W}^i\bar{W}_i + \bar{W}^{ij}\bar{W}_{ij},
\nonumber
\end{eqnarray}
and $H(k)$ is smooth in $k\in S^2$. Therefore, the system is indeed strongly hyperbolic for $\alpha\beta > 0$.

In order to analyze under which conditions the system is symmetric hyperbolic we notice that because of rotational and parity invariance the most general $k$-independent symmetrizer must have the form
\begin{displaymath}
u^* H u = a (E^i)^* E_i + b (W^{[ij]})^* W_{[ij]} + c(\hat{W}^{ij})^*\hat{W}_{ij} 
+ d\, W^* W,
\end{displaymath}
with strictly positive constants $a$, $b$, $c$ and $d$, where $\hat{W}_{ij} := W_{(ij)} - \delta_{ij} W/3$ denotes the symmetric, trace-free part of $W_{ij}$ and $W := W^j{}_j$ its trace. Then,
\begin{eqnarray}
u^* H P_0(ik) u &=& i a (E^i)^*\left[ (\alpha+2) k^j W_{[ij]} + \alpha k^j\hat{W}_{ij}
 - \frac{2\alpha}{3} k_i W \right] 
\nonumber\\
 &+& i b(W^{[ij]})^* E_i k_j
 - i c(\hat{W}^{ij})^* E_i k_j
 - i d\left( 1 + \frac{3\beta}{2} \right)W^* k^i E_i.
\nonumber
\end{eqnarray}
For $H$ to be a symmetrizer, the expression on the right-hand side must be purely imaginary. This is the case if and only if $a(\alpha + 2) = b$, $-a\alpha = c$ and $2a\alpha/3 = d( 1 + 3\beta/2)$. Since $a$, $b$, $c$ and $d$ are positive, these equalities can be satisfied if and only if $-2 < \alpha < 0$ and $\beta < -2/3$. Therefore, if either $\alpha$ and $\beta$ are both positive or $\alpha$ and $\beta$ are both negative and $\alpha\leq -2$ or $\beta \geq -2/3$, then the system~(\ref{Eq:FatMaxwell1}, \ref{Eq:FatMaxwell2}) is strongly but not symmetric hyperbolic.
\end{example}

\subsubsection{Second order systems}
\label{SubSubSec:SecondOrderCC}

Another class of important systems in physics are wave problems. In the linear, constant coefficient case, they are described by an equation of the form
\begin{equation}
v_{tt} = \sum\limits_{j,k=1}^n A^{jk}\frac{\partial^2}{\partial x^j\partial x^k} v
 + \sum\limits_{j=1}^n 2B^j\frac{\partial}{\partial x^j} v_t 
 + \sum\limits_{j=1}^n C^j\frac{\partial}{\partial x^j} v + D v_t + E v,
\qquad x\in\Real^n,\quad t\geq 0,
\label{Eq:LinearCCWave}
\end{equation}
where $v = v(t,x)\in\Complex^m$ is the state vector, and $A^{ij} = A^{ji},B^j,C^j,D,E$ denote complex $m\times m$ matrices. In order to apply the theory described so far, we reduce this equation to a system which is first order in time. This is achieved by introducing the new variable $w := v_t - \sum\limits_{j=1}^n B^j\frac{\partial}{\partial x^j} v$.\epubtkFootnote{Here, the advection term $\sum\limits_{j=1}^n B^j\frac{\partial}{\partial x^j} v$ is subtracted from $v_t$ for convenience only.} With this redefinition one obtains a system of the form~(\ref{Eq:LinearCCPDE}) with $u = (v,w)^T$ and
\begin{displaymath}
P(\partial/\partial x) = \sum\limits_{j=1}^n B^j\frac{\partial}{\partial x^j} +
 \left( \begin{array}{cc}
  0 & I \\
 \sum\limits_{j,k=1}^n (A^{jk} + B^j B^k)\frac{\partial^2}{\partial x^j\partial x^k} 
 + \sum\limits_{j=1}^n(C^j + D B^j)\frac{\partial}{\partial x^j} + E & D 
\end{array} \right)
\end{displaymath}
Now we could apply the matrix theorem, Theorem~\ref{Thm:MatrixTheorem}, to the corresponding symbol $P(ik)$ and analyze under which conditions on the matrix coefficients $A^{ij},B^j,C^j,D,E$ the Cauchy problem is well posed. However, since our problem originates from a second order equation, it is convenient to rewrite the symbol in a slightly different way: instead of taking the Fourier transform of $v$ and $w$ directly, we multiply $\hat{v}$ by $|k|$ and write the symbol in terms of the variable $\hat{U} := (|k|\hat{v},\hat{w})^T$. Then, the $L^2$-norm of $\hat{U}$ controls, through Parseval's identity, the $L^2$-norms of the first partial derivatives of $v$, as is the case for the usual energies for second order systems. In terms of $\hat{U}$ the system reads
\begin{equation}
\hat{U}_t = Q(i k)\hat{U},\qquad
t\geq 0,\quad k\in\Real^n,
\label{Eq:LinearCCWaveFourier}
\end{equation}
in Fourier space, where
\begin{equation}
Q(i k) = i|k|\sum\limits_{j=1}^n B^j\hat{k}_j
 + \left( \begin{array}{cc}
 0 & |k| I \\
 -|k|\sum\limits_{j,k=1}^n(A^{jk} + B^j B^k)\hat{k}_j\hat{k}_k 
 + i\sum\limits_{j=1}^n(C^j + D B^j)\hat{k}_j + \frac{1}{|k|} E & D 
\end{array} \right)
\end{equation}
with $\hat{k}_j := k_j/|k|$. As for first order systems, we can split $Q(ik)$ into its principal part,
\begin{equation}
Q_0(i k) := i|k|\sum\limits_{j=1}^n B^j\hat{k}_j
+ |k|\left( \begin{array}{cc} 0 & I \\
 -\sum\limits_{j,k=1}^n(A^{jk} + B^j B^k)\hat{k}_j\hat{k}_k & 0
\end{array} \right),
\end{equation}
which dominates for $|k|\to\infty$, and the remaining, lower order terms. Because of the homogeneity of $Q_0(ik)$ in $k$ we can restrict ourselves to values of $k\in S^{n-1}$ on the unit sphere, like for first order systems. Then, it follows as a consequence of the matrix theorem that the problem is well posed if and only if there exists a symmetrizer $H(k)$ and a constant $M > 0$ satisfying
\begin{displaymath}
M^{-1} I \leq H(k) \leq M I,\qquad H(k) Q_0(ik) + Q_0(ik)^* H(k) = 0
\end{displaymath}
for all such $k$. Necessary and sufficient conditions under which such a symmetrizer exists have been given in Ref.~\cite{hKoO02} for the particular case in which the mixed second order derivative term in Equation~(\ref{Eq:LinearCCWave}) vanishes; that is, when $B^j = 0$. This result can be generalized in a straightforward manner to the case where the matrices $B^j = \beta^j I$ are proportional to the identity:

\begin{theorem}
\label{Thm:SecondOrder}
Suppose $B^j = \beta^j I$, $j=1,2,\ldots,n$. (Note that this condition is trivially satisfied if $m=1$.) Then, the Cauchy problem for the equation~(\ref{Eq:LinearCCWave}) is well posed if and only if the symbol
\begin{displaymath}
R(k) := \sum\limits_{i,j=1}^n (A^{ij} + B^i B^j) k_i k_j,\qquad k\in S^{n-1},
\end{displaymath}
has the following properties: there exist constants $M > 0$ and $\delta > 0$ and a family $h(k)$ of Hermitian $m\times m$ matrices such that
\begin{equation}
M^{-1} I \leq h(k) \leq M I,\qquad
h(k) R(k) = R(k)^* h(k) \geq \delta I
\label{Eq:SecondOrderSymmetrizer}
\end{equation}
for all $k\in S^{n-1}$.
\end{theorem}

\proof Since for $B^j = \beta^j I$ the advection term $i|k|\sum\limits_{j=1}^n B^j\hat{k}_j$ commutes with any Hermitian matrix $H(k)$, it is sufficient to prove the theorem for $B^j = 0$, in which case the principal symbol reduces to
\begin{displaymath}
Q_0(i k) := \left( \begin{array}{cc} 0 & I \\
 -R(k) & 0 \end{array} \right),\qquad
 k\in S^{n-1}.
\end{displaymath}
We write the symmetrizer $H(k)$ in the following block form,
\begin{displaymath}
H(k) = \left( \begin{array}{cc} H_{11}(k) & H_{12}(k) \\
 H_{12}(k)^* & H_{22}(k) \end{array} \right),
\end{displaymath}
where $H_{11}(k)$, $H_{22}(k)$ and $H_{12}(k)$ are complex $m\times m$ matrices, the first two being Hermitian. Then,
\begin{displaymath}
H(k) Q_0(ik) + Q_0(ik)^* H(k) 
= \left( \begin{array}{cc} -H_{12}(k) R(k) - R(k)^* H_{12}(k)^*
 & H_{11}(k) - R(k)^* H_{22}(k) \\
 H_{11}(k) - H_{22}(k) R(k) & H_{12}(k) + H_{12}(k)^* \end{array} \right).
\end{displaymath}
Now, suppose $h(k)$ satisfies the conditions~(\ref{Eq:SecondOrderSymmetrizer}). Then, choosing $H_{12}(k) := 0$, $H_{22}(k) := h(k)$ and $H_{11}(k) := h(k) R(k)$ we find that $H(k)Q_0(ik) + Q_0(ik)^* H(k) = 0$. Furthermore, $M^{-1} I \leq H_{22}(k) \leq M I$ and $\delta I \leq H_{11}(k) = h(k) R(k) \leq M C I$ where
\begin{displaymath}
C := \sup\{  |R(k) u| : k\in S^{n-1}, u\in\Complex^m, |u| = 1 \}
\end{displaymath}
is finite because $R(k) u$ is continuous in $k$ and $u$. Therefore, $H(k)$ is a symmetrizer for $Q_0(ik)$, and the problem is well posed.

Conversely, suppose that the problem is well posed with symmetrizer $H(k)$. Then, the vanishing of $H(k) Q_0(ik) + Q_0(ik)^* H(k)$ yields the conditions $H_{11}(k) = H_{22}(k) R(k) = R(k)^* H_{22}(k)$ and the conditions~(\ref{Eq:SecondOrderSymmetrizer}) are satisfied for $h(k) := H_{22}(k)$.
\qed\\

\textbf{Remark}: The conditions~(\ref{Eq:SecondOrderSymmetrizer}) imply that $R(k)$ is symmetric and positive with respect to the scalar product defined by $h(k)$. Hence it is diagonalizable, and all its eigenvalues are positive. A practical way of finding $h(k)$ is to construct $T(k)$ which diagonalizes $R(k)$, $T(k)^{-1} R(k) T(k) = P(k)$ with $P(k)$ diagonal and positive. Then, $h(k) := (T(k)^{-1})^* T(k)^{-1}$ is the candidate for satisfying the conditions~(\ref{Eq:SecondOrderSymmetrizer}).\\

Let us give some examples and applications:

\begin{example}
The Klein--Gordon equation $v_{tt} = \Delta v - m^2 v$ on flat spacetime. In this case, $A^{ij} = \delta^{ij}$ and $B^j=0$, and $R(k) = |k|^2$ trivially satisfies the conditions of Theorem~\ref{Thm:SecondOrder}.
\end{example}

\begin{example}
\label{Example:KleinGordon} In anticipation of the following subsection, where linear problems with variable coefficients are treated, let us generalize the previous example on a curved spacetime $(M,g)$. We assume that $(M,g)$ is globally hyperbolic such that it can be foliated by space-like hypersurfaces $\Sigma_t$. In the ADM decomposition, the metric in adapted coordinates assumes the form
\begin{displaymath}
g = -\alpha^2 dt\otimes dt + \gamma_{ij}(dx^i + \beta^i dt)\otimes (dx^j + \beta^j dt),
\end{displaymath}
with $\alpha > 0$ the lapse, $\beta^i$ the shift vector which is tangent to $\Sigma_t$ and $\gamma_{ij} dx^i \otimes dx^j$ the induced three-metric on the spacelike hypersurfaces $\Sigma_t$. The inverse of the metric is given by
\begin{displaymath}
g^{-1} = -\frac{1}{\alpha^2}
\left( \frac{\partial}{\partial t} - \beta^i\frac{\partial}{\partial x^i} \right)\otimes
\left( \frac{\partial}{\partial t} - \beta^j\frac{\partial}{\partial x^j} \right)
 + \gamma^{ij}\frac{\partial}{\partial x^i}\otimes\frac{\partial}{\partial x^j},
\end{displaymath}
where $\gamma^{ij}$ are the components of the inverse three-metric. The Klein--Gordon equation on $(M,g)$ is
\begin{displaymath}
g^{\mu\nu}\nabla_\mu\nabla_\nu v 
 = \frac{1}{\sqrt{-\det(g)}}\partial_\mu\left( \sqrt{-\det(g)} g^{\mu\nu}\partial_\nu v \right)
 = m^2 v, 
\end{displaymath}
which, in the constant coefficient case, has the form of Equation~(\ref{Eq:LinearCCWave}) with
\begin{displaymath}
A^{jk} = \alpha^2\gamma^{jk} - \beta^j\beta^k,\qquad
B^j = \beta^j.
\end{displaymath}
Hence, $R(k) = \alpha^2 \gamma^{ij} k_i k_j$, and the conditions of Theorem~\ref{Thm:SecondOrder} are satisfied with $h(k) = 1$ since $\alpha > 0$ and $\gamma^{ij}$ is symmetric positive definite.
\end{example}

%===================================================================
\subsection{Linear problems with variable coefficients}
\label{SubSec:LPVC}
%===================================================================

Next, we generalize the theory to linear evolution problems with variable coefficients. That is, we consider equations of the following form:
\begin{equation}
u_t = P(t,x,\partial/\partial x) u
  \equiv \sum\limits_{|\nu| \leq p} A_\nu(t,x) D_\nu u,
\qquad x\in \Real^n, \quad t \geq 0,
\label{Eq:LinearVCPDE}
\end{equation}
where now the complex $m\times m$ matrices $A_\nu(t,x)$ may depend on $t$ and $x$. For simplicity, we assume that each matrix coefficient of $A_\nu$ belongs to the class $C_b^\infty([0,\infty)\times\Real^n)$ of bounded, $C^\infty$-functions with bounded derivatives. Unlike the constant coefficient case, the different $k$-modes couple when performing a Fourier transformation, and there is no simple explicit representation of the solutions through the exponential of the symbol. Therefore, Definition~\ref{Def:WP} of well posedness needs to be altered. Instead of giving an operator-based definition, let us define well posedness by the basic requirements a Cauchy problem should satisfy:

\begin{definition}
\label{Def:WPVC}
The Cauchy problem
\begin{eqnarray}
u_t(t,x) = P(t,x,\partial/\partial x) u(t,x),
&& x\in \Real^n, \quad t \geq 0,
\label{Eq:LinearVCPDECauchy1}\\
u(0,x) = f(x),
&& x\in\Real^n,
\label{Eq:LinearVCPDECauchy2}
\end{eqnarray}
is well posed, if any $f\in C^\infty_0(\Real^n)$ gives rise to a unique $C^\infty$-solution $u(t,x)$, and if there are constants $K\geq 1$ and $\alpha\in\Real$ such that
\begin{equation}
\| u(t,\cdot) \| \leq K e^{\alpha t} \| f \|
\label{Eq:WPInequality}
\end{equation}
for all $f\in C^\infty_0(\Real^n)$ and all $t\geq 0$.
\end{definition}

Before we proceed and analyze under which conditions on the operator $P(t,x,\partial/\partial x)$ the Cauchy problem~(\ref{Eq:LinearVCPDECauchy1}, \ref{Eq:LinearVCPDECauchy2}) is well posed, let us make the following observations:
\begin{itemize}
\item In the constant coefficient case, the inequality~(\ref{Eq:WPInequality}) is equivalent to the inequality~(\ref{Eq:WPEstimate}), and in this sense Definition~\ref{Def:WPVC} is a generalization of Definition~\ref{Def:WP}.
\item If $u_1$ and $u_2$ are the solutions corresponding to the initial data $f_1,f_2\in C^\infty_0(\Real^n)$, then the difference $u = u_2 - u_1$ satisfies the Cauchy problem~(\ref{Eq:LinearVCPDECauchy1}, \ref{Eq:LinearVCPDECauchy2}) with $f = f_2 - f_1$ and
the estimate~(\ref{Eq:WPInequality}) implies that
\begin{equation}
\| u_2(t,\cdot) - u_1(t,\cdot) \| \leq K e^{\alpha t} \| f_2 - f_1 \|,\qquad t\geq 0.
\label{Eq:WPInequalityDifference}
\end{equation}
In particular, this implies that $u_2(t,\cdot)$ converges to $u_1(t,\cdot)$ if $f_2$ converges to $f_1$ in the $L^2$-sense. In this sense, \emph{the solution depends continuously on the initial data}. This property is important for the convergence of a numerical approximation, as discussed in Section~\ref{sec:num_stability}. 
\item The estimate~(\ref{Eq:WPInequality}) also implies \emph{uniqueness} of the solution, because for two solutions $u_1$ and $u_2$ with the same initial data $f_1=f_2\in C^\infty_0(\Real^n)$ the inequality~(\ref{Eq:WPInequalityDifference}) implies $u_1 = u_2$.
\item As in the constant coefficient case, it is possible to extend the solution concept to weak ones by taking sequences of $C^\infty$-elements. This defines a propagator
$U(t,s): L^2(\Real^n)\to L^2(\Real^n)$ which maps the solution at time $s\geq 0$ to the solution at time $t\geq s$ and satisfies similar properties than the ones described in Section~\ref{SubSubSec:ExtensionOfSolutions}: (i) $U(t,t) = I$ for all $t\geq 0$, (ii) $U(t,s) U(s,r) = U(t,r)$ for all $t\geq s\geq r\geq 0$, (iii) for $f\in C^\infty_0(\Real^n)$, $U(t,0) f$ is the unique solution of the Cauchy problem~(\ref{Eq:LinearVCPDECauchy1}, \ref{Eq:LinearVCPDECauchy2}), (iv) $\| U(t,s) f \| \leq K e^{\alpha(t-s)} \| fÊ\|$ for all $f\in L^2(\Real)$ and all $t\geq s\geq 0$. Furthermore, the Duhamel formula~(\ref{Eq:DuhamelFormula}) holds with the replacement $U(t-s)\mapsto U(t,s)$.
\end{itemize}

\subsubsection{The localization principle}

Like in the constant coefficient case, we would like to have a criterion for well posedness that is based on the coefficients $A_\nu(t,x)$ of the differential operator alone. As we have seen in the constant coefficient case, well posedness is essentially a statement about high frequencies. Therefore, we are led to consider solutions with very high frequency or, equivalently, with very short wavelength. In this regime we can consider small neighborhoods and since the coefficients $A_\nu(t,x)$ are smooth, they are approximately constant in such neighborhoods. Therefore, intuitively, the question of well posedness for the variable coefficient problem can be reduced to a frozen coefficient problem, where the values of the matrix coefficients $A_\nu(t,x)$ are frozen to their values at a given point.

In order to analyze this more carefully, and for the sake of illustration, let us consider a first order linear system with variable coefficients
\begin{equation}
u_t = P(t,x,\partial/\partial x) u
 \equiv \sum\limits_{j=1}^n A^j(t,x)\frac{\partial}{\partial x^j} u + B(t,x) u,
\qquad x\in \Real^n, \quad t \geq 0,
\label{Eq:LinearFOSVC}
\end{equation}
where $A^1, \ldots ,A^n,B$ are complex $m\times m$ matrices whose coefficients belong to the class $C_b^\infty([0,\infty)\times\Real^n)$ of bounded, $C^\infty$-functions with bounded derivatives. As mentioned above, the Fourier transform of this operator does not yield a simple, algebraic symbol like in the constant coefficient case.\epubtkFootnote{The Fourier transform of the function $v(t,\cdot) = P(t,\cdot,\partial/\partial x)u(t,\cdot)$ in space is formally given by
\begin{displaymath}
\hat{v}(t,\cdot) = \frac{1}{(2\pi)^{n/2}}\left[ 
 \sum\limits_{j=1}^n \hat{A}^j(t,\cdot) * i k_j\hat{u}(t,\cdot) + \hat{B}(t,\cdot) * \hat{u}(t,\cdot)
\right],
\end{displaymath}
where $\hat{A}^j(t,\cdot)$, $\hat{B}(t,\cdot)$ denote the Fourier transform of $A^j(t,\cdot)$ and $B(t,\cdot)$, respectively, and where the star denotes the convolution operator. Unless $A^j$ and $B$ are independent of $x$, the different $k$-modes couple to each other.} However, given a specific point $p_0 = (t_0,x_0)\in [0,\infty) \times \Real^n$, we may zoom into a very small neighborhood of $p_0$. Since the coefficients $A^j(t,x)$ and $B(t,x)$ are smooth, they will be approximately constant in this neighborhood and we may freeze the coefficients of $A^j(t,x)$ and $B(t,x)$ to their values at the point $p_0$. More precisely, let $u(t,x)$ be a smooth solution of the equation~(\ref{Eq:LinearFOSVC}). Then, we consider the formal expansion
\begin{equation}
u(t_0 + \varepsilon t,x_0 + \varepsilon x) = u(t_0,x_0) 
+ \varepsilon u^{(1)}(t,x) + \varepsilon^2 u^{(2)}(t,x) + \ldots,\qquad\varepsilon > 0.
\label{Eq:ZoomIn}
\end{equation}
As a consequence of Equation~(\ref{Eq:LinearFOSVC}) one obtains
\begin{eqnarray}
u^{(1)}_t(t,x) + \varepsilon u^{(2)}_t(t,x) + \ldots 
 &=& \sum\limits_{j=1}^n A^j(t_0 + \varepsilon t,x_0 + \varepsilon x)\left[
 \frac{\partial u^{(1)}}{\partial x^j}(t,x) 
 + \varepsilon\frac{\partial u^{(2)}}{\partial x^j}(t,x) + \ldots \right]
\nonumber\\
 &+& B(t_0 + \varepsilon t,x_0 + \varepsilon x)
 \left[ u(t_0,x_0) + \varepsilon u^{(1)}(t,x) + \ldots \right].
\nonumber
\end{eqnarray}
Taking the pointwise limit $\varepsilon\to 0$ on both sides of this equation we obtain
\begin{displaymath}
u^{(1)}_t(t,x) = \sum\limits_{j=1}^n A^j(t_0,x_0) 
\frac{\partial u^{(1)}}{\partial x^j}(t,x) + F_0
 = P_0(t_0,x_0,\partial/\partial x) u^{(1)}(t,x) + F_0,
\end{displaymath}
where $F_0 := B(t_0,x_0) u(t_0,x_0)$. \emph{Therefore, if $u$ is a solution of the variable coefficient equation $u_t = P(t,x,\partial/\partial x)u$, then, $u^{(1)}$ satisfies the linear constant coefficient problem $u^{(1)}_t(t,x) = P_0(t_0,x_0,\partial/\partial x) u^{(1)} + F_0$ obtained by freezing the coefficients in the principal part of $P(t,x,\partial/\partial x)u$ to their values at the point $p_0$ and by replacing the lower order term $B(t,x) u$ by the forcing term $F_0$.} By adjusting the scaling of $t$, a similar conclusion can be obtained when $P(t,x,\partial/\partial x)$ is a higher-derivative operator.

This leads us to the following statement: a necessary condition for the linear, variable coefficient Cauchy problem for the equation $u_t = P(t,x,\partial/\partial x)u$ to be well posed is that all the corresponding problems for the frozen coefficient equations $v_t = P_0(t_0,x_0,\partial/\partial x)v$ are well posed. For a rigorous proof of this statement for the case in which $P(t,x,\partial/\partial x)$ is time-independent, see Ref.~\cite{gS66}. We stress that it is important to replace $P(t,x,\partial/\partial x)$ by its principal part $P_0(t,x,\partial/\partial x)$ when freezing the coefficients. The statement is false if lower order terms are retained, see Refs.~\cite{KL89,gS66} for counterexamples. 

Now it is natural to ask whether or not the converse statement is true: suppose that the Cauchy problems for all frozen coefficient equations $v_t = P_0(t_0,x_0,\partial/\partial x)v$ are well posed; is the original, variable coefficient problem also well posed? It turns out this \textbf{localization principle} is valid in many cases under additional smoothness requirements. In order to formulate the latter, let us go back to the first order equation~(\ref{Eq:LinearFOSVC}). We define its principal symbol as
\begin{equation}
P_0(t,x,ik) := i\sum\limits_{j=1}^n A^j(t,x) k_j.
\end{equation}
In analogy to the constant coefficient case we define:
\begin{definition}
\label{Def:HyperbolicityVC}
The first order system~(\ref{Eq:LinearFOSVC}) is called
\begin{enumerate}
\item[(i)] \textbf{weakly hyperbolic} if all the eigenvalues of its principal symbol $P_0(t,x,ik)$ are purely imaginary.
\item[(ii)] \textbf{strongly hyperbolic} if there exist $M > 0$ and a family of positive definite, Hermitian $m\times m$ matrices $H(t,x,k)$, $(t,x,k)\in\Omega\times S^{n-1}$, whose coefficients belong to the class $C_b^\infty(\Omega\times S^{n-1})$, such that
\begin{equation}
M^{-1} I \leq H(t,x,k) \leq M I,\qquad
H(t,x,k)P_0(t,x,ik) + P_0(t,x,ik)^* H(t,x,k) = 0,
\label{Eq:SymmetrizerConditionsVC}
\end{equation}
for all $(t,x,k)\in \Omega\times S^{n-1}$, where $\Omega := [0,\infty) \times \Real^n$.
\item[(iii)] \textbf{symmetric hyperbolic} if it is strongly hyperbolic and the symmetrizer $H(t,x,k)$ can be chosen independent of $k$.
\end{enumerate}
\end{definition}
We see that these definitions are straight extrapolations of the corresponding definitions (see Definition~\ref{Def:Hyperbolicity}) in the constant coefficient case, \emph{except for the smoothness requirements for the symmetrizer $H(t,x,k)$.}\epubtkFootnote{These smoothness requirements are sometimes omitted in the numerical relativity literature.} There are examples of ill posed Cauchy problems for which a Hermitian, positive definite symmetrizer $H(t,x,k)$ exists but is not smooth, see Ref.~\cite{gS66}, showing that these requirements are necessary in general.

The smooth symmetrizer is used in order to construct a pseudo-differential operator
\begin{displaymath}
[H(t) v](x) := \frac{1}{(2\pi)^{n/2}}\int H(t,x,k/|k|) e^{i k\cdot x}\hat{v}(k) d^n k,\qquad
\hat{v}(k) =  \frac{1}{(2\pi)^{n/2}}\int e^{-i k\cdot x} v(x) d^n x,
\end{displaymath}
from which one defines a scalar product $(\cdot,\cdot)_{H(t)}$ which, for each $t$ is equivalent to the $L^2$ product. This scalar product has the property that a solution $u$ to the equation~(\ref{Eq:LinearFOSVC}) satisfies an inequality of the form
\begin{displaymath}
\frac{d}{dt} (u,u)_{H(t)} \leq b(T) (u,u)_{H(t)},\qquad
0\leq t\leq T,
\end{displaymath}
see, for instance, Ref.~\cite{Taylor96b}. Upon integration this yields an estimate of the form of Equation~(\ref{Eq:WPInequality}). In the symmetric hyperbolic case, we have simply $[H(t) v] = H(t,x) v(x)$ and the scalar product is given by
\begin{displaymath}
(u,v)_{H(t)} := \int u(x)^* H(t,x) v(x) d^n x,\qquad
u,v\in L^2(\Real^n).
\end{displaymath}
We will return to the application of this scalar product for deriving energy estimates below. Let us state the important result:

\begin{theorem}
If the first order system (\ref{Eq:LinearFOSVC}) is strongly or symmetric hyperbolic in the sense of Definition~\ref{Def:HyperbolicityVC}, then the Cauchy problem (\ref{Eq:LinearVCPDECauchy1}, \ref{Eq:LinearVCPDECauchy2}) is well posed in the sense of Definition~\ref{Def:WPVC}.
\end{theorem}

For a proof of this theorem, see for instance Proposition 7.1 and the comments following its formulation in chapter 7 of Ref.~\cite{Taylor96b}. Let us look at some examples:

\begin{example}
For a given, stationary fluid field, the non-relativistic, ideal magnetohydrodynamic equations reduce to the simple system~\cite{mClLoR08}
\begin{equation}
B_t = \nabla\wedge (v\wedge B)
\label{Eq:IdealMHD}
\end{equation}
for the magnetic field $B$, where $v$ is the fluid velocity. The principal symbol for this equation is given by
\begin{displaymath}
P_0(x,ik) B = ik\wedge (v(x)\wedge B) = (ik\cdot B) v(x) - (ik\cdot v(x)) B.
\end{displaymath}
In order to analyze it, it is convenient to introduce an orthonormal frame $e_1,e_2,e_3$ such that $e_1$ is parallel to $k$. With respect to this, the matrix corresponding to $P_0(x,ik)$ is
\begin{displaymath}
i|k|\left(\begin{array}{ccc} 
0 & 0 & 0 \\ v_2(x) & -v_1(x) & 0 \\ v_3(x) & 0 &-v_1(x)
\end{array} \right),
\end{displaymath}
with purely imaginary eigenvalues $0$, $-i|k| v_1(x)$. However, the symbol is not diagonalizable when $k$ is orthogonal to the fluid velocity, $v_1(x)=0$, and so the system is only weakly hyperbolic.

One can still show that the system is well posed if one takes into account the constraint $\nabla\cdot B = 0$, which is preserved by the evolution equation~(\ref{Eq:IdealMHD}). In Fourier space, this constraint forces $B_1=0$, which eliminates the first row and column in the principal symbol, and yields a strongly hyperbolic symbol. However, at the numerical level, this means that special care needs to be taken when discretizing the system~(\ref{Eq:IdealMHD}) since any discretization which does not preserve $\nabla\cdot B = 0$ will push the solution away from the constraint manifold in which case the system is weakly hyperbolic. For numerical schemes which explicitly preserve (divergence-transport) or enforce (divergence-cleaning) the constraints, see Refs.~\cite{jHcE88} and \cite{aDfKdKcMtSmW02}, respectively. For alternative formulations which are strongly hyperbolic without imposing the constraint, see Ref.~\cite{mClLoR08}.
\end{example}

\begin{example}
The localization principle can be generalized to a certain class of second order systems \cite{hKoO02}\cite{gNoOoR04}: For example, we may consider a second order linear equation of the form
\begin{equation}
v_{tt} = \sum\limits_{j,k=1}^n A^{jk}(t,x)\frac{\partial^2}{\partial x^j\partial x^k} v
 + \sum\limits_{j=1}^n 2B^j(t,x)\frac{\partial}{\partial x^j} v_t 
 + \sum\limits_{j=1}^n C^j(t,x)\frac{\partial}{\partial x^j} v + D(t,x) v_t + E(t,x) v,
\label{Eq:LinearVCWave}
\end{equation}
$x\in\Real^n$, $t\geq 0$, where now the $m\times m$ matrices $A^{jk}$, $B^j$, $C^j$, $D$ and $E$ belong to the class $C_b^\infty([0,\infty)\times\Real^n)$ of bounded, $C^\infty$-functions with bounded derivatives. Zooming into a very small neighborhood of a given point $p_0 = (t_0,x_0)$ by applying the expansion in Equation~(\ref{Eq:ZoomIn}) to $v$, one obtains, in the limit $\varepsilon\to 0$,
the constant coefficient equation
\begin{equation}
v^{(2)}_{tt}(t,x)
 = \sum\limits_{j,k=1}^n A^{jk}(t_0,x_0)\frac{\partial^2 v^{(2)}}{\partial x^j\partial x^k}(t,x)
 + \sum\limits_{j=1}^n 2B^j(t_0,x_0)\frac{\partial v^{(2)}_t}{\partial x^j}(t,x) + F_0,
\label{Eq:SecondOrderFrozen}
\end{equation}
with
\begin{displaymath}
F_0 :=  \sum\limits_{j=1}^n C^j(t_0,x_0)\frac{\partial v}{\partial x^j}(t_0,x_0)
 + D(t_0,x_0) v_t(t_0,x_0) + E(t_0,x_0) v(t_0,x_0),
\end{displaymath}
where we have used the fact that $v^{(1)}(t,x) = t v_t(t_0,x_0) + \sum\limits_{j=1}^n x^j\frac{\partial v}{\partial x^j}(t_0,x_0)$. Equation~(\ref{Eq:SecondOrderFrozen}) can
be rewritten as a first order system in Fourier space for the variable
\begin{displaymath}
\hat{U} = \left( \begin{array}{ll} |k|\hat{v} \\ 
 \hat{v}_t - i\sum\limits_{j=1}^n B^j(t_0,x_0) k_j\hat{v} \end{array}\right),
\end{displaymath}
see Section~\ref{SubSubSec:SecondOrderCC}. Now Theorem~\ref{Thm:SecondOrder} implies that the problem is well posed if there exist constants $M > 0$ and $\delta > 0$ and a family of positive definite $m\times m$ Hermitian matrices $h(t,x,k)$, $(t,x,k)\in \Omega\times S^{n-1}$, which is $C^\infty$-smooth in all its arguments such that $M^{-1} I \leq h(t,x,k) \leq M I$ and $h(t,x,k) R(t,x,k) = R(t,x,k)^* h(t,x,k)\geq \delta I$ for all $(t,x,k)\in\Omega\times S^{n-1}$, where $R(t,x,k) := \sum\limits_{i,j=1}^n (A^{ij}(t,x) + B^i(t,x) B^j(t,x)) k_i k_j$.

In particular, it follows that the Cauchy problem for the Klein--Gordon equation on a globally hyperbolic spacetime $M = [0,\infty) \times \Real^n$ with $\alpha,\beta^i,\gamma_{ij}\in C_b^\infty([0,\infty)\times \Real^n)$, is well posed provided that $\alpha^2\gamma^{ij}$ is uniformly positive definite, see Example~\ref{Example:KleinGordon}.
\end{example}

\subsubsection{Characteristic speeds and fields}

Consider a first order linear system of the form~(\ref{Eq:LinearFOSVC}) which is strongly hyperbolic. Then, for each $t\geq 0$, $x\in\Real^n$ and $k\in S^{n-1}$ the principal symbol $P_0(t,x,ik)$ is diagonalizable and has purely complex eigenvalues. In the constant coefficient case with no lower order terms ($B=0$) an eigenvalue $i\mu(k)$ of $P_0(ik)$ with corresponding eigenvector $a(k)$ gives rise to the plane wave solution
\begin{equation}
u(t,x) = a(k) e^{i\mu(k) t + ik\cdot x},\qquad t\geq 0, x\in \Real^n.
\end{equation}
If lower order terms are present and the matrix coefficients $A^j(t,x)$ are not constant one can look for approximate plane wave solutions which have the form
\begin{equation}
u(t,x) = a_\varepsilon(t,x) e^{i\varepsilon^{-1}\psi(t,x)}, \qquad
t\geq 0, x\in \Real^n,
\label{Eq:GeometricOpticLimit}
\end{equation}
where $\varepsilon > 0$ is a small parameter, $\psi(t,x)$ a smooth phase function and $a_\varepsilon(t,x) = a_0(t,x) + \varepsilon a_1(t,x) + \varepsilon^2 a_2(t,x) + \ldots$ a slowly varying amplitude. Introducing the ansatz (\ref{Eq:GeometricOpticLimit}) into Equation~(\ref{Eq:LinearFOSVC}) and taking the limit $\varepsilon\to 0$ yields the problem
\begin{equation}
i\psi_t a_0 = P_0(t,x,i\nabla\psi) a_0
 = i\sum\limits_{j=1}^n A^j(t,x)\frac{\partial \psi}{\partial x^j} a_0.
\end{equation}
Setting $\omega(t,x) := \psi_t(t,x)$ and $k(t,x) := \nabla\psi(t,x)$, a nontrivial solution exists if and only if the eikonal equation
\begin{equation}
\det\left[ i\omega I - P_0(t,x,ik) \right] = 0
\label{Eq:Eikonal}
\end{equation}
is satisfied. Its solutions provide the phase function $\psi(t,x)$ whose level sets have conormal $\omega dt + k\cdot dx$. The phase function and $a_0$ determine approximate plane wave solutions of the form~(\ref{Eq:GeometricOpticLimit}). For this reason we call $\omega(k)$ the \textbf{characteristic speed} in the direction $k\in S^{n-1}$, and $a_0$ a corresponding \textbf{characteristic mode}. For a strongly hyperbolic system, the solution at each point $(t,x)$ can be expanded in terms of the characteristic modes $e_j(t,x,k)$ with respect to a given direction $k\in S^{n-1}$,
\begin{displaymath}
u(t,x) = \sum\limits_{j=1}^m u^{(j)}(t,x,k) e_j(t,x,k).
\end{displaymath}
The corresponding coefficients $u^{(j)}(t,x,k)$ are called the \textbf{characteristic fields}.

\begin{example}
\label{Example:KleinGordonChar}
Consider the Klein--Gordon equation on a hyperbolic spacetime, like in Example~\ref{Example:KleinGordon}. In this case the eikonal equation is
\begin{displaymath}
0 = \det\left[ i\omega I - Q_0(ik) \right]
 = \det\left( \begin{array}{cc}
 i(\omega - \beta^j k_j) & |k| \\
 -\alpha^2\gamma^{ij} k_i k_j/|k| & i(\omega - \beta^j k_j)
 \end{array} \right)
  = -(\omega -\beta^j k_j)^2 + \alpha^2\gamma^{ij} k_i k_j,
\end{displaymath}
which yields $\omega_\pm(k) = \beta^j k_j \pm \alpha\sqrt{\gamma^{ij} k_i k_j}$. The corresponding conormals $\omega_{\pm}(k) dt + k_j dx^j$ is null; hence the surfaces of constant phase are null surfaces. The characteristic modes and fields are
\begin{displaymath}
e_{\pm}(k) = \left( \begin{array}{c} i|k| \\ \mp\alpha\sqrt{\gamma^{ij} k_i k_j} 
 \end{array} \right), \qquad
u^{(\pm)}(k) = \frac{1}{2}\left( \frac{U_1}{i |k|} 
 \mp \frac{U_2}{\alpha\sqrt{\gamma^{ij} k_i k_j}} \right),
\end{displaymath}
where $U = (U_1,U_2) = (|k| v, v_t - i\beta^j k_j v)$ and $v$ is the Klein--Gordon field.
\end{example}

\begin{example}
\label{Example:FatMaxwellChar}
In the formulation of Maxwell's equations discussed in Example~\ref{Example:FatMaxwell}, the characteristic speeds are $0$, $\pm\sqrt{\alpha\beta}$ and $\pm 1$, and the corresponding characteristic fields are the components of the vector on the right-hand side of Equation~(\ref{Eq:FatMaxwellCharVars}).
\end{example}

\subsubsection{Energy estimates and finite speed of propagation}
\label{sec:EnergyEstimate}

Here we focus our attention on first order linear systems which are symmetric hyperbolic. In this case it is not difficult to derive \emph{a priori energy estimates} based on integration by parts. Such estimates assume the existence of a sufficiently smooth solution and bound an appropriate norm of the solution at some time $t > 0$ in terms of the same norm of the solution at the initial time $t=0$. As we will illustrate here, such estimates already yield quite a lot of information on the qualitative behavior of the solutions. In particular, they give uniqueness, continuous dependence on the initial data and finite speed of propagation.

The word ``energy'' stems from the fact that for many problems the squared norm satisfying the estimate is directly or indirectly related to the physical energy of the system, although for many other problems the squared norm does not have a physical interpretation of any kind.

For first order symmetric hyperbolic linear systems, an a priori energy estimate can be constructed from the symmetrizer $H(t,x)$ in the following way. For a given smooth solution $u(t,x)$ of Equation~(\ref{Eq:LinearFOSVC}), define the vector field $J$ on $\Omega = [0,\infty)\times \Real^n$ by its components
\begin{equation}
J^\mu(t,x) := -u(t,x)^* H(t,x) A^\mu(t,x) u(t,x),\qquad
\mu = 0,1,2,\ldots,n,
\end{equation}
where $A^0(t,x) := -I$. By virtue of the evolution equation, $J$ satisfies
\begin{equation}
\partial_\mu J^\mu(t,x) \equiv 
\frac{\partial}{\partial t} J^0(t,x) + \sum\limits_{k=1}^n\frac{\partial}{\partial x^k} J^k(t,x)
 = u(t,x)^* K(t,x) u(t,x),
\label{Eq:PseudoConservationLaw}
\end{equation}
where the Hermitian $m\times m$ matrix $K(t,x)$ is defined as
\begin{displaymath}
K(t,x) := H(t,x) B(t,x) + B(t,x)^* H(t,x) + H_t(t,x) - \sum\limits_{k=1}^n\frac{\partial}{\partial x^k}\left( H(t,x) A^k(t,x) \right).
\end{displaymath}
If $K=0$, Equation~(\ref{Eq:PseudoConservationLaw}) formally looks like a conservation law for the current density $J$. If $K\neq 0$ we obtain, instead of a conserved quantity, an energy-like expression whose growth can be controlled by its initial value. For this, we first notice that our assumptions on the matrices $H(t,x)$, $B(t,x)$ and $A^k(t,x)$ imply that $K(t,x)$ is bounded on $\Omega$. In particular, since $H(t,x)$ is uniformly positive, there is a constant $\alpha > 0$ such that
\begin{equation}
K(t,x) \leq 2\alpha H(t,x) ,\qquad (t,x)\in\Omega.
\label{Eq:UniformLBound}
\end{equation}
Let $\Omega_T = \bigcup_{0\leq t \leq T} \Sigma_t$ be a tubular region obtained by piling up open subsets $\Sigma_t$ of $t=const$ hypersurfaces. This region is enclosed by the initial surface $\Sigma_0$, the final surface $\Sigma_T$ and the boundary surface ${\cal T} := \bigcup_{0\leq t \leq T} \partial\Sigma_t$ which is assumed to be smooth. Integrating Equation~(\ref{Eq:PseudoConservationLaw}) over $\Omega_T$ and using Gauss' theorem, one obtains
\begin{equation}
\int\limits_{\Sigma_T} J^0(t,x) d^n x = \int\limits_{\Sigma_0} J^0(t,x) d^n x
 - \int\limits_{\cal T} e_\mu J^\mu(t,x) dS
 + \int\limits_{\Omega_T} u(t,x)^* K(t,x) u(t,x) dt d^n x,
\label{Eq:IntegratedPseudoConservationLaw}
\end{equation}
where $e_\mu$ is the unit outward normal covector to ${\cal T}$ and $dS$ the volume element on that surface. Defining the ``energy'' contained in the surface $\Sigma_t$ by
\begin{equation}
E(\Sigma_t) := \int\limits_{\Sigma_t} J^0(t,x) d^n x
 = \int\limits_{\Sigma_t} u(t,x)^* H(t,x) u(t,x) d^n x
\label{Eq:EnergyDef}
\end{equation}
and assuming for the moment that the ``flux'' integral over ${\cal T}$ is positive or zero one obtains the estimate
\begin{eqnarray}
E(\Sigma_T) &\leq& E(\Sigma_0) + \int\limits_0^T\left( 
 \int\limits_{\Sigma_t} u(t,x)^* K(t,x) u(t,x) d^n x \right) dt
 \nonumber\\
 &\leq& E(\Sigma_0) 
 + 2\alpha\int\limits_0^T E(\Sigma_t)  dt,
\label{Eq:FirstEnergyEstimate}
\end{eqnarray}
where we have used the inequality (\ref{Eq:UniformLBound}) and the definition of $E(\Sigma_t)$ in the last step. Defining the function $h(T) := \int_0^T E(\Sigma_t) dt$ this inequality can be rewritten as
\begin{displaymath}
\frac{d}{dt}\left( h(t) e^{-2\alpha t} \right) \leq E(\Sigma_0) e^{-2\alpha t},
\qquad 0\leq t \leq T,
\end{displaymath}
which yields $\alpha h(T) \leq E(\Sigma_0)( e^{2\alpha T} - 1 )$ upon integration. This together with (\ref{Eq:FirstEnergyEstimate}) gives
\begin{equation}
E(\Sigma_t) \leq e^{2\alpha t} E(\Sigma_0),
\qquad 0 \leq t \leq T,
\label{Eq:EnergyEstimate}
\end{equation}
which bounds the energy at any time $t\in [0,T]$ in terms of the initial energy.

In order to analyze the conditions under which the flux integral is positive or zero, we examine the sign of the integrand $e_\mu J^\mu(t,x)$. Decomposing $e_\mu dx^\mu = N[a\, dt + s_1 dx^1 + \ldots + s_2 dx^n]$ where $s = (s_1,\ldots,s_n)$ is a unit vector and $N > 0$ a positive normalization constant, we have
\begin{displaymath}
e_\mu J^\mu(t,x) = N(t,x)\,u(t,x)^* [ a(t,x) H(t,x) - H(t,x) P_0(t,x,s) ] u(t,x), 
\end{displaymath}
where $P_0(t,x,s) = \sum\limits_{j=1}^n A^j(t,x) s_j$ is the principal symbol in the direction of the unit vector $s$. This is guaranteed to be positive if the boundary surface ${\cal T}$ is such that $a(t,x)$ is greater than or equal to all the eigenvalues of the \textbf{boundary matrix} $P_0(t,x,s)$, for each $(t,x)\in {\cal T}$. This is equivalent to the condition
\begin{equation}
a(t,x) \geq \sup\limits_{u\in\Complex^m, u\neq 0} \frac{u^* H(t,x) P_0(t,x,s) u}{u^* H(t,x) u}
\qquad\hbox{for all $(t,x)\in {\cal T}$}.
\label{Eq:OutflowBoundaryCondition}
\end{equation}
Since $H(t,x) P_0(t,x,s)$ is symmetric, the supremum is equal to the maximum eigenvalue of $P_0(t,x,s)$. Therefore, condition~(\ref{Eq:OutflowBoundaryCondition}) is equivalent to the requirement that $a(t,x)$ be greater than or equal to the maximum characteristic speed in the direction of the unit outward normal $s$.

With these arguments, we arrive at the following conclusions and remarks:
\begin{itemize}
\item \textbf{Finite speed of propagation}.
Let $p_0 = (t_0,x_0)\in\Omega$ be a given event, and set
\begin{displaymath}
v(t_0) := \sup\left\{ \frac{u^* H(t,x) P_0(t,x,s) u}{u^* H(t,x) u} : 0\leq t \leq t_0, x\in\Real^n, s\in S^{n-1}, u\in\Complex^m, u\neq 0 \right\}.
\end{displaymath}
Define the past cone at $p_0$ as\epubtkFootnote{In principle, the maximum propagation speed $v(t_0)$ could be infinite. Finiteness can be guaranteed, for instance, by requiring the principal symbol $P_0(t,x,s)$ to be independent of $(t,x)$ for $|x| > R$ outside a large ball of radius $R > 0$.}
\begin{equation}
C^-(p_0) := \{ (t,x)\in \Omega : |x| \leq v(t_0)(t_0 - t) \}.
\label{Eq:PastLightCone}
\end{equation}
The unit outward normal to its boundary is $e_\mu dx^\mu = N[ v(t_0) dt + x\cdot dx/|x| ]$ which satisfies the condition~(\ref{Eq:OutflowBoundaryCondition}). It follows from the estimate (\ref{Eq:EnergyEstimate}) applied to the domain $\Omega_T = C^-(p_0)$ that the solution $u$ is zero on $C^-(p_0)$ if the initial data is zero on the intersection of the cone $C^-(p_0)$ with the initial surface $t=0$. In other words, a perturbation in the initial data \emph{outside} the ball $|x| \leq v(t_0) t_0$ does not alter the solution \emph{inside} the cone $C^-(p_0)$. Using this argument, it also follows that if $f$ has compact support, the corresponding solution $u(t,\cdot)$ also has compact support for all $t > 0$.
\item \textbf{Continuous dependence on the initial data}. Let $f\in C^\infty_0(\Real^n)$ be smooth initial data with compact support. As we have seen above, the corresponding smooth solution $u(t,\cdot)$ also has compact support for each $t\geq 0$. Therefore, applying the estimate~(\ref{Eq:EnergyEstimate}) to the case $\Sigma_t := \{ t \} \times \Real^n$, the boundary integral vanishes and we obtain
\begin{displaymath}
E(\Sigma_t) \leq e^{2\alpha t} E(\Sigma_0), \qquad t \geq 0.
\end{displaymath}
In view of the definition of $E(\Sigma_t)$, see Equation~(\ref{Eq:EnergyDef}), and the properties~(\ref{Eq:SymmetrizerConditionsVC}) of the symmetrizer it follows that
\begin{equation}
\| u(t,\cdot) \| \leq M e^{\alpha t} \| f \|,\qquad t\geq 0,
\label{Eq:WPInequalityBis}
\end{equation}
which is of the required form, see Definition~\ref{Def:WPVC}. In particular, we have uniqueness and continuous dependence on the initial data.
\item The statements about finite speed of propagation and continuous dependence on the data can easily be generalized to the case of a first order symmetric hyperbolic inhomogeneous equation $u_t = P(t,x,\partial/\partial x) u + F(t,x)$, with $F: \Omega \to \Complex^m$ a bounded, $C^\infty$-function with bounded derivatives. In this case, the inequality (\ref{Eq:WPInequalityBis}) is replaced by
\begin{equation}
\| u(t,\cdot) \| \leq M e^{\alpha t}\left[ \| f \| 
 + \int\limits_0^t e^{-\alpha s} \| F(s,\cdot) \| ds \right],
\qquad t\geq 0.
\label{Eq:WPInequalityInhomogeneous}
\end{equation}
\item If the boundary surface ${\cal T}$ does not satisfy the condition~(\ref{Eq:OutflowBoundaryCondition}) for the boundary integral to be positive, then suitable boundary conditions need to be specified in order to control the sign of this term. This will be discussed in Section~\ref{section:MaxDiss}.
\item Although different techniques have to be used to prove them, very similar results hold for strongly hyperbolic systems, see Ref.~\cite{oR04}.
\end{itemize}
 
\begin{example}
We have seen that for the Klein--Gordon equation propagating on a globally hyperbolic spacetime, the characteristic speeds are the speed of light. Therefore, in the case of a constant metric (i.e., Minkowksi space), the past cone $C^-(p_0)$ defined in Equation~(\ref{Eq:PastLightCone}) coincides with the past \emph{light} cone at the event $p_0$. A slight refinement of the above argument shows that the statement remains true for a Klein--Gordon field propagating on any hyperbolic spacetime. 
\end{example}

\begin{example}
In Example~\ref{Example:FatMaxwellChar} we have seen that the characteristic speeds of the system given in Example~\ref{Example:FatMaxwell} are $0$, $\pm\sqrt{\alpha\beta}$ and $\pm 1$, where $\alpha\beta > 0$ is assumed for strong hyperbolicity. Therefore, the past cone $C^-(p_0)$ corresponds to the past light cone \emph{provided that $0 < \alpha\beta \leq 1$.} For $\alpha\beta > 1$, the formulation has superluminal constraint-violating modes, and an initial perturbation emanating from a region \emph{outside} the past light cone at $p_0$ could affect the solution at $p_0$. In this case, the past light cone at $p_0$ is a proper subset of $C^-(p_0)$. 
\end{example}

%===================================================================
\subsection{Quasilinear equations}
\label{SubSec:QLP}
%===================================================================

Next, we generalize the theory one more step and consider evolution systems which are described by quasilinear partial differential equations, that is, by nonlinear partial differential equations which are linear in their highest order derivatives. This already covers most of the interesting physical systems, including the Yang--Mills and the Einstein equations. Restricting ourselves to the first order case, such equations have the form
\begin{equation}
u_t = \sum\limits_{j=1}^n A^j(t,x,u)\frac{\partial}{\partial x^j} u + F(t,x,u),
\qquad 0\leq t\leq T,\quad x\in \Real^n,
\label{Eq:QuasiLinearFOSVC}
\end{equation}
where all the coefficients of the complex $m\times m$ matrices $A^1(t,x,u)$, \ldots , $A^n(t,x,u)$ and the nonlinear source term $F(t,x,u)\in\Complex^m$ belong to the class $C_b^\infty([0,T]\times\Real^n\times\Complex^m)$ of bounded, $C^\infty$-functions with bounded derivatives. Compared to the linear case, there are two new features the solutions may exhibit:
\begin{itemize}
\item The nonlinear term $F(t,x,u)$ may induce \textbf{blowup of the solutions in finite time}. This is already the case for the simple example where $m=1$, all the matrices $A^j$ vanish identically and $F(t,x,u) = u^2$, in which case Equation~(\ref{Eq:QuasiLinearFOSVC}) reduces to $u_t = u^2$.
\item In contrast to the linear case, the matrix functions $A^j$ in front of the derivative operator now depend pointwise on the state vector $u$ itself, which implies, in particular, that the characteristic speeds and fields depend on $u$. This can lead to the formation of \textbf{shocks} where characteristics cross each other, like in the simple example of Burger's equation $u_t = u u_x$ corresponding to the case $m=n=1$, $A^1(t,x,u) = u$ and $F(t,x,u) = 0$.
\end{itemize}
For these reasons, one cannot expect global existence of smooth solutions from smooth initial data with compact support in general, and the best one can hope for is existence of a smooth solution on some finite time interval $[0,T]$, where $T$ might depend on the initial data.

Under such restrictions, it is possible to prove well posedness of the Cauchy problem. The idea is to linearize the problem and to apply Banach's fixed point theorem. This is is discussed next.

\subsubsection{The principle of linearization}

Suppose $u^{(0)}(t,x)$ is a $C^\infty$ (reference) solution of Equation~(\ref{Eq:QuasiLinearFOSVC}), corresponding to initial data $f(x) = u^{(0)}(0,x)$. Assuming this solution to be uniquely determined by the initial data $f$, we may ask if a unique solution $u$ also exists for the perturbed problem
\begin{eqnarray}
u_t(t,x) = \sum\limits_{j=1}^n A^j(t,x,u)\frac{\partial}{\partial x^j} u(t,x) 
 + F(t,x,u) + \delta F(t,x), && x\in \Real^n,\quad 0\leq t\leq T,
\label{Eq:QuasiLinearFOSVCPert}\\
u(0,x) = f(x) + \delta f(x), && x\in\Real^n,
\label{Eq:QuasiLinearFOSVCPertID}
\end{eqnarray}
where the perturbations $\delta F(t,x)$ and $\delta f(x)$ belong to the class of bounded, $C^\infty$-functions with bounded derivatives. This leads to the following definition:

\begin{definition}
\label{Def:WPNonLinear}
Consider the nonlinear Cauchy problem given by Equation~(\ref{Eq:QuasiLinearFOSVC}) and prescribed initial data for $u$ at $t=0$. Let $u^{(0)}$ be a $C^\infty$-solution to this problem which is uniquely determined by its initial data $f$. Then, the problem is called well posed at $u^{(0)}$, if there are normed vector spaces $X$, $Y$, and $Z$ and constants $K > 0$, $\varepsilon > 0$ such that for all sufficiently smooth perturbations $\delta f$ and $\delta F$ lying in $Y$ and $Z$, respectively, with
\begin{displaymath}
\| \delta f \|_Y + \| \delta F \|_Z < \varepsilon,
\end{displaymath}
the perturbed problem~(\ref{Eq:QuasiLinearFOSVCPert}, \ref{Eq:QuasiLinearFOSVCPertID}) is also uniquely solvable and the corresponding solution $u$ satisfies $u - u^{(0)}\in X$ and the estimate
\begin{equation}
\| u - u^{(0)} \|_X \leq K\left( \| \delta f \|_Y + \| \delta F \|_Z \right).
\label{Eq:WPNonLinear}
\end{equation}
\end{definition}
Here, the norms $X$ and $Y$ appearing on both sides of Equation~(\ref{Eq:WPNonLinear}) are different from each other because $\| u - u^{(0)} \|_X$ controls the function $u - u^{(0)}$ over the spacetime region $[0,T]\times\Real^n$ while $\| \delta f \|_Y$
is a norm controlling the function $\delta f$ on $\Real^n$.

If the problem is well posed at $u^{(0)}$ we may consider a one-parameter curve $f_\varepsilon$ of initial data lying in $C^\infty_0(\Real^n)$ that goes through $f$ and assume that there is a corresponding solution $u_\varepsilon(t,x)$ for each small enough $|\varepsilon|$ which lies close to $u^{(0)}$ in the sense of inequality~(\ref{Eq:WPNonLinear}). Expanding
\begin{displaymath}
u_\varepsilon(t,x) = u^{(0)}(t,x) + \varepsilon v^{(1)}(t,x) 
 + \varepsilon^2 v^{(2)}(t,x) + \ldots
\end{displaymath}
and plugging into the Equation~(\ref{Eq:QuasiLinearFOSVC}) we find, to first order in $\varepsilon$,
\begin{equation}
v^{(1)}_t
  = \sum\limits_{j=1}^n A_0^j(t,x)\frac{\partial}{\partial x^j} v^{(1)} + B_0(t,x) v^{(1)},
\label{Eq:LinearFOSVCBis}
\end{equation}
with
\begin{displaymath}
A_0^j(t,x) = A^j(t,x,u^{(0)}(t,x)),\qquad
B_0(t,x) = \frac{\partial A^j}{\partial u}(t,x,u^{(0)}(t,x))\frac{\partial u^{(0)}}{\partial x^j}
 + \frac{\partial F}{\partial u}(t,x,u^{(0)}(t,x)).
\end{displaymath}
Equation~(\ref{Eq:LinearFOSVCBis}) is a first order linear equation with variable coefficients for the first variation, $v^{(1)}$, for which we can apply the theory described in Section~\ref{SubSec:LPVC}. Therefore, it is reasonable to assume that the linearized problem is strongly hyperbolic for any smooth function $u^{(0)}(t,x)$. In particular, if we generalize the definitions of strongly and symmetric hyperbolicity given in Definition~\ref{Def:HyperbolicityVC} to the quasilinear case by requiring that the symmetrizer $H(t,x,k,u)$ has coefficients in $C^\infty_b(\Omega\times S^{n-1}\times\Complex^m)$, it follows that the linearized problem is well posed provided that the quasilinear problem is strongly or symmetric hyperbolic.

The \textbf{linearization principle} states that the converse is also true: the nonlinear problem is well posed at $u^{(0)}$ if all the linear problems which are obtained by linearizing Equation~(\ref{Eq:QuasiLinearFOSVC}) at functions in a suitable neighborhood of $u^{(0)}$ are well posed. To prove that this principle holds, one sets up the following iteration. We define the sequence $u^{(k)}$ of functions by iteratively solving the linear problems
\begin{eqnarray}
u^{(k+1)}_t = \sum\limits_{j=1}^n A^j(t,x,u^{(k)})\frac{\partial}{\partial x^j} u^{(k+1)} 
 + F(t,x,u^{(k)}) + \delta F(t,x), && x\in \Real^n,\quad 0\leq t\leq T,
\label{Eq:CauchyIteration1}\\
u^{(k+1)}(0,x) = f(x) + \delta f(x), && x\in\Real^n,
\label{Eq:CauchyIteration2}
\end{eqnarray}
for $k=0,1,2,\ldots$ starting with the reference solution $u^{(0)}$. If the linearized problems are well posed in the sense of Definition~\ref{Def:WPVC} for functions lying in a neighborhood of $u^{(0)}$, one can solve each Cauchy problem~(\ref{Eq:CauchyIteration1}, \ref{Eq:CauchyIteration2}), at least for small enough time $T_k$. The key point then, is to prove that $T_k$ does not shrink to zero when $k\to\infty$ and to show that the sequence $u^{(k)}$ of functions converges to a solution of the perturbed problem~(\ref{Eq:QuasiLinearFOSVCPert}, \ref{Eq:QuasiLinearFOSVCPertID}). This is, of course, a nontrivial task which requires controlling $u^{(k)}$ and its derivatives in an appropriate way. For particular examples where this program is carried through, see~\cite{KL89}. For general results on quasilinear symmetric hyperbolic systems, see~\cite{tK75a,aFjM72,Taylor96c}.

%===================================================================
\subsection{Abstract evolution operators}
\label{SubSec:SemiGroups}
%===================================================================

A general framework for treating evolution problems is based on methods from functional analysis. Here, one considers a linear operator $A: D(A) \subset X \to X$ with dense domain, $\overline{D(A)} = X$, in a Banach space $X$ and asks under which conditions the Cauchy problem
\begin{eqnarray}
&& u_t(t) = A u(t),\qquad t\geq 0,
\label{Eq:AbstractCP1}\\
&& u(0) = f,
\label{Eq:AbstractCP2}
\end{eqnarray}
possesses a unique solution curve, i.e., a continuously differentiable map $u: [0,\infty) \to D(A)\subset X$ satisfying Equations~(\ref{Eq:AbstractCP1}, \ref{Eq:AbstractCP2}) for each $f\in D(A)$. Under a mild assumption on $A$ this turns out to be the case if and only if\epubtkFootnote{See Theorem~4.1.3 in Ref.~\cite{Pazy-Book}.} the operator $A$ is the infinitesimal generator of a strongly continuous semigroup $P(t)$, that is, a map $P: [0,\infty) \to {\cal L}(X)$, with ${\cal L}(X)$ denoting the space of bounded, linear operators on $X$, with the properties that
\begin{enumerate}
\item[(i)] $P(0) = I$,
\item[(ii)] $P(t + s) = P(t) P(s)$ for all $t,s\geq 0$,
\item[(iii)] $\lim\limits_{t\to 0} P(t) u = u$ for all $u\in X$,
\item[(iv)] $D(A) = \left\{ u\in X : \lim\limits_{t\to 0} \frac{1}{t}\left[ P(t) u - u \right] \hbox{ exists in $X$} \right\}$ and $A u = \lim\limits_{t\to 0} \frac{1}{t}\left[ P(t) u - u \right]$, $u\in D(A)$.
\end{enumerate}
In this case, the solution curve $u$ of the Cauchy problem~(\ref{Eq:AbstractCP1}, \ref{Eq:AbstractCP2}) is given by $u(t) = P(t) f$, $t\geq 0$, $f\in D(A)$. One can show \cite{Pazy-Book,hB05} that $P(t)$ always possesses constants $K \geq 1$ and $\alpha\in\Real$ such that
\begin{equation}
\| P(t) \| \leq K e^{\alpha t}, \qquad t\geq 0,
\end{equation}
which implies that $\| u(t) \| \leq K e^{\alpha t} \| f \|$ for all $f\in D(A)$ and all $t\geq 0$. Therefore, the semigroup $P(t)$ gives existence, uniqueness and continuous dependence on the initial data.

There are several results giving necessary and sufficient conditions for the linear operator $A$ to generate a strongly continuous semigroup, see, for instance, Refs.~\cite{Pazy-Book,hB05}. One useful result, which we formulate for Hilbert spaces, is the following:

\begin{theorem}[Lumer--Phillips]
\label{Thm:LumerPhillips}
Let $X$ be a Hilbert space with scalar product $(\cdot,\cdot)$, and let $A: D(A)\subset X \to X$ be a linear operator. Let $\alpha\in\Real$. Then, the following statements are equivalent:
\begin{enumerate}
\item[(i)] $A$ is the infinitesimal generator of a strongly continuous semigroup $P(t)$ such that $\| P(t) \| \leq e^{\alpha t}$ for all $t\geq 0$.
\item[(ii)] $A - \alpha I$ is dissipative, that is, $\re(u,Au - \alpha u) \leq 0$ for all $u\in D(A)$, and the range of $A - \lambda I$ is equal $X$ for some $\lambda > \alpha$.
\end{enumerate}
\end{theorem}

\begin{example}
As a simple example consider the Hilbert space $X = L^2(\Real^n)$ with the linear operator $A: D(A)\subset X\to X$ defined by
\begin{eqnarray}
&& D(A) := \{ u\in X : (1 + |k|^2){\cal F} u\in L^2(\Real^n) \},
\nonumber\\
&& A u := \Delta u = -{\cal F}^{-1}( |k|^2 {\cal F} u),\qquad u\in D(A),
\nonumber
\end{eqnarray}
where ${\cal F}$ denotes the Fourier-Plancharel operator, see Section~\ref{section:notation}. Using Parseval's identity, we find
\begin{displaymath}
\re(u,A u) = \re({\cal F} u, -|k|^2 {\cal F} u) = -\| (|k| {\cal F}u) \|^2 \leq 0,
\end{displaymath}
hence $A$ is dissipative. Furthermore, let $v\in L^2(\Real^n)$, then
\begin{displaymath}
u := {\cal F}^{-1}\left( \frac{{\cal F}v}{1 + |k|^2} \right)
\end{displaymath}
defines an element in $D(A)$ satisfying $(I - A) u = u - \Delta u = v$. Therefore, the range of $A - I$ is equal to $X$, and Theorem~\ref{Thm:LumerPhillips} implies that $A = \Delta$ generates a strongly continuous semigroup $P(t)$ on $X$ such that $\| P(t) \| \leq 1$ for all $t\geq 0$. The curves $u(t) := P(t) f$, $t\geq 0$, $f\in L^2(\Real^n)$ are the weak solutions to the heat equation on $\Real^n$, see Section~\ref{SubSubSec:ExtensionOfSolutions}.
\end{example}

In general, the requirement for $A - \alpha I$ to be dissipative is equivalent to finding an energy estimate for the squared norm $E := \| u \|^2$ of $u$. Indeed, setting $u(t) := P(t) f$ and using $u_t = A P(t) f$ we find
\begin{displaymath}
\frac{d}{dt} E(t) = \frac{d}{dt} \| u(t) \|^2 = 2\re(u(t), A u(t) ) \leq 2\alpha \| u(t) \|^2
 = 2\alpha E(t)
\end{displaymath}
for all $t\geq 0$ and $f\in D(A)$, which yields the estimate
\begin{equation}
\| u(t) \| \leq e^{\alpha t} \| f \|,\qquad t\geq 0,
\end{equation}
for all $f\in D(A)$. Given the dissipativity of $A - \alpha I$, the second requirement, that the range of $A - \lambda I$ is $X$ for some $\lambda > \alpha$, is equivalent to demanding that the linear operator $A - \lambda I : D(A) \to X$ be invertible. Therefore, proving this condition requires solving the linear equation
\begin{displaymath}
A u - \lambda u = v
\end{displaymath}
for given $v\in X$. This condition is important for the existence of solutions, and shows that for general evolution problems requiring an energy estimate is not sufficient. This statement is rather obvious, because given that $A - \alpha I$ is dissipative on $D(A)$, one could just make $D(A)$ smaller, and still have an energy estimate. However, if $D(A)$ is too small, the Cauchy problem is over-determined and a solution might not exist. We will encounter explicit examples for this phenomena in Section~\ref{section:ibvp}, when discussing boundary conditions.

Finding the correct domain $D(A)$ for the infinitesimal generator $A$ is not always a trivial task, especially for equations involving singular coefficients. Fortunately, there are weaker versions of the Lumer--Phillips theorem which only require checking conditions on a subspace $D\subset D(A)$ which is dense in $X$. It is also possible to formulate the Lumer--Phillips theorem on Banach spaces. See Refs.~\cite{Pazy-Book, EngelNagel-Book, hB05} for more details.

The semigroup theory can be generalized to time-dependent operators $A(t)$, and to quasilinear equations where $A(u)$ depends on the solution $u$ itself. We refer the reader to Ref.~\cite{hB05} for these generalizations and for applications to examples from mathematical physics including general relativity. The theory of strongly continuous semigroups has also been used for formulating well posed initial-boundary value formulations for the Maxwell equations~\cite{oRoS05} and the linearized Einstein equations~\cite{gNoS06} with elliptic gauge conditions.

\newpage
%===================================================================
%===================================================================
\section{Initial Value Formulations for Einstein's Equations}
\label{section:IVFEinstein}
%===================================================================
%===================================================================

In this section we apply the theory discussed in the previous section to well posed Cauchy formulations of Einstein's vacuum equations. The first such formulation dates back to the fifties \cite{cB52} and will be discussed in Section~\ref{SubSec:Harmonic}. Since then, there has been a plethora of new formulations which distinguish themselves by the choice of variables (metric vs.\ tetrad, Christoffel symbols vs.\ connection coefficients, inclusion or not of curvature components as independent variables, etc.), the choice of gauges and the use of the constraint equations in order to modify the evolution equations off the constraint surface. Many of these new formulations have been motivated by numerical calculations which try to solve a given physical problem in a stable way.

By far the most successful formulations for numerically evolving binary compact objects have been the harmonic system, which is based on the original work of Ref.~\cite{cB52}, and the Baumgarte-Shapiro-Shibata-Nakamura (BSSN) one~\cite{mStN95,tBsS98}. For this reason, we review these two formulations in detail in Sections~\ref{SubSec:Harmonic} and~\ref{SubSec:BSSN}, respectively. In Section~\ref{SubSec:ADM} we also review the Arnotwitt-Deser-Misner (ADM) formulation~\cite{ADM62}, which is based on a Hamiltonian approach to general relativity and serves as a starting point for many hyperbolic systems, including the BSSN one. A list of references for hyperbolic reductions of Einstein's equations not discussed here is given in Section~\ref{SubSec:HypForm}.

%===================================================================
\subsection{The harmonic formulation}
\label{SubSec:Harmonic}
%===================================================================

We start by discussing the harmonic formulation of Einstein's field equations. Like in the potential formulation of electromagnetism, where the Lorentz gauge $\nabla_\mu A^\mu = 0$ allows to cast Maxwell's equations into a system of wave equations, it was observed early in Ref.~\cite{DeDonder-Book} that Einstein's equations reduce to a system of wave equations when harmonic coordinates,
\begin{equation}
\nabla^\mu \nabla_\mu x^\nu = 0, \qquad \nu = 0,1,2,3,
\label{Eq:HarmonicCoordinates}
\end{equation}
are used. There are many straightforward generalizations of these gauge conditions; one of them is to replace the right-hand side of Equation~(\ref{Eq:HarmonicCoordinates}) by given source functions $H^\nu$ \cite{hF96, dG02}.

In order keep general covariance, we follow \cite{HawkingEllis-Book} and choose a \emph{fixed} smooth background metric $\gz_{\alpha\beta}$ with corresponding Levi-Civita connection $\nablaz$, Christoffel symbols $\Gammaz^\mu{}_{\alpha\beta}$, and curvature tensor $\Rz^\alpha{}_{\beta\mu\nu}$. Then, the generalized harmonic gauge condition can be rewritten as\epubtkFootnote{Geometrically, this means that the identity map $\phi : (M,g) \to (M,\gz)$, $p\mapsto p$, satisfies the inhomogeneous harmonic wave map equation
\begin{equation}
\nabla^\mu\nabla_\mu\phi^D + \Gammaz^D{}_{AB}\frac{\partial\phi^A}{\partial x^\mu}
\frac{\partial\phi^B}{\partial x^\nu} g^{\mu\nu} = H^D,
\label{Eq:HarmonicWaveMap}
\end{equation}
where $x^\mu$ and $x^A$ are local coordinates on $(M,g)$ and $(M,\gz)$, respectively.}
\begin{equation}
C^\mu := g^{\alpha\beta}\left( \Gamma^\mu{}_{\alpha\beta} 
 - \Gammaz^\mu{}_{\alpha\beta} \right) + H^\mu = 0.
\label{Eq:HarmonicConstraint}
\end{equation}
In the particular case where $H^\mu = 0$ and where the background metric is Minkowski in standard Cartesian coordinates, $\Gammaz^\mu{}_{\alpha\beta}$ vanishes, and the condition $C^\mu = 0$ reduces to the harmonic coordinate expression~(\ref{Eq:HarmonicCoordinates}). However, unlike the condition~(\ref{Eq:HarmonicCoordinates}), Equation~(\ref{Eq:HarmonicConstraint}) yields a coordinate-independent condition for any given vector field $H^\mu$ on spacetime since the difference $C^\mu{}_{\alpha\beta} := \Gamma^\mu{}_{\alpha\beta} - \Gammaz^\mu{}_{\alpha\beta}$ between two connections is a tensor field. In terms of the difference, $h_{\alpha\beta} := g_{\alpha\beta} - \gz_{\alpha\beta}$, between the dynamical and background metric, this tensor field can be expressed as
\begin{equation}
C^\mu{}_{\alpha\beta} = \frac{1}{2} g^{\mu\nu}\left( 
\nablaz_\alpha h_{\beta\nu} + \nablaz_\beta h_{\alpha\nu} - \nablaz_\nu h_{\alpha\beta} \right).
\label{Eq:Christoffel}
\end{equation}
Of course, the coordinate-independence is now traded for the introduction of a background metric $\gz_{\alpha\beta}$, and the question remains of how to choose $\gz_{\alpha\beta}$ and the vector field $H^\mu$. A simple possibility is to choose $H^\mu = 0$ and $\gz_{\alpha\beta}$ equal to the initial data for the metric, such that $h_{\mu\nu} = 0$ initially.

Einstein's field equations in the gauge $C^\mu = 0$ are equivalent to the wave system
\begin{eqnarray}
g^{\mu\nu}\nablaz_\mu\nablaz_\nu h_{\alpha\beta} 
&=& 2\, g_{\sigma\tau} g^{\mu\nu} C^\sigma{}_{\alpha\mu} C^\tau{}_{\beta\nu} 
 + 4\, C^\mu{}_{\nu(\alpha} g_{\beta)\sigma} C^\sigma{}_{\mu\tau} g^{\nu\tau}
 - 2\, g^{\mu\nu} \Rz^\sigma{}_{\mu\nu(\alpha} g_{\beta)\sigma}
\nonumber\\
&+& 16\pi G_N
  \left( T_{\alpha\beta} - \frac{1}{2} g_{\alpha\beta} g^{\mu\nu} T_{\mu\nu} \right)
 - 2\,\nabla_{(\alpha} H_{\beta)}\, ,
\label{Eq:EinsteinWave}
\end{eqnarray}
where $T_{\alpha\beta}$ is the stress-energy tensor and $G_N$ Newton's constant. This system is subject to the harmonic constraint
\begin{equation}
0 = C^\mu = g^{\mu\nu} g^{\alpha\beta}\left(
 \nablaz_\alpha h_{\beta\nu} - \frac{1}{2}\nablaz_\nu h_{\alpha\beta} \right) + H^\mu.
\label{Eq:HarmonicConstraintBis}
\end{equation}

\subsubsection{Hyperbolicity}

For any given smooth stress-energy tensor $T_{\alpha\beta}$, the equations~(\ref{Eq:EinsteinWave}) constitute a quasilinear system of ten coupled wave equations for the ten coefficients of the difference metric $h_{\alpha\beta}$ (or equivalently, for the ten components of the dynamical metric $g_{\alpha\beta}$) and, therefore, we can apply the results of Section \ref{section:ivp} to formulate a (local in time) well-posed Cauchy problem for the wave system~(\ref{Eq:EinsteinWave}) with initial conditions
\begin{equation}
h_{\alpha\beta}(0,x) = h^{(0)}_{\alpha\beta}(x),\qquad
\frac{\partial h_{\alpha\beta}}{\partial t}(0,x) = k^{(0)}_{\alpha\beta}(x),
\label{Eq:EinsteinWaveID}
\end{equation}
where $h^{(0)}_{\alpha\beta}$ and $k^{(0)}_{\alpha\beta}$ are two sufficiently smooth symmetric tensor fields defined on the initial slice $t=0$ satisfying the requirement that $g_{\alpha\beta}(0,x) = \gz_{\alpha\beta}(0,x) + h^{(0)}_{\alpha\beta}$ has Lorentz signature such that $g^{00}(0,x) < 0$ and the induced metric $g_{ij}(0,x)$, $i,j,=1,2,3$, on $t=0$ is positive definite \epubtkFootnote{As indicated above, given initial data $g^{(0)}_{\alpha\beta}$ and $k^{(0)}_{\alpha\beta}$ it is always possible to adjust the background metric $\gz_{\alpha\beta}$ such that the initial data for $h_{\alpha\beta}$ is trivial; just replace $\gz_{\alpha\beta}(t,x)$ by $\gz_{\alpha\beta}(t,x) + h^{(0)}_{\alpha\beta}(x) + t k^{(0)}_{\alpha\beta}(x)$.}. For detailed well posed Cauchy formulations we refer the reader to the original work in \cite{cB52}, see also \cite{cB62}, \cite{aFjM72}, and Ref.~\cite{tHtKjM77}, which presents an improvement of the results in the previous references due to weaker smoothness assumptions on the initial data.

An alternative way of establishing the hyperbolicity of the system~(\ref{Eq:EinsteinWave}) is to cast it into first order symmetric hyperbolic form \cite{aFjM72, kA02, lLmSlKrOoR06}. There are several ways of constructing such a system; the simplest one is obtained \cite{aFjM72} by introducing the first partial derivatives of $g_{\alpha\beta}$ as new variables,
\begin{displaymath}
k_{\alpha\beta} := \frac{\partial g_{\alpha\beta}}{\partial t},\qquad
D_{j\alpha\beta} := \frac{\partial g_{\alpha\beta}}{\partial x^j},\qquad j=1,2,3.
\end{displaymath}
Then, the second order wave system~(\ref{Eq:EinsteinWave}) can be rewritten in the form
\begin{eqnarray}
\frac{\partial g_{\alpha\beta}}{\partial t} &=& k_{\alpha\beta},
\label{Eq:EinsteinWaveFOSH1}\\
\frac{\partial D_{j\alpha\beta}}{\partial t} 
 &=& \frac{\partial k_{\alpha\beta}}{\partial x^j}
\label{Eq:EinsteinWaveFOSH2}\\
\frac{\partial k_{\alpha\beta}}{\partial t} 
 &=& -2\frac{g^{0j}}{g^{00}}\frac{\partial k_{\alpha\beta}}{\partial x^j}
  - \frac{g^{ij}}{g^{00}}\frac{\partial D_{i\alpha\beta}}{\partial x^j} + l.o.,
\label{Eq:EinsteinWaveFOSH3}
\end{eqnarray}
where $l.o.$ are lower order terms not depending on any derivatives of
the state vector $u =
(g_{\alpha\beta},k_{\alpha\beta},D_{j\alpha\beta})$. The system of
equations~(\ref{Eq:EinsteinWaveFOSH1}, \ref{Eq:EinsteinWaveFOSH2}, \ref{Eq:EinsteinWaveFOSH3}) constitutes a quasilinear first order symmetric hyperbolic system for $u$ with symmetrizer given by the quadratic form
\begin{equation}
u^* H(u) u = \sum\limits_{\alpha,\beta=0}^3 \left( 
  g_{\alpha\beta}^* g_{\alpha\beta}
 + |g^{00}| k_{\alpha\beta}^* k_{\alpha\beta}
 + g^{ij} D_{i\alpha\beta}^* D_{j\alpha\beta} \right).
\end{equation}
However, it should be noted that the symmetrizer is only positive definite if $g^{ij}$ is; that is, only if the time evolution vector field $\partial_t$ is time-like. In many situations, this requirement might be too restrictive. Inside a Schwarzschild black hole, for example, the asymptotically time-like Killing field $\partial_t$ is space-like.

However, as indicated above, the first order symmetric hyperbolic
reduction (\ref{Eq:EinsteinWaveFOSH1}, \ref{Eq:EinsteinWaveFOSH2},
\ref{Eq:EinsteinWaveFOSH3}) is not unique. A different reduction is
based on the variables $\tilde{u} =
(h_{\alpha\beta},\Pi_{\alpha\beta},\Phi_{j\alpha\beta})$, where
$\Pi_{\alpha\beta} := n^\mu\nablaz_\mu h_{\alpha\beta}$ is the
derivative of $g_{\alpha\beta}$ in the direction of the
future-directed unit normal $n^\mu$ to the time-slices $t=const$, and
$\Phi_{j\alpha\beta} := \nablaz_j h_{\alpha\beta}$. This yields a
first order system which is symmetric hyperbolic as long as the
$t=const$ slices are space-like, independent of whether or not
$\partial_t$ is time-like~\cite{kA02, lLmSlKrOoR06}.

\subsubsection{Constraint propagation and damping}

The hyperbolicity results described above guarantee that unique solutions of the nonlinear wave system~(\ref{Eq:EinsteinWave}) exist, at least for short times, and that they depend continuously on the initial data $h^{(0)}_{\alpha\beta}$, $k^{(0)}_{\alpha\beta}$. However, in order to obtain a solution of Einstein's field equations one has to ensure that the harmonic constraint~(\ref{Eq:HarmonicConstraint}) is identically satisfied.

The system~(\ref{Eq:EinsteinWave}) is equivalent to the modified Einstein equations
\begin{equation}
R^{\alpha\beta} + \nabla^{(\alpha} C^{\beta)}
 = 8\pi G_N\left( T^{\alpha\beta} - \frac{1}{2} g^{\alpha\beta} g_{\mu\nu} T^{\mu\nu} \right),
\label{Eq:ReducedEinsteinWave}
\end{equation}
where $R^{\alpha\beta}$ denotes the Ricci tensor, and where $C^\mu = 0$ if the harmonic constraint holds. From the twice contracted Bianchi identities $2\nabla_\beta R^{\alpha\beta} - \nabla^\alpha(g_{\mu\nu} R^{\mu\nu}) = 0$ one obtains the following equation for the constraint variable $C^\alpha$,
\begin{equation}
g^{\mu\nu}\nabla_\mu\nabla_\nu C^\alpha + R^{\alpha}{}_\beta C^\beta 
 = -16\pi G_N\nabla_\beta T^{\alpha\beta}.
\label{Eq:EinsteinWaveCP}
\end{equation}
This system describes the propagation of constraint violations which are present if $C^\mu$ is nonzero. For this reason, we call it the \textbf{constraint propagation system}, or \textbf{auxiliary system}. Provided the stress-energy tensor is divergence free, $\nabla_\beta T^{\alpha\beta} = 0$, this is a linear, second order hyperbolic equation for $C^\alpha$.\epubtkFootnote{Notice that the condition of $T^{\alpha\beta}$ being divergence-free depends on the metric $g_{\alpha\beta}$ itself, which is not known before actually solving the nonlinear wave equation~(\ref{Eq:EinsteinWave}), and the latter in turn depends on $T^{\alpha\beta}$. Therefore, one cannot specify $T^{\alpha\beta}$ by hand, except in the vacuum case, $T^{\alpha\beta} = 0$, or in the case $T^{\alpha\beta} = -\Lambda g^{\alpha\beta}$ with $\Lambda$ the cosmological constant. In the more general case, the stress-energy tensor has to be computed from a diffeomorphism-independent action for the matter fields and one has to consistently solve the coupled Einstein-matter system.} Therefore, it follows from the uniqueness properties of such hyperbolic problems that $C^\alpha = 0$ provided the initial data $h^{(0)}_{\alpha\beta}$, $k^{(0)}_{\alpha\beta}$ satisfies the initial constraints
\begin{equation}
C^\alpha(0,x) = 0, \qquad
\frac{\partial C^\alpha}{\partial t}(0,x) = 0.
\label{Eq:EinsteinWaveConstraints}
\end{equation}
This turns out to be equivalent to solving $C^\alpha(0,x) = 0$ plus
the usual Hamiltonian and momentum constraints, see Refs.~\cite{cB52, lLmSlKrOoR06}. Summarizing, specifying initial data $h^{(0)}_{\alpha\beta}$, $k^{(0)}_{\alpha\beta}$ satisfying the constraints~(\ref{Eq:EinsteinWaveConstraints}), the corresponding unique solution to the nonlinear wave system~(\ref{Eq:EinsteinWave}) yields a solution to the Einstein equations.

However, in numerical calculations, one cannot assume that the initial constraints~(\ref{Eq:EinsteinWaveConstraints}) are satisfied exactly, due to truncation and roundoff errors. The propagation of these errors is described by the constraint propagation system~(\ref{Eq:EinsteinWaveCP}), and hyperbolicity guarantees that for each fixed time $t > 0$ of existence, these errors converge to zero if the initial constraint violation converges to zero, which is usually the case when resolution is increased. On the other hand, due to limited computer resources, one cannot reach the limit of infinite resolution, and from a practical point of view one does not want the constraint errors to grow rapidly in time for fixed resolution. Therefore, one would like to design an evolution scheme in which the constraint violations are damped in time, such that the constraint hypersurface is an attractor set in phase space. A general method for damping constraints violations in the context of first order symmetric hyperbolic formulations of Einstein's field equations was given in Ref.~\cite{oBsFpHoR99}. This method was then adapted to the harmonic formulation in Ref.~\cite{cGjMgCiH05}. The procedure proposed in Ref.~\cite{cGjMgCiH05} consists in adding lower order friction terms in Equation~(\ref{Eq:ReducedEinsteinWave}) which damp constraint violations. Explicitly, the modified system reads
\begin{equation}
R^{\alpha\beta} + \nabla^{(\alpha} C^{\beta)}
 - \kappa\left( n^{(\alpha} C^{\beta)}
 - \frac{1}{2}(1 + \rho) g^{\alpha\beta} n_\mu C^\mu \right)
 = 8\pi G_N\left( T^{\alpha\beta} - \frac{1}{2} g^{\alpha\beta} g_{\mu\nu} T^{\mu\nu} \right),
\label{Eq:ReducedEinsteinWaveDamped}
\end{equation}
with $n^\mu$ the future directed unit normal to the $t=const$ surfaces, and $\kappa$ and $\rho$ real constants, where $\kappa > 0$ determines the timescale on which the constraint violations $C^\mu$ are damped. 

With this modification the constraint propagation system reads
\begin{equation}
g^{\mu\nu}\nabla_\mu\nabla_\nu C^\alpha + R^{\alpha}{}_\beta C^\beta 
 - \kappa\nabla_\beta\left( 2n^{(\alpha} C^{\beta)} 
 + \rho g^{\alpha\beta} n_\mu C^\mu\right)
 = -16\pi G_N\nabla_\beta T^{\alpha\beta}.
\label{Eq:EinsteinWaveCPDamped}
\end{equation}
A mode analysis for linear vacuum perturbations of the Minkowski metric reveals \cite{cGjMgCiH05} that for $\kappa > 0$ and $\rho > -1$ all modes except those which are  constant in space are damped. Numerical codes based on the modified system~(\ref{Eq:ReducedEinsteinWaveDamped}) or similar systems have been used in the context of binary black hole evolutions~\cite{fP05,fP06,lLmSlKrOoR06,mShPlLlKoRsT06,mBbSjW06b,bSdPlRjTjW07,cPmAlLsLdN09}, the head-on collision of boson stars~\cite{cPiOlLsL07} and the evolution of black strings in five-dimensional gravity~\cite{lLfP10}, among other references.

\subsubsection{Geometric issues}
\label{SubSubSec:ivpGeometric}

The results described so far guarantee the local in time unique
existence of solutions to Einstein's equations in harmonic
coordinates, given a sufficiently smooth initial data set
$(h^{(0)},k^{(0)})$. However, since general relativity is a
diffeomorphism invariant theory, some questions remain. The first
issue is whether or not the harmonic gauge is sufficiently general
such that any solution of the field equations can be obtained by this
method, at least for short enough time. The answer is
affirmative~\cite{cB52, aFjM72}. Namely, let $(M,g)$, $M = (-\varepsilon,\varepsilon) \times \Real^3$, be a smooth spacetime satisfying Einstein's field equations such that the initial surface $t=0$ is spacelike with respect to $g$. Then, we can find a diffeomorphism $\phi: M \to M$ in a neighborhood of the initial surface which leaves it invariant and casts the metric into the harmonic gauge. For this, one solves the harmonic wave map equation~(\ref{Eq:HarmonicWaveMap}) with initial data
\begin{displaymath}
\phi^0(0,x) = 0,\qquad
\frac{\partial \phi^0}{\partial t}(0,x) = 1,\qquad
\phi^i(0,x) = x^i,\qquad
\frac{\partial \phi^i}{\partial t}(0,x) = 0.
\end{displaymath}
Since equation~(\ref{Eq:HarmonicWaveMap}) is a second order hyperbolic one, a unique solution exists, at least on some sufficiently small time interval $(-\varepsilon',\varepsilon')$. Furthermore, choosing $\varepsilon' > 0$ small enough, $\phi : (-\varepsilon',\varepsilon')\times \Real^3 \to M$ describes a diffeomorphism when restricted to its image. By construction, $\bar{g} := (\phi^{-1})^* g$ satisfies the harmonic gauge condition~(\ref{Eq:HarmonicConstraint}). 

The next issue is the question about geometric uniqueness. Let $g^{(1)}$ and $g^{(2)}$ be two solutions of Einstein's equations with the same initial data on $t=0$, i.e., $g^{(1)}_{\alpha\beta}(0,x) = g^{(2)}_{\alpha\beta}(0,x)$, $\partial_t g^{(1)}_{\alpha\beta}(0,x) = \partial_t g^{(2)}_{\alpha\beta}(0,x)$. Are these solutions related, at least for small time, by a diffeomorphism? Again, the answer is affirmative \cite{cB52, aFjM72} because one can transform both solutions to harmonic coordinates using the above diffeomorphism $\phi$ without changing their initial data. It then follows by the uniqueness property of the nonlinear wave system~(\ref{Eq:EinsteinWave}) that the transformed solutions must be identical, at least on some sufficiently small time interval. Note that this geometric uniqueness property also implies that the solutions are, at least locally, independent of the background metric. For further results on geometric uniqueness involving only the first and second fundamental forms of the initial surface see Ref.~\cite{cBrG69}, where it is shown that every such initial data set satisfying the Hamiltonian and momentum constraints possesses a unique maximal Cauchy development.

Finally, we mention that results about the nonlinear stability of Minkowski spacetime with respect to vacuum and vacuum-scalar perturbations have been established based on the harmonic system~\cite{hLiR05, hLiR04}, offering an alternative proof to the one of Ref.~\cite{ChristodoulouKlainerman-Book}.

%===================================================================
\subsection{The ADM formulation}
\label{SubSec:ADM}
%===================================================================

In the usual 3+1 decomposition of Einstein's field equations (see, for example  \cite{Gourgoulhon}, for a through discussion of it) one evolves the three metric and the extrinsic curvature (the first and second fundamental forms) relative to a foliation $\Sigma_t$ of spacetime by spacelike hypersurfaces. The motivation for this formulation stems from the Hamiltonian description of general relativity (see, for instance, Appendix~E in~\cite{Wald-Book}) where the ``$q$'' variables are the three metric $\gamma_{ij}$ and the associated canonical momenta $\pi^{ij}$ (the ``$p$'' variables) are related to the extrinsic curvature $K_{ij}$ according to
\begin{displaymath}
\pi^{ij} = -\sqrt{\gamma}\left( K^{ij} - \gamma^{ij} K \right),
\end{displaymath}
where $\gamma = \det(\gamma_{ij})$ denotes the determinant of the three-metric and $K = \gamma^{ij} K_{ij}$ the trace of the extrinsic curvature.

In York's formulation \cite{jY79} of the 3+1 decomposed Einstein equations, the evolution equations are
\begin{eqnarray}
\partial_0\gamma_{ij} &=& -2K_{ij},
\label{Eq:EinsteinADM1}\\
\partial_0 K_{ij} &=& R^{(3)}_{ij} - \frac{1}{\alpha} D_i D_j\alpha 
+ K K_{ij} - 2K_i{}^l K_{lj} 
- 8\pi G_N\left[ \sigma_{ij} + \frac{1}{2}\gamma_{ij}(\rho - \sigma) \right].
\label{Eq:EinsteinADM2}
\end{eqnarray}
Here, the operator $\partial_0$ is defined as $\partial_0 := \alpha^{-1}(\partial_t - \pounds_\beta)$ with $\alpha$ and $\beta^i$ denoting lapse and shift, respectively. It is equal to the Lie derivative along the future-directed unit normal $n$ to the time slices when acting on covariant tensor fields orthogonal to $n$. Next, $R^{(3)}_{ij}$ and $D_j$ are the Ricci tensor and covariant derivative operator belonging to the three metric $\gamma_{ij}$, and $\rho := n^\alpha n^\beta T_{\alpha\beta}$ and $\sigma_{ij} := T_{ij}$ are the energy density and the stress tensor as measured by observers moving along the future-directed unit normal $n$ to the time slices. Finally, $\sigma := \gamma^{ij} T_{ij}$ denotes the trace of the stress tensor. The evolution system~(\ref{Eq:EinsteinADM1}, \ref{Eq:EinsteinADM2}) is subject to the Hamiltonian and momentum constraints,
\begin{eqnarray}
H &:=& \frac{1}{2}
 \left( \gamma^{ij} R^{(3)}_{ij} + K^2 - K^{ij} K_{ij} \right) = 8\pi G_N\rho,
\label{Eq:EinsteinADMHam}\\
M_i &:=& D^j K_{ij} - D_i K = 8\pi G_N j_i,
\label{Eq:EinsteinADMMom}
\end{eqnarray}
where $j_i: = -n^\beta T_{i\beta}$ is the flux density.

\subsubsection{Algebraic gauge conditions}

One issue with the evolution equations~(\ref{Eq:EinsteinADM1}, \ref{Eq:EinsteinADM2}) is the principle part of the Ricci tensor belonging to the three-metric,
\begin{equation}
R^{(3)}_{ij} = \frac{1}{2}\gamma^{kl}\left( -\partial_k\partial_l\gamma_{ij}
 - \partial_i\partial_j\gamma_{kl} 
 + \partial_i\partial_k\gamma_{lj} + \partial_j\partial_k\gamma_{li} \right) + l.o.,
\label{Eq:ThreeRicciTensor}
\end{equation}
which does not define a positive definite operator. This is due to the fact that the linearized Ricci tensor is invariant with respect to infinitesimal coordinate transformations $\gamma_{ij}\mapsto \gamma_{ij} + 2\partial_{(i} \xi_{j)}$ generated by a vector field $\xi = \xi^i\partial_i$. This has the following implications for the evolution equations~(\ref{Eq:EinsteinADM1},\ref{Eq:EinsteinADM2}), assuming for the moment that lapse and shift are fixed, a priori specified functions, in which case the system is equivalent to the second order system $\partial_0^2\gamma_{ij} = -2R^{(3)}_{ij} + l.o.$ for the three metric. Linearizing and localizing as described in Section~\ref{section:ivp} one obtains a linear, constant coefficient problem of the form~(\ref{Eq:LinearCCWave}) which can be brought into first order form via the reduction in Fourier space described in Section~\ref{SubSubSec:SecondOrderCC}. The resulting first order system has the form of Equation~(\ref{Eq:LinearCCWaveFourier}) with the symbol
\begin{displaymath}
Q(ik)= i|k|\sum\limits_{j=1}^n\betaz^j\hat{k}_j
+ |k|\left( \begin{array}{cc} 0 & I \\
 -\alphaz^2 R(\hat{k}) & 0 
\end{array} \right),
\end{displaymath}
where $R(k)$ is, up to a factor $2$, the principal symbol of the Ricci operator,
\begin{displaymath}
R(\hat{k})\gamma_{ij} = \gammaz^{lm}\left( \hat{k}_l\hat{k}_m\gamma_{ij} 
+ \hat{k}_i\hat{k}_j\gamma_{lm}
- \hat{k}_i\hat{k}_l\gamma_{mj} - \hat{k}_j\hat{k}_l\gamma_{mi} \right).
\end{displaymath}
Here, $\alphaz$, $\betaz^i$ and $\gammaz_{ij}$ refer to the frozen lapse, shift and three-metric, respectively. According to Theorem~\ref{Thm:SecondOrder},  the problem is well posed if and only there is a uniformly positive and bounded symmetrizer $h(\hat{k})$ such that $h(\hat{k}) R(\hat{k})$ is symmetric and uniformly positive for $\hat{k}\in S^2$. Although $R(\hat{k})$ is diagonalizable and its eigenvalues are not negative, some of them are zero since $R(\hat{k})\gamma_{ij} = 0$ for $\gamma_{ij}$ of the form $\gamma_{ij} = 2\hat{k}_{(i}\xi_{j)}$ with an arbitrary one-form $\xi_j$, so $h(k) R(k)$ cannot be positive.

These arguments were used in~\cite{gNoOoR04} to show that the evolution system~(\ref{Eq:EinsteinADM1},  \ref{Eq:EinsteinADM2}) with fixed lapse and shift is weakly but not strongly hyperbolic. The results in~\cite{gNoOoR04} also analyze modifications of the equations for which the lapse is densitized and the Hamiltonian constraint is used to modify the trace of Equation~(\ref{Eq:EinsteinADM2}). The conclusion is that such changes cannot make the evolution equations~(\ref{Eq:EinsteinADM1}, \ref{Eq:EinsteinADM2}) strongly hyperbolic. Therefore, these equations, with given shift and densitized lapse, are not suited for numerical evolutions. \epubtkFootnote{Weak hyperbolicity of the system~(\ref{Eq:EinsteinADM1}, \ref{Eq:EinsteinADM2}) with given shift and densitized lapse has also been pointed out in~\cite{lKmSsT01} based on a reduction which is first order in time and space. However, there are several inequivalent such reductions, and so it is not sufficient to show that a particular one is weakly hyperbolic in order to infer that the second order in space system~(\ref{Eq:EinsteinADM1}, \ref{Eq:EinsteinADM2}) is weakly hyperbolic.}

\subsubsection{Dynamical gauge conditions leading to a well-posed formulation}
 
The results obtained so far often lead to the popular statement ``The ADM equations are not strongly hyperbolic.''. However, consider the possibility of determining the lapse and shift through evolution equations. A natural choice, motivated by the discussion in Section~\ref{SubSec:Harmonic}, is to impose the harmonic gauge constraint~(\ref{Eq:HarmonicConstraint}). Assuming that the background metric $\gz_{\alpha\beta}$ is Minkowski in Cartesian coordinates for simplicity, this yields the following equations for the 3+1 decomposed variables,
\begin{eqnarray}
(\partial_t - \beta^j\partial_j)\alpha &=& -\alpha^2 f K + \alpha^3 H^t,
\label{Eq:ADMHarmonicGauge1}\\
(\partial_t - \beta^j\partial_j)\beta^i &=&
 -\alpha\gamma^{ij}\partial_j\alpha + \alpha^2\gamma^{ij}\gamma^{kl}\left(
 \partial_k\gamma_{jl} - \frac{1}{2}\partial_j\gamma_{kl} \right) 
 + \alpha^2(H^i + \beta^i H^t).
\label{Eq:ADMHarmonicGauge2}
\end{eqnarray}
with $f$ a constant which is equal to one for the harmonic time coordinate $t$. Let us analyze the hyperbolicity of the evolution system~(\ref{Eq:ADMHarmonicGauge1}, \ref{Eq:ADMHarmonicGauge2}, \ref{Eq:EinsteinADM1}, \ref{Eq:EinsteinADM2}) for the fields $u = (\alpha,\beta^i,\gamma_{ij},K_{ij})$, where for generality and later use we do not necessarily assume $f=1$ in Equation~(\ref{Eq:ADMHarmonicGauge1}). Since this is a mixed first/second order system, we base our analysis on the first order pseudodifferential reduction discussed in Section~\ref{SubSubSec:SecondOrderCC}. After linearizing and localizing, we obtain the constant coefficient linear problem
\begin{eqnarray}
(\partial_t - \betaz^k\partial_k)\alpha &=& -\alphaz^2 f K,
\label{Eq:ADMLinLapse}\\
(\partial_t - \betaz^k\partial_k)\beta^i &=& -\alphaz\gammaz^{ij}\partial_j\alpha
 + \alphaz^2\gammaz^{ij}\gammaz^{kl}\left(
 \partial_k\gamma_{jl} - \frac{1}{2}\partial_j\gamma_{kl} \right),
 \label{Eq:ADMLinShift}\\
(\partial_t - \betaz^k\partial_k)\gamma_{ij} 
 &=& 2\gammaz_{k(i}\partial_{j)}\beta^k - 2\alphaz K_{ij},
\label{Eq:ADMLinThreeMetric}\\
(\partial_t - \betaz^k\partial_k) K_{ij} 
 &=&  - \partial_i\partial_j\alpha
 + \frac{\alphaz}{2}\gammaz^{kl}\left( -\partial_k\partial_l\gamma_{ij}
 - \partial_i\partial_j\gamma_{kl} 
 + \partial_i\partial_k\gamma_{lj} + \partial_j\partial_k\gamma_{li} \right),
\label{Eq:ADMLinExtrinsicCurvature}
\end{eqnarray}
where $\alphaz$, $\betaz^k$ and $\gammaz_{ij}$ refer to the quantities corresponding to $\alpha$, $\beta^k$, $\gamma_{ij}$ of the background metric when frozen at a given point. In order to rewrite this in first order form, we perform a Fourier transformation in space and introduce the variables $\hat{U} = (a,b_i,l_{ij},p_{ij})$ with
\begin{displaymath}
a := |k|\hat{\alpha}/\alphaz,\qquad
b_i := |k|\gammaz_{ij}\hat{\beta}^j/\alphaz,\qquad
l_{ij} := |k|\hat{\gamma}_{ij},\qquad
p_{ij} := 2i\hat{K}_{ij},
\end{displaymath}
where $|k|:=\sqrt{\gammaz^{ij} k_i k_j}$ and the hatted quantities refer to their Fourier transform. With this, we obtain the first order system $\hat{U}_t = P(ik)\hat{U}$ where the symbol has the form $P(ik) = i\betaz^s k_s I + \alphaz Q(ik)$ with
\begin{displaymath}
Q(ik) \left( \begin{array}{c} a \\ b_i \\ l_{ij} \\ p_{ij} \end{array} \right)
 = i |k|\left( \begin{array}{c} \frac{f}{2} p \\
 -\hat{k}_i a + \hat{k}^j l_{ij} - \frac{1}{2}\hat{k}_i l \\
 2\hat{k}_{(i} b_{j)} + p_{ij} \\
 2\hat{k}_i\hat{k}_j a + l_{ij} + \hat{k}_i\hat{k}_j l - 2\hat{k}^s\hat{k}_{(i} l_{j)s}  
 \end{array} \right),
\end{displaymath}
where $\hat{k}_i := k_i/|k|$, $\hat{k}^i := \gammaz^{ij}\hat{k}_j$, $l:=\gammaz^{ij} l_{ij}$, and $p:=\gammaz^{ij} p_{ij}$. In order to determine the eigenfields $S(k)^{-1}\hat{U}$ such that $S(k)^{-1} P(ik) S(k)$ is diagonal, we decompose
\begin{eqnarray}
&& b_i = \bar{b}\hat{k}_i + \bar{b}_i,\nonumber\\
&& l_{ij} = \bar{l}\hat{k}_i\hat{k}_j + 2\hat{k}_{(i}\bar{l}_{j)} + \hat{l}_{ij} 
 + \frac{1}{2}(\gammaz_{ij} - \hat{k}_i\hat{k}_j)\bar{l}',\qquad
p_{ij} = \bar{p}\hat{k}_i\hat{k}_j + 2\hat{k}_{(i}\bar{p}_{j)} + \hat{p}_{ij} 
 + \frac{1}{2}(\gammaz_{ij} - \hat{k}_i\hat{k}_j)\bar{p}'
\nonumber
\end{eqnarray}
into pieces parallel and orthogonal to $\hat{k}_i$, similar as in Example~\ref{Example:FatMaxwell}. Then, the problem decouples into a tensor sector, involving $(\hat{l}_{ij},\hat{p}_{ij})$, into a vector sector, involving $(\bar{b}_i,\bar{l}_i,\bar{p}_i)$ and a scalar sector involving $(a,\bar{b},\bar{l},\bar{p},\bar{l}',\bar{p}')$. In the tensor sector, we have
\begin{displaymath}
Q^{(tensor)}(ik)\left( \begin{array}{c} \hat{l}_{ij} \\ \hat{p}_{ij} \end{array} \right)
 = i|k|\left( \begin{array}{c} \hat{p}_{ij} \\ \hat{l}_{ij} \end{array} \right),
\end{displaymath}
which has the eigenvalues $\pm i|k|$ with corresponding eigenfields $\hat{l}_{ij} \pm \hat{p}_{ij}$. In the vector sector, we have
\begin{equation}
Q^{(vector)}(ik)\left( \begin{array}{c} \bar{b}_j \\ \bar{l}_j \\ \bar{p}_j \end{array} \right)
 = i|k|\left( \begin{array}{c} \bar{l}_j \\ \bar{b}_j + \bar{p}_j \\ 0 \end{array} \right),
\label{Eq:ADMVectorBlock}
\end{equation}
which is also diagonalizable  with eigenvalues $0$, $\pm i|k|$ and corresponding eigenfields $\bar{p}_j$ and $\bar{l}_j \pm (\bar{b}_j + \bar{p}_j)$. Finally, in the scalar sector we have
\begin{displaymath}
Q^{(scalar)}(ik)\left( \begin{array}{c} 
a \\ \bar{b} \\ \bar{l} \\ \bar{p} \\ \bar{l}' \\ \bar{p}' 
\end{array} \right) = i|k|\left( \begin{array}{c}
 \frac{f}{2}(\bar{p} + \bar{p}') \\ -a + \frac{1}{2}(\bar{l} - \bar{l}') \\
 2\bar{b} + \bar{p} \\ 2a + \bar{l}' \\ \bar{p}' \\ \bar{l}' \end{array} \right).
\end{displaymath}
It turns out $Q^{(scalar)}(ik)$ is diagonalizable with purely imaginary values if and only if $f > 0$ and $f\neq 1$. In this case, the eigenvalues and corresponding eigenfields are $\pm i|k|$, $\pm i|k|$, $\pm i\sqrt{f}|k|$ and $\bar{l}' \pm \bar{p}'$, $\bar{l} \pm (2\bar{b} + \bar{p})$, $a + f\bar{l}'/(f-1) \pm \sqrt{f}[\bar{p} + (f+1)/(f-1)\bar{p}']/2$, respectively. A symmetrizer for $P(ik)$ which is smooth in $k\in S^2$, $\alphaz$, $\betaz^k$ and $\gammaz_{ij}$ can be constructed from the eigenfields as in Example~\ref{Example:FatMaxwell}.

\textbf{Remarks:}
\begin{itemize}
\item If instead of imposing the dynamical shift condition~(\ref{Eq:ADMHarmonicGauge2}), $\beta$ is a priori specified, then the resulting evolution system, consisting of Equations~(\ref{Eq:ADMHarmonicGauge1}, \ref{Eq:EinsteinADM1}, \ref{Eq:EinsteinADM2}), is weakly hyperbolic for any choice of $f$. Indeed, in that case the symbol~(\ref{Eq:ADMVectorBlock}) in the vector sector reduces to the Jordan block
\begin{displaymath}
Q^{(vector)}(ik)\left( \begin{array}{c} \bar{l}_j \\ \bar{p}_j \end{array} \right)
 = i|k|\left( \begin{array}{cc} 0 & 1 \\ 0 & 0 \end{array} \right)
 \left( \begin{array}{c} \bar{l}_j \\ \bar{p}_j \end{array} \right),
\end{displaymath}
which cannot be diagonalized.
\item When linearized about Minkowski spacetime, it is possible to classify the characteristic fields into physical, constraint-violating and gauge fields, see Ref.~\cite{gCjPoRoSmT03}. For the system~(\ref{Eq:ADMLinLapse}--\ref{Eq:ADMLinExtrinsicCurvature}) the physical fields are the ones in the tensor sector, $\hat{l}_{ij} \pm \hat{p}_{ij}$, the constraint-violating ones are $\bar{p}_j$ and $\bar{l}'\pm\bar{p}'$, and the gauge fields are the remaining characteristic variables. Observe that the constraint-violating fields are governed by a strongly hyperbolic system (see also Section~\ref{SubSubSec:ADMCP} below), and that in this particular formulation of the ADM equations the gauge fields are coupled to the constraint-violating ones. This coupling is one of the properties which make it possible to cast the system as a strongly hyperbolic one. 
\end{itemize}

We conclude that the evolution system~(\ref{Eq:ADMHarmonicGauge1}, \ref{Eq:ADMHarmonicGauge2}, \ref{Eq:EinsteinADM1}, \ref{Eq:EinsteinADM2}) is strongly hyperbolic if and only if $f > 0$ and $f\neq 1$. Although the full harmonic gauge condition~(\ref{Eq:HarmonicConstraint}) is excluded from these restrictions,\epubtkFootnote{Notice that even when $f=1$ the evolution system~(\ref{Eq:ADMHarmonicGauge1}, \ref{Eq:ADMHarmonicGauge2}, \ref{Eq:EinsteinADM1}, \ref{Eq:EinsteinADM2}) is not equivalent to the harmonic system discussed in the previous subsection. In the former case, the harmonic constraint is exactly enforced in order to obtain evolution equations for the lapse and shift; while in the latter case first derivatives of the harmonic constraint are combined with the Hamiltonian and momentum constraints, see Equation~(\ref{Eq:ReducedEinsteinWave}).} there is still a large family of evolution equations for the lapse and shift that give rise to a strongly hyperbolic problem together with the standard evolution equations~(\ref{Eq:EinsteinADM1}, \ref{Eq:EinsteinADM2}) from the 3+1 decomposition.

\subsubsection{Elliptic gauge conditions leading to a well-posed formulation}
\label{SubSubSec:ADMEllipticGauge}
  
Rather than fixing the lapse and shift algebraically or dynamically, an alternative which has been considered in the literature is to fix them according to elliptic equations. A natural restriction on the extrinsic geometry of the time slices $\Sigma_t$ is to require that their mean curvature, $c = -K/3$, vanishes or is constant \cite{lSjY78}. Taking the trace of Equation~(\ref{Eq:EinsteinADM2}) and using the Hamiltonian constraint to eliminate the trace of $R^{(3)}_{ij}$ yields the following equation for the lapse,
\begin{equation}
\left[ -D^j D_j + K^{ij} K_{ij} + 4\pi G_N(\rho + \sigma) \right]\alpha = \partial_t K, 
\label{Eq:CMCLapse}
\end{equation}
which is a second order linear elliptic equation. The operator inside the square parenthesis is formally positive if the strong energy condition, $\rho + \sigma\geq 0$, holds, and so it is invertible when defined on appropriate function spaces. See also Ref.~\cite{dGcG99} for generalizations of this condition. Concerning the shift, one choice which is motivated by eliminating the ``bad'' terms in the expression for the Ricci tensor, Equation~(\ref{Eq:ThreeRicciTensor}), is the spatial harmonic gauge \cite{lAvM03}. In terms of a fixed (possibly time-dependent) background metric $\gammaz_{ij}$ on $\Sigma_t$, this gauge is defined as (cf. Equation~(\ref{Eq:HarmonicConstraint}))
\begin{equation}
0 = V^k := \gamma^{ij}\left( \Gamma^k{}_{ij} - \Gammaz^k{}_{ij} \right) 
 = \gamma^{ij}\gamma^{kl}\left( \Dz_k\gamma_{lj} - \frac{1}{2}\Dz_j\gamma_{kl} \right),
\label{Eq:SpatialHarmonicGauge}
\end{equation}
where $\Dz$ is the Levi-Civita connection with respect to $\gammaz$ and $\Gammaz^k{}_{ij}$ denote the corresponding Christoffel symbols. The main importance of this gauge is that it permits to rewrite the Ricci tensor belonging to the three metric in the form
\begin{displaymath}
R^{(3)}_{ij} = -\frac{1}{2}\gamma^{kl}\Dz_k\Dz_l\gamma_{ij} + D_{(i} V_{j)} + l.o.,
\end{displaymath}
where $\Dz_k$ denotes the covariant derivative with respect to the background metric $\gammaz$ and where the lower order terms ``l.o.'' depend only on $\gamma_{ij}$ and its first derivatives $\Dz_k\gamma_{ij}$. When $V^k = 0$ The operator on the right-hand side is second order quasilinear elliptic, and with this, the evolution system~(\ref{Eq:EinsteinADM1}, \ref{Eq:EinsteinADM2}) has the form of a nonlinear wave equation for the three-metric $\gamma_{ij}$. However, the coefficients and source terms in this equation still depend on the lapse and shift. For constant mean curvature slices the lapse satisfies the elliptic scalar equation~(\ref{Eq:CMCLapse}), and with the spatial harmonic gauge the shift is determined by the requirement that Equation~(\ref{Eq:SpatialHarmonicGauge}) is preserved throughout evolution, which yields an elliptic vector equation for it. In Ref.~\cite{lAvM03} it was shown that the coupled hyperbolic-elliptic system consisting of the evolution equations~(\ref{Eq:EinsteinADM1}, \ref{Eq:EinsteinADM2}) with the Ricci tensor $R^{(3)}_{ij}$ rewritten in elliptic form using the condition $V^k = 0$, the constant mean curvature condition~(\ref{Eq:CMCLapse}), and this elliptic equation for $\beta^i$, gives rise to a well posed Cauchy problem in vacuum. Besides eliminating the ``bad'' terms in the Ricci tensor, the spatial harmonic gauge has also other nice properties which were exploited in the well-posed formulation of Ref.~\cite{lAvM03}. For example, the covariant Laplacian of a function $f$ is
\begin{displaymath}
D^k D_k f = \gamma^{ij}\Dz_i\Dz_j f - V^k\Dz_k f,
\end{displaymath}
which does not contain any derivatives of the three metric $\gamma^{ij}$ if $V^k = 0$. For applications of the hyperbolic-elliptic formulation in \cite{lAvM03} to the global existence of expanding vacuum cosmologies, see Refs.~\cite{lAvM04,lAvM09}.

Other methods for specifying the shift have been proposed in \cite{lSjY78}, with the idea of minimizing a functional of the type
\begin{equation}
I[\beta] = \int\limits_{\Sigma_t} \Theta^{ij}\Theta_{ij} \sqrt{\gamma} d^3 x,
\label{Eq:StrainFunctional}
\end{equation}
where $\Theta_{ij} := \partial_t\gamma_{ij}/2 = -\alpha K_{ij} + D_{(i}\beta_{(j)}$ is the strain tensor. Therefore, the functional $I[\beta]$ minimizes time changes in the three metric in an averaged sense. In particular, $I[\beta]$ attains its absolute minimum (zero) if $\partial_t$ is a Killing vector field. Therefore, one expects the resulting gauge condition to minimize the time dependence of the coordinate components of the three metric. An alternative is to replace the strain by its trace-free part on the right-hand side of Equation~(\ref{Eq:StrainFunctional}), giving rise to the minimal distortion gauge. Both conditions yield a second order elliptic equation for the shift vector which has unique solutions provided suitable boundary conditions are specified. For generalizations and further results on these type of gauge conditions, see Refs.~\cite{pBjCkT98,dGcG99,dGcGjInM00}. However, it seems to be currently unknown whether or not these elliptic shift conditions, together with the evolution system~(\ref{Eq:EinsteinADM1}, \ref{Eq:EinsteinADM2}) and an appropriate condition on the lapse, lead to a well posed Cauchy problem.

\subsubsection{Constraint propagation}
\label{SubSubSec:ADMCP}

The evolution equations~(\ref{Eq:EinsteinADM1}, \ref{Eq:EinsteinADM2}) are equivalent to the components of the Einstein equations corresponding the spatial part of the Ricci tensor,
\begin{equation}
R_{ij} = 8\pi G_N\left( T_{ij} - \frac{1}{2}\gamma_{ij} g^{\mu\nu} T_{\mu\nu} \right),
\label{Eq:ReducedEinsteinADM}
\end{equation}
and in order to obtain a solution of the full Einstein equations one also needs to solve the constraints $H = 8\pi G_N\rho$ and $M_i = 8\pi G_N j_i$. As in the previous subsection, the constraint propagation system can be obtained from the twice contracted Bianchi identities which, in the 3+1 decomposition, read
\begin{eqnarray}
&& \partial_0 H + \frac{1}{\alpha^2} D^j\left( \alpha^2 M_j \right) - 2K H 
 - \left( K^{ij} - K\gamma^{ij} \right) R_{ij} = 0,
\label{Eq:TwiceContractedBianchi1}\\
&& \partial_0 M_i + \frac{1}{\alpha^2} D_i\left( \alpha^2 H \right) - K M_i 
 + \frac{1}{\alpha}D^j\left( \alpha R_{ij} - \alpha \gamma_{ij}\gamma^{kl} R_{kl} \right) = 0.
\label{Eq:TwiceContractedBianchi2}
\end{eqnarray}
The condition of the stress-energy tensor being divergence-free leads to similar evolution equations for $\rho$ and $j_i$. Therefore, the equations~(\ref{Eq:ReducedEinsteinADM}) lead to the following symmetric hyperbolic system \cite{sF97,jY98} for the constraint variables ${\cal H} := H - 8\pi G_N\rho$ and ${\cal M}_i := M_i - 8\pi G_N j_i$,
\begin{eqnarray}
\partial_0 {\cal H} &=& -\frac{1}{\alpha^2} D^j\left( \alpha^2 {\cal M}_j \right) + 2K {\cal H},
\label{Eq:ADMConsProp1}\\
\partial_0 {\cal M}_i &=& -\frac{1}{\alpha^2} D_i\left( \alpha^2 {\cal H} \right) + K {\cal M}_i.
\label{Eq:ADMConsProp2}
\end{eqnarray}
As has also been observed in \cite{sF97}, the constraint propagation system associated to the standard ADM equations, where Equation~(\ref{Eq:ReducedEinsteinADM}) is replaced by its trace-reversed version $R_{ij} - \gamma_{ij} g^{\mu\nu} R_{\mu\nu}/2 = 8\pi G_N T_{ij}$ is
\begin{eqnarray}
\partial_0 {\cal H} &=& -\frac{1}{\alpha^2} D^j\left( \alpha^2 {\cal M}_j \right) + K {\cal H},
\nonumber\\
\partial_0 {\cal M}_i &=& -\frac{D_i\alpha}{\alpha} {\cal H} + K {\cal M}_i,
\nonumber
\end{eqnarray}
which is only weakly hyperbolic. Therefore, it is much more difficult to control the constraint fields in the standard ADM case than in York's formulation of the 3+1 equations.

%===================================================================
\subsection{The BSSN formulation}
\label{SubSec:BSSN}
%===================================================================

The BSSN formulation is based on the 3+1 decomposition of Einstein's field equations. Unlike the harmonic formulation, which has been motivated by the mathematical structure of the equations and the understanding of the Cauchy formulation in general relativity, this system has been mainly developed and improved based on its capability of numerically evolving spacetimes containing compact objects in a stable way. Interestingly, it turns out that in spite of the fact that the BSSN formulation is based on an entirely different motivation, mathematical questions like the well posedness of its Cauchy problem can be answered, at least for most gauge conditions.

In the BSSN formulation, the three metric $\gamma_{ij}$ and the extrinsic curvature $K_{ij}$ are decomposed according to
\begin{eqnarray}
\gamma_{ij} &=& e^{4\phi}\tilde{\gamma}_{ij}\; ,\\
K_{ij} &=& e^{4\phi}\left( \tilde{A}_{ij} 
 + \frac{1}{3}\tilde{\gamma}_{ij} K \right) .
\end{eqnarray}
Here, $K = \gamma^{ij} K_{ij}$ and $\tilde{A}_{ij}$ are the trace and the trace-less part, respectively, of the conformally rescaled extrinsic curvature. The conformal factor $e^{2\phi}$ is determined by the requirement for the conformal metric $\tilde{\gamma}_{ij}$ to have unit determinant. Aside from these variables one also evolves the lapse ($\alpha$), the shift ($\beta^i$) and its time derivative ($B^i$), and the variable
\begin{displaymath}
\tilde{\Gamma}^i := -\partial_j\tilde{\gamma}^{ij}.
\end{displaymath}
In terms of the operator $\hat{\partial}_0 = \partial_t - \beta^j\partial_j$ the BSSN evolution equations are
\begin{eqnarray}
\hat{\partial}_0 \alpha &=& -\alpha^2 f(\alpha,\phi,x^\mu) (K - K_0(x^\mu)),
\label{Eq:BSSN1}\\
%%%%%%%%%%%%%%%%%%%
\hat{\partial}_0 K &=& -e^{-4\phi}\left[ 
 \tilde{D}^i\tilde{D}_i \alpha + 2\partial_i\phi \cdot\tilde{D}^i\alpha \right]
 + \alpha\left( \tilde{A}^{ij}\tilde{A}_{ij} + \frac{1}{3} K^2 \right)
 - \alpha S,
\label{Eq:BSSN2}\\
%%%%%%%%%%%%%%%%%%%
\hat{\partial}_0 \beta^i &=& \alpha^2 G(\alpha,\phi,x^\mu) B^i,
\label{Eq:BSSN3}\\
%%%%%%%%%%%%%%%%%%%
\hat{\partial}_0 B^i &=& e^{-4\phi} H(\alpha,\phi,x^\mu)
  \hat{\partial}_0\tilde{\Gamma}^i - \eta^i(B^i,\alpha,x^\mu)
\label{Eq:BSSN4}\\
%%%%%%%%%%%%%%%%%%%
\hat{\partial}_0 \phi &=& -\frac{\alpha}{6}\, K + \frac{1}{6}\partial_k\beta^k,
\label{Eq:BSSN5}\\
%%%%%%%%%%%%%%%%%%%
\hat{\partial}_0 \tilde{\gamma}_{ij} &=& -2\alpha\tilde{A}_{ij} 
 + 2\tilde{\gamma}_{k(i}\partial_{j)}\beta^k 
 - \frac{2}{3}\tilde{\gamma}_{ij}\partial_k\beta^k ,
\label{Eq:BSSN6}\\
%%%%%%%%%%%%%%%%%%%
\hat{\partial}_0 \tilde{A}_{ij} &=& e^{-4\phi}\left[ 
 \alpha\tilde{R}_{ij} + \alpha R^\phi_{ij} - \tilde{D}_i\tilde{D}_j\alpha 
  + 4\partial_{(i}\phi\cdot\tilde{D}_{j)}\alpha\right]^{TF}
\nonumber\\
 && {} + \alpha K\tilde{A}_{ij} - 2\alpha\tilde{A}_{ik}\tilde{A}^k_{\; j}
  + 2\tilde{A}_{k(i}\partial_{j)}\beta^k 
  - \frac{2}{3}\tilde{A}_{ij}\partial_k\beta^k
  - \alpha e^{-4\phi} \hat{S}_{ij} ,
\label{Eq:BSSN7}\\
%%%%%%%%%%%%%%%%%%%
\hat{\partial}_0\tilde{\Gamma}^i &=& 
 \tilde{\gamma}^{kl}\partial_k\partial_l\beta^i
 + \frac{1}{3} \tilde{\gamma}^{ij}\partial_j\partial_k\beta^k 
 + \partial_k\tilde{\gamma}^{kj} \cdot \partial_j\beta^i
 - \frac{2}{3}\partial_k\tilde{\gamma}^{ki} \cdot \partial_j\beta^j\nonumber\\
 && {} - 2\tilde{A}^{ij}\partial_j\alpha 
 + 2\alpha\left[ (m-1)\partial_k\tilde{A}^{ki} - \frac{2m}{3}\tilde{D}^i K
    + m(\tilde{\Gamma}^i_{\; kl}\tilde{A}^{kl} + 6\tilde{A}^{ij}\partial_j\phi)
\right] - S^i,
\label{Eq:BSSN8}
\end{eqnarray}
Here, quantities with a tilde refer to the conformal three metric $\tilde{\gamma}_{ij}$, which is also used in order to raise and lower indices. In particular, $\tilde{D}_i$ and $\tilde{\Gamma}^k{}_{ij}$ denote the covariant derivative and the Christoffel symbols, respectively, with respect to $\tilde{\gamma}_{ij}$. Expressions with a superscript $TF$ refer to their trace-less part with respect to the conformal metric. Next, the sum $\tilde{R}_{ij} + R^\phi_{ij}$ represents the Ricci tensor associated to the physical three metric $\gamma_{ij}$, where
\begin{eqnarray}
\tilde{R}_{ij} 
 &=& -\frac{1}{2} \tilde{\gamma}^{kl}\partial_k\partial_l\tilde{\gamma}_{ij} 
  + \tilde{\gamma}_{k(i}\partial_{j)}\tilde{\Gamma}^k
  - \tilde{\Gamma}_{(ij)k}\partial_l\tilde{\gamma}^{lk} 
  + \tilde{\gamma}^{ls}\left( 2\tilde{\Gamma}^k{}_{l(i}\tilde{\Gamma}_{j)ks} 
  + \tilde{\Gamma}^k{}_{is}\tilde{\Gamma}_{klj} \right),
\label{Eq:BSSNRtildeij}\\
R^\phi_{ij} &=& -2\tilde{D}_i\tilde{D}_j\phi 
  - 2\tilde{\gamma}_{ij} \tilde{D}^k\tilde{D}_k\phi
  + 4\tilde{D}_i\phi\, \tilde{D}_j\phi 
  - 4\tilde{\gamma}_{ij}\tilde{D}^k\phi\, \tilde{D}_k\phi.
\label{Eq:BSSNRphiij}
\end{eqnarray}
The term $\hat{\partial}_0\tilde{\Gamma}^i$ in Equation~(\ref{Eq:BSSN4}) is set equal to the right-hand side of Equation~(\ref{Eq:BSSN8}). The parameter $m$ in the latter equation modifies the evolution flow off the constraint surface by adding the momentum constraint to the evolution equation for the variable $\tilde{\Gamma}^i$. This parameter was first introduced in Ref.~\cite{mAgAbBeSwS00} in order to compare the stability properties of the BSSN evolution equations with those of the ADM formulation.

The gauge conditions which are imposed on the lapse and shift in Equations~(\ref{Eq:BSSN1}, \ref{Eq:BSSN3}, \ref{Eq:BSSN4}) were introduced in \cite{hBoS04} and generalize the Bona--Mass\'o condition~\cite{cBjMeSjS95} and the hyperbolic Gamma driver condition \cite{mAbBpDmKdPeSrT03}. It is assumed that the functions $f(\alpha,\phi,x^\mu)$, $G(\alpha,\phi,x^\mu)$ and $H(\alpha,\phi,x^\mu)$ are strictly positive and smooth in their arguments, and that $K_0(x^\mu)$ and $\eta^i(B^j,\alpha,x^\mu)$ are smooth functions of their arguments. The choice
\begin{eqnarray}
&& m = 1,\qquad 
f(\alpha,\phi,x^\mu) = \frac{2}{\alpha}\; ,\qquad
K_0(x^\mu) = 0,
\label{Eq:StandardBSSNGaugeChoice1}\\
&& G(\alpha,\phi,x^\mu) = \frac{3}{4\alpha^2}\; ,\qquad
H(\alpha,\phi,x^\mu) = e^{4\phi}\; ,\qquad
\eta^i(B^j,\alpha,x^\mu) = \eta B^i,
\label{Eq:StandardBSSNGaugeChoice2}
\end{eqnarray}
with $\eta$ a positive constant, corresponds to the evolution system used in many black hole simulations based on $1+\log$ slicing and the moving puncture technique (see, for instance, Ref. \cite{vMjBmKdC06} and references therein). Finally, the source terms $S$, $\hat{S}_{ij}$ and $S^i$ are defined in the following way: denoting by $R^{(3)}_{ij}$ and $R^{(4)}_{ij}$ the Ricci tensors belonging to the three-metric $\gamma_{ij}$ and the spacetime metric, respectively, and introducing the constraint variables
\begin{eqnarray}
H &:=& \frac{1}{2}\left( \gamma^{ij}\ R^{(3)}_{ij} 
  + \frac{2}{3} K^2 - \tilde{A}^{ij}\tilde{A}_{ij} \right),
\label{Eq:BSSNCons1}\\
M_i &:=& \tilde{D}^j \tilde{A}_{ij} 
  - \frac{2}{3} \tilde{D}_i K + 6\tilde{A}_{ij} \tilde{D}^j\phi,
\label{Eq:BSSNCons2}\\
C^i &:=& \tilde{\Gamma}^i + \partial_j\tilde{\gamma}^{ij},
\label{Eq:BSSNCons3}
\end{eqnarray}
the source terms are defined as
\begin{displaymath}
S := \gamma^{ij} R^{(4)}_{ij} - 2H,\qquad
\hat{S}_{ij} := \left[ R^{(4)}_{ij} + \tilde{\gamma}_{k(i}\partial_{j)} C^k \right]^{TF},\quad
S^i := 2\alpha\, m\,\tilde{\gamma}^{ij} M_j - \hat{\partial}_0 C^i.
\end{displaymath}
For vacuum evolutions one sets $S=0$, $\hat{S}_{ij} = 0$ and $S^i = 0$. When matter fields are present, the Einstein field equations are equivalent to the evolution equations
(\ref{Eq:BSSN1}--\ref{Eq:BSSN8}) setting $S = -4\pi G_N(\rho + \sigma)$, $\hat{S}_{ij} = 8\pi G_N\sigma_{ij}^{TF}$, $S^i = 16\pi G_N m\alpha\tilde{\gamma}^{ik} j_k$ and the constraints $H = 8\pi G_N\rho$, $M_i = 8\pi G_N j_i$ and $C^i = 0$.

When comparing Cauchy evolutions in different spatial coordinates, it is very convenient to reformulate the BSSN system such that it is covariant with respect to spatial coordinate transformations. This is indeed possible, see Refs.~\cite{dB09a,dBetal12}. One way of achieving this is to fix a smooth background three-metric $\gammaz_{ij}$, similarly as in Section~\ref{SubSec:Harmonic}, and to replace the fields $\phi$ and $\tilde{\Gamma}^i$ by the scalar and vector fields
\begin{displaymath}
\phi := \frac{1}{12}\log\left(\frac{\gamma}{\gammaz}\right),\qquad
\tilde{\Gamma}^i := -\Dz_j\tilde{\gamma}^{ij},
\end{displaymath}
where $\gamma$ and $\gammaz$ denote the determinants of $\gamma_{ij}$ and $\gammaz_{ij}$, and $\Dz_j$ is the covariant derivative associated to the latter. If $\gammaz_{ij}$ is flat\epubtkFootnote{If $\gammaz_{ij}$ is time-independent but not flat, additional curvature terms appear in the equations, see Appendix A in Ref.~\cite{dBetal12}.} and time-independent, the corresponding BSSN equations are obtained by replacing $\partial_k\mapsto \Dz_k$ and $\tilde{\Gamma}^k{}_{ij}\mapsto \tilde{\Gamma}^k{}_{ij} - \Gammaz^k{}_{ij}$ in Equations~(\ref{Eq:BSSN1}--\ref{Eq:BSSN8},\ref{Eq:BSSNRtildeij},\ref{Eq:BSSNRphiij},\ref{Eq:BSSNCons1}--\ref{Eq:BSSNCons3}).

\subsubsection{The hyperbolicity of the BSSN evolution equations}
\label{SubSubSec:BSSNHyp}

In fact, the ADM formulation in the spatial harmonic gauge described in Section~\ref{SubSubSec:ADMEllipticGauge} and the BSSN formulation are based on some common ideas. In the covariant reformulation of BSSN just mentioned, the variable $\tilde{\Gamma}^i$ is just the quantity $V^i$ defined in Equation~(\ref{Eq:SpatialHarmonicGauge}), where $\gamma_{ij}$ is replaced by the conformal metric $\gammaz_{ij}$. Instead of requiring $\tilde{\Gamma}^i$ to vanish, which would convert the operator on the right-hand side of Equation~(\ref{Eq:BSSNRtildeij}) into a quasilinear elliptic operator, one promotes this quantity to an independent field satisfying the evolution equation~(\ref{Eq:BSSN8}) (see also the discussion below Equation~(2.18) in Ref.~\cite{mStN95}). In this way, the $\tilde{\gamma}_{ij}-\tilde{A}_{ij}$-block of the evolution equations forms a wave system. However, this system is coupled through its principal terms to the evolution equations of the remaining variables, and so one needs to analyze the complete system. As follows from the discussion below, it is crucial to add the momentum constraint to Equation~(\ref{Eq:BSSN8}) with an appropriate factor $m$ in order to obtain a hyperbolic system.

The hyperbolicity of the BSSN evolution equations was first analyzed  in a systematic way in Ref.~\cite{oSgCjPmT02}, where it was established that for fixed shift and densitized lapse,
\begin{equation}
\alpha = e^{12\sigma\phi}
\label{Eq:DensitizedLapse}
\end{equation}
the evolution system~(\ref{Eq:BSSN2}, \ref{Eq:BSSN5}--\ref{Eq:BSSN8}) is strongly hyperbolic for $\sigma > 0$ and $m > 1/4$ and symmetric hyperbolic for $m > 1$ and $6\sigma = 4m - 1$. This was shown by introducing new variables and enlarging the system to a strongly or symmetric hyperbolic first order one. In fact, similar first order reductions were obtained in~\cite{sFoR99,hFaR00}. However, in Ref.~\cite{oSgCjPmT02} it was shown that the first order enlargements are equivalent to the original system if the extra constraints associated to the definition of the new variables are satisfied, and that these extra constraints propagate \emph{independently} of the BSSN constraints $H=0$, $M_i=0$ and $C^i = 0$. This establishes the well posedness of the Cauchy problem for the system~(\ref{Eq:DensitizedLapse}, \ref{Eq:BSSN2}, \ref{Eq:BSSN5}--\ref{Eq:BSSN8}) under the aforementioned conditions on $\sigma$ and $m$. Based on the same method, a symmetric hyperbolic first order enlargement of the evolution equations~(\ref{Eq:BSSN1}, \ref{Eq:BSSN2}, \ref{Eq:BSSN5}--\ref{Eq:BSSN8}) and fixed shift was obtained in Ref.~\cite{hBoS04} under the conditions $f > 0$ and $4m = 3f + 1$ and used to construct boundary conditions for BSSN. First order strongly hyperbolic reductions for the full system~(\ref{Eq:BSSN1}--\ref{Eq:BSSN8}) have also been recently analyzed in Ref.~\cite{dBetal12}.

An alternative and efficient method for analyzing the system consists in reducing it to a first order pseudodifferential system, as described in Section~\ref{SubSubSec:SecondOrderCC}. This method has been applied in~\cite{gNoOoR04} to derive a strongly hyperbolic system very similar to BSSN with fixed, densitized lapse and fixed shift. This system is then shown to yield a well posed Cauchy problem. In~\cite{hBoS04} the same method was applied to the evolution system~(\ref{Eq:BSSN1}--\ref{Eq:BSSN8}). Linearizing and localizing, one obtains a first order system of the form $\hat{U}_t = P(ik)\hat{U} = i\betaz^s k_s\hat{U} + \alphaz Q(ik)\hat{U}$. The eigenvalues of $Q(ik)$ are $0$, $\pm i$, $\pm i\sqrt{m}$, $\pm i\sqrt{\mu}$, $\pm\sqrt{f}$, $\pm\sqrt{GH}$, $\pm\sqrt{\kappa}$, where we have defined $\mu := (4m-1)/3$ and $\kappa := 4GH/3$. The system is weakly hyperbolic provided that
\begin{displaymath}
f > 0,\qquad \mu > 0,\qquad \kappa > 0,
\end{displaymath}
and it is strongly hyperbolic if, in addition, the parameter $m$ and the functions $f$, $G$, and $H$ can be chosen such that the functions
\begin{displaymath}
\frac{\kappa}{f - \kappa}\; , \qquad
\frac{m-1}{\mu - \kappa}\; , \qquad
\frac{6(m-1)\kappa}{4m-3\kappa}
\end{displaymath}
are bounded and smooth. In particular, this requires that the nominators converge to zero at least as fast as the denominators when $f\to \kappa$, $\mu\to \kappa$ or $3\kappa\to 4m$, respectively. Since $\kappa > 0$, the boundedness of $\kappa/(f-\kappa)$ requires that $f\neq \kappa$. For the standard choice $m=1$, the conditions on the gauge parameters leading to strong hyperbolicity are, therefore, $f > 0$, $\kappa > 0$ and $f\neq\kappa$. Unfortunately, for the choice~(\ref{Eq:StandardBSSNGaugeChoice1}, \ref{Eq:StandardBSSNGaugeChoice2}) used in binary black hole simulations these conditions reduce to
\begin{equation}
e^{4\phi} \neq 2\alpha.
\label{Eq:StandardBSSNConstraint}
\end{equation}
which is typically violated at some two-surface, since asymptotically, $\alpha\to 1$ and $\phi\to 0$ while near black holes $\alpha$ is small and $\phi$ positive. It is currently not known whether or not the Cauchy problem is well posed if the system is strongly hyperbolic everywhere except at points belonging to a set of zero measure, such as a two-surface. Although numerical simulations based on finite difference discretizations with the standard choice~(\ref{Eq:StandardBSSNGaugeChoice1}, \ref{Eq:StandardBSSNGaugeChoice2}) show no apparent sign of instabilities near such surfaces, the well posedness for the Cauchy problem for the BSSN system~(\ref{Eq:BSSN1}--\ref{Eq:BSSN8}) with the choice~(\ref{Eq:StandardBSSNGaugeChoice1}, \ref{Eq:StandardBSSNGaugeChoice2}) for the gauge source functions remains an open problem when the condition~(\ref{Eq:StandardBSSNConstraint}) is violated. However, a well posed problem could be formulated by modifying the choice for the functions $G$ and $H$ such that $f\neq\kappa$ and $f,\kappa > 0$ are guaranteed to hold everywhere.

Yet a different approach to analyzing the hyperbolicity of BSSN has been given in~\cite{cGjM04a,cGjM04b} based on a new definition of strongly and symmetric hyperbolicity for evolution systems which are first order in time and second order in space. Based on this definition, it has been verified that the BSSN system~(\ref{Eq:DensitizedLapse}, \ref{Eq:BSSN2}, \ref{Eq:BSSN5}--\ref{Eq:BSSN8}) is strongly hyperbolic for $\sigma > 0$ and $m > 1/4$ and symmetric hyperbolic for $6\sigma = 4m-1 > 0$. (Note that this generalizes the original result in~\cite{oSgCjPmT02} where, in addition, $m > 1$ was required.) The results in~\cite{cGjM04b} also discuss more general 3+1 formulations including the one in~\cite{gNoOoR04} and construct constraint-preserving boundary conditions. The relation between the different approaches to analyzing hyperbolicity of evolution systems which are first order in time and second order in space has been analyzed in~\cite{cGjM06a}.

Strong hyperbolicity for different versions of the gauge evolution equations~(\ref{Eq:BSSN1}, \ref{Eq:BSSN3}, \ref{Eq:BSSN4}), where the normal operator $\hat{\partial}_0$ is sometimes replaced by $\partial_t$, has been analyzed in~\cite{cGjM06b}. See Table~I in that reference for a comparison between the different versions and the conditions they are subject to in order to satisfy strong hyperbolicity. It should be noted that when $m=1$ and $\hat{\partial}_0$ is replaced by $\partial_t$, additional conditions restricting the magnitude of the shift appear in addition to $f > 0$ and $f\neq\kappa$.

\subsubsection{Constraint propagation}

As mentioned above, the BSSN evolution equations~(\ref{Eq:BSSN1}--\ref{Eq:BSSN8}) are only equivalent to Einstein's field equation if the constraints
\begin{displaymath}
{\cal H} := H - 8\pi G_N\rho = 0,\qquad
{\cal M}_i := M_i - 8\pi G_N j_i = 0,\qquad
C^i = 0
\end{displaymath}
are satisfied. Using the twice contracted Bianchi identities in their 3+1 decomposed form, Equations~(\ref{Eq:TwiceContractedBianchi1}, \ref{Eq:TwiceContractedBianchi2}), and assuming that the stress-energy tensor is divergence free, it is not difficult to show that the equations~(\ref{Eq:BSSN1}--\ref{Eq:BSSN8}) imply the following 
evolution system for the constraint fields \cite{hBoS04,cGjM04b}:
\begin{eqnarray}
\hat{\partial}_0 {\cal H} &=& -\frac{1}{\alpha}\, D^j(\alpha^2 {\cal M}_j) 
- \alpha e^{-4\phi}\tilde{A}^{ij}\tilde{\gamma}_{ki}\partial_j C^k + \frac{2\alpha}{3}\, K {\cal H},
\label{Eq:BSSNConsProp1}\\
\hat{\partial}_0 {\cal M}_j
  &=& \frac{\alpha^3}{3} D_j( \alpha^{-2} {\cal H} ) 
+ \alpha K {\cal M}_j + {\cal M}_i \partial_j\beta^i 
   + D^i\left( \alpha\left[ \tilde{\gamma}_{k(i}\partial_{j)}C^k \right]^{TF} \right),
\label{Eq:BSSNConsProp2}\\
\hat{\partial}_0 C^i &=& 2\alpha\, m\, \tilde{\gamma}^{ij} {\cal M}_j.
\label{Eq:BSSNConsProp3}
\end{eqnarray}
This is the constraint propagation system for BSSN, which describes the propagation of constraint violations which are usually present in numerical simulations due to truncation and roundoff errors. There are at least three reasons for establishing the well posedness of its Cauchy problem. The first reason is to show that the unique solution of the system~(\ref{Eq:BSSNConsProp1}--\ref{Eq:BSSNConsProp3}) with zero initial data is the trivial solution. This implies that it is sufficient to solve the constraints at the initial time $t=0$. Then, any smooth enough solution of the BSSN evolution equations with such data satisfies the constraint propagation system with ${\cal H} = 0$, ${\cal M}_j = 0$ and $C^i = 0$, and it follows from the uniqueness property of this system that the constraints must hold everywhere and at each time. In this way, one obtains a solution to Einstein's field equations. However, in numerical calculations, the initial constraints are not exactly satisfied due to numerical errors. This brings us to the second reason for having a well posed problem at the level of the constraint propagation system; namely, the continuous dependence on the initial data. Indeed, the initial constraint violations give rise to constraint violating solutions; but if these violations are governed by a well posed evolution system the norm of the constraint violations is controlled by those of the initial violations for each fixed time $t > 0$. In particular, the constraint violations must converge to zero if the initial constraint violations do. Since the initial constraint errors go to zero when resolution is increased (provided a stable numerical scheme is used to solve the constraints), this guarantees convergence to a constraint-satisfying solution.\epubtkFootnote{However, one should mention that this convergence is usually not sufficient for obtaining accurate solutions. If the constraint manifold is unstable, small departure from it may grow exponentially in time and even though the constraint errors converge to zero they remain large for finite resolutions, see Ref.~\cite{lKmSsT01} for an example of this effect.} Finally, the third reason for establishing well posedness for the constraint propagation system is the construction of constraint-preserving boundary conditions which will be explained in detail in Section~\ref{section:ibvpEinstein}.

The hyperbolicity of the constraint propagation system~(\ref{Eq:BSSNConsProp1}--\ref{Eq:BSSNConsProp3}) has been analyzed in Refs.~\cite{cGjM04b, hBoS04, dBoSeSmTpDiHdP07, dBpDoSeSmT09}, and \cite{dNoS10} and shown to be reducible to a symmetric hyperbolic first order system for $m > 1/4$. Furthermore, there are no superluminal characteristic fields if $1/4 < m \leq 1$. Because of finite speed of propagation, this means that BSSN with $1/4 < m \leq 1$ (which includes the standard choice $m=1$) does not possess superluminal constraint-violating modes. This is an important property, for it shows that constraint violations that originate inside black hole regions (which usually dominate the constraint errors due to high gradients at the punctures or stuffing of the black hole singularities in the turducken approach~\cite{zEjFyLsStB07, dBoSeSmTpDiHdP07, dBpDoSeSmT09}) cannot propagate to the exterior region.

In~\cite{oR04} a general result is derived, showing that under a mild assumption on the form of the constraints, strong hyperbolicity of the main evolution system implies strong hyperbolicity of the constraint propagation system, with the characteristic speeds of the latter being a subset of those of the former. The result does not hold in general if "strong" is replaced by "symmetric", since there are known examples for which the main evolution system is symmetric hyperbolic while the constraint propagation system is only strongly hyperbolic~\cite{gCoS03}.

 %===================================================================
\subsection{Other hyperbolic formulations}
\label{SubSec:HypForm}
%===================================================================

There exist many other hyperbolic reductions of Einstein's field equations. In particular, there has been a large amount of work on casting the evolution equations into first order symmetric~\cite{aAaAyCjY95, hF96, sFoR96, aAaAyCjY97, Anderson:1997jn, fErRhW97, mIeLoR97, gYhS99, aAjY99, oBsFpHoR99, sH00, lKmSsT01, mSlKlLhPsT02, oSmT02, kA02, lLmS03, lLmS03, lBjB03} and strongly hyperbolic~\cite{cBjMeSjS95, cBjMeSjS97, Alcubierre:1999wj, Bona:2002fq, Bona:2003fj, Alcubierre:2003ti, Bona:2004yp, Rinne:2005sk, cGjM06b, Brown:2008cca, Bona:2010is,dBetal12} form, see~\cite{hF96, oR98, hFaR00, oR04} for reviews. For systems involving wave equations for the extrinsic curvature, see~\cite{yCtR83, aAaAyCjY95}, see also Ref.~\cite{mPdE96} and Refs.~\cite{aAaAcL98, oBmHoS00, oSmHoB01, oSeW01, eWoS02} for applications to perturbation theory and the linear stability of solitons and hairy black holes.

Recently, there has also been work deriving strongly or symmetric hyperbolic formulations from an action principle~\cite{dB11, Bona:2010is, Hilditch:2010wp}.

\newpage
%===================================================================
%===================================================================
\section{Boundary Conditions: The Initial-Boundary Value Problem}
\label{section:ibvp}
%===================================================================
%===================================================================

In Section~\ref{section:ivp} we discussed the general Cauchy problem for quasilinear hyperbolic evolution equations on the unbounded domain $\Real^n$. However, in the numerical modeling of such problems one is faced with the finiteness of computer resources. A common approach for dealing with this problem is to truncate the domain via an artificial boundary, thus forming a finite computational domain with outer boundary. Absorbing boundary conditions must then be specified at the boundary such that the resulting initial-boundary value problem (IBVP) is well posed and such that the amount of spurious reflection is minimized.

Therefore, we examine in this section quasilinear hyperbolic evolution equations on a finite, open domain $\Sigma\subset \Real^n$ with $C^\infty$-smooth boundary $\partial\Sigma$. Let $T > 0$. We are considering an IBVP of the following form,
\begin{eqnarray}
& u_t = \sum\limits_{j=1}^n A^j(t,x,u)\frac{\partial}{\partial x^j} u + F(t,x,u),
& x\in\Sigma,\quad t\in [0,T],
\label{Eq:QLIBVP1}\\
& u(0,x) = f(x),& x\in\Sigma,
\label{Eq:QLIBVP2}\\
& b(t,x,u) u = g(t,x), & x\in\partial\Sigma,\quad t\in [0,T],
\label{Eq:QLIBVP3}
\end{eqnarray}
where $u(t,x)\in\Complex^m$ is the state vector, $A^1(t,x,u), \ldots ,A^n(t,x,u)$ are complex $m\times m$ matrices, $F(t,x,u)\in\Complex^m$, and $b(t,x,u)$ is a complex $r\times m$ matrix. As before, we assume for simplicity that all coefficients belong to the class $C^\infty_b([0,T]\times\Sigma\times\Complex^m)$ of bounded, smooth functions with bounded derivatives. The data consists of the initial data $f\in C_b^\infty(\Sigma,\Complex^m)$ and the boundary data $g\in C_b^\infty([0,T]\times\partial\Sigma,\Complex^r)$. 

Compared to the initial-value problem discussed in Section~\ref{section:ivp} the following new issues and difficulties appear when boundaries are present:
\begin{itemize}
\item For a smooth solution to exist, the data $f$ and $g$ must satisfy appropriate \textbf{compatibility conditions} at the intersection $S := \{ 0 \} \times \partial\Sigma$ between the initial and boundary surface~\cite{jRfM74}. Assuming that $u$ is continuous, for instance, Equations~(\ref{Eq:QLIBVP2}, \ref{Eq:QLIBVP3}) imply that $g(0,x) = b(0,x,f(x)) f(x)$ for all $x\in\partial\Sigma$. If $u$ is continuously differentiable, then taking a time derivative of Equation~(\ref{Eq:QLIBVP3}) and using Equations~(\ref{Eq:QLIBVP1}, \ref{Eq:QLIBVP2}) leads to
\begin{displaymath}
g_t(0,x) = c(x)\left[ \sum\limits_{j=1}^n A^j(0,x,f(x))\frac{\partial f}{\partial x^j}(x) 
 + F(0,x,f(x)) \right] + b_t(0,x,f(x))f(x),\qquad
 x\in\partial\Sigma
\end{displaymath}
where $c(x)$ is the complex $r\times m$ matrix with coefficients
\begin{displaymath}
c(x)^A{}_B = b(0,x,f(x))^A{}_B 
 + \sum\limits_{C=1}^m\frac{\partial b^A{}_C}{\partial u^B}(0,x,f(x)) f(x)^C,\qquad
 A = 1,\ldots,r,\quad B = 1,\ldots,m.
\end{displaymath}
Assuming higher regularity of $u$, one obtains additional compatibility conditions by taking further time derivatives of Equation~(\ref{Eq:QLIBVP3}). In particular, for an infinitely differentiable solution $u$, one has an infinite family of such compatibility conditions at $S$, and one must make sure that the data $f$, $g$ satisfies each of them if the solution $u$ is to be reproduced by the IBVP. If an exact solution $u^{(0)}$ of the partial differential equation~(\ref{Eq:QLIBVP1}) is known, a convenient way of satisfying these conditions is to choose the data such that in a neighborhood of $S$, $f$ and $g$ agree with the corresponding values for $u^{(0)}$, i.e such that $f(x) = u^{(0)}(0,x)$ and $g(t,x) = b(t,x,u^{(0)}(t,x))u^{(0)}(t,x)$ for $(t,x)$ in a neighborhood of $S$. Depending on the problem at hand, this might be too restrictive, however.
\item The next issue is the question of what class of boundary conditions~(\ref{Eq:QLIBVP3}) leads to a well-posed problem. In particular, one would like to know which are the restrictions on the matrix $b(t,x,u)$ implying existence of a unique solution, provided the compatibility conditions hold. In order to illustrate this issue on a very simple example, consider the advection equation $u_t = u_x$ on the interval $[-1,1]$. The most general solution has the form $u(t,x) = h(t + x)$ for some differentiable function $h: (-1,\infty)\to\Complex$. The function $h$ is determined on the interval $[-1,1]$ by the initial data alone, and so the initial data alone fixes the solution on the strip $-1-t \leq x \leq 1-t$. Therefore, one is not allowed to specify any boundary conditions at $x=-1$, whereas data must be specified for $u$ at $x=1$ in order to uniquely determine the function $h$ on the interval $(1,\infty)$. 
\item Additional difficulties appear when the system has constraints, like in the case of electromagnetism and general relativity. In the previous section we saw in the case of Einstein's equations that it is usually sufficient to solve these constraints on an initial Cauchy surface, since the Bianchi identities and the evolution equations imply that the constraints propagate. However, in the presence of boundaries one can only guarantee that the constraints remain satisfied inside the future domain of dependence of the initial surface $\Sigma_0 := \{ 0 \} \times \Sigma$ unless the boundary conditions are chosen with care. Methods for constructing \textbf{constraint-preserving boundary conditions} which make sure that the constraints propagate correctly on the whole spacetime domain $[0,T]\times\Sigma$ will be discussed in the next section.
\end{itemize}

There are two common techniques for analyzing an IBVP. The first, discussed in Section~\ref{section:Laplace}, is based on the linearization and localization principles, and reduces the problem to linear, constant coefficient IBVPs which can be explicitly solved using Fourier transformations, similarly to the case without boundaries. This approach, called the Laplace method, is very useful for finding necessary conditions for the well posedness of linear, constant coefficient IBVPs. Likely, these conditions are also necessary for the quasilinear IBVP since small amplitude high frequency perturbations are essentially governed by the corresponding linearized, frozen coefficient problem. Based on the Kreiss symmetrizer construction \cite{hK70} and the theory of pseudo-differential operators, the Laplace method also gives sufficient conditions for the linear, variable coefficient problem to be well posed; however, the general theory is rather technical. For a discussion and interpretation of this approach in terms of wave propagation we refer to~\cite{rHReview86}.

The second method, which is discussed in Section~\ref{section:MaxDiss}, is based on energy inequalities obtained from integration by parts and does not require the use of pseudo-differential operators. It provides a class of boundary conditions, called maximal dissipative, which leads to a well posed IBVP. Essentially, these boundary conditions specify data to the incoming normal characteristic fields, or to an appropriate linear combination of the in- and outgoing normal characteristic fields. Although technically less involved than the Laplace one, this method requires the evolution equations~(\ref{Eq:QLIBVP1}) to be symmetric hyperbolic in order to be applicable, and it gives sufficient but not necessary conditions for well posedness.

In Section~\ref{section:Absorbing} we also discuss absorbing boundary conditions, which are designed to minimize spurious reflections from the boundary surface.

%===================================================================
\subsection{The Laplace method}
\label{section:Laplace}
%===================================================================

Upon linearization and localization, the IBVP~(\ref{Eq:QLIBVP1}, \ref{Eq:QLIBVP2}, \ref{Eq:QLIBVP3}) reduces to a linear, constant coefficient problem of the following form,
\begin{eqnarray}
& u_t = \sum\limits_{j=1}^n A^j\frac{\partial}{\partial x^j} u + F_0(t,x),
& x\in\Sigma,\quad t\geq 0,
\label{Eq:QLIBVP1Lin}\\
& u(0,x) = f(x),& x\in\Sigma,
\label{Eq:QLIBVP2Lin}\\
& b u = g(t,x),& x\in\partial\Sigma,\quad t\geq 0,
\label{Eq:QLIBVP3Lin}
\end{eqnarray}
where $A^j = A^j(t_0,x_0,u^{(0)}(t_0,x_0))$, $b = b(t_0,x_0,u^{(0)}(t_0,x_0))$ denote the matrix coefficients corresponding to $A^j(t,x,u)$ and $b(t,x,u)$ linearized about a solution $u^{(0)}$ and frozen at the point $p_0 = (t_0,x_0)$, and where for generality we include the forcing term $F_0(t,x)$ with components in the class $C_b^\infty([0,\infty)\times\Sigma)$. Since the freezing process involves a zoom into a very small neighborhood of $p_0$, we may replace $\Sigma$ by $\Real^n$ for all points $p_0$ lying \emph{inside} the domain $\Sigma$. We are then back into the case of the previous section, and we conclude that a necessary condition for the IBVP~(\ref{Eq:QLIBVP1}, \ref{Eq:QLIBVP2}, \ref{Eq:QLIBVP3}) to be well posed at $u^{(0)}$, is that all linearized, frozen coefficient Cauchy problems corresponding to $p_0\in\Sigma$ are well posed. In particular, the equation~(\ref{Eq:QLIBVP1Lin}) must be strongly hyperbolic.

Now let us consider a point $p_0\in\partial\Sigma$ at the boundary. Since $\partial\Sigma$ is assumed to be smooth, it will be mapped to a plane during the freezing process. Therefore, taking points $p_0\in\partial\Sigma$, it is sufficient to consider the linear, constant coefficient IBVP~(\ref{Eq:QLIBVP1Lin}, \ref{Eq:QLIBVP2Lin}, \ref{Eq:QLIBVP3Lin}) on the half space
\begin{displaymath}
\Sigma := \{ (x_1,x_2,\ldots,x_n)\in \Real^n : x_1 > 0 \},
\end{displaymath}
say. This is the subject of this subsection. Because we are dealing with a constant coefficient problem on the half-space, we can reduce the problem to an ordinary differential boundary problem on the interval $[0,\infty)$ by employing Fourier transformation in the directions $t$ and $y := (x_2,\ldots,x_n)$ tangential to the boundary. More precisely, we first exponentially damp the function $u(t,x)$ in time by defining for $\eta > 0$ the function
\begin{eqnarray}
u_\eta(t,x) :=
\left\{ \begin{array}{ll} 
 e^{-\eta t}\,u(t,x)  & \hbox{for $t\geq 0$, $x\in\Sigma$,} \\ 
 0                         & \hbox{for $t < 0$, $x\in\Sigma$.}
\end{array}\, \right.
\label{Eq:ueta}
\end{eqnarray}
We denote by $\tilde{u}_\eta(\xi,x_1,k)$ the Fourier transformation of $u_\eta(t,x_1,y)$ with respect to the directions $t$, and $y$ tangential to the boundary and define the Laplace--Fourier transformation of $u$ by
\begin{displaymath}
\tilde{u}(s,x_1,k) := \hat{u}_\eta(\xi,x_1,k) 
 = \frac{1}{(2\pi)^{n/2}}\int e^{-s t - i k\cdot y} u(t,x_1,y) dt d^{n-1} y, \qquad
s := \eta + i\xi,
\end{displaymath}
Then, $\tilde{u}$ satisfies the following boundary value problem,
\begin{eqnarray}
& A\frac{\partial}{\partial x_1}\tilde{u} = B(s,k)\tilde{u} + \tilde{F}(s,x_1,k),
& x_1 > 0,
\label{Eq:IBVPLaplace}\\
& b\tilde{u} = \tilde{g}(s,k)
& x_1 = 0,
\label{Eq:IBVPLaplaceBC}
\end{eqnarray}
where for notational simplicity we set $A := A^1$ and $B^j := A^j$, $j=2,\ldots,n$, and where $B(s,k) := s I - i B^2 k_2 - \ldots - i B^n k_n$. Here, $\tilde{F}(s,x_1,k) = \tilde{F}_0(s,x_1,k) + \hat{f}(x_1,k)$ with $\tilde{F}_0$ and $\hat{f}$ denoting the Laplace-Fourier and Fourier transform, respectively, of $F_0$ and $f$, and $\tilde{g}(s,k)$ is the Laplace-Fourier transform of the boundary data $g$.

In the following, we assume for simplicity that the \textbf{boundary matrix} $A$ is invertible, and that the equation~(\ref{Eq:QLIBVP1Lin}) is strongly hyperbolic. An interesting example with a singular boundary matrix is mentioned in Example~\ref{Example:FatMaxwellBC} below. If $A$ can be inverted, then we rewrite Equation~(\ref{Eq:IBVPLaplace}) as the linear ordinary differential equation
\begin{equation}
\frac{\partial}{\partial x_1}\tilde{u} = M(s,k)\tilde{u} + A^{-1}\tilde{F}(s,x_1,k),
\qquad x_1 > 0,
\label{Eq:IBVPLaplaceODE}
\end{equation}
where $M(s,k) := A^{-1} B(s,k)$. We solve this equation subject to the boundary conditions~(\ref{Eq:IBVPLaplaceBC}) and the requirement that $\tilde{u}$ vanishes as $x_1\to\infty$. For this it is useful to have information about the eigenvalues of $M(s,k)$.

\begin{lemma}[\cite{hK70,KL89,GKO95}]
Suppose the equation~(\ref{Eq:QLIBVP1Lin}) is strongly hyperbolic and the boundary matrix $A$ has $q$ negative and $m-q$ positive eigenvalues. Then, $M(s,k)$ has precisely $q$ eigenvalues with negative real part and $m-q$ eigenvalues with positive real part. (The eigenvalues are counted according to their algebraic multiplicity.) Furthermore, there is a constant $\delta > 0$ such that the eigenvalues $\kappa$ of $M(s,k)$ satisfy the estimate
\begin{equation}
|\re(\kappa)| \geq \delta \re(s),
\label{Eq:kappaEstimate}
\end{equation}
for all $\re(s) > 0$ and $k\in\Real^{n-1}$.
\end{lemma}

\proof Let $\re(s) > 0$, $\beta\in\Real$ and $k\in\Real^{n-1}$. Then
\begin{displaymath}
M(s,k) - i\beta I = A^{-1}\left[ s I - i\beta A - i k_j B^j \right] 
= A^{-1}\left[ sI - P_0(i\beta,ik) \right].
\end{displaymath}
Since the equation~(\ref{Eq:QLIBVP1Lin}) is strongly hyperbolic there is a constant $K$ and matrices $S(\beta,k)$ such that (see the comments below Definition~\ref{Def:Hyperbolicity})
\begin{displaymath}
|S(\beta,k)| + |S(\beta,k)^{-1}| \leq K,\qquad
S(\beta,k)^{-1} P_0(i\beta,ik) S(\beta,k) = i\Lambda(\beta,k),
\end{displaymath}
for all $(\beta,k)\in\Real^n$, where $\Lambda(\beta,k)$ is a real, diagonal matrix. Hence,
\begin{displaymath}
M(s,k) - i\beta I = A^{-1} S(\beta,k)\left[ s I - i\Lambda(\beta,k) \right] S(\beta,k)^{-1},
\end{displaymath}
and since $s I - i\Lambda(\beta,k)$ is diagonal and its diagonal entries have real part greater than or equal to $\re(s)$, it follows that
\begin{displaymath}
| [ M(s,k) - i\beta I ]^{-1} | \leq 
 |A| |S(\beta,k)| |S(\beta,k)^{-1}| | [s I - i\Lambda(\beta,k)]^{-1}|
\leq \frac{1}{\delta\re(s)},
\end{displaymath}
with $\delta := (K^2 |A|)^{-1}$. Therefore, the eigenvalues $\kappa$ of $M(s,k)$ must satisfy
\begin{displaymath}
|\kappa - i\beta| \geq \delta \re(s)
\end{displaymath}
for all $\beta\in\Real$. Choosing $\beta := \im(\kappa)$ proves the inequality~(\ref{Eq:kappaEstimate}). Furthermore, since the eigenvalues $\kappa = \kappa(s,k)$ can be chosen to be continuous functions of $(s,k)$ \cite{Kato-Book}, and since for $k=0$, $M(s,0) = s A^{-1}$, the number of eigenvalues $\kappa$ with positive real part is equal to the number of positive eigenvalues of $A$.
\qed

According to this lemma, the Jordan normal form of the matrix $M(s,k)$ has the following form:
\begin{equation}
M(s,k) = T(s,k)\left[ D(s,k) + N(s,k) \right] T(s,k)^{-1},
\end{equation}
with $T(s,k)$ a regular matrix, $N(s,k)$ is nilpotent ($N(s,k)^m = 0$) and
\begin{displaymath}
D(s,k) = \diag(\kappa_1,\ldots,\kappa_q,\kappa_{q+1},\ldots,\kappa_m)
\end{displaymath}
is the diagonal matrix with the eigenvalues of $M(s,k)$, where $\kappa_1,\ldots,\kappa_q$ have negative real part. Furthermore, $N(s,k)$ commutes with $D(s,k)$. Transforming to the variable $\tilde{v}(s,x,k) := T(s,k)^{-1}\tilde{u}(s,x,k)$ the boundary value problem~(\ref{Eq:IBVPLaplace}, \ref{Eq:IBVPLaplaceBC}) simplifies to
\begin{eqnarray}
& \frac{\partial}{\partial x_1}\tilde{v} = \left[ D(s,k) + N(s,k) \right]\tilde{v} 
 + T(s,k)^{-1}A^{-1}\tilde{F}(s,x_1,k),
& x_1 > 0,
\label{Eq:IBVPLaplaceTrans}\\
& b T(s,k)\tilde{v} = \tilde{g}(s,k)
& x_1 = 0.
\label{Eq:IBVPLaplaceBCTrans}
\end{eqnarray}

\subsubsection{Necessary conditions for well posedness and the Lopatinsky condition}
\label{SubSubSec:LaplaceNecessary}

Having cast the IBVP into the ordinary differential system~(\ref{Eq:IBVPLaplaceTrans}, \ref{Eq:IBVPLaplaceBCTrans}), we are ready to obtain a simple necessary condition for well posedness. For this, we consider the problem for $\tilde{F} = 0$ and split $\tilde{v} = (\tilde{v}_-,\tilde{v}_+)$ where $\tilde{v}_- := (\tilde{v}_1,\ldots,\tilde{v}_q)$ and $\tilde{v}_+ := (\tilde{v}_{q+1},\ldots,\tilde{v}_m)$ are the variables corresponding to the eigenvalues of $M(s,k)$ with negative and positive real parts, respectively. Accordingly, we split
\begin{displaymath}
D(s,k) = \left( \begin{array}{cc} D_-(s,k) & 0 \\ 0 & D_+(s,k) \end{array} \right),\qquad
N(s,k) = \left( \begin{array}{cc} N_-(s,k) & 0 \\ 0 & N_+(s,k) \end{array} \right),
\end{displaymath}
and $bT(s,k) = (b_-(s,k),b_+(s,k))$. When $\tilde{F} = 0$ the most general solution of Equation~(\ref{Eq:IBVPLaplaceTrans}) is
\begin{eqnarray}
\tilde{v}_-(s,x_1,k) &=& e^{D_-(s,k) x_1} e^{N_-(s,k) x_1}\sigma_-(s,k),
\nonumber\\
\tilde{v}_+(s,x_1,k) &=& e^{D_+(s,k) x_1} e^{N_+(s,k) x_1}\sigma_+(s,k),
\nonumber
\end{eqnarray}
with constant vectors $\sigma_-(s,k)\in\Complex^q$ and $\sigma_+(s,k)\in\Complex^{m-q}$. The expression for $\tilde{v}_+$ describes modes that grow exponentially in $x_1$ and do not satisfy the required boundary condition at $x_1\to\infty$ unless $\sigma_+(s,k) = 0$; hence we set $\sigma_+(s,k) = 0$. In view of the boundary conditions~(\ref{Eq:IBVPLaplaceBCTrans}) we then obtain the algebraic equation
\begin{equation}
b_-(s,k)\sigma_-(s,k) = \tilde{g}.
\label{Eq:BoundaryEquation}
\end{equation}
Therefore, a necessary condition for existence and uniqueness is that the $r\times q$ matrix $b_-(s,k)$ be a square matrix, i.e.\ $r=q$, and
that
\begin{equation}
\det( b_-(s,k) ) \neq 0
\label{Eq:LopatinskyCondition}
\end{equation}
for all $\re(s) > 0$ and $k\in\Real^{n-1}$. Let us make the following observations:
\begin{itemize}
\item The condition~(\ref{Eq:LopatinskyCondition}) implies that we must specify exactly as many linearly independent boundary conditions as there are incoming characteristic fields, since $q$ is the number of negative eigenvalues of the boundary matrix $A = A^1$.
\item The violation of condition~(\ref{Eq:LopatinskyCondition}) at some $(s_0,k_0)$ with $\re(s_0) > 0$ and $k\in\Real^{n-1}$ gives rise to the simple wave solutions
\begin{equation}
u(t,x_1,y) = e^{s_0 t + i k_0\cdot y}\tilde{u}(s_0,x_1,k_0),\qquad
t\geq 0,\quad (x_1,y)\in\Sigma,
\label{Eq:LaplacePlaneWave}
\end{equation}
where $\tilde{u}(s_0,\cdot,k_0) = T(s,k)\tilde{v}(s_0,\cdot,k_0)\in L^2(0,\infty)$ is a nontrivial solution of the problem~(\ref{Eq:IBVPLaplaceTrans}, \ref{Eq:IBVPLaplaceBCTrans}) with homogeneous data $\tilde{F}=0$ and $\tilde{g}=0$. Therefore, an equivalent necessary condition for well posedness is that no such simple wave solutions exist. This is known as the \textbf{Lopatinsky condition}.
\item If such a simple wave solution exists for some $(s_0,k_0)$, then the homogeneity of the problem implies the existence of a whole family,
\begin{equation}
u_\alpha(t,x_1,y)
 = e^{\alpha (s_0t + i k_0\cdot y)}\tilde{u}(\alpha s_0,\alpha x_1,\alpha k_0),
\qquad t\geq 0,\quad (x_1,y)\in\Sigma,
\label{Eq:LaplacePlaneWaveFamily}
\end{equation}
of such solutions parametrized by $\alpha > 0$. In particular, it follows that
\begin{displaymath}
|u_\alpha(t,x_1,y)| = e^{\alpha\re(s) t} |\tilde{u}(\alpha s_0,\alpha x_1,\alpha k_0)| 
 = e^{\alpha\re(s) t} |u_\alpha(0,x_1,y)|,
\end{displaymath}
such that
\begin{displaymath}
\frac{|u_\alpha(t,x_1,y)|}{|u_\alpha(0,x_1,y)|} = e^{\alpha\re(s) t} \to \infty
\end{displaymath}
for all $t > 0$, as $\alpha\to\infty$. Therefore, one has solutions growing exponentially in time with an arbitrarily large rate.\epubtkFootnote{However, it should be noted that these solutions are not square integrable, due to their harmonic dependency in $y$. This problem can be remedied by truncation of $u_\alpha$, see Section 3.2 in~\cite{rHReview86}.} 
\end{itemize}

\begin{example}
\label{Example:2DDiracIBVP}
Consider the IBVP for the massless Dirac equation in two spatial dimensions (cf. Section 8.4.1 in \cite{KL89}),
\begin{eqnarray}
& u_t = \left( \begin{array}{rr}
 1 & 0 \\
 0 & -1 \end{array} \right) u_x
 + \left( \begin{array}{rr}
 0 & 1 \\
 1 & 0 \end{array} \right) u_y,
& t \geq 0,\quad x\geq 0,\quad y\in\Real,\qquad
u = \left( \begin{array}{c} u_1 \\ u_2 \end{array} \right)
\label{Eq:2DDiracIBVP}\\
& u(0,x,y) = f(x,y),
& x\geq 0,\quad y\in\Real,
\label{Eq:2DDiracIBVPID}\\
& a u_1 + b u_2 = g(t,y),
& t\geq 0,\quad y\in\Real,
\label{Eq:2DDiracIBVPBC}
\end{eqnarray} 
where $a$ and $b$ are two complex constants to be determined. Assuming $f=0$, Laplace--Fourier transformation leads to the boundary value problem
\begin{eqnarray}
& \tilde{u}_x = M(s,k)\tilde{u},
& x > 0,\qquad
M(s,k) = \left( \begin{array}{rr}
 s & -i k \\
 i k & -s \end{array} \right)
\label{Eq:IBVPLaplaceDirac}\\
& a\tilde{u}_1 + b\tilde{u}_2 = \tilde{g}(s,k), & x=0.
\label{Eq:IBVPLaplaceBCDirac}
\end{eqnarray}
The eigenvalues and corresponding eigenvectors of the matrix $M(s,k)$ are $\kappa_{\pm} = \pm\lambda$ and $e_{\pm} = (ik,s \mp \lambda)^T$, with $\lambda := \sqrt{s^2 + k^2}$, where the root is chosen such that $\re(\lambda) > 0$ for $\re(s) > 0$. The solution which is square integrable on $[0,\infty)$ is the one associated with $\kappa_-$; that is,
\begin{displaymath}
\tilde{u}(s,x,k) = \sigma e^{-\lambda x} e_-,
\end{displaymath}
with $\sigma$ a constant. Introduced into the boundary condition~(\ref{Eq:IBVPLaplaceBCDirac}) leads to the condition
\begin{displaymath}
\left[ i k a + (s + \lambda) b \right]\sigma = \tilde{g}(s,k),
\end{displaymath}
and the Lopatinsky condition is satisfied if and only if the expression inside the square brackets on the left-hand side is different from zero for all $\re(s) > 0$ and $k\in\Real$. Clearly, this implies $b\neq 0$ since otherwise this expression is zero for $k=0$. Assuming $b\neq 0$ and $k\neq 0$, we then obtain the condition
\begin{displaymath}
z + \sqrt{z^2+1} \pm i\frac{a}{b}\neq 0,
\end{displaymath}
for all $z := s/|k|$ with $\re(z) > 0$, which is the case if and only if $|a/b|\leq 1$ or $a/b\in\Real$, see Figure~\ref{Fig:Complex}. The particular case $a=0$, $b=1$ corresponds to fixing the incoming normal characteristic field $u_2$ to $g$ at the boundary.

\epubtkImage{}{
\begin{figure}[htbp]
\centerline{\includegraphics[width=6cm]{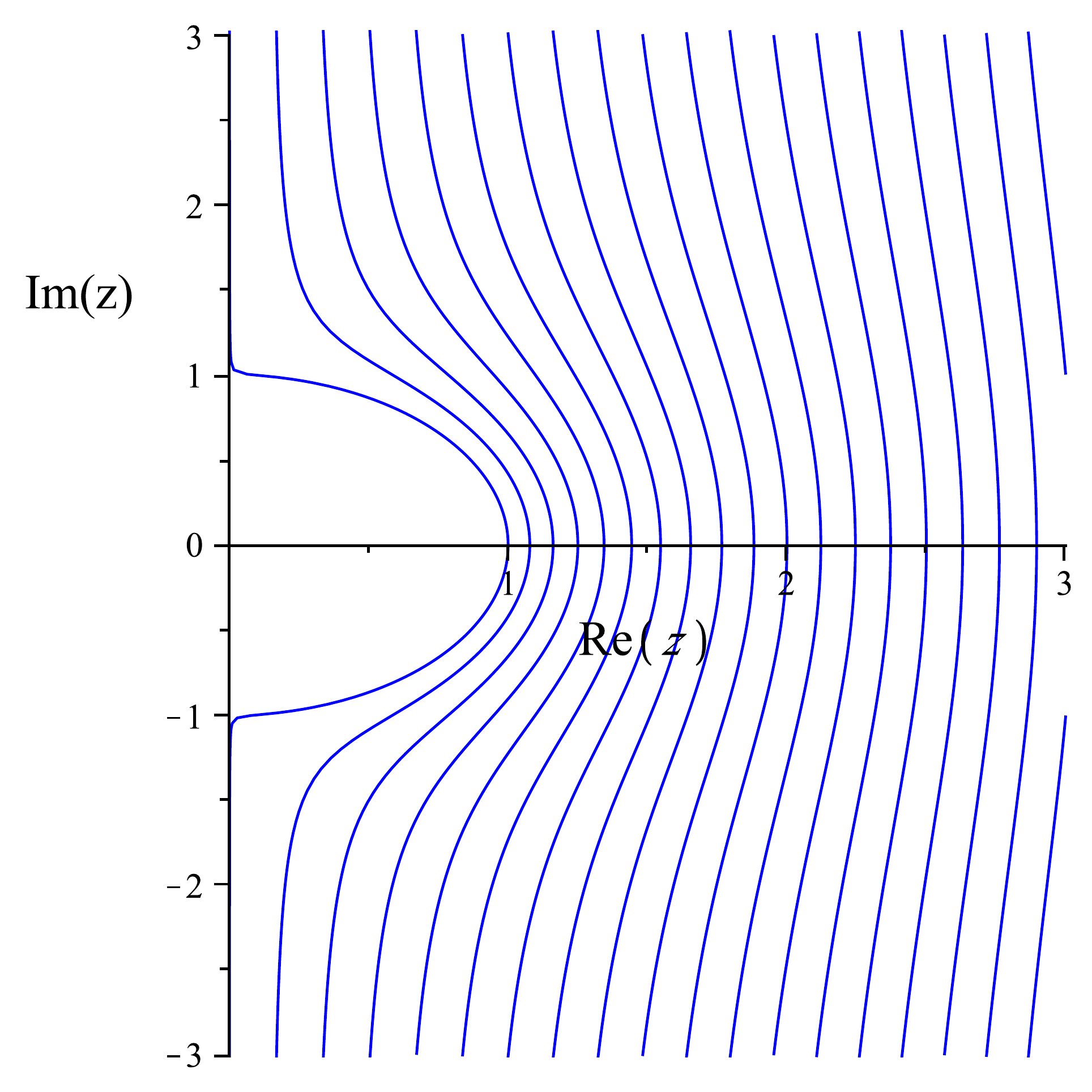}}
\caption{Image of the lines $\re(z)=const > 0$ under the map
  $\Complex\to\Complex$, $z\mapsto z+ \sqrt{z^2 + 1}$.}
\label{Fig:Complex}
\end{figure}}

\end{example}

\begin{example}
\label{Example:FatMaxwellBC}
We consider the Maxwell evolution equations of Example~\ref{Example:FatMaxwell} on the half-space $x_1 > 0$, and freeze the incoming normal characteristic fields to zero at the boundary. These fields are the ones defined in Equation~(\ref{Eq:FatMaxwellCharVars}) which correspond to negative eigenvalues and $k = \partial_x$;\epubtkFootnote{Alternatively, if $k=-\partial_x$ is taken to be the unit outward normal, then the incoming normal characteristic fields are the ones with positive characteristic speeds with respect to $k$. This more geometrical definition will be the one taken in Section~\ref{section:MaxDiss}.} hence
\begin{equation}
E_1 + \frac{\mu}{\beta}(W_{22} + W_{33}) = 0,\qquad
E_A + W_{1A} -  (1 + \alpha) W_{A1} = 0,\qquad
x_1 = 0,\quad x_A\in\Real,\quad t\geq 0,
\label{Eq:FatMaxwellBC}
\end{equation}
where $A=2,3$ label the coordinates tangential to the boundary, and where we recall that $\mu = \sqrt{\alpha\beta}$, assuming that $\alpha$ and $\beta$ have the same sign such that the evolution system~(\ref{Eq:FatMaxwell1}, \ref{Eq:FatMaxwell2}) is strongly hyperbolic. In this example, we apply the Lopatinsky condition in order to find necessary conditions for the resulting IBVP to be well posed. For simplicity, we assume that $\mu = \sqrt{\alpha\beta} = 1$, which implies that the system is strongly hyperbolic for all values of $\alpha\neq 0$, but symmetric hyperbolic only if $-3/2 < \alpha < 0$, see Example~\ref{Example:FatMaxwell}.

In order to analyze the system, it is convenient to introduce the variables $U_1 := W_{22} + W_{33}$, $U_A := W_{1A} - (1 + \alpha) W_{A1}$, $Z:=\beta W_{11} - (1 + \beta/2)U_1$, and $\bar{W}_{AB} := W_{AB} - \delta_{AB} U_1/2$ which are motivated by the form of the characteristic fields with respect to the direction $k = -\partial_1$ normal to the boundary $x_1=0$, see Example~\ref{Example:FatMaxwell}. With these assumptions and definitions, Laplace--Fourier transformation of the system~(\ref{Eq:FatMaxwell1}, \ref{Eq:FatMaxwell2}) yields
\begin{eqnarray}
s\tilde{E}_1 &=& -\alpha\partial_1\tilde{U}_1 
 + i k^A\left[ (1+\alpha)\tilde{U}_A + \alpha(2+\alpha)\tilde{W}_{A1} \right],
\nonumber\\
s\tilde{E}_A &=& -\partial_1\tilde{U}_A
 - i k^B\left[ \tilde{\bar W}_{BA} - (1+\alpha)\tilde{\bar W}_{AB} \right]
 - \alpha i k_A\left[ \alpha\tilde{Z} + (1+\alpha)\tilde{U}_1 \right],
\nonumber\\
s\tilde{U}_1 &=& -\frac{1}{\alpha}\left[ \partial_1\tilde{E}_1 
 + (1+\alpha) i k^A\tilde{E}_A \right],
\nonumber\\
s\tilde{U}_A &=& -\partial_1\tilde{E}_A + (1+\alpha) i k_A\tilde{E}_1,
\nonumber\\
s\tilde{Z} &=& \frac{3+2\alpha}{2\alpha} i k^A\tilde{E}_A,
\nonumber\\
s\tilde{W}_{A1} &=& -i k_A\tilde{E}_1,
\nonumber\\
s\tilde{\bar W}_{AB} &=& -i k_A\tilde{E}_B + \frac{i}{2}\delta_{AB} k^C\tilde{E}_C,
\nonumber
\end{eqnarray}
where we have used $\beta = 1/\alpha$ since $\mu=1$. The last three equations are purely algebraic and can be used to eliminate the zero speed fields $\tilde{Z}$, $\tilde{W}_{A1}$ and $\tilde{\bar W}_{AB}$ from the remaining equations. The result is the ordinary differential system
\begin{eqnarray}
\partial_1\tilde{E}_1 &=& -\alpha s\tilde{U}_1 - (1+\alpha) i k^A\tilde{E}_A,
\nonumber\\
\partial_1\tilde{U}_1 &=& -\left[ \frac{s}{\alpha} 
 - (2+\alpha)\frac{|k|^2}{s} \right]\tilde{E}_1 
 + \frac{1+\alpha}{\alpha} i k^A\tilde{U}_A,
\nonumber\\
\partial_1\tilde{E}_A &=& -s\tilde{U}_A + (1+\alpha) i k_A\tilde{E}_1,
\nonumber\\
\partial_1\tilde{U}_A &=& -\left[ s + \frac{|k|^2}{s} \right]\tilde{E}_A
 + (1+\alpha)^2\frac{k_A k^B}{s}\tilde{E}_B - \alpha(1+\alpha)i k_A\tilde{U}_1.
\nonumber
\end{eqnarray}
In order to diagonalize this system, we decompose $\tilde{E}_A$ and $\tilde{U}_A$ into their components parallel and orthogonal to $k$; if $\hat{k} := k/|k|$ and $\hat{l}$ form and orthonormal basis of the boundary $x_1=0$,\epubtkFootnote{$\hat{k}$ is not well-defined if $k=0$; however, in this case the scalar block comprising $(\tilde{E}_1,\tilde{U}_1)$ decouples from the vector block comprising $(\tilde{E}_A,\tilde{U}_A)$ and it is simple to verify that the resulting system does not possess nontrivial simple wave solutions.} then these are defined as
\begin{displaymath}
\tilde{E}_{||} := \hat{k}^A\tilde{E}_A,\qquad
\tilde{E}_{\perp} := \hat{l}^A\tilde{E}_A,\qquad
\tilde{U}_{||} := \hat{k}^A\tilde{U}_A,\qquad
\tilde{U}_{\perp} := \hat{l}^A\tilde{U}_A.
\end{displaymath}
Then, the system decouples into two blocks, one comprising the transverse quantities $(\tilde{E}_{\perp},\tilde{U}_{\perp})$ and the other the quantities $(\tilde{E}_1,\tilde{U}_1,\tilde{E}_{||},\tilde{U}_{||})$. The first block gives
\begin{displaymath}
\partial_1\left( \begin{array}{c} \tilde{E}_{\perp} \\ \tilde{U}_{\perp} \end{array} \right) 
 = \left( \begin{array}{cc} 0 & -s \\ -\left[ s + \frac{|k|^2}{s} \right] & 0 \end{array} \right)
 \left( \begin{array}{c} \tilde{E}_{\perp} \\ \tilde{U}_{\perp} \end{array} \right),
\end{displaymath}
and the corresponding solutions with exponential decay at $x_1\to\infty$ have the form
\begin{equation}
\left( \begin{array}{c} \tilde{E}_{\perp}(s,x_1,k) \\ \tilde{U}_{\perp}(s,x_1,k) \end{array} \right)
 = \sigma_0 e^{-\lambda x_1}\left( \begin{array}{c} s \\ \lambda \end{array} \right),
\label{Eq:FatMaxwellTransverseBlock}
\end{equation}
where $\sigma_0$ is a complex constant, and where we have defined $\lambda := \sqrt{s^2 + |k|^2}$ with the root chosen such that $\re(\lambda) > 0$ for $\re(s) > 0$. The second block is
\begin{displaymath}
\partial_1\left( \begin{array}{c} 
 \tilde{E}_1 \\ \tilde{U}_1 \\ \tilde{E}_{||} \\ \tilde{U}_{||} \end{array} \right) = 
\left( \begin{array}{cccc} 
 0 & -\alpha s & -i(1+\alpha)|k| & 0 \\ 
 -\frac{s}{\alpha} + (2+\alpha)\frac{|k|^2}{s} & 0 & 0 & i\frac{1+\alpha}{\alpha} |k| \\
 i(1+\alpha)|k| & 0 & 0 & -s \\
0 & -i\alpha(1+\alpha)|k| & -s + \alpha(2+\alpha)\frac{|k|^2}{s} & 0 \end{array} \right)
\left( \begin{array}{c} 
 \tilde{E}_1 \\ \tilde{U}_1 \\ \tilde{E}_{||} \\ \tilde{U}_{||} \end{array} \right) ,
\end{displaymath}
with corresponding decaying solutions
\begin{equation}
\left( \begin{array}{c} 
 \tilde{E}_1(s,x_1,k) \\ \tilde{U}_1(s,x_1,k) \\ \tilde{E}_{||}(s,x_1,k) \\ \tilde{U}_{||}(s,x_1,k)
\end{array} \right)  
 = \sigma_1 e^{-\lambda x_1}
 \left( \begin{array}{c} i|k|s \\ -i|k| \lambda \\ s\lambda \\ s^2 - \alpha |k|^2  
 \end{array} \right)
 + \sigma_2 e^{-\lambda x_1}
 \left( \begin{array}{c} i s\lambda \\ i(s^2/\alpha - |k|^2) \\ |k|s \\ -\alpha|k|\lambda 
 \end{array} \right),
\label{Eq:FatMaxwellParallelBlock}
\end{equation}
with complex constants $\sigma_1$ and $\sigma_2$.

On the other hand, Laplace--Fourier transformation of the boundary conditions~(\ref{Eq:FatMaxwellBC}) leads to
\begin{displaymath}
\tilde{E}_1 + \alpha\tilde{U}_1 = 0,\quad
\tilde{E}_A + \tilde{U}_A = 0,\qquad
x_1 = 0.
\end{displaymath}
Introducing into this the solutions~(\ref{Eq:FatMaxwellTransverseBlock}, \ref{Eq:FatMaxwellParallelBlock}) gives
\begin{displaymath}
(s + \lambda)\sigma_0 = 0
\end{displaymath}
and 
\begin{displaymath}
\left( \begin{array}{cc} 
 |k|(s - \alpha\lambda) & s\lambda + s^2 - \alpha|k|^2) \\
 s\lambda + s^2 - \alpha|k|^2 & |k|(s - \alpha\lambda)
\end{array} \right)
\left( \begin{array}{c} \sigma_1 \\ \sigma_2 \end{array} \right) = 0.
\end{displaymath}
In the first case, since $\re(s+\lambda) \geq \re(s) > 0$, we obtain $\sigma_0 = 0$ and there are no simple wave solutions in the transverse sector. In the second case, the determinant of the system is
\begin{displaymath}
-s^2\left[ (s+\lambda)^2 - (1+\alpha)^2|k|^2 \right],
\end{displaymath}
which is different from zero if and only if $z + \sqrt{z^2 + 1} \neq \pm (1+\alpha)$ for all $\re(z) > 0$, where $z:=s/|k|$. Since $\alpha$ is real this is the case if and only if $-2\leq \alpha \leq 0$, see Figure~\ref{Fig:Complex}.

We conclude that the strongly hyperbolic evolution system~(\ref{Eq:FatMaxwell1}, \ref{Eq:FatMaxwell2}) with $\alpha\beta = 1$ and incoming normal characteristic fields set to zero at the boundary does not give rise to a well posed IBVP when $\alpha > 0$ or $\alpha < -2$. This excludes the parameter range $-3/2 < \alpha < 0$ for which the system is symmetric hyperbolic. This case is covered by the results in Section~\ref{section:MaxDiss} below which utilize energy estimates and show that symmetric hyperbolic problems with zero incoming normal characteristic fields are well posed.
\end{example}

\subsubsection{Sufficient conditions for well posedness and boundary stability}
\label{SubSubSec:LaplaceSufficient}

Next, let us discuss sufficient conditions for the linear, constant coefficient IBVP~(\ref{Eq:QLIBVP1Lin}, \ref{Eq:QLIBVP2Lin}, \ref{Eq:QLIBVP3Lin}) to be well posed. For this, we first transform the problem to trivial initial data by replacing $u(t,x)$ with $u(t,x) - e^{-t} f(x)$. Then, we obtain the IBVP
\begin{eqnarray}
& u_t = \sum\limits_{j=1}^n A^j\frac{\partial}{\partial x^j} u + F(t,x),
& x\in\Sigma,\quad t\geq 0,
\label{Eq:QLIBVP1LinBis}\\
& u(0,x) = 0,& x\in\Sigma,
\label{Eq:QLIBVP2LinBis}\\
& b u = g(t,x),& x\in\partial\Sigma,\quad t\geq 0,
\label{Eq:QLIBVP3LinBis}
\end{eqnarray}
with $F(t,x) = F_0(t,x) -e^{-t}[ f(x) + \sum\limits_{j=1}^n A^j\frac{\partial}{\partial x^j} f(x) ]$ and $g(t,x)$ replaced by $g(t,x) + e^{-t} b f(x)$. By applying the Laplace--Fourier transformation to it, one obtains the boundary value problem~(\ref{Eq:IBVPLaplace}, \ref{Eq:IBVPLaplaceBC}) which could be solved explicitly, provided the Lopatinsky condition holds. However, in view of the generalization to variable coefficients, one would like to have a method which does not rely on the explicit representation of the solution in Fourier space.

In order to formulate the next definition, let $\Omega := [0,\infty) \times \bar{\Sigma}$ be the bulk and ${\cal T} := [0,\infty) \times \partial\Sigma$ the boundary surface, and introduce the associated norms $\| \cdot \|_{\eta,0,\Omega}$ and $\| \cdot \|_{\eta,0,{\cal T}}$ defined by
\begin{eqnarray}
\| u \|^2_{\eta,0,\Omega}
 &:=& \int\limits_\Omega e^{-2\eta t} |u(t,x_1,y) |^2 dt\, dx_1 d^{n-1} y
 = \int\limits_{\Real^{n+1}} |u_\eta(t,x)|^2 dt\, d^n x, 
\nonumber\\
\| u \|^2_{\eta,0,{\cal T}} &:=& \int\limits_{\cal T} e^{-2\eta t} |u(t,0,y) |^2 dt\, d^{n-1} y
= \int\limits_{\Real^n}  |u_\eta(t,0,y)|^2 dt\, d^{n-1} y,
\nonumber
\end{eqnarray}
where we have used the definition of $u_\eta$ as  in Equation~(\ref{Eq:ueta}). Using Parseval's identities we may also rewrite these norms as
\begin{eqnarray}
\| u \|^2_{\eta,0,\Omega} &:=& 
 \int\limits_{\Real} \left[ \int\limits_0^\infty \left(\; \int\limits_{\Real^{n-1}}
  |\tilde{u}(\eta + i\xi,x_1,k)|^2 d^{n-1} k\right) dx_1 \right] d\xi,
\label{Eq:BulkNorm}\\
\| u \|^2_{\eta,0,{\cal T}} &:=& 
 \int\limits_{\Real}\left( \;
  \int\limits_{\Real^{n-1}} |\tilde{u}(\eta + i\xi,0,k)|^2 d^{n-1} k \right) d\xi.
\label{Eq:BoundaryNorm}
\end{eqnarray}

The relevant concept of well posedness is the following one:
\begin{definition}\cite{hK70}
\label{Def:SWPGS}
The IBVP~(\ref{Eq:QLIBVP1LinBis}, \ref{Eq:QLIBVP2LinBis}, \ref{Eq:QLIBVP3LinBis}) is called \textbf{strongly well posed in the generalized sense} if there is a constant $K > 0$ such that each compatible data $F\in C_0^\infty(\Omega)$ and $g\in C_0^\infty({\cal T})$ gives rise to a unique solution $u$ satisfying the estimate
\begin{equation}
\eta\, \| u \|^2_{\eta,0,\Omega} + \| u \|^2_{\eta,0,{\cal T}}
\leq K^2 \left( \frac{1}{\eta} \| F \|^2_{\eta,0,\Omega} + \| g \|^2_{\eta,0,{\cal T}} \right),
\label{Eq:SWPGSEstimate}
\end{equation}
for all $\eta > 0$.
\end{definition}

The inequality~(\ref{Eq:SWPGSEstimate}) implies that both the bulk norm $\| \cdot \|_{\eta,0,\Omega}$ and the boundary norm $\| \cdot \|_{\eta,0,{\cal T}}$ of $u$ are bounded by the corresponding norms of $F$ and $g$. For a trivial source term, $F=0$, the inequality~(\ref{Eq:SWPGSEstimate}) implies, in particular,
\begin{equation}
\| u \|_{\eta,0,{\cal T}} \leq K \| g \|_{\eta,0,{\cal T}},\qquad \eta > 0,
\label{Eq:BoundaryEstimate}
\end{equation}
which is an estimate for the solution at the boundary in terms of the norm of the boundary data $g$. In view of Equation~(\ref{Eq:BoundaryNorm}) this is equivalent to the following requirement:

\begin{definition}\cite{KL89,hKjW06} \label{Def:BoundaryStable}
The boundary problem~(\ref{Eq:QLIBVP1LinBis}, \ref{Eq:QLIBVP2LinBis}, \ref{Eq:QLIBVP3LinBis}) is called \textbf{boundary stable} if there is a constant $K > 0$ such that all solutions $\tilde{u}(s,\cdot,k)\in L^2(0,\infty)$ of Equations~(\ref{Eq:IBVPLaplace}, \ref{Eq:IBVPLaplaceBC}) with $\tilde{F}=0$  satisfy
\begin{equation}
|\tilde{u}(s,0,k)| \leq K |\tilde{g}(s,k)|
\label{Eq:BoundaryStability}
\end{equation}
for all $\re(s) > 0$ and $k\in\Real^{n-1}$.
\end{definition}

Since boundary stability only requires considering solutions for trivial source terms, $F=0$, it is a much simpler condition than Equation~(\ref{Eq:SWPGSEstimate}). Clearly, well posedness in the generalized sense implies boundary stability. The main result is that, modulo technical assumptions, the converse is also true: boundary stability implies strong well posedness in the generalized sense.

\begin{theorem}\cite{hK70, jR71}
\label{Thm:FrozenIBVP}
Consider the linear, constant coefficient IBVP~(\ref{Eq:QLIBVP1LinBis}, \ref{Eq:QLIBVP2LinBis}, \ref{Eq:QLIBVP3LinBis}) on the half space $\Sigma = \{ (x_1,x_2,\ldots,x_n)\in\Real^n : x_1 > 0 \}$. Assume that equation~(\ref{Eq:QLIBVP1LinBis}) is
strictly hyperbolic, meaning that the eigenvalues of the principal symbol $P_0(ik)$ are distinct for all $k\in S^{n-1}$. Assume that the boundary matrix $A = A^1$ is invertible. Then, the problem is strongly well posed in the generalized sense if and only if it is boundary stable.
\end{theorem}

Maybe the importance of Theorem~\ref{Thm:FrozenIBVP} is not so much its statement, which concerns only the linear, constant coefficient case for which the solutions can also be constructed explicitly, but rather the method for its proof, which is based on the construction of a smooth symmetrizer symbol, and which is amendable to generalizations to the variable coefficient case using pseudo-differential operators.

In order to formulate the result of this construction, define $\rho:=\sqrt{|s|^2 + |k|^2}$, $s' := s/\rho$, $k':= k/\rho$, such that $(s',k')\in S_+^n$ lies on the half sphere $S_+^n := \{ (s',k')\in \Complex \times \Real^n : |s'|^2 + |k'|^2 = 1, \re(s') > 0 \}$ for $\re(s) > 0$ and $k\in\Real^{n-1}$. Then, we have,

\begin{theorem}\cite{hK70}
\label{Thm:IBVPSymmetrizer}
Consider the linear, constant coefficient IBVP~(\ref{Eq:QLIBVP1LinBis}, \ref{Eq:QLIBVP2LinBis}, \ref{Eq:QLIBVP3LinBis}) on the half space $\Sigma$. Assume that equation~(\ref{Eq:QLIBVP1LinBis}) is strictly hyperbolic, that the boundary matrix $A = A^1$ is invertible, and that the problem is boundary stable. Then, there exists a family of complex $m\times m$ matrices $H(s',k)$, $(s',k)\in S_+^n$, whose coefficients belong to the class $C^\infty(S_+^n)$, with the following properties:
\begin{enumerate}
\item[(i)] $H(s',k') = H(s',k')^*$ is Hermitian.
\item[(ii)] $H(s',k') M(s',k') + M(s',k')^*H(s',k') \geq 2\re(s') I$ for all $(s',k')\in S^n_+$.
\item[(iii)] There is a constant $C > 0$ such that
\begin{displaymath}
\tilde{u}^* H(s',k')\tilde{u} + C | b\tilde{u} |^2 \geq |\tilde{u}|^2
\end{displaymath}
for all $\tilde{u}\in\Complex^m$ and all $(s',k')\in S^+_n$.
\end{enumerate}
Furthermore, $H$ can be chosen to be a smooth function of the matrix coefficients of $A^j$ and $b$.
\end{theorem}

Let us show how the existence of the symmetrizer $H(s',k')$ implies the estimate~(\ref{Eq:SWPGSEstimate}). First, using Equation~(\ref{Eq:IBVPLaplaceODE}) and properties (i) and (ii) we have
\begin{eqnarray}
\frac{\partial}{\partial x_1}\left[ \tilde{u}^* H(s',k')\tilde{u} \right]
 &=& \left( \frac{\partial\tilde{u}}{\partial x_1} \right)^* H(s',k')\tilde{u} 
  + \tilde{u}^* H(s',k')\frac{\partial\tilde{u}}{\partial x_1}
\nonumber\\
 &=& \rho \tilde{u}^* \left[ H(s',k') M(s',k') + M(s',k')^* H(s',k') \right]\tilde{u} 
 + 2\re\left(\tilde{u}^* H(s',k')A^{-1}\tilde{F}\right)
\nonumber\\
 &\geq& 2\re(s) |\tilde{u}|^2 
  - C_1 |\tilde{u}|^2 - \frac{1}{C_1} | H(s',k')A^{-1}\tilde{F} |^2,
\nonumber
\end{eqnarray}
where we have used the fact that $M(s,k) = \rho M(s',k')$ in the second step, and the inequality $2\re(a^* b) \leq 2|a| |b| \leq C_1 |a|^2 + C_1^{-1} |b|^2$ for complex numbers $a$ and $b$ and any positive constant $C_1 > 0$ in the third step. Integrating both sides from $x_1 = 0$ to $\infty$ and choosing $C_1 = \re(s)$, we obtain, using (iii),
\begin{eqnarray}
\re(s)\int\limits_0^\infty |\tilde{u}|^2 dx_1 
&\leq& -\left[ \tilde{u}^* H(s',k')\tilde{u} \right]_{x_1=0} 
 + \frac{1}{\re(s)}\int\limits_0^\infty | HA^{-1}\tilde{F} |^2 dx_1
\nonumber\\
 &\leq& -\left. |\tilde{u} |^2 \right|_{x_1=0} + C |\tilde{g}|^2
 + \frac{1}{\re(s)}\int\limits_0^\infty |H A^{-1}\tilde{F}|^2 dx_1.
\label{Eq:SWPELaplace}
\end{eqnarray}
Since $H$ is bounded, there exists a constant $C_2 > 0$ such that $|H A^{-1}\tilde{F}| \leq C_2 |\tilde{F}|$ for all $(s',k')\in S^n_+$. Integrating over $\xi = \im(s)\in\Real$ and $k\in\Real^{n-1}$ and using Parseval's identity, we obtain from this
\begin{displaymath}
\eta\, \| u \|^2_{\eta,0,\Omega} + \| u \|^2_{\eta,0,{\cal T}}
\leq  \frac{C_2^2}{\eta} \| F \|^2_{\eta,0,\Omega} + C\| g \|^2_{\eta,0,{\cal T}},
\end{displaymath}
and the estimate~(\ref{Eq:SWPGSEstimate}) follows with $K^2:=\max\{ C_2^2,C\}$.

\begin{example}
Let us go back to the Example~\ref{Example:2DDiracIBVP} of the 2D Dirac equation on the halfspace with boundary condition~(\ref{Eq:2DDiracIBVPBC}) at $x=0$. The solution of Equations~(\ref{Eq:IBVPLaplaceDirac},\ref{Eq:IBVPLaplaceBCDirac}) at the boundary is given by $\tilde{u}(s,0,k) = \sigma(ik,s+\lambda)^T$, where $\lambda = \sqrt{s^2 + k^2}$, and
\begin{displaymath}
\sigma = \frac{\tilde{g}(s,k)}{ik a + (s+\lambda)b}.
\end{displaymath}
Therefore, the IBVP is boundary stable if and only if there exists a constant $K > 0$ such that
\begin{displaymath}
\frac{\sqrt{ k^2 + |s + \lambda|^2}}{| i k a + (s+\lambda)b |} \leq K
\end{displaymath}
for all $\re(s) > 0$ and $k\in\Real^{n-1}$. We may assume $b\neq 0$, otherwise the Lopatinsky condition is violated. For $k=0$ the left-hand side is $1/|b|$. For $k\neq 0$ we can rewrite the condition as
\begin{displaymath}
\frac{1}{|b|}\frac{\sqrt{1 + |\psi(z)|^2}}{|\psi(z) \pm i\frac{a}{b}|} \leq K,
\end{displaymath}
for all $\re(z) > 0$, where $\psi(z) := z + \sqrt{z^2+1}$ and $z := s/|k|$. This is satisfied if and only if the function $|\psi(z) \pm i\frac{a}{b}|$ is bounded away from zero, which is the case if and only if $|a/b| < 1$, see Figure~\ref{Fig:Complex}.

This, together with the results obtained in Example~\ref{Example:2DDiracIBVP}, yields the following conclusions: the IBVP~(\ref{Eq:2DDiracIBVP},\ref{Eq:2DDiracIBVPID},\ref{Eq:2DDiracIBVPBC}) gives rise to an ill posed problem if $b=0$ or if $|a/b| > 1$ and $a/b\notin\Real$ and to a problem which is strongly well posed in the generalized sense if $b\neq 0$ and $|a/b| < 1$. The case $|a|=|b|\neq 0$ is covered by the energy method discussed in Section~\ref{section:MaxDiss} below. For the case $|a/b| > 1$ with $a/b\in\Real$ see Section 10.5 in~\cite{GKO95}.
\end{example}

Before discussing second order systems, let us make a few remarks concerning Theorem~\ref{Thm:FrozenIBVP}:
\begin{itemize}
\item The boundary stability condition~(\ref{Eq:BoundaryStability}) is often called the \textbf{Kreiss condition}. Provided the eigenvalues of the matrix $M(s,k)$ are suitably normalized it can be shown \cite{hK70,GKO95,rHReview86} that the determinant $\det(b_-(s,k))$ in Equation~(\ref{Eq:LopatinskyCondition}) can be extended to a continuous function defined for all $\re(s)\geq 0$ and $k\in\Real^{n-1}$, and condition~(\ref{Eq:BoundaryStability}) can be restated as the following algebraic condition:
\begin{equation}
\det( b_-(s,k) ) \neq 0
\label{Eq:KreissCondition}
\end{equation}
for all $\re(s)\geq 0$ and $k\in\Real^{n-1}$. This is a strengthened version of the Lopatinsky condition, since it requires the determinant to be different from zero also for $s$ on the imaginary axis.
\item As anticipated above, the importance of the symmetrizer construction in Theorem~\ref{Thm:IBVPSymmetrizer} relies on the fact that, based on the theory of pseudo-differential operators, it can be used to treat the linear, variable coefficient IBVP~\cite{hK70}. Therefore, the localization principle holds: if all the frozen coefficient IBVPs are boundary stable and satisfy the assumptions of Theorem~\ref{Thm:FrozenIBVP}, then the variable coefficient problem is strongly well posed in the generalized sense.
\item If the problem is boundary stable, it is also possible to estimate higher order derivatives of the solutions. For example, if we multiply both sides of the inequality~(\ref{Eq:SWPELaplace}) by $|k|^2$, integrate over $\xi = \im(s)$ and $k$ and use Parseval's identity as before, we obtain the estimate ~(\ref{Eq:SWPGSEstimate}) with $u$, $F$ and $g$ replaced by their tangential derivatives $u_y$, $F_y$ and $g_y$, respectively. Similarly, one obtains the estimate ~(\ref{Eq:SWPGSEstimate}) with $u$, $F$ and $g$ replaced by their time derivatives $u_t$, $F_t$ and $g_t$ if we multiply both sides of the inequality~(\ref{Eq:SWPELaplace}) by $|s|^2$ and assume that $u_t(0,x) = 0$ for all $x\in\Sigma$.\epubtkFootnote{One can always assume that $u(0,x) = u_t(0,x) = 0$ for all $x\in\Sigma$ by a suitable redefinition of $u$, $F$ and $g$.} Then, a similar estimate follows for the partial derivative, $\partial_1 u$, in the $x_1$-direction using the evolution equation~(\ref {Eq:QLIBVP1Lin}) and the fact that the boundary matrix $A^1$ is invertible. Estimates for higher order derivatives of $u$ follow by an analogous process.
\item Theorem~\ref{Thm:FrozenIBVP} assumes that the initial data $f$ is trivial, which is not an important restriction since one can always achieve $f=0$ by transforming the source term $F$ and the boundary data $g$, as described below Equation~(\ref{Eq:QLIBVP3LinBis}). Since the transformed $F$ involves derivatives of $f$, this means that derivatives of $f$ would appear on the right-hand side of the inequality~(\ref{Eq:SWPGSEstimate}), and at first sight it looks like one ``loses a derivative'' in the sense that one needs to control the derivatives of $f$ to one degree higher than the ones of $u$. However, the results in Refs.~\cite{jR72a,jR73} improve the statement of Theorem~\ref{Thm:FrozenIBVP} by allowing nontrivial initial data and by showing that the same hypotheses lead to a stronger concept of well posedness (strong well posedness, defined below in Definition~\ref{Def:WPIBVP} as opposed to strong well posedness in the generalized sense).
\item The results mentioned so far assume strict hyperbolicity and an invertible boundary matrix, which are too restrictive conditions for many applications. Unfortunately, there does not seem to exist a general theory which removes these two assumptions. Partial results include Ref.~\cite{mA72}, which treats strongly hyperbolic problems with an invertible boundary matrix that are not necessarily strictly hyperbolic, and Ref.~\cite{aMsO75}, which discusses symmetric hyperbolic problems with a singular boundary matrix.
\end{itemize}

\subsubsection{Second order systems}
\label{SubSubSec:LaplaceSecondOrder}

It has been shown in Ref.~\cite{hKjW06} that certain systems of wave problems can be reformulated in such a way that they satisfy the hypotheses of Theorem~\ref{Thm:IBVPSymmetrizer}. In order to illustrate this, we consider the IBVP for the wave equation on the half-space $\Sigma := \{ (x_1,x_2,\ldots,x_n)\in \Real^n : x_1 > 0 \}$, $n\geq 1$,
\begin{eqnarray}
v_{tt} = \Delta v + F(t,x), && x\in\Sigma,\quad t\geq 0,
\label{Eq:IBVPWaveEq}\\
v(0,x) = 0,\quad v_t(0,x) = 0, && x\in\Sigma,
\label{Eq:IBVPWaveID}\\
L v = g(t,x), && x\in\partial\Sigma,\quad t\geq 0,
\label{Eq:IBVPWaveBC}
\end{eqnarray}
where $F\in C_0^\infty([0,\infty)\times\Sigma)$ and $g\in C_0^\infty([0,\infty)\times\partial\Sigma)$, and where $L$ is a first order linear differential operator of the form
\begin{displaymath}
L := a\frac{\partial}{\partial t} - b\frac{\partial}{\partial x_1}
- \sum\limits_{j=2}^n c_j\frac{\partial}{\partial x_j},
\end{displaymath}
where $a$, $b$, $c_2$, \ldots , $c_n$ are real constants. We ask under which conditions on these constants the IBVP~(\ref{Eq:IBVPWaveEq}, \ref{Eq:IBVPWaveID}, \ref{Eq:IBVPWaveBC}) is strongly well posed in the generalized sense. Since we are dealing with a second order system, the estimate~(\ref{Eq:SWPGSEstimate}) in Definition~\ref{Def:SWPGS} has to be replaced with
\begin{equation}
\eta\, \| v \|^2_{\eta,1,\Omega} + \| v \|^2_{\eta,1,{\cal T}}
\leq K^2 \left( \frac{1}{\eta} \| F \|^2_{\eta,0,\Omega} + \| g \|^2_{\eta,0,{\cal T}} \right),
\label{Eq:SWPGSEstimateSO}
\end{equation}
where the norms $\| \cdot \|^2_{\eta,1,\Omega}$ and $\| \cdot \|^2_{\eta,1,{\cal T}}$
control the first partial derivatives of $v$,
\begin{eqnarray}
\| v \|^2_{\eta,1,\Omega} &:=& \int\limits_\Omega e^{-2\eta t} 
\sum\limits_{\mu=0}^n \left| \frac{\partial v}{\partial x^\mu}(t,x_1,y) 
\right|^2 dt\, dx_1 d^{n-1} y,
\nonumber\\
\| v \|^2_{\eta,1,{\cal T}} &:=& \int\limits_{\cal T} e^{-2\eta t}
\sum\limits_{\mu=0}^n \left| \frac{\partial v}{\partial x^\mu}(t,0,y) 
\right|^2 dt\, d^{n-1} y,
\nonumber
\end{eqnarray}
with $(x^\mu) = (t,x_1,x_2,\ldots,x_n)$. Likewise, the inequality~(\ref{Eq:BoundaryStability}) in the definition of boundary stability needs to be replaced by
\begin{equation}
|\tilde{u}(s,0,k)| \leq K\frac{|\tilde{g}(s,k)|}{\sqrt{|s|^2 + |k|^2}}.
\label{Eq:BoundaryStability2ndOrder}
\end{equation}

Laplace-Fourier transformation of Equations~(\ref{Eq:IBVPWaveEq}, \ref{Eq:IBVPWaveBC}) leads to the second order differential problem
\begin{eqnarray}
\frac{\partial^2}{\partial x_1^2}\tilde{v} = (s^2 + |k|^2)\tilde{v} - \tilde{F},
&& x_1 > 0,\\
b\frac{\partial}{\partial x_1}\tilde{v} = (a s - i c(k))\tilde{v} - \tilde{g},
&& x_1 = 0,
\end{eqnarray}
where we have defined $c(k) := \sum\limits_{j=2}^n c_j k_j$ and where $\tilde{F}$ and $\tilde{g}$ denote the Laplace--Fourier transformations of $F$ and $g$, respectively. In order to apply the theory described in the previous subsection, we rewrite this system in first order pseudo-differential form. Defining
\begin{displaymath}
\tilde{u} := \left( \begin{array}{c} \rho\tilde{v} \\ 
 \frac{\partial\tilde{v}}{\partial x_1} \end{array} \right),
\qquad \tilde{f} := -\left( \begin{array}{c} 0 \\ \tilde{F} \end{array} \right),
\end{displaymath}
where $\rho := \sqrt{|s|^2 + |k|^2}$, we find
\begin{eqnarray}
\frac{\partial}{\partial x_1}\tilde{u} = M(s,k)\tilde{u} + \tilde{f},
&& x_1 > 0,
\label{Eq:IBVPLaplaceWave}\\
L(s,k)\tilde{u} = \tilde{g},
&& x_1 = 0,
\label{Eq:IBVPLaplaceWaveBC}
\end{eqnarray}
where we have defined
\begin{displaymath}
M(s,k) := \rho\left( \begin{array}{cc}
 0  & 1 \\ s'^2 + |k'|^2 & 0
\end{array} \right), \qquad
L(s,k) := \left( a s' - i c(k'), -b \right),
\end{displaymath}
with $s' := s/\rho$, $k' := k/\rho$. This system has the same form as the one described by  Equations~(\ref{Eq:IBVPLaplaceODE}, \ref{Eq:IBVPLaplaceBC}), and the eigenvalues of the matrix $M(s,k)$ are distinct for $\re(s) > 0$ and $k\in\Real^{n-1}$. Therefore, we can construct a symmetrizer $H(s',k')$ according to Theorem~\ref{Thm:IBVPSymmetrizer} provided that the problem is boundary stable. In order to check boundary stability, we diagonalize $M(s,k)$ and consider the solution of Equation~(\ref{Eq:IBVPLaplaceWave}) for $\tilde{f}=0$ which decays exponentially as $x_1\to\infty$,
\begin{displaymath}
\tilde{u}(s,x_1,k)
  = \sigma_- e^{-\lambda x_1}\left( \begin{array}{c} \rho \\ -\lambda \end{array} \right),
\end{displaymath}
where $\sigma_-$ is a complex constant and $\lambda := \sqrt{s^2 + |k|^2}$ with the root chosen such that $\re(\lambda) > 0$ for $\re(s) > 0$. Introduced into the boundary condition~(\ref{Eq:IBVPLaplaceWaveBC}) this gives
\begin{displaymath}
\left[ a s' + b\lambda' - i c(k') \right]\sigma_- = \frac{\tilde{g}}{\rho},
\end{displaymath}
and the system is boundary stable if and only if the expression inside the square parenthesis is different from zero for all $\re(s')\geq 0$ and $k'\in\Real^{n-1}$ with $|s'|^2 + |k'|^2 = 1$. In the one-dimensional case, $n=1$, this condition reduces to $(a+b)s' = 0$ with $|s'|=1$, and the system is boundary stable if and only if $a + b\neq 0$, that is, if and only if the boundary vector field $L$ is not proportional to the \emph{ingoing} null vector at the boundary surface,
\begin{displaymath}
\frac{\partial}{\partial t} + \frac{\partial}{\partial x_1}.
\end{displaymath}
Indeed, if $a + b = 0$, $L u = a(u_t + u_{x_1})$ is proportional to the \emph{outgoing} characteristic field, for which it is not permitted to specify boundary data since it is completely determined by the initial data.

When $n\geq 2$ it follows that $b$ must be different from zero since otherwise the square parenthesis is zero for purely imaginary $s'$ satisfying $a s' = i c(k')$. Therefore, one can choose $b=1$ without loss of generality. It can then be shown that the system is boundary stable if and only if $a > 0$ and $\sum\limits_{j=2}^n |c_j|^2 < a^2$, see Ref.~\cite{hKjW06}, which is equivalent to the condition that the boundary vector field $L$ is pointing outside the domain, and that its orthogonal projection onto the boundary surface ${\cal T}$,
\begin{displaymath}
T := a\frac{\partial}{\partial t} - \sum\limits_{j=2}^n c_j\frac{\partial}{\partial x_j},
\end{displaymath}
is future-directed time-like. This includes as a particular case the ``Sommerfeld'' boundary condition $u_t - u_{x_1} = 0$ for which $L$ is the null vector obtained from the sum of the time evolution vector field $\partial_t$ and the normal derivative $N = -\partial_{x_1}$. While $N$ is uniquely determined by the boundary surface ${\cal T}$, $\partial_t$ is not unique since one can transform it to an arbitrary future-directed time-like vector field $T$ which is tangent to ${\cal T}$ by means of an appropriate Lorentz transformation. Since the wave equation is Lorentz-invariant, it is clear that the new boundary vector field $\hat{L} = T + N$ must also give rise to a well-posed IBVP, which explains why there is so much freedom in the choice of $L$.

For a more geometric derivation of these results based on estimates derived from the stress-energy tensor associated to the scalar field $v$, which shows that the above construction for $L$ is sufficient for strong well-posedness, see Appendix B in Ref.~\cite{hKoRoSjW07}. For a generalization to the shifted wave equation, see Ref.~\cite{mRoRoS07}.

As pointed out in~\cite{hKjW06}, the advantage of obtaining a strong well-posedness estimate~(\ref{Eq:SWPGSEstimateSO}) for the scalar wave problem is the fact that it allows the treatment of \emph{systems} of wave equations where the boundary conditions can be coupled in a certain way through terms involving first derivatives of the fields. In order to illustrate this with a simple example, consider a system of two wave equations,
\begin{displaymath}
\left( \begin{array}{c} v_1 \\ v_2 \end{array} \right)_{tt} 
 = \Delta\left( \begin{array}{c} v_1 \\ v_2 \end{array} \right)
  + \left( \begin{array}{c} F_1(t,x) \\ F_2(t,x) \end{array} \right),
\qquad x\in\Sigma,\quad t\geq 0,
\end{displaymath}
which is coupled through the boundary conditions
\begin{equation}
\left( \frac{\partial}{\partial t} - \frac{\partial}{\partial x_1} \right)
\left( \begin{array}{c} v_1 \\ v_2 \end{array} \right)
=  N\left( \begin{array}{c} v_1 \\ v_2 \end{array} \right)
+ \left( \begin{array}{c} g_1(t,x) \\ g_2(t,x) \end{array} \right),
\qquad x\in\partial\Sigma,\quad t\geq 0,
\label{Eq:CoupledSommerFeldBC}
\end{equation}
where $N$ has the form
\begin{displaymath}
N = \left( \begin{array}{cc} 0 & 0 \\ 0 & X \end{array} \right),\qquad
X = X^0\frac{\partial}{\partial t} + X^1\frac{\partial}{\partial x_1}
+ \ldots + X^n\frac{\partial}{\partial x_n},
\end{displaymath}
with $(X^0,X^1,\ldots,X^n)\in\Complex^{n+1}$ any vector. Since the wave equation and boundary condition for $v_1$ decouples from the one for $v_2$, we can apply the estimate~(\ref{Eq:SWPGSEstimateSO}) to $v_1$, obtaining
\begin{equation}
\eta\, \| v_1 \|^2_{\eta,1,\Omega} + \| v_1 \|^2_{\eta,1,{\cal T}}
\leq K^2 \left( \frac{1}{\eta} \| F_1 \|^2_{\eta,0,\Omega} 
 + \| g_1 \|^2_{\eta,0,{\cal T}} \right).
\label{Eq:SWPGSEstimateSOv1}
\end{equation}
If we set $g_3(t,x) := g_2(t,x) + X v_1(t,x)$, $t\geq 0$, $x\in\partial\Sigma$, we have a similar estimate for $v_2$,
\begin{equation}
\eta\, \| v_2 \|^2_{\eta,1,\Omega} + \| v_2 \|^2_{\eta,1,{\cal T}}
\leq K^2 \left( \frac{1}{\eta} \| F_2 \|^2_{\eta,0,\Omega} 
 + \| g_3 \|^2_{\eta,0,{\cal T}} \right).
\label{Eq:SWPGSEstimateSOv2}
\end{equation}
However, since the boundary norm of $v_1$ is controlled by the estimate~(\ref{Eq:SWPGSEstimateSOv1}), one also controls
\begin{displaymath}
\| g_3 \|^2_{\eta,0,{\cal T}} 
 \leq 2\| g_2 \|^2_{\eta,0,{\cal T}} + C^2\| v_1 \|^2_{\eta,1,{\cal T}}
 \leq \frac{(C K)^2}\eta \| F_1 \|^2_{\eta,0,\Omega} 
 + (C K)^2 \| g_1 \|^2_{\eta,0,{\cal T}} + 2\| g_2 \|^2_{\eta,0,{\cal T}}
\end{displaymath}
with some constant $C > 0$ depending only on the vector field $X$. Therefore, the inequalities~(\ref{Eq:SWPGSEstimateSOv1},\ref{Eq:SWPGSEstimateSOv2}) together yield an estimate of the form~(\ref{Eq:SWPGSEstimateSO}) for $v=(v_1,v_2)$, $F=(F_1,F_2)$ and $g=(g_1,g_2)$, which shows strong well posedness in the generalized sense for the coupled system. Notice that the key point which allows the coupling of $v_1$ and $v_2$ through the boundary matrix operator $N$ is the fact that one controls the \emph{boundary norm} of $v_1$ in the estimate~(\ref{Eq:SWPGSEstimateSOv1}). The result can be generalized to larger systems of wave equations, where the matrix operator $N$ is in triangular form with zero on the diagonal, or where it can be brought into this form by an appropriate transformation \cite{hKjW06,hKoRoSjW09}.

\begin{example}
\label{Example:MaxwellLorentzIBVP}
As an application of the theory for systems of wave equations which are coupled through the boundary conditions, we discuss Maxwell's equations in their potential formulation on the half space $\Sigma$ \cite{hKjW06}. In the Lorentz gauge and the absence of sources, this system is described by four wave equations $\partial^\mu\partial_\mu A_\nu = 0$ for the components $(A_t,A_x,A_y,A_z)$ of the vector potential $A_\mu$, which are subject to the constraint $C:=\partial^\mu A_\mu = 0$, where we use the Einstein summation convention.

As a consequence of the wave equation for $A_\nu$, the constraint variable $C$ also satisfies the wave equation, $\partial^\mu\partial_\mu C = 0$. Therefore, the constraint is correctly propagated if the initial data is chosen such that $C$ and its first time derivative vanish, and if $C$ is set to zero at the boundary. Setting $C=0$ at the boundary amounts in the following condition for $A_\nu$ at $x=0$:
\begin{displaymath}
\frac{\partial A_t}{\partial t} = \frac{\partial A_x}{\partial x}
 + \frac{\partial A_y}{\partial y} + \frac{\partial A_z}{\partial z},
\end{displaymath}
which can be rewritten as
\begin{equation}
\left( \frac{\partial}{\partial t} - \frac{\partial}{\partial x} \right)(A_t + A_x)
 = -\left( \frac{\partial}{\partial t} + \frac{\partial}{\partial x} \right)(A_t - A_x)
 + 2\frac{\partial}{\partial y} A_y + 2\frac{\partial}{\partial z} A_z.
\label{Eq:LorentzConstraint}
\end{equation}
Together with the boundary conditions
\begin{eqnarray}
\left( \frac{\partial}{\partial t} - \frac{\partial}{\partial x} \right)(A_t - A_x) &=& 0,
\nonumber\\
\left( \frac{\partial}{\partial t} - \frac{\partial}{\partial x} \right) A_y &=&
\frac{\partial}{\partial y}(A_t - A_x)
\nonumber\\
\left( \frac{\partial}{\partial t} - \frac{\partial}{\partial x} \right) A_z &=&
\frac{\partial}{\partial z}(A_t - A_x),
\nonumber
\end{eqnarray}
this yields a system of the form of Equation~(\ref{Eq:CoupledSommerFeldBC}) with $N$ having the required triangular form, where $v$ is the four component vector function $v = (A_t - A_x,A_y,A_z,A_t + A_x)$. Notice that the Sommerfeld-like boundary conditions on $A_y$ and $A_z$ set the gauge-invariant quantities $E_y + B_z$ and $E_z - B_y$ to zero, where $E$ and $B$ are the electric and magnetic fields, which is compatible with an outgoing plane wave traveling in the normal direction to the boundary.
\end{example}

For a recent development based on the Laplace method which allows the treatment of second order IBVPs  with more general classes of boundary conditions, including those admitting boundary phenomena like glancing and surface waves, see~\cite{hKoOnP10}.

%===================================================================
\subsection{Maximal dissipative boundary conditions}
\label{section:MaxDiss}
%===================================================================

An alternative technique for specifying boundary conditions which does not require Laplace-Fourier transformation and the use of pseudo-differential operators when generalizing to variable coefficients is based on energy estimates. In order to understand this, we go back to Section~\ref{sec:EnergyEstimate}, where we discussed such estimates for linear, first order symmetric hyperbolic evolution equations with symmetrizer $H(t,x)$. We obtained the estimate~(\ref{Eq:EnergyEstimate}), bounding the energy $E(\Sigma_t) = \int_{\Sigma_t} J^0(t,x) d^n x$ at any time $t\in [0,T]$ in terms of the initial energy $E(\Sigma_0)$, provided that the flux integral
\begin{displaymath}
\int\limits_{\cal T} e_\mu J^\mu(t,x) dS,\qquad
J^\mu(t,x) := -u(t,x)^* H(t,x) A^\mu(t,x) u(t,x)
\end{displaymath}
was nonnegative. Here, the boundary surface is ${\cal T} = [0,T]\times\partial\Sigma$, and its unit outward normal $e = (0,s_1,\ldots,s_n)$ is determined by the unit outward normal $s$ to $\partial\Sigma$. Therefore, the integral is nonnegative if
\begin{equation}
u(t,x)^* H(t,x) P_0(t,x,s) u(t,x) \leq 0,\qquad
(t,x)\in {\cal T},
\label{Eq:BoundaryIntegrandCond}
\end{equation}
where $P_0(t,x,s) = \sum\limits_{j=1}^n A^j(t,x) s_j$ is the principal symbol in the direction of the unit normal $s$. Hence, the idea is to specify homogeneous boundary conditions, $b(t,x) u = 0$ at ${\cal T}$, such that the condition~(\ref{Eq:BoundaryIntegrandCond}) is satisfied.\epubtkFootnote{The restriction to homogeneous boundary conditions $g=0$ in the IBVP~(\ref{Eq:QLIBVP1}, \ref{Eq:QLIBVP2}, \ref{Eq:QLIBVP3}) is not severe, since it can always be achieved by redefining $u$, $f$ and $F$.} In this case, one obtains an a priori energy estimate as in Section~\ref{sec:EnergyEstimate}. Of course, there are many possible choices for $b(t,x)$ which fulfill the condition~(\ref{Eq:BoundaryIntegrandCond}); however, an additional requirement is that one should not overdetermine the IBVP. For example, 
setting all the components of $u$ to zero at the boundary does not lead to a well posed problem if there are outgoing modes, as discussed in Section~\ref{SubSubSec:LaplaceNecessary} for the constant coefficient case. Correct boundary conditions turn out to be a minimal condition on $u$ for which the inequality~(\ref{Eq:BoundaryIntegrandCond}) holds. In other words, at the boundary surface, $u$ has to be restricted to a space for which Equation~(\ref{Eq:BoundaryIntegrandCond}) holds and which cannot be extended. The precise definition which captures this idea is:

\begin{definition}
\label{Def:MaxDiss}
Denote for each boundary point $p = (t,x)\in {\cal T}$ the boundary space
\begin{displaymath}
V_p := \{ u\in\Complex^m : b(t,x) u = 0 \} \subset \Complex^m
\end{displaymath}
of state vectors satisfying the homogeneous boundary condition. $V_p$ is called \textbf{maximal nonpositive} if
\begin{enumerate}
\item[(i)] $u^* H(t,x) P_0(t,x,s) u \leq 0$ for all $u\in V_p$,
\item[(ii)] $V_p$ is maximal with respect to condition (i); that is, if $W_p\supset V_p$ is a linear subspace of $\Complex^m$ containing $V_p$ which satisfies (i), then $W_p = V_p$.
\end{enumerate}
The boundary condition $b(t,x) u = g(t,x)$ is called \textbf{maximal dissipative} if the associated boundary spaces $V_p$ are maximal nonpositive for all $p\in {\cal T}$.
\end{definition}

Maximal dissipative boundary conditions were proposed in Refs.~\cite{kF58,pLrP60} in the context of symmetric positive operators, which include symmetric hyperbolic operators as a special case. With such boundary conditions, the IBVP is well posed in the following sense:

\begin{definition}
\label{Def:WPIBVP}
Consider the linearized version of the IBVP~(\ref{Eq:QLIBVP1}, \ref{Eq:QLIBVP2}, \ref{Eq:QLIBVP3}), where the matrix functions $A^j(t,x)$ and $b(t,x)$ and the vector function $F(t,x)$ do not depend on $u$. It is called \textbf{well posed} if there are constants $K = K(T)$ and $\varepsilon = \varepsilon(T) \geq 0$ such that each compatible data $f\in C_b^\infty(\Sigma,\Complex^m)$ and $g\in C_b^\infty([0,T]\times\partial\Sigma,\Complex^r)$ gives rise to a unique $C^\infty$-solution $u$ satisfying the estimate
\begin{equation}
\| u(t,\cdot) \|_{L^2(\Sigma)}^2 
 + \varepsilon \int\limits_0^t \| u(s,\cdot) \|_{L^2(\partial\Sigma)}^2 ds
 \leq K^2\left[ \| f \|_{L^2(\Sigma)}^2
 + \int\limits_0^t\left( \| F(s,\cdot) \|_{L^2(\Sigma)}^2 
 + \| g(s,\cdot) \|_{L^2(\partial\Sigma)}^2 \right) ds\right],
\label{Eq:SWPEstimate}
\end{equation}
for all $t\in [0,T]$. If, in addition, the constant $\varepsilon > 0$ can be chosen strictly positive, the problem is called \textbf{strongly well posed}.
\end{definition}

This definition strengthens the corresponding definition in the Laplace analysis, where trivial initial data was assumed and only a time-integral of the $L^2(\Sigma)$-norm of the solution could be estimated (see Definition~\ref{Def:SWPGS}). The main result of the theory of maximal dissipative boundary conditions is:

\begin{theorem}
\label{Thm:MaxDiss}
Consider the linearized version of the IBVP~(\ref{Eq:QLIBVP1}, \ref{Eq:QLIBVP2}, \ref{Eq:QLIBVP3}), where the matrix functions $A^j(t,x)$ and $b(t,x)$ and the vector function $F(t,x)$ do not depend on $u$. Suppose the system is symmetric hyperbolic, and that the boundary conditions~(\ref{Eq:QLIBVP3}) are maximal dissipative. Suppose, furthermore, that the rank of the boundary matrix $P_0(t,x,s)$ is constant in $(t,x)\in {\cal T}$.

Then, the problem is well posed in the sense of Definition~\ref{Def:WPIBVP}. Furthermore, it is strongly well posed if the boundary matrix $P_0(t,x,s)$ is invertible.
\end{theorem}

This theorem was first proven in \cite{kF58,pLrP60,jRfM74} for the case where the boundary surface ${\cal T}$ is non-characteristic, that is, the boundary matrix $P_0(t,x,s)$ is invertible for all $(t,x)\in {\cal T}$. A difficulty with the characteristic case is the loss of derivatives of $u$ in the normal direction to the boundary (see Ref.~\cite{mT72}). This case was studied in \cite{aMsO75,jR85,pS96a}, culminating with the regularity theorem in \cite{pS96a} which is based on special function spaces which control the $L^2$-norms of $2k$ tangential derivatives and $k$ normal derivatives at the boundary (see also Ref.~\cite{pS00}). For generalizations of Theorem~\ref{Thm:MaxDiss} to the quasilinear case, see Refs.~\cite{oG90,pS96b}.

A more practical way of characterizing maximal dissipative boundary conditions is the following. Fix a boundary point $(t,x)\in {\cal T}$, and define the scalar product $(\cdot,\cdot)$ by $(u,v) := u^* H(t,x) v$, $u,v\in\Complex^m$. Since the boundary matrix $P_0(t,x,s)$ is Hermitian with respect to this scalar product, there exists a basis $e_1,e_2,\ldots,e_m$ of eigenvectors of $P_0(t,x,s)$ which are orthonormal with respect to $(\cdot,\cdot)$. Let $\lambda_1,\lambda_2,\ldots,\lambda_m$ be the corresponding eigenvalues, where we might assume that the first $r$ of these eigenvalues are strictly positive, and the last $s$ are strictly negative. We can expand any vector $u\in\Complex^m$ as $u = \sum\limits_{j=1}^m u^{(j)} e_j$, the coefficients $u^{(j)}$ being the characteristic fields with associated speeds $\lambda_j$. Then, the condition~(\ref{Eq:BoundaryIntegrandCond}) at the point $p$ can be written as
\begin{equation}
0\geq (u, P_0(t,x,s) u ) = \sum\limits_{j=1}^m \lambda_j |u^{(j)}|^2
 = \sum\limits_{j=1}^r \lambda_j |u^{(j)}|^2 
 - \sum\limits_{j=m-s+1}^m |\lambda_j| |u^{(j)}|^2,
\label{Eq:BoundaryIntegrandCondBis}
\end{equation}
where we have used the fact that $\lambda_1,\ldots,\lambda_r > 0$, $\lambda_{m-s+1},\ldots,\lambda_m < 0$  and the remaining $\lambda_j$'s are zero. Therefore, a maximal dissipative boundary condition must have the form
\begin{equation}
u_+ = q u_-,\qquad
u_+ := \left( \begin{array}{c} u^{(1)} \\ \ldots \\ u^{(r)} \end{array} \right),\qquad
u_- := \left( \begin{array}{c} u^{(m-s+1)} \\ \ldots \\ u^{(m)} \end{array} \right),
\label{Eq:MaxDissBC}
\end{equation}
with $q$ a complex $r\times s$ matrix, since $u_-=0$ must imply $u_+=0$. Furthermore, the matrix $q$ has to be small enough such that the inequality~(\ref{Eq:BoundaryIntegrandCondBis}) holds. There can be no further conditions since an additional, independent condition on $u$ would violate the maximality of the boundary space $V_p$.

In conclusion, a maximal dissipative boundary condition must have the form of Equation~(\ref{Eq:MaxDissBC}) which describes a linear coupling of the outgoing characteristic fields $u_-$ to the incoming ones, $u_+$. In particular, there are exactly as many independent boundary conditions as there are incoming fields, in agreement with the Laplace analysis in Section~\ref{SubSubSec:LaplaceNecessary}. Furthermore, the boundary conditions must not involve the zero speed fields. The simplest choice for $q$ is the trivial one, $q=0$, in which case data for the incoming fields is specified. A nonzero value of $q$ would be chosen if the boundary is to incorporate some reflecting properties, like the case of a perfect conducting surface in electromagnetism, for example.

\begin{example}
\label{Example:KleinGordonBC}
Consider the first order reformulation of the Klein--Gordon equation for the variables $u = (\Phi,\Phi_t,\Phi_x,\Phi_y)$, see Example~\ref{Example:KleinGordonFlat}. Suppose the spatial domain is $x > 0$, with the boundary located at $x=0$. Then, $s = (-1,0)$ and the boundary matrix is
\begin{displaymath}
P_0(s) = -\left( \begin{array}{llll}
 0 & 0 & 0 & 0 \\
 0 & 0 & 1 & 0 \\
 0 & 1 & 0 & 0 \\
 0 & 0 & 0 & 0 \end{array} \right).
\end{displaymath}
Therefore, the characteristic fields and speeds are $\Phi$, $\Phi_y$ (zero speed fields, $\lambda=0$), $\Phi_t - \Phi_x$ (incoming field with speed $\lambda=1$) and $\Phi_t + \Phi_x$ (outgoing field with speed $\lambda=-1$). It follows from Equations~(\ref{Eq:BoundaryIntegrandCondBis}, \ref{Eq:MaxDissBC}) that the class of maximal dissipative boundary conditions is
\begin{displaymath}
(\Phi_t - \Phi_x) = q(t,y) (\Phi_t + \Phi_x) + g(t,y),
\qquad t\geq 0,\quad y\in\Real,
\end{displaymath}
where the function $q$ satisfies $|q(t,y)|\leq 1$ and $g$ is smooth boundary data. Particular cases are:
\begin{itemize}
\item $q=0$: Sommerfeld boundary condition,
\item $q=-1$: Dirichlet boundary condition,
\item $q=1$: Neumann boundary condition.
\end{itemize}
\end{example}

\begin{example}
For Maxwell's equations on a domain $\Sigma\subset\Real^3$ with $C^\infty$-boundary $\partial\Sigma$, the boundary matrix is given by
\begin{displaymath}
P_0(s)\left( \begin{array}{c} E \\ B \end{array} \right)
 = \left( \begin{array}{c} +s\wedge B \\ -s\wedge E \end{array} \right),
\end{displaymath}
see Example~\ref{Example:Maxwell}. In terms of the components $E_{||}$ of $E$ parallel to the boundary surface $\partial\Sigma$, and the ones $E_{\perp}$ which are orthogonal to it (and, hence, parallel to $s$) the characteristic speeds and fields are,
\begin{eqnarray}
0: && E_\perp,\quad B_\perp,
\nonumber\\
\pm 1: && E_{||} \pm s\wedge B_{||}
\nonumber.
\end{eqnarray}
Therefore, maximal dissipative boundary conditions have the form
\begin{equation}
(E_{||} + s\wedge B_{||}) = q(E_{||} - s\wedge B_{||}) + g_{||},
\label{Eq:MaxwellMaxDiss}
\end{equation}
with $g_{||}$ some smooth vector-valued function at the boundary which is parallel to $\partial\Sigma$, and $q$ a matrix-valued function satisfying the condition $|q|\leq 1$. Particular cases are:
\begin{itemize}
\item $q=-1$, $g_{||}=0$: The boundary condition $E_{||}=0$ describes a perfectly conducting boundary surface.
\item $q=0$, $g_{||}=0$: This is a Sommerfeld-type boundary condition, which, locally, is transparent to \emph{outgoing} plane waves traveling in the normal direction $s$,
\begin{displaymath}
E(t,x) = {\cal E} e^{i(\omega t - k\cdot x)},
\qquad
B(t,x) = s\wedge E(t,x),
\end{displaymath}
where $\omega$ the frequency, $k = \omega s$ the wave vector, and ${\cal E}$ the polarization vector which is orthogonal to $k$. The generalization of this boundary condition to inhomogeneous data $g_{||}\neq 0$ allows to specify data on the \emph{incoming} field $E_{||} + s\wedge B_{||}$ at the boundary surface, which is equal to $2{\cal E} e^{i\omega t}$ for the plane waves traveling in the normal inward direction $-s$.
\end{itemize}
Recall that the constraints $\nabla\cdot E = \rho$ and $\nabla\cdot B = 0$ propagate along the time evolution vector field $\partial_t$, $(\nabla\cdot E - \rho)_t = 0$, $(\nabla\cdot B)_t = 0$, provided the continuity equations holds. Since $\partial_t$ is tangent to the boundary, no additional conditions controlling the constraints must be specified at the boundary; the constraints are automatically satisfied everywhere provided they are satisfied on the initial surface. 
\end{example}

\begin{example}
\label{Example:MaxwellLorentz}
Commonly, one writes Maxwell's equations as a system of wave equations for the electromagnetic potential $A_\mu$ in the Lorentz gauge, as discussed in Example~\ref{Example:MaxwellLorentzIBVP}. By reducing the problem to a first order symmetric hyperbolic system, one may wonder if it is possible to apply the theory of maximal dissipative boundary conditions and obtain a well posed IBVP, as in the previous example. As we shall see in Section~\ref{SubSubSec:MaxDissWaveSystems} the answer is affirmative, but the correct application of the theory is not completely straightforward. In order to illustrate why this is the case, introduce the new independent fields $D_{\mu\nu} := \partial_\mu A_\nu$. Then, the set of wave equations can be rewritten as the first order system for the $20$-component vector $(A_\nu,D_{t\nu},D_{j\nu})$, $j=x,y,z$,
\begin{displaymath}
\partial_t A_\nu = D_{t\nu},\qquad
\partial_t D_{t\nu} = \partial^j D_{j\nu},\qquad
\partial_t D_{j\nu} = \partial_j D_{t\nu},
\end{displaymath}
which is symmetric hyperbolic. The characteristic fields with respect to the unit outward normal $s = (-1,0,0)$ at the boundary are
\begin{eqnarray}
D_{t\nu} - D_{x\nu} = (\partial_t - \partial_x)A_\nu &&  \mbox{(incoming field)}
\nonumber\\
D_{t\nu} + D_{x\nu} = (\partial_t + \partial_x)A_\nu &&  \mbox{(outgoing field)}
\nonumber\\
D_{y\nu} = \partial_y A_\nu  && \mbox{(zero speed field)}
\nonumber\\
D_{z\nu} = \partial_z A_\nu  && \mbox{(zero speed field)}
\nonumber
\end{eqnarray}
According to Equation~(\ref{Eq:LorentzConstraint}) we can rewrite the Lorentz constraint in the following way:
\begin{displaymath}
(D_{tt} - D_{xt}) + (D_{tx} - D_{xx}) 
 = -(D_{tt} + D_{xt}) + (D_{tx} + D_{xx}) + 2D_{yy} + 2D_{zz}.
\end{displaymath}
The problem is that when written in terms of the characteristic fields, the Lorentz constraint not only depends on the in- and outgoing fields, but also on the zero speed fields $D_{yy}$ and $D_{zz}$. Therefore, imposing the constraint on the boundary in order to guarantee constraint preservation leads to a boundary condition which couples the incoming fields to outgoing and zero speed fields,\epubtkFootnote{Instead of imposing the constraint itself on the boundary one might try to set some linear combination of its normal and time derivatives to zero, obtaining a constraint-preserving boundary condition that does not involve zero speed fields. Unfortunately, this trick only seems to work for reflecting boundaries, see Refs.~\cite{bSjW03} and \cite{gCjPoRoSmT03} for the case of general relativity. In our example, such boundary conditions are given by $\partial_t A_t = \partial_x A_x = \partial_t A_y = \partial_t A_z = 0$ which imply $\partial_t(\partial^\nu A_\nu) = 0$.} which does not fall in the class of admissaadmissibleble boundary conditions.

At this point, one might ask why we were able to formulate a well-posed IBVP based on the second order formulation in Example~\ref{Example:MaxwellLorentzIBVP}, while the first order reduction discussed here fails. As we shall see, the reason for this is that there exist many first order reductions which are inequivalent to each other, and a slightly more sophisticated reduction works, while the simplest choice adopted here does not. See also Refs.~\cite{oRoS05,aA07} for well posed formulations of the IBVP in electromagnetism based on the potential formulation in a different gauge.
\end{example}

\begin{example}
\label{Example:Weyl}
A generalization of Maxwell's equations is the evolution system
\begin{eqnarray}
\partial_t E_{ij} &=& -\varepsilon_{kl(i}\partial^k B^l{}_{j)},
\label{Eq:WeylSystem1}\\
\partial_t B_{ij} &=& +\varepsilon_{kl(i}\partial^k E^l{}_{j)},
\label{Eq:WeylSystem2}
\end{eqnarray}
for the symmetric, trace-free tensor fields $E_{ij}$ and $B_{ij}$, where here we use the Einstein summation convention, the indices $i,j,k,l$ run over $1,2,3$, $(ij)$ denotes symmetrization over $ij$, and $\varepsilon_{ijk}$ is the totally antisymmetric tensor with $\varepsilon_{123} = 1$. Notice that the right-hand sides of Equations~(\ref{Eq:WeylSystem1}, \ref{Eq:WeylSystem2}) are symmetric and trace-free, such that one can consistently assume that $E^i{}_i = B^i{}_i = 0$. The evolution system~(\ref{Eq:WeylSystem1}, \ref{Eq:WeylSystem2}) which is symmetric hyperbolic with respect to the trivial symmetrizer describes the propagation of the electric and magnetic parts of the Weyl tensor for linearized gravity on a Minkowski background, see for instance Ref.~\cite{hF96}.

Decomposing $E_{ij}$ into its parts parallel and orthogonal to the unit outward normal $s$,
\begin{displaymath}
E_{ij} = \bar{E}\left( s_i s_j - \frac{1}{2}\gamma_{ij} \right) 
 + 2 s_{(i}\bar{E}_{j)} + \hat{E}_{ij},
\end{displaymath}
where $\gamma_{ij} := \delta_{ij} - s_i s_j$, $\bar{E} := s^i s^j E_{ij}$, $\bar{E}_i := \gamma_i{}^k E_{kj} s^j$, $\hat{E}_{ij} := (\gamma_i{}^k\gamma_j^l - \gamma_{ij}\gamma^{kl}/2) E_{kl}$, and similarly for $B_{ij}$, the eigenvalue problem $\lambda u = P_0(s) u$ for the boundary matrix is
\begin{eqnarray*}
\lambda\bar{E} &=& 0,\\
\lambda\bar{B} &=& 0,\\
\lambda\bar{E}_i &=& -\frac{1}{2}\varepsilon_{kli} s^k\bar{B}^l,\\
\lambda\bar{B}_i &=& +\frac{1}{2}\varepsilon_{kli} s^k\bar{E}^l, \\ 
\lambda\hat{E}_{ij} &=& -\varepsilon_{kl(i} s^k \hat{B}^l{}_{j)}, \\
\lambda\hat{B}_{ij} &=& +\varepsilon_{kl(i} s^k \hat{E}^l{}_{j)},
\end{eqnarray*}
from which one obtains the following characteristic speeds and fields,
\begin{eqnarray}
0: && \bar{E},\quad \bar{B},
\nonumber\\
\pm \frac{1}{2}: && \bar{E}_i \mp \varepsilon_{kli} s^k\bar{B}^l,
\nonumber\\
\pm 1: && \hat{E}_{ij} \mp \varepsilon_{kl(i} s^k \hat{B}^l{}_{j)}
\nonumber.
\end{eqnarray}
Similarly to the Maxwell case, the boundary condition $\hat{E}_{ij} - \varepsilon_{kl(i} s^k \hat{B}^l{}_{j)} = 0$ on the incoming, symmetric trace-free characteristic field is, locally, transparent to outgoing linear gravitational plane waves traveling in the normal direction $s$. In fact, this condition is equivalent to setting to zero at the boundary surface the complex Weyl scalar $\Psi_0$, computed from the adapted, complex null tetrad $K:=\partial_t + s$, $L:=\partial_t - s$, $Q$, $\bar{Q}$.\epubtkFootnote{This relation is explicitly given in terms of the Weyl tensor $C$, namely $\Psi_0 = C_{\alpha\beta\gamma\delta} K^\alpha Q^\beta K^\gamma Q^\delta = 2(E_{\alpha\beta} - \varepsilon_{\gamma\delta(\alpha} s^\gamma B^\delta{}_{\beta)})Q^\alpha Q^\beta$, where $E_{\alpha\beta} = C_{\alpha\gamma\beta\delta} T^\gamma T^\delta$ and $B_{\alpha\beta} = *C_{\alpha\gamma\beta\delta} T^\gamma T^\delta$ are the electric and magnetic parts of $C$ with respect to the timelike vector field $T = \partial_t$.} Variants of this condition have been proposed in the literature in the context of the IBVP for Einstein's field equations in order to approximately control the incoming gravitational radiation, see Refs.~\cite{hFgN99,jBlB02,lKlLmSlBhP05,oSmT05,oR06,gNoS06,lLmSlKrOoR06,mShPlLlKoRsT06,oRlLmS07}.

However, one also needs to control the incoming field $\bar{E}_i - \varepsilon_{kli} s^k\bar{B}^l$ at the boundary. This field, which propagates with speed $1/2$, is related to the constraints in the theory. Like in electromagnetism, the fields $E_{ij}$ and $B_{ij}$ are subject to the divergence constraints $P_j := \partial^i E_{ij} = 0$, $Q_j := \partial^i B_{ij} = 0$. Unlike the Maxwell case, however, these constraints do not propagate trivially. As a consequence of the evolution equations~(\ref{Eq:WeylSystem1}, \ref{Eq:WeylSystem2}), the constraint fields $P_j$ and $Q_j$ obey
\begin{displaymath}
\partial_t P_j = -\frac{1}{2}\varepsilon_{jkl}\partial^k Q^l,\qquad
\partial_t Q_j = +\frac{1}{2}\varepsilon_{jkl}\partial^k P^l,
\end{displaymath}
which is equivalent to Maxwell's equations except that the propagation speed for the transverse modes is $1/2$ instead of $1$. Therefore, guaranteeing constraint propagation requires specifying homogeneous maximal dissipative boundary conditions for this system, which have the form of Equation~(\ref{Eq:MaxwellMaxDiss}) with $E \mapsto P$, $B \mapsto -Q$ and $g_{||}=0$. A problem is that this yields conditions involving \emph{first derivatives} of the fields $E_{ij}$ and $B_{ij}$, when rewritten as a boundary condition for the main system~(\ref{Eq:WeylSystem1}, \ref{Eq:WeylSystem2}). Except in some particular cases involving totally reflecting boundaries, it is not possible to cast these conditions into maximal dissipative form.

A solution to this problem has been presented in Refs.~\cite{hF95} and \cite{hFgN99}, where a similar system appears in the context of the IBVP for Einstein's field equations for solutions with anti-de~Sitter asymptotics, or for solutions with an artificial boundary, respectively. The method consists in modifying the evolution system~(\ref{Eq:WeylSystem1}, \ref{Eq:WeylSystem2}) by using the constraint equations $P_j = Q_j = 0$ in such a way that the constraint fields for the resulting boundary adapted system propagate along $\partial_t$ at the boundary surface. In order to describe this system, extend $s$ to a smooth vector field on $\Sigma$ with the property that $|s|\leq 1$. Then, the boundary adapted system reads:
\begin{eqnarray}
\partial_t E_{ij} &=& -\varepsilon_{kl(i}\partial^k B^l{}_{j)}
 + s_{(i}\varepsilon_{j)kl} s^k Q^l,
\label{Eq:WeylSystemModified1}\\
\partial_t B_{ij} &=& +\varepsilon_{kl(i}\partial^k E^l{}_{j)}
 - s_{(i}\varepsilon_{j)kl} s^k P^l.
\label{Eq:WeylSystemModified2}
\end{eqnarray}
This system is symmetric hyperbolic, and the characteristic fields in the normal direction are identical to the unmodified system with the important difference that the fields $\bar{E}_i \mp \varepsilon_{kli} s^k\bar{B}^l$ now propagate with zero speed. The induced evolution system for the constraint fields is symmetric hyperbolic, and has a trivial boundary matrix. As a consequence, the constraints propagate tangentially to the boundary surface, and no extra boundary conditions for controlling the constraints must be specified.
\end{example}

\subsubsection{Application to systems of wave equations}
\label{SubSubSec:MaxDissWaveSystems}

As anticipated in Example~\ref{Example:MaxwellLorentz}, the theory of symmetric hyperbolic first order equations with maximal dissipative boundary conditions can also be used to formulate well posed IBVP for systems of wave equations which are coupled through the boundary conditions, as already discussed in Section~\ref{SubSubSec:LaplaceSecondOrder} based on the Laplace method. Again, the key idea is to show strong well posedness; that is, an a priori estimate which controls the first derivatives of the fields in the bulk \emph{and at the boundary}.

In order to explain how this is performed, we consider the simple case of the Klein-Gordon equation $\Phi_{tt} = \Delta\Phi - m^2\Phi$ on the half plane $\Sigma := \{ (x,y)\in\Real^2 : x > 0 \}$. In Example~\ref{Example:KleinGordonFlat} we reduced the problem to a first order symmetric hyperbolic system for the variables $u = (\Phi,\Phi_t,\Phi_x,\Phi_y)$ with symmetrizer $H = \diag(m^2,1,1,1)$, and in Example~\ref{Example:KleinGordonBC} we determined the class of maximal dissipative boundary conditions for this first order reduction. Consider the particular case of Sommerfeld boundary conditions, where $\Phi_t = \Phi_x$ is specified at $x=0$. Then, Equation~(\ref{Eq:IntegratedPseudoConservationLaw}) gives the following conservation law,
\begin{equation}
E(\Sigma_T) = E(\Sigma_0) 
 + \int\limits_0^T\int\limits_\Real \left. u^* H P_0(s) u \right|_{x=0} dy dt,
\label{Eq:ConservationLawKG}
\end{equation}
where $E(\Sigma_t) = \int_{\Sigma_t} u^* H u\, dx dy = \int_{\Sigma_t}\left( m^2|\Phi|^2 + |\Phi_t|^2 + |\Phi_x|^2 + |\Phi_y|^2 \right) dx dy$, and $u^* H P_0(s) u = -2\re(\Phi_t^*\Phi_x)$, see Example~\ref{Example:KleinGordonBC}. Using the Sommerfeld boundary condition, we may rewrite $-2\re(\Phi_t^*\Phi_x) = -(|\Phi_t|^2 + |\Phi_x|^2)$, and obtain the energy equality
\begin{displaymath}
E(\Sigma_T) 
 + \int\limits_0^T\int\limits_\Real \left[ |\Phi_t|^2 + |\Phi_x|^2 \right]_{x=0} dy dt
 = E(\Sigma_0),
\end{displaymath}
controlling the derivatives of $\Phi_t$ and $\Phi_x$ at the boundary surface. However, a weakness of this estimate is that it does not control the zero speed fields $\Phi$ and $\Phi_y$ at the boundary, and so one does not obtain strong well posedness.

On the other hand, the first order reduction is not unique, and as we show now, different reductions may lead to stronger estimates. For this, we choose a real constant $b$ such that $0 < b\leq 1/2$ and define the new fields $\bar{u} := (\Phi,\Phi_t - b\Phi_x,\Phi_x,\Phi_y)$, which yield the symmetric hyperbolic system
\begin{displaymath}
\bar{u}_t = \left( \begin{array}{rrrr}
 b & 0 & 0 & 0 \\
 0 & -b & 1-b^2 & 0 \\
 0 & 1 & b & 0 \\
 0 & 0 & 0 & b \end{array} \right)\bar{u}_x
 + \left( \begin{array}{rrrr}
 0 & 0 & 0 & 0 \\
 0 & 0 & 0 & 1 \\
 0 & 0 & 0 & 0 \\
 0 & 1 & 0 & 0 \end{array} \right)\bar{u}_y
 +  \left( \begin{array}{rrrr}
 0   & 1 & 0 & 0 \\
-m^2 & 0 & 0 & 0 \\
 0   & 0 & 0 & 0 \\
 0   & 0 & 0 & 0 \end{array} \right)\bar{u},
\end{displaymath}
with symmetrizer $\bar{H} = \diag(m^2,1,1-b^2,1)$. The characteristic fields in terms of $\Phi$ and its derivatives are $\Phi$, $\Phi_y$, $\Phi_t + \Phi_x$, and $\Phi_t - \Phi_x$, as before. However, the fields now have characteristic speeds $-b,-b,-1,+1$, respectively, whereas in the previous reduction they were $0,0,-1,+1$. Therefore, the effect of the new reduction versus the old one is to shift the speeds of the zero speed fields, and to convert them to outgoing fields with speed $-b$. Notice that the Sommerfeld boundary condition $\Phi_t = \Phi_x$ is still maximal dissipative with respect to the new reduction. Repeating the energy estimates again leads to a conservation law of the form~(\ref{Eq:ConservationLawKG}), but where now the energy and flux quantities are $E(\Sigma_t) = \int_{\Sigma_t} \bar{u}^* \bar{H} \bar{u}\, dx dy = \int_{\Sigma_t}\left( m^2|\Phi|^2 + |\Phi_t - b\Phi_x|^2 + (1-b^2)|\Phi_x|^2 + |\Phi_y|^2 \right) dx dy$ and
\begin{displaymath}
\bar{u}^* \bar{H} P_0(s)\bar{u}
 = -b\left[ m^2|\Phi|^2 + |\Phi_t|^2 + |\Phi_x|^2 + |\Phi_y|^2 \right]
 + 2b\left[ |\Phi_t|^2 + |\Phi_x|^2 \right] - 2\re(\Phi_t^*\Phi_x).
\end{displaymath}
Imposing the boundary condition $\Phi_t = \Phi_x$ at $x=0$ and using $2b\leq 1$ leads to the energy estimate
\begin{displaymath}
E(\Sigma_T) + b\int\limits_0^T\int\limits_\Real 
 \left[ m^2|\Phi|^2 + |\Phi_t|^2 + |\Phi_x|^2 + |\Phi_y|^2 \right]_{x=0} dy dt
 \leq E(\Sigma_0),
\end{displaymath}
controlling $\Phi$ and all its first derivatives at the boundary surface.

Summarizing, we have seen that the most straightforward first order reduction of the Klein--Gordon equation does not lead to strong well-posedness. However, strong well-posedness can be obtained by choosing a more sophisticated reduction, in which the time-derivative of $\Phi$ is replaced by its derivative $\Phi_t - b\Phi_x$ along the time-like vector $(1,-b)$ which is pointing \emph{outside} the domain at the boundary surface. In fact, it is possible to obtain a symmetric hyperbolic reduction leading to strong well posedness for any future-directed time-like vector field $u$ which is pointing outside the domain at the boundary. Based on the geometric definition of first order symmetric hyperbolic systems in~\cite{rG96}, it is possible to generalize this result to systems of quasilinear wave equations on curved backgrounds~\cite{hKoRoSjW09}. 

In order to describe the result in~\cite{hKoRoSjW09}, let $\pi: E \to M$ be a vector bundle over $M = [0,T]\times \Sigma $ with fibre $\Real^N$, let $\nabla_\mu$ be a fixed, given connection on $E$ and let $g_{\mu\nu} = g_{\mu\nu}(\Phi)$ be a Lorentz metric on $M$ with inverse $g^{\mu\nu}(\Phi)$ which depends pointwise and smoothly on a vector-valued function $\Phi = \{ \Phi^A \}_{A=1,2,\ldots,N}$, parameterizing a local section of $E$. Assume that each time-slice $\Sigma_t = \{ t \} \times \Sigma$ is space-like and that the boundary ${\cal T} = [0,T] \times \partial\Sigma$ is time-like with respect to $g_{\mu\nu}(\Phi)$. We consider a system
of quasilinear wave equations of the form
\begin{equation}
g^{\mu\nu}(\Phi) \nabla_\mu\nabla_\nu\Phi^A = F^A(\Phi,\nabla\Phi), 
\label{Eq:WaveSystemEq}
\end{equation}
where $F^A(\Phi,\nabla\Phi)$ is a vector-valued function which depends pointwise and smoothly on its arguments. The wave system~(\ref{Eq:WaveSystemEq}) is subject to the initial conditions
\begin{equation}
\left. \Phi^A \right|_{\Sigma_0} = \Phi^A_0\; , \qquad
\left. n^\mu\nabla_\mu\Phi^A \right|_{\Sigma_0} = \Pi^A_0\; ,
\label{Eq:WaveSystemID}
\end{equation}
where $\Phi^A_0$ and $\Pi^A_0$ are given vector-valued functions on $\Sigma_0$, and where $n^\mu = n^\mu(\Phi)$ denotes the future-directed unit normal to $\Sigma_0$ with respect to $g_{\mu\nu}$. In order to describe the boundary conditions, let $T^\mu = T^\mu(p,\Phi)$, $p\in {\cal T}$, be a future-directed vector field on ${\cal T}$ which is normalized with respect to $g_{\mu\nu}$ and let $N^\mu = N^\mu(p,\Phi)$ be the unit outward normal to ${\cal T}$ with respect to the metric $g_{\mu\nu}$. We consider boundary conditions on ${\cal T}$ of the following form
\begin{equation}
\left. \left[ T^\mu + \alpha N^\mu \right]\nabla_\mu\Phi^A \right|_{\cal T}
 = c^{\mu\, A}{}_B\left. \nabla_\mu\Phi^B \right|_{\cal T} 
 + d^A{}_B\left. \Phi^B \right|_{\cal T} + G^A,
\label{Eq:WaveSystemBC}
\end{equation}
where $\alpha = \alpha(p,\Phi) > 0$ is a strictly positive, smooth function, $G^A = G^A(p)$ is a given, vector-valued function on ${\cal T}$ and the matrix coefficients $c^{\mu\, A}{}_B = c^{\mu\, A}{}_B(p,\Phi)$ and $d^A{}_B = d^A{}_B(p,\Phi)$ are smooth functions of their arguments. Furthermore, we assume that $c^{\mu\, A}{}_B$ satisfies the following property. Given a local trivialization $\varphi: U \times \Real^N \mapsto \pi^{-1}(U)$ of $E$ such that $\bar{U}\subset M$ is compact and contains a portion ${\cal U}$ of the boundary ${\cal T}$, and given $\varepsilon > 0$, there exists a smooth map $J: U \to GL(N,\Real), p \mapsto (J^A{}_B(p))$ such that the transformed matrix coefficients
\begin{displaymath}
\tilde{c}^{\mu\, A}{}_B := J^A{}_C c^{\mu\, C}{}_D \left( J^{-1}\right)^D{}_B
\end{displaymath}
are in upper triangular form with zeroes on the diagonal, that is
\begin{displaymath}
\tilde{c}^{\mu\, A}{}_B = 0,\qquad B \leq A.
\end{displaymath}

\begin{theorem}\cite{hKoRoSjW09}
\label{Thm:IBVPWaveSystems}
The IBVP~(\ref{Eq:WaveSystemEq}, \ref{Eq:WaveSystemID}, \ref{Eq:WaveSystemBC}) is well posed. Given $T > 0$ and sufficiently small and smooth initial and boundary data $\Phi_0^A$, $\Pi_0^A$ and $G^A$ satisfying the compatibility conditions at the edge $S = \{ 0 \} \times \partial\Sigma$, there exists a unique smooth solution on $M$ satisfying the evolution equation~(\ref{Eq:WaveSystemEq}), the initial condition~(\ref{Eq:WaveSystemID}) and the boundary condition~(\ref{Eq:WaveSystemBC}). Furthermore, the solution depends continuously on the initial and boundary data.
\end{theorem}

Theorem~\ref{Thm:IBVPWaveSystems} provides the general framework for treating wave systems with constraints, such as Maxwell's equations in the Lorentz gauge and, as we will see in Section~\ref{SubSec:HarmonicBC}, Einstein's field equations with artificial outer boundaries.

\subsubsection{Existence of weak solutions and the adjoint problem}

Here, we show how to prove the existence of weak solutions for linear, symmetric hyperbolic equations with variable coefficients and maximal dissipative boundary conditions. The method can also be applied to a more general class of linear symmetric operators with maximal dissipative boundary conditions, see Refs.~\cite{kF58,pLrP60}. The proof below will shed some light on the maximality condition for the boundary space $V_p$.

Our starting point is an IBVP of the form~(\ref{Eq:QLIBVP1}, \ref{Eq:QLIBVP2}, \ref{Eq:QLIBVP3}) where the matrix functions $A^j(t,x)$ and $b(t,x)$ do not depend on $u$, and where $F(t,x,u)$ is replaced by $B(t,x) u + F(t,x)$, such that the system is linear. Furthermore, we can assume that the initial and boundary data is trivial, $f=0$, $g=0$. We require the system to be symmetric hyperbolic with symmetrizer $H(t,x)$ satisfying the conditions in Definition~\ref{Def:HyperbolicityVC}(iii), and assume the boundary conditions~(\ref{Eq:QLIBVP3}) are maximal dissipative. We rewrite the IBVP on $\Omega_T := [0,T]\times\Sigma$ as the abstract linear problem
\begin{equation}
-L u = F,
\label{Eq:SymHypStrongForm}
\end{equation}
where $L: D(L)\subset X\to X$ is the linear operator on the Hilbert space $X:=L^2(\Omega_T)$ defined by the evolution equation and the initial and boundary conditions:
\begin{eqnarray}
&& D(L) := \{ u\in C^\infty_b(\Omega_T) : u(p) = 0\hbox{ for all $p\in\Sigma_0$ and }
u(p)\in V_p\hbox{ for all $p\in{\cal T}$} \},
\nonumber\\
&& L u := \sum\limits_{\mu=0}^n A^\mu(t,x)\frac{\partial u}{\partial x^\mu} 
+ B(t,x) u,\qquad u\in D(L),
\nonumber
\end{eqnarray}
where we have defined $A^0 := -I$ and $x^0 := t$, where $V_p = \{ u\in\Complex^m : b(t,x) u = 0 \}$ is the boundary space, and where $\Sigma_0 := \{ 0 \} \times \Sigma$, $\Sigma_T := \{ T \} \times \Sigma$ and ${\cal T} := [0,T]\times\partial\Sigma$ denote the initial, the final and the boundary surface, respectively.

For the following, the adjoint IBVP plays an important role. This problem is defined as follows. First, the symmetrizer defines a natural scalar product on $X$,
\begin{displaymath}
\la v, u \ra_H := \int\limits_{\Omega_T} v^*(t,x) H(t,x) u(t,x)\, dt d^n x,\qquad
u,v\in X,
\end{displaymath}
which, because of the properties of $H$, is equivalent to the standard scalar product on $L^2(\Omega_T)$. In order to obtain the adjoint problem, we take $u\in D(L)$ and $v\in C^\infty_b(\Omega_T)$, and use Gauss's theorem to find
\begin{equation}
\la v,Lu \ra_H = \la L^*v, u \ra_H 
 + \int\limits_{\Sigma_0} v^* H u\, d^n x - \int\limits_{\Sigma_T} v^* H u\, d^n x 
 + \int\limits_{\cal T} v^* H P_0(t,x,s) u\, dS,
\label{Eq:GreenIdentity}
\end{equation}
where we have defined the formal adjoint $L^*: D(L^*)\subset X\to X$ of $L$ by
\begin{displaymath}
L^* v := -\sum\limits_{\mu=0}^n A^\mu(t,x)\frac{\partial v}{\partial x^\mu} 
- H(t,x)^{-1}\sum\limits_{\mu=0}^n \frac{\partial( H(t,x) A^\mu(t,x))}{\partial x^\mu} v 
+ H(t,x)^{-1}B(t,x)^* H(t,x) v.
\end{displaymath}
In order for the integrals on the right-hand side of Equation~(\ref{Eq:GreenIdentity}) to vanish, such that $\la v,Lu \ra_H = \la L^*v, u \ra_H$, we first notice that the integral over $\Sigma_0$ vanishes, because $u=0$ on $\Sigma_0$. The integral over $\Sigma_T$ also vanishes if we require $v=0$ on $\Sigma_T$. The last term also vanishes if we require $v$ to lie in the dual boundary space
\begin{displaymath}
V_p^* := \{ v\in\Complex^m : v^* H P_0(t,x,s) u = 0 \hbox{ for all $u\in V_p$} \}.
\end{displaymath}
for each $p\in {\cal T}$. Therefore, if we define
\begin{displaymath}
D(L^*) := \{ v\in C^\infty_b(\Omega_T) : v(p) = 0\hbox{ for all $p\in\Sigma_T$ and }
v(p)\in V_p^*\hbox{ for all $p\in{\cal T}$} \},
\end{displaymath}
we have $\la v,Lu \ra_H = \la L^*v, u \ra_H$ for all $u\in D(L)$ and $v\in D(L^*)$, that is, the operator $L^*$ is adjoint to $L$. There is the following nice relation between the boundary spaces $V_p$ and $V_p^*$:

\begin{lemma}
Let $p\in{\cal T}$ be a boundary point. Then, $V_p$ is maximal nonpositive if and only if $V_p^*$ is maximal nonnegative.
\end{lemma}

\proof
Fix a boundary point $p=(t,x)\in{\cal T}$ and define the matrix ${\cal B} := H(t,x) P_0(t,x,s)$ with $s$ the unit outward normal to $\partial\Sigma$ at $x$. Since the system is symmetric hyperbolic, ${\cal B}$ is Hermitian. We decompose $\Complex^m = E_+ \oplus E_- \oplus E_0$ into orthogonal subspaces $E_+$, $E_-$, $E_0$ on which ${\cal B}$ is positive, negative and zero, respectively. We equip $E_\pm$ with the scalar products $(\cdot,\cdot)_\pm$ which are defined by
\begin{displaymath}
(u_\pm,v_\pm)_\pm := \pm u_\pm^*{\cal B} v_\pm,\qquad
u_\pm,v_\pm\in E_\pm.
\end{displaymath}
In particular, we have $u^*{\cal B} u = ( u_+,u_+)_+ - (u_-,u_-)_-$ for all $u\in\Complex^m$. Therefore, if $V_p$ is maximal nonpositive, there exists a linear transformation $q: E_-\to E_+$ satisfying $|q u_-|_+ \leq |u_-|_-$ for all $u_-\in E_-$ such that (cf. Equation~(\ref{Eq:MaxDissBC}))
\begin{displaymath}
V_p = \{ u\in\Complex^m : u_+ = q u_- \}.
\end{displaymath}
Let $v\in V_p^*$. Then,
\begin{displaymath}
0 = v^* {\cal B} u = (v_+,u_+)_+ - (v_-,u_-)_- = (v_+,q u_-)_+ - (v_-,u_-)_-
 = (q^\dagger v_+,u_-)_- - (v_-,u_-)_-
\end{displaymath}
for all $u\in V_p$, where $q^\dagger : E_+\to E_-$ is the adjoint of $q$ with respect to the scalar products $(\cdot,\cdot)_\pm$ defined on $E_\pm$. Therefore, $v_- = q^\dagger v_+$, and
\begin{displaymath}
V_p^* = \{ v\in\Complex^m : v_- = q^\dagger v_+ \}.
\end{displaymath}
Since $q^\dagger$ has the same norm as $q$, which is one, it follows that $V_p^*$ is maximal nonnegative. The converse statement follows in an analogous way.
\qed

The lemma implies that solving the original problem $-L u = F$ with $u\in D(L)$ is equivalent to solving the adjoint problem $L^* v = F$ with $v\in D(L^*)$, which, since $v(T,x)=0$ is held fixed at $\Sigma_T$, corresponds to the time-reversed problem with the adjoint boundary conditions. From the energy a priori estimates we obtain:

\begin{lemma}
\label{Lem:LowerBoundness}
There is a constant $\delta = \delta(T)$ such that
\begin{equation}
\| L u \|_H \geq \delta \| u \|_H,\qquad
\| L^* v \|_H \geq \delta \| v \|_H
\label{Eq:LLstarIneq}
\end{equation}
for all $u\in D(L)$ and $v\in D(L^*)$, where $\|\cdot\|_H$ is the norm induced by the scalar product $\la\cdot,\cdot\ra_H$.
\end{lemma}

\proof
Let $u\in D(L)$ and set $F:=-L u$. From the energy estimates in Section~\ref{sec:EnergyEstimate} one easily obtains
\begin{displaymath}
E(\Sigma_t) \leq C \| F \|_H^2,\qquad 0 \leq t \leq T,
\end{displaymath}
for some positive constants $C$ depending on $T$. Integrating both sides from $t=0$ to $t=T$ gives
\begin{displaymath}
\| u \|_H^2 \leq C T \| F \|_H^2 = C T \| L u \|_H^2,
\end{displaymath}
which yields the statement for $L$ setting $\delta := (C T)^{-1/2}$. The estimate for $L^*$ follows from a similar energy estimate for the adjoint problem, or it follows from the established result for $L$ using $\la v,Lu \ra_H = \la L^*v, u \ra_H$ for all $u\in D(L)$ and $v\in D(L^*)$.
\qed

In particular, Lemma~\ref{Lem:LowerBoundness} implies that (strong) solutions to the IBVP and its adjoint are unique. Since $L$ and $L^*$ are closable operators~\cite{ReedSimon80I}, their closures $\overline{L}$ and $\overline{L^*}$ satisfy the same inequalities as in Equation~(\ref{Eq:LLstarIneq}). Now we are ready to define weak solutions and to prove their existence:

\begin{definition}
$u\in X$ is called a \textbf{weak solution} of the problem~(\ref{Eq:SymHypStrongForm}) if
\begin{displaymath}
\la L^*v, u \ra_H = -\la v, F \ra_H
\end{displaymath}
for all $v\in D(L^*)$. 
\end{definition}

In order to prove the existence of such $u\in X$, we introduce the linear space $Y = D(\overline{L^*})$ and equip it with the scalar product $\la\cdot,\cdot\ra_Y$ defined by
\begin{displaymath}
\la v,w\ra_Y := \la \overline{L^*}v,\overline{L^*}w \ra_H,\qquad v,w\in Y.
\end{displaymath}
The positivity of this product is a direct consequence of Lemma~\ref{Lem:LowerBoundness}, and since $\overline{L^*}$ is closed, $Y$ defines a Hilbert space. Next, we define the linear form $J: Y \to \Complex$ on $Y$ by
\begin{displaymath}
J(v) := -\la F , v \ra_H.
\end{displaymath}
This form is bounded, according to Lemma~\ref{Lem:LowerBoundness}, $|J(v)| \leq \| F \|_H \| v \|_H \leq \delta^{-1} \| F \|_H \| \overline{L^*} v \|_H = \delta^{-1} \| F \|_H \| v \|_Y$ for all $v\in Y$. Therefore, according to the Riesz representation lemma there exists a unique $w\in Y$ such that $\la w, v \ra_Y = J(v)$ for all $v\in Y$. Setting $u:=\overline{L^*} w\in X$ gives a weak solution of the problem.

If $u\in X$ is a weak solution which is sufficiently smooth, it follows from the Green type identity~(\ref{Eq:GreenIdentity}) that $u$ has vanishing initial data and that it satisfies the required boundary conditions, and hence is a solution to the original IBVP~(\ref{Eq:SymHypStrongForm}). The difficult part is to show that a weak solution is indeed sufficiently regular for this conclusion to be made. See Refs.~\cite{kF58,pLrP60,jRfM74,jR85,pS96a} for such ``weak=strong'' results.

%===================================================================
\subsection{Absorbing boundary conditions}
\label{section:Absorbing}
%===================================================================

When modeling isolated systems, the boundary conditions have to be chosen such that they minimize spurious reflections from the boundary surface. This means that inside the computational domain, the solution of the IBVP should lie as close as possible to the true solution of the Cauchy problem on the unbounded domain. In this sense, the dynamics outside the computational domain is replaced by appropriate conditions on a finite, artificial boundary. Clearly, this can only work in particular situations, where the solutions outside the domain are sufficiently simple so that they can be computed and used to construct boundary conditions which are, at least, approximately compatible with them. Boundary conditions which give rise to a well-posed IBVP and achieve this goal are called \textbf{absorbing}, \textbf{non-reflecting} or \textbf{radiation} boundary conditions in the literature, and there has been a substantial amount of work on the construction of such conditions for wave problems in acoustics, electromagnetism, meteorology, and solid geophysics (see \cite{dG91} for a review). Some recent applications to General Relativity are mentioned in Sections~\ref{section:ibvpEinstein} and \ref{SubSubSec:AbsorbingNum}.

One approach in the construction of absorbing boundary conditions is based on suitable series or Fourier expansions of the solution, and derives a hierarchy of \emph{local} boundary conditions with increasing order of accuracy~\cite{bEaM77,aBeT80,rH86}. Typically, such higher order local boundary conditions involve solving differential equations at the boundary surface, where the order of the differential equation is increasing with the order of the accuracy. This problem can be dealt with by introducing auxiliary variables at the boundary surface~\cite{dG01,dGbN03}.

The starting point for a slightly different approach is an \emph{exact nonlocal} boundary condition which involves the convolution with an appropriate integral kernel. A method based on an efficient approximation of this integral kernel is then implemented, see for instance Ref.~\cite{bAlGtH2000,bAlGtH2002} for the case of the 2D and 3D flat wave equations and Refs.~\cite{sL04a,sL04b,sL05} for the Regge--Wheeler~\cite{tRjW57} and Zerilli~\cite{fZ70a} equations describing linear gravitational waves on a Schwarzschild background. Although this method is robust, very accurate and stable, it is based on detailed knowledge of the solutions which might not always be available in more general situations.

In the following, we illustrate some aspects of the problem of constructing absorbing boundary conditions on some simple examples~\cite{oS07}. Specifically, we construct local absorbing boundary conditions for the wave equation with a spherical outer boundary at radius $R > 0$.

\subsubsection{The one-dimensional wave equation}

Consider first the one-dimensional case,
\begin{equation}
u_{tt} - u_{xx} = 0,\qquad
|x| < R,\quad t > 0.
\label{Eq:1DWaveEq}
\end{equation}
The general solution is a superposition of a left- and a right-moving solution,
\begin{displaymath}
u(t,x) = f_{\nwarrow}(x+t) + f_{\nearrow}(x-t).
\end{displaymath}
Therefore, the boundary conditions
\begin{equation}
(b_- u)(t,-R) = 0, \qquad
(b_+ u)(t,+R) = 0, \qquad 
b_\pm := \frac{\partial}{\partial t} \pm \frac{\partial}{\partial x},
\qquad t > 0,
\label{Eq:1DWaveBC}
\end{equation}
are perfectly absorbing according to our terminology. Indeed, the operator $b_+$ has as its kernel the right-moving solutions $f_{\nearrow}(x-t)$, hence the boundary condition $(b_+ u)(t,R) = 0$ at $x=R$ is transparent to these solutions. On the other hand, $b_+ f_{\nwarrow}(t+x) = 2f'_{\nwarrow}(t+x)$, which implies that at $x=R$, the boundary condition requires that $f_{\nwarrow}(v) = f_{\nwarrow}(1)$ is constant for advanced time $v = t+x > R$. A similar argument shows that the left boundary condition $(b_-u)(t,-R) = 0$ implies that $f_{\nearrow}(-u) = f_{\nearrow}(-R)$ is constant for retarded time $u = t - x > R$. Together with initial conditions for $u$ and its time derivative at $t=0$ satisfying the compatibility conditions, Equations~(\ref{Eq:1DWaveEq}, \ref{Eq:1DWaveBC}) give rise to a well-posed IBVP. In particular, the solution is identically zero after one crossing time $t\geq 2R$ for initial data which are compactly supported inside the interval $(-R,R)$.

\subsubsection{The three-dimensional wave equation}

Generalizing the previous example to higher dimensions is a nontrivial task. This is due to the fact that there are infinitely many propagation directions for outgoing waves, and not just two as in the one-dimensional case. Ideally, one would like to control all the propagation directions $k$ which are outgoing at the boundary ($k\cdot n > 0$, where $n$ is the unit outward normal to the boundary), but this is obviously difficult. Instead, one can try to control \emph{specific} directions (starting with the one that is normal to the outer boundary). Here, we illustrate the method of Ref.~\cite{aBeT80} on the three-dimensional wave equation,
\begin{equation}
u_{tt} - \Delta u = 0,\qquad
|x| < R,\quad t > 0.
\label{Eq:3DWaveEq}
\end{equation}
The general solution can be decomposed into spherical harmonics $Y^{\ell m}$ according to
\begin{displaymath}
u(t,r,\vartheta,\varphi)
 = \frac{1}{r}\sum\limits_{\ell=0}^\infty\sum\limits_{m = -\ell}^{\ell}
    u_{\ell m}(t,r) Y^{\ell m}(\vartheta,\varphi)
\end{displaymath}
which yields the family of reduced equations
\begin{equation}
\left[ \frac{\partial^2}{\partial t^2} - \frac{\partial^2}{\partial r^2} 
 + \frac{\ell(\ell+1)}{r^2} \right] u_{\ell m}(t,r) = 0, \qquad
0 < r < R,\quad t > 0.
\label{Eq:3DWaveReduced}
\end{equation}
For $\ell=0$ this equation reduces to the one-dimensional wave equation, for which the general solution is $u_{00}(t,r) = U_{00\nearrow}(r - t) + U_{00\nwarrow}(r + t)$ with $U_{00\nearrow}$ and $U_{00\nwarrow}$ two arbitrary functions. Therefore, the boundary condition
\begin{equation}
{\cal B}_0: \qquad \left. b(r u) \right|_{r=R} = 0,\qquad
b := r^2\left( \frac{\partial}{\partial t} + \frac{\partial}{\partial r} \right),\qquad t > 0,
\end{equation}
is perfectly absorbing for spherical waves. For $\ell\geq 1$, exact solutions can be generated from the solutions for $\ell=0$ by applying suitable differential operators to $u_{00}(t,r)$. For this, we define the operators~\cite{wB71}
\begin{displaymath}
a_\ell \equiv \frac{\partial}{\partial r} + \frac{\ell}{r}\, ,\qquad
a_\ell^\dagger \equiv -\frac{\partial}{\partial r} + \frac{\ell}{r}
\end{displaymath}
which satisfy the operator identities
\begin{displaymath}
a_{\ell+1} a_{\ell+1}^\dagger = a_\ell^\dagger a_\ell 
 = -\frac{\partial^2}{\partial r^2} + \frac{\ell(\ell+1)}{r^2}\; .
\end{displaymath}
As a consequence, for each $\ell=1,2,3,\ldots$, we have
\begin{eqnarray}
\left[ \frac{\partial^2}{\partial t^2}
 - \frac{\partial^2}{\partial r^2} + \frac{\ell(\ell+1)}{r^2} \right]
a_\ell^\dagger a_{\ell-1}^\dagger \ldots a_1^\dagger
 &=& \left[ \frac{\partial^2}{\partial t^2} + a_{\ell}^\dagger a_{\ell} \right]
   a_\ell^\dagger a_{\ell-1}^\dagger \ldots a_1^\dagger
\nonumber\\
 &=& a_\ell^\dagger \left[ \frac{\partial^2}{\partial t^2} 
 + a_{\ell-1}^\dagger a_{\ell-1} \right] a_{\ell-1}^\dagger \ldots a_1^\dagger
\nonumber\\
 &=& a_\ell^\dagger a_{\ell-1}^\dagger \ldots a_1^\dagger
     \left[ \frac{\partial^2}{\partial t^2} - \frac{\partial^2}{\partial r^2} \right].
\nonumber
\end{eqnarray}
Therefore, we have the explicit in- and outgoing solutions
\begin{eqnarray}
u_{\ell m\nwarrow}(t,r)
 &=& a_\ell^\dagger a_{\ell-1}^\dagger \ldots a_1^\dagger V_{\ell m}(r+t)
  = \sum\limits_{j=0}^\ell (-1)^j\frac{(2\ell-j)!}{(\ell-j)!\, j!} 
   (2r)^{j-\ell} V_{\ell m}^{(j)}(r+t),
\nonumber\\
u_{\ell m\nearrow}(t,r)
 &=& a_\ell^\dagger a_{\ell-1}^\dagger \ldots a_1^\dagger U_{\ell m}(r-t)
  = \sum\limits_{j=0}^\ell (-1)^j\frac{(2\ell-j)!}{(\ell-j)!\, j!} 
   (2r)^{j-\ell} U_{\ell m}^{(j)}(r-t),
\label{Eq:OutgoingFlatSolution}
\end{eqnarray}
where $V_{\ell m}$ and $U_{\ell m}$ are arbitrary smooth functions with $j$'th derivatives $V_{\ell m}^{(j)}$ and $U_{\ell m}^{(j)}$, respectively. In order to construct boundary conditions which are perfectly absorbing for $u_{\ell m}$, one first notices the following identity:
\begin{equation}
b^{\ell+1} a_\ell^\dagger a_{\ell-1}^\dagger \ldots a_1^\dagger 
U(r-t) = 0
\end{equation}
for all $\ell=0,1,2,\ldots$ and all sufficiently smooth functions $U$. This identity follows easily from Equation~(\ref{Eq:OutgoingFlatSolution}) and the fact that $b^{\ell+1}(r^{k}) = k(k+1)\cdot\cdot\cdot (k+\ell) r^{k+\ell+1} = 0$ if $k\in \{ 0,-1,-2,\ldots,-\ell \}$. Therefore, given $L\in \{ 1,2,3,\ldots\}$, the boundary condition
\begin{equation}
{\cal B}_L: \qquad \left. b^{L+1}(r u) \right|_{r=R} = 0,
\label{Eq:BCBL}
\end{equation}
leaves the outgoing solutions with $\ell \leq L$ unaltered. Notice that this condition is \emph{local} in the sense that its formulation does not require the decomposition of $u$ into spherical harmonics. Based on the Laplace method, it was proven in \cite{aBeT80} (see also Ref.~\cite{mRoRoS07}) that each boundary condition ${\cal B}_L$ yields a well posed IBVP. By uniqueness this implies that initial data corresponding to a purely outgoing solution with $\ell \leq L$ yields a purely outgoing solution (without reflections). In this sense, the condition ${\cal B}_L$ is \emph{perfectly absorbing for waves with $\ell \leq L$}. For waves with $\ell > L$, one obtains spurious reflections; however, for monochromatic radiation with wave number $k$, the corresponding amplitude reflection coefficients can be calculated to decay as $(k R)^{-2(L+1)}$ in the wave zone $k R \gg 1$~\cite{lBoS06}. Furthermore, in most scenarios with smooth solutions, the amplitudes corresponding to the lower few $\ell$'s will dominate over the ones with high $\ell$ so that reflections from high $\ell$'s are unimportant. For a numerical implementation of the boundary condition ${\cal B}_2$ via spectral methods and a possible application
to General Relativity see Ref. \cite{jNsB04}.

\subsubsection{The wave equation on a curved background}

When the background is curved, it is not always possible to construct in- and outgoing solutions explicitly, as in the previous example. Therefore, it is not even clear how a hierarchy of absorbing boundary conditions should be formulated. However, in many applications the spacetime is asymptotically flat, and if the boundary surface is placed sufficiently far from the strong field region, one can assume that the metric is a small deformation of the flat, Minkowski metric. To first order in $M/R$ with $M$ the ADM mass and $R$ the areal radius of the outer boundary, these correction terms are given by those of the Schwarzschild metric, and approximate in- and outgoing solutions for all $(\ell , m)$ modes can again be computed~\cite{oS07}.\epubtkFootnote{A systematic derivation of exact solutions including the correction terms in $M/R$ of arbitrarily high order in the Schwarzschild metric was given in Ref.~\cite{jBwP73} in a different context. However, a generic asymptotically flat spacetime is not expected to be Schwarzschild beyond the order of $M/R$, since higher-correction terms involve, for example, the total angular momentum. See Ref.~\cite{eDjS11} for the behavior of linearized gravitational perturbations of the Kerr spacetime near future null infinity.} The $M/R$ terms in the background metric induce two kind of corrections in the in- and outgoing solutions $u_{\ell m\nwarrow}$. The first is a curvature correction term which just adds $M/R$ terms to the coefficients in the sum of Equation~(\ref{Eq:OutgoingFlatSolution}). This term is local and still obeys Huygens' principle. The second term is fast decaying (it decays as $R/r^{\ell+1}$) and describes the backscatter off the curvature of the background. As a consequence, it is nonlocal (it depends on the past history of the unperturbed solution) and violates Huygens' principle.

By construction, the boundary conditions ${\cal B}_L$ are perfectly absorbing for outgoing waves with angular momentum number $\ell\leq L$, including their curvature corrections to first order in $M/R$. If the first order correction terms responsible for the backscatter are taken into account, then ${\cal B}_L$ are not perfectly absorbing anymore, but the spurious reflections arising from these correction terms have  been estimated in \cite{oS07} to decay at least as fast as $(M/R) (kR)^{-2}$ for monochromatic waves with wave number $k$ satisfying $M\ll k^{-1} \ll R$.

The well-posedness of higher order absorbing boundary conditions for wave equations on a curved background can be established by assuming the localization principle and the Laplace method~\cite{mRoRoS07}. Some applications to General Relativity are discussed in Sections~\ref{section:ibvpEinstein} and~\ref{SubSubSec:AbsorbingNum}.

\newpage

%===================================================================
%===================================================================
\section{Boundary Conditions for Einstein's Equations}
\label{section:ibvpEinstein}
%===================================================================
%===================================================================

The subject of this section is the discussion of the initial-boundary value problem (IBVP) for Einstein's field equations. There are, at least, three difficulties when formulating Einstein's equations on a finite domain with artificial outer boundaries. First, as we have seen in Section~\ref{section:IVFEinstein}, the evolution equations are subject to constraints which in general propagate with nontrivial characteristic speeds. As a consequence, in general there are incoming constraint fields at the boundary that need to be controlled in order to make sure that the constraints propagate correctly, i.e., that constraint satisfying initial data yields a solution of the evolution equations \emph{and} the constraints on the complete computational domain, and not just on its domain of dependence. The control of these incoming constraint fields leads to constraint-preserving boundary conditions, and a nontrivial task is to fit these conditions into one of the admissible boundary conditions discussed in the previous section, for which well posedness can be shown.

A second issue is the construction of absorbing boundary conditions. Unlike the simple examples considered in Section~\ref{section:Absorbing}, for which the fields evolve on a \emph{fixed background} and in- and outgoing solutions can be represented explicitly, or at least characterized precisely, in General Relativity it is not even clear how to define in- and outgoing gravitational radiation since there are no local expressions for the gravitational energy density and flux. Therefore, the best one can hope for is to construct boundary conditions which approximately control the incoming gravitational radiation in certain regimes, like, for example, in the weak field limit where the field equations can be linearized around, say, a Schwarzschild or Minkowski spacetime. 

Finally, the third issue is related to the diffeomorphism invariance of the theory. Ideally, one would like to formulate a geometric version of the IBVP, for which the data given on the initial and boundary surfaces $\Sigma_0$ and ${\cal T}$ can be characterized in terms of geometric quantities such as the first and second fundamental forms of these surfaces as embedded in the yet unknown spacetime $(M,g)$. In particular, this means that one should be able to identify equivalent data sets, i.e., those which are related to each other by a diffeomorphism of $M$ leaving $\Sigma_0$ and ${\cal T}$ invariant, by local transformations on $\Sigma_0$ and ${\cal T}$, without knowing the solution $(M,g)$. It is currently not even clear if such a geometric uniqueness property does exist, see Refs.~\cite{hF09,oRoS10} for further discussions on these points.

A well posed IBVP for Einstein's vacuum field equations was first formulated by Friedrich and Nagy~\cite{hFgN99} based on a tetrad formalism which incorporates the Weyl curvature tensor as an independent field. This formulation exploits the freedom of choosing local coordinates and the tetrad orientation in order to impose very precise gauge conditions which are adapted to the boundary surface ${\cal T}$ and tailored to the IBVP. These gauge conditions, together with a suitable modification of the evolution equations for the Weyl curvature tensor using the constraints (cf.\ Example~\ref{Example:Weyl}), lead to a first order symmetric hyperbolic system in which all the constraint fields propagate tangentially to ${\cal T}$ at the boundary. As a consequence, no constraint-preserving boundary conditions need to be specified, and the only incoming fields are related to the gravitational radiation, at least in the context of the approximations mentioned above. With this, the problem can be shown to be well posed using the techniques described in Section~\ref{section:MaxDiss}.

After the pioneering work of Ref.~\cite{hFgN99}, there has been much effort in formulating a well posed IBVP for metric formulations of general relativity, on which most numerical calculations are based. However, with the exception of particular cases in spherical symmetry~\cite{mIoR02}, the linearized field equations~\cite{gNoS06} or the restriction to flat, totally reflecting boundaries~\cite{bSbSjW02,bSjW03,gCjPoRoSmT03,gC-PhD-03,cGjM04a,cGjM04b,nT-PhD-05,dAnT06,aA08}, not much progress had been made towards obtaining a manifestly well posed IBVP with nonreflecting, constraint-preserving boundary conditions. The difficulties encountered were similar to those described in Examples~\ref{Example:MaxwellLorentz} and \ref{Example:Weyl}. Namely, controlling the incoming constraint fields usually resulted in boundary conditions for the main system involving either derivatives of its characteristic fields or fields propagating with zero speed, when it was written in first order symmetric hyperbolic form. Therefore, the theory of maximal dissipative boundary conditions could not be applied in these attempts. Instead, boundary conditions controlling the incoming characteristic constraint fields were specified and combined with more or less ad hoc conditions controlling the gauge and gravitational degrees of freedom and verified to satisfy the Lopatinsky condition~(\ref{Eq:LopatinskyCondition}) using the Laplace method, see Refs.~\cite{jS98,gCoS03,oSmT05,cGjM04b,oR06,mRdHsB10}. 

The breakthrough in the metric case came with the work by Kreiss and Winicour~\cite{hKjW06} who formulated a well-posed IBVP for the linearized Einstein vacuum field equations with harmonic coordinates. Their method is based on the pseudo-differential first order reduction of the wave equation described in Section~\ref {SubSubSec:LaplaceSecondOrder}, which, when combined with Sommerfeld boundary conditions, yields a problem which is strongly well posed in the generalized sense and, when applied to systems of equations, allows a certain hierarchical coupling in the boundary conditions. This work was then generalized to shifted wave equations and higher order absorbing boundary conditions in~\cite{mRoRoS07}. Later, it was recognized that the results in~\cite{hKjW06} could also be established based on the usual a priori energy estimates based on integration by parts~\cite{hKoRoSjW07}. Finally, it was found that the boundary conditions imposed were actually maximal dissipative for a specific nonstandard class of first order symmetric hyperbolic reduction of the wave system, see Section~\ref{SubSubSec:MaxDissWaveSystems}. Unlike the reductions considered in earlier work, such nonstandard class has the property that the boundary surface is noncharacteristic, which implies that no zero speed fields are present, and yields a strong well posed system. Based on this reduction and the theory of quasilinear symmetric hyperbolic formulations with maximal dissipative boundary conditions~\cite{oG90,pS96b}, it was possible to extend the results in~\cite{hKjW06,hKoRoSjW07} and formulate a well-posed IBVP for quasilinear systems of wave equations \cite{hKoRoSjW09} with a certain class of boundary conditions (see Theorem~\ref{Thm:IBVPWaveSystems}) which was sufficiently flexible to treat the Einstein equations. Furthermore, the new reduction also offers the interesting possibility to extend the proof to the discretized case using finite difference operators  satisfying the \emph{summation by parts} property, discussed in Sections~\ref{sec:sbp} and \ref{sec:gauss_and_sbp}.

In order to parallel the presentation in Section~\ref{section:IVFEinstein}, here we focus on the IBVP for Einstein's equations in generalized harmonic coordinates and the IBVP for the BSSN system. The first case, which is discussed in Section~\ref{SubSec:HarmonicBC}, is an application of Theorem~\ref{Thm:IBVPWaveSystems}. In the BSSN case, only partial results have been obtained so far, but since the BSSN system is widely used, we nevertheless present some of these results in Section~\ref{SubSec:BSSNBC}. In Section~\ref{SubSec:GeometricUniqueness} we discuss some of the problems encountered when trying to formulate a geometric uniqueness theorem and, finally, in Section~\ref{SubSec:OutToInfinity} we briefly mention alternative approaches to the IBVP which do not require an artificial boundary.

For an alternative approach to treating the IBVP which is based on the imposition of the Gauss-Codazzi equations at ${\cal T}$, see Refs.~\cite{sFrG03a,sFrG03b,sFrG04a,sFrG04b}. For numerical studies, see Refs.~\cite{mIoR02,gClLmT02,jBlB02,bSbSjW02,bSjW03,gC-PhD-03,lLmSlKhPdSsT04,mHlLrOhPmSlK04,oSmT05,lKlLmSlBhP05,cBtLcPmZ05,oR-PhD-05,mBbSjW06a,mBhKjW06,mRdHsB10,cBcB10a,cBcB10b}, especially Refs.~\cite{oRlLmS07} and \cite{mRoRoS07} for a comparison between different boundary conditions used in numerical relativity and Ref.~\cite{oRlBmShP09} for a numerical implementation of higher absorbing boundary conditions. For review articles on the IBVP in general relativity see Refs.~\cite{oS07,oRoS10,jW12}.

At present, there are no numerical simulations which are based directly on the well posed IBVP for the tetrad formulation~\cite{hFgN99} or the well posed IBVP for the harmonic formulation~\cite{hKjW06,hKoRoSjW07,hKoRoSjW09} described in Section~\ref{SubSec:HarmonicBC}, nor is there a numerical implementation of the constraint-preserving boundary conditions for the BSSN system presented in Section~\ref{SubSec:BSSNBC}. The closest example is the harmonic approach described in~\cite{lLmSlKrOoR06,oR06,oRlLmS07} which has been shown to be well posed in the generalized sense in the high frequency limit~\cite{mRoRoS07}. However, as mentioned above, the well posed IBVP in~\cite{hKoRoSjW09} opens the door for a numerical discretization based on the energy method which can be proven to be stable, at least in the linearized case.

%===================================================================
\subsection{The harmonic formulation}
\label{SubSec:HarmonicBC}
%===================================================================

Here, we discuss the IBVP formulated in Ref.~\cite{hKoRoSjW09} for the Einstein vacuum equations in generalized harmonic coordinates. The starting point is a manifold of the form $M = [0,T]\times \Sigma$, with $\Sigma$ a three-dimensional compact manifold with $C^\infty$-boundary $\partial\Sigma$, and a given, fixed smooth background metric $\gz_{\alpha\beta}$ with corresponding Levi-Civita connection $\nablaz$, as in Section~\ref{SubSec:Harmonic}. We assume that the time slices $\Sigma_t := \{ t \} \times \Sigma$ are space-like and that the boundary surface ${\cal T} := [0,T]\times\partial\Sigma$ is time-like with respect to $\gz_{\alpha\beta}$.

In order to formulate the boundary conditions, we first construct a null tetrad $\{ K^\mu,L^\mu,Q^\mu,\bar{Q}^\mu \}$ which is adapted to the boundary. This null tetrad is based on the choice of a future-directed time-like vector field $T^\mu$ tangent to ${\cal T}$ which is normalized such that $g_{\mu\nu} T^\mu T^\nu = -1$. One possible choice is to tie $T^\mu$ to the foliation $\Sigma_t$, and the define it in the direction orthogonal to the cross sections $\partial\Sigma_t = \{ t \} \times \partial\Sigma$ of the boundary surface. A more geometric choice has been proposed in~\cite{hF09}, where instead $T^\mu$ is chosen as a distinguished future-directed time-like eigenvector of the second fundamental form of ${\cal T}$, as embedded in $(M,g)$. Next, we denote by $N^\mu$ the unit outward normal to ${\cal T}$ with respect to the metric $g_{\mu\nu}$ and complete $T^\mu$ and $N^\mu$ to an orthonormal basis $\{ T^\mu,N^\mu,V^\mu,W^\mu \}$ of $T_p M$ at each point $p\in{\cal T}$. Then, we define the complex null tetrad by
\begin{displaymath}
K^\mu := T^\mu + N^\mu, \qquad
L^\mu := T^\mu - N^\mu, \qquad
Q^\mu := V^\mu + i\, W^\mu, \qquad
\bar{Q}^\mu := V^\mu - i\, W^\mu,
\end{displaymath}
where $i = \sqrt{-1}$. Notice that the construction of these vectors is implicit, since it  depends on the dynamical metric $g_{\alpha\beta}$ which is yet unknown. However, the dependency is algebraic, and does not involve any derivatives of $g_{\alpha\beta}$. We also note that the complex null vector $Q^\mu$ is not unique since it can be rotated by an angle $\varphi\in\Real$, $Q^\mu\mapsto e^{i\varphi} Q^\mu$. Finally, we define a radial function $r$ on ${\cal T}$ as the areal radius of the cross sections $\partial\Sigma_t$ with respect to the background metric. 

Then, the boundary conditions which were proposed in~\cite{hKoRoSjW09} for the harmonic system~(\ref{Eq:EinsteinWave}) are:
\begin{eqnarray}
\left. \nablaz_K h_{KK} + \frac{2}{r} h_{KK} \right|_{\cal T} &=& q_K,
\label{Eq:KKK}\\
\left. \nablaz_K h_{KL} + \frac{1}{r} (h_{KL} + h_{Q\bar{Q}}) \right|_{\cal T} 
&=& q_L,
\label{Eq:KKL}\\
\left. \nablaz_K h_{KQ} + \frac{2}{r} h_{KQ} \right|_{\cal T} &=& q_Q,
\label{Eq:KKQ}\\
\left. \nablaz_K h_{QQ} - \nablaz_Q h_{QK} \right|_{\cal T} &=& q_{QQ},
\label{Eq:KQQ-QQK}\\
\left. \nablaz_K h_{Q\bar{Q}} + \nablaz_L h_{KK} - \nablaz_Q h_{K\bar{Q}}
 - \nablaz_{\bar{Q}} h_{KQ} \right|_{\cal T} &=& \left. -2H_K \right|_{\cal T},
\label{Eq:KQQbar}\\
\left. \nablaz_K h_{LQ} + \nablaz_L h_{KQ} - \nablaz_Q h_{KL}
 - \nablaz_{\bar{Q}} h_{QQ} \right|_{\cal T} &=& \left. -2H_Q \right|_{\cal T},
\label{Eq:KLQ}\\
\left. \nablaz_K h_{LL} + \nablaz_L h_{Q\bar{Q}} - \nablaz_Q h_{L\bar{Q}}
 - \nablaz_{\bar{Q}} h_{LQ} \right|_{\cal T} &=& \left. -2H_L \right|_{\cal T},
\label{Eq:KLL}
\end{eqnarray}
where $\nablaz_K h_{LQ} := K^\mu L^\alpha Q^\beta \nablaz_\mu h_{\alpha\beta}$, $h_{KL} := K^\alpha L^\beta h_{\alpha\beta}$, $H_K := K^\mu H_\mu$, etc., and where $q_K$ and $q_L$ are real-valued given smooth functions on ${\cal T}$ and $q_Q$ and $q_{QQ}$ are complex-valued given smooth functions on ${\cal T}$. Since $Q$ is complex, these constitute ten real boundary conditions for the metric coefficients $h_{\alpha\beta}$. The content of the boundary conditions~(\ref{Eq:KKK}, \ref{Eq:KKL}, \ref{Eq:KKQ}, \ref{Eq:KQQ-QQK}) can be clarified by considering linearized gravitational waves on a Minkowski background with a spherical boundary. The analysis in~\cite{hKoRoSjW09} shows that in this context the four real conditions~(\ref{Eq:KKK}),(\ref{Eq:KKL}, \ref{Eq:KKQ}) are related to the gauge freedom; and the two conditions~(\ref{Eq:KQQ-QQK}) control the gravitational radiation. The remaining conditions~(\ref{Eq:KQQbar}, \ref{Eq:KLQ}, \ref{Eq:KLL}) enforce the constraint $C^\mu=0$ on the boundary, see Equation~(\ref{Eq:HarmonicConstraintBis}), and so together with the constraint propagation system~(\ref{Eq:EinsteinWaveCP}) and the initial constraints~(\ref{Eq:EinsteinWaveConstraints}) they guarantee that the constraints are correctly propagated. Based on these observations, it is expected that these boundary conditions yield small spurious reflections in the case of a nearly spherical boundary in the wave zone of an asymptotically flat curved spacetime.

\subsubsection{Well posedness of the IBVP}

The IBVP consisting of the harmonic Einstein equations~(\ref{Eq:EinsteinWave}), initial data~(\ref{Eq:EinsteinWaveID}) and the boundary conditions~(\ref{Eq:KKK}--\ref{Eq:KLL}) can be shown to be well posed as an application of Theorem~\ref{Thm:IBVPWaveSystems}. For this, we first notice that the evolution equations~(\ref{Eq:EinsteinWave}) have the required form of Equation~(\ref{Eq:WaveSystemEq}), where $E$ is the vector bundle of symmetric, covariant tensor fields $h_{\mu\nu}$ on $M$. Next, the boundary conditions can be written in the form of Equation~(\ref{Eq:WaveSystemBC}) with $\alpha = 1$. In order to compute the matrix coefficients $c^{\mu\, A}{}_B$, it is convenient to decompose $h_{\mu\nu} = h^A e_{A\, \mu\nu}$ in terms of the basis vectors
\begin{eqnarray}
&& e_{1\, \alpha\beta} := K_\alpha K_\beta,\quad
     e_{2\, \alpha\beta} := -2 K_{(\alpha}\bar{Q}_{\beta)},\quad
     e_{3\, \alpha\beta} := -2 K_{(\alpha} Q_{\beta)},\quad
     e_{4\, \alpha\beta} := 2 Q_{(\alpha}\bar{Q}_{\beta)},     
\nonumber\\
&& e_{5\, \alpha\beta}:= \bar{Q}_\alpha\bar{Q}_\beta,\quad
     e_{6\, \alpha\beta}:= Q_\alpha Q_\beta,
\nonumber\\
&& e_{7\, \alpha\beta} := -2 L_{(\alpha}\bar{Q}_{\beta)},\quad
     e_{8\, \alpha\beta}:= -2 L_{(\alpha} Q_{\beta)},\quad
     e_{9\, \alpha\beta}:= 2 K_{(\alpha} L_{\beta)},\quad
     e_{10\, \alpha\beta} := L_\alpha L_\beta,
\nonumber
\end{eqnarray}
with $h^1 = h_{LL}/4$, $h^2 = \bar{h}^3 = h_{LQ}/4$, $h^4 = h_{Q\bar{Q}}/4$, $h^5 = \bar{h}^6 = h_{QQ}/4$, $h^7 = \bar{h}^8 = h_{KQ}/4$, $h^9 = h_{KL}/4$, $h^{10} = h_{KK}/4$. With respect to this basis, the only nonzero matrix coefficients are
\begin{eqnarray}
&& c^{\mu\, 1}{}_2 = \bar{Q}^\mu,\quad
     c^{\mu\, 1}{}_3 = Q^\mu,\quad
     c^{\mu\, 1}{}_4 = -L^\mu, 
\nonumber\\
&& c^{\mu\, 2}{}_5 = \bar{Q}^\mu,\quad
     c^{\mu\, 2}{}_7 = -L^\mu,\quad
     c^{\mu\, 2}{}_9 = Q^\mu, 
\nonumber\\
&& c^{\mu\, 3}{}_6 = Q^\mu,\quad
     c^{\mu\, 3}{}_8 = -L^\mu,\quad
     c^{\mu\, 3}{}_9 = \bar{Q}^\mu, 
\nonumber\\
&& c^{\mu\, 4}{}_7 = \bar{Q}^\mu,\quad
     c^{\mu\, 4}{}_8 = Q^\mu,\quad
     c^{\mu\, 4}{}_{10} = -L^\mu, 
\nonumber\\
&& c^{\mu\, 5}{}_7 = Q^\mu,\quad 
      c^{\mu\, 6}{}_8 = \bar{Q}^\mu,
\nonumber
\end{eqnarray}
which has the required upper triangular form with zeroes in the diagonal. Therefore, the hypothesis of Theorem~\ref{Thm:IBVPWaveSystems} are verified and one obtains a well posed IBVP for Einstein's equations in harmonic coordinates.

This result also applies the the modified system~(\ref{Eq:ReducedEinsteinWaveDamped}), since the constraint damping terms which are added do not modify the principal part of the main evolution system nor the one of the constraint propagation system.

%===================================================================
\subsection{Boundary conditions for BSSN}
\label{SubSec:BSSNBC}
%===================================================================

Here we discuss boundary conditions for the BSSN system~(\ref{Eq:BSSN1}--\ref{Eq:BSSN8}), which is used extensively in numerical calculations of spacetimes describing dynamical black holes and neutron stars. Unfortunately, to date, this system lacks an initial-boundary value formulation for which well posedness in the full nonlinear case has been proven. Without doubt the reason for this relies in the structure of the evolution equations, which are mixed first/second order in space and whose principal part is much more complicated than the harmonic case, where one deals with a system of wave equations.

A first step towards formulating a well posed IBVP for the BSSN system was performed in~\cite{hBoS04}, where the evolution equations~(\ref{Eq:BSSN1}, \ref{Eq:BSSN2}, \ref{Eq:BSSN5}--\ref{Eq:BSSN8}) with a fixed shift and the relation $f=\mu\equiv (4m - 1)/3$ were reduced to a first order symmetric hyperbolic system. Then, a set of six boundary conditions consistent with this system could be formulated based on the theory of maximal dissipative boundary conditions. Although this gives rise to a well posed IBVP, the boundary conditions specified in~\cite{hBoS04} are not compatible with the constraints, and therefore, one does not necessarily obtain a solution to the full set of Einstein's equations beyond the domain of dependence of the initial data surface. In a second step, constraint-preserving boundary conditions for BSSN with a fixed shift were formulated in~\cite{cGjM04b}, and cast into maximal dissipative form for the linearized system (see also Ref.~\cite{aA08}). However, even at the linearized level, these boundary conditions are too restrictive because they constitute a combination of Dirichlet and Neumann boundary conditions on the metric components, and in this sense they are totally reflecting instead of absorbing. More general constraint-preserving boundary conditions were also considered in~\cite{cGjM04b} and, based on the Laplace method, they were shown to satisfy the Lopatinsky condition~(\ref{Eq:LopatinskyCondition}).

Radiative-type constraint-preserving boundary conditions for the BSSN system~(\ref{Eq:BSSN1}--\ref{Eq:BSSN8}) with dynamical lapse and shift were formulated in~\cite{dNoS10} and shown to yield a well posed IBVP in the linearized case. The assumptions on the parameters in this formulation are $m=1$, $f > 0$, $\kappa = 4GH/3 > 0$, $f\neq\kappa$, which guarantee that the BSSN system is strongly hyperbolic, and as long as $e^{4\phi}\neq 2\alpha$ they allow for the gauge conditions~(\ref{Eq:StandardBSSNGaugeChoice1}, \ref{Eq:StandardBSSNGaugeChoice2}) used in recent numerical calculations, where $f = 2/\alpha$ and $\kappa = e^{4\phi}/\alpha^2$, see Section~\ref{SubSubSec:BSSNHyp}. In the following, we describe this IBVP in more detail. First, we notice that the analysis in Section~\ref{SubSubSec:BSSNHyp} reveals that for the standard choice $m=1$ the characteristic speeds with respect to the  unit outward normal $s_i$ to the boundary are
\begin{displaymath}
\beta^s, \qquad
\beta^s \pm \alpha, \qquad
\beta^s \pm \alpha\,\sqrt{f}, \qquad
\beta^s \pm \alpha\,\sqrt{G H},\qquad
\beta^s \pm \alpha\,\sqrt{\kappa},
\end{displaymath}
where $\beta^s = \beta^i s_i$ is the normal component of the shift. According to the theory described in Section~\ref{section:ibvp} it is the sign of these speeds which determines the number of incoming fields and boundary conditions that must be specified. Namely, the number of boundary conditions is equal to the number of characteristic fields with \emph{positive} speed. Assuming $|\beta^s|$ is small enough such that $|\beta^s/\alpha| < \min\{ 1,\sqrt{f},\sqrt{GH},\sqrt{\kappa} \}$, which is satisfied asymptotically if $\beta^s\to 0$ and $\alpha\to 1$, it is the sign of the normal component of the shift which determines the number of boundary conditions. Therefore, in order keep the number of boundary conditions fixed throughout evolution\epubtkFootnote{Notice that this is one of the requirements of Theorem~\ref{Thm:MaxDiss}, for instance, where the boundary matrix must have constant rank in order for the theorem to be applicable.} one has to ensure that either $\beta^s > 0$ or $\beta^s \leq 0$ at the boundary surface. If the condition $\beta^s\to 0$ is imposed asymptotically, the most natural choice is to set the normal component of the shift to zero at the boundary, $\beta^s = 0$ at ${\cal T}$. The analysis in~\cite{hBoS04} then reveals that there are precisely nine incoming characteristic fields at the boundary, and thus, nine conditions have to be imposed at the boundary. These nine boundary conditions are as follows:
\begin{itemize}
\item \textbf{Boundary conditions on the gauge variables}\\
There are four conditions that must be imposed on the gauge functions, namely the lapse and shift. These conditions are motivated by the linearized analysis, where the gauge propagation system, consisting of the evolution equations for lapse and shift obtained from the BSSN equations~(\ref{Eq:BSSN1}--\ref{Eq:BSSN4}, \ref{Eq:BSSN8}), decouples from the remaining evolution equations. Surprisingly, this gauge propagation system can be cast into symmetric hyperbolic form \cite{dNoS10}, for which maximal dissipative boundary conditions can be specified, as described in Section~\ref{section:MaxDiss}. It is remarkable that the gauge propagation system has such a nice mathematical structure, since the equations~(\ref{Eq:BSSN1}, \ref{Eq:BSSN3}, \ref{Eq:BSSN4}) have been specified by hand and mostly motivated by numerical experiments instead of mathematical analysis.

In terms of the operator $\Pi^i{}_j = \delta^i{}_j - s^i\,s_j$  projecting onto vectors tangential to the boundary, the four conditions on the gauge variables can be written as
\begin{eqnarray}
&& s^i\partial_i\alpha = 0,
\label{Eq:BSSNBC1}\\
&& \beta^s = 0,
\label{Eq:BSSNBC2}\\
&& \Pi^i{}_j\,\left(\partial_t + \frac{\sqrt{3\kappa}}2 s^k\partial_k\right)\beta^j
 =\frac{\kappa}{f-\kappa}\Pi^i{}_j\,\tilde{\gamma}^{jk}\partial_k\alpha.
\label{Eq:BSSNBC3}
\end{eqnarray}
Equation~(\ref{Eq:BSSNBC1}) is a Neumann boundary condition on the lapse, and Equation~(\ref{Eq:BSSNBC2}) sets the normal component of the shift to zero, as explained above. Geometrically, this implies that the boundary surface ${\cal T}$ is orthogonal to the time slices $\Sigma_t$. The other two conditions in Equation~(\ref{Eq:BSSNBC3}) are Sommerfeld-like boundary conditions involving the tangential components of the shift and the tangential derivatives of the lapse; they arise from the analysis of the characteristic structure of the gauge propagation system. An alternative to Equation~(\ref{Eq:BSSNBC3}) also described in~\cite{dNoS10} is to set the tangential components of the shift to zero, which, together with Equation~(\ref{Eq:BSSNBC2}) is equivalent to setting $\beta^i = 0$ at the boundary. This alternative may be better suited for IBVP with non-smooth boundaries, such as cubes, where additional compatibility conditions must be enforced at the edges.

\item \textbf{Constraint-preserving boundary conditions}\\
Next, there are three conditions requiring that the momentum constraint be satisfied at the boundary. In terms of the BSSN variables this implies
\begin{equation}
\tilde{D}^j \tilde{A}_{ij} - \frac{2}{3} \tilde{D}_i K + 6\tilde{A}_{ij} \tilde{D}^j\phi 
 = 8\pi G_N j_i.
\label{Eq:BSSNBC4}
\end{equation}
As shown in~\cite{dNoS10}, Equations~(\ref{Eq:BSSNBC4}) yields homogeneous maximal dissipative boundary conditions for a symmetric hyperbolic first order reduction of the constraint propagation system~(\ref{Eq:BSSNConsProp1}, \ref{Eq:BSSNConsProp2}, \ref{Eq:BSSNConsProp3}). Since this system is also linear and its boundary matrix has constant rank if $\beta^s = 0$, it follows from Theorem~\ref{Thm:MaxDiss} that the propagation of constraint violations is governed by a well posed IBVP. This implies, in particular, that solutions whose initial data satisfy the constraints exactly automatically satisfy the constraints on each time slice $\Sigma_t$. Furthermore, small initial constraint violations which are usually present in numerical applications yield solutions for which the growth of the constraint violations can be bounded in terms of the initial violations.

\item \textbf{Radiation controlling boundary conditions}\\
Finally, the last two boundary conditions are intended to control the incoming gravitational radiation, at least approximately, and specify the complex Weyl scalar $\Psi_0$, cf.\ Example~\ref{Example:Weyl}. In order to describe this boundary condition we first define the quantities ${\bar{\cal E}}_{ij} := \tilde{R}_{ij} + R^\phi_{ij} + e^{4\phi}\left( \frac{1}{3}\, K\tilde{A}_{ij} - \tilde{A}_{il}\,\tilde{A}^l_j\right) - 4\pi G_N\sigma_{ij}$ and ${\bar{\cal B}}_{kij} := e^{4\,\phi}\,\left[\tilde{D}_k\,\tilde{A}_{ij} - 4\,\left(\tilde{D}_{(i}\,\phi\right)\tilde{A}_{j)k}\right]$ which determine the electric and magnetic parts of the Weyl tensor through $E_{ij} = {\bar{\cal E}}_{ij} - \frac{1}{3}\gamma_{ij}\gamma^{kl}{\bar{\cal E}}_{kl}$ and $B_{ij} = \varepsilon_{kl(i} {\bar{\cal B}}^{kl}{}_{j)}$, respectively. Here, $\varepsilon_{kij}$ denotes the volume form with respect to the three metric $\gamma_{ij}$. In terms of the operator $P^{ij}{}_{lm} = \Pi^i{}_{(l}\Pi^j{}_{m)} - \frac{1}{2}\Pi^{ij}\Pi_{lm}$ projecting onto symmetric trace-less tangential tensors to the boundary, the boundary condition reads
\begin{equation}
P^{ij}{}_{lm}{\bar{\cal E}}_{ij}
 + \left(s^k P^{ij}{}_{lm} - s^i P^{kj}{}_{lm} \right){\bar{\cal B}}_{kij} = P^{ij}{}_{lm} G_{ij},
\label{Eq:BSSNBC5}
\end{equation}
with $G_{ij}$ a given smooth tensor field on the boundary surface ${\cal T}$. The relation between $G_{ij}$ and $\Psi_0$ is the following: if $n = \alpha^{-1}(\partial_t - \beta^i\partial_i)$ denotes the future-directed unit normal to the time slices, we may construct an adapted Newman-Penrose null tetrad $\{ K,L,Q,\bar{Q} \}$ at the boundary by defining $K := n + s$, $L := n - s$, and by choosing $Q$ to be a complex null vector orthogonal to $K$ and $L$, normalized such that $Q^\mu\bar{Q}_\mu = 2$. Then, we have $\Psi_0 = (E_{kl} - i B_{kl}) Q^k Q^l = G_{kl} Q^k Q^l$. For typical applications involving the modeling of isolated systems one may set $G_{ij}$ to zero. However, this in general is not compatible with the initial data (see the discussion in Section~\ref{sec:boundary_numrel}), an alternative is then to freeze $G_{ij}$ its value to the one computed from the initial data.

The boundary condition~(\ref{Eq:BSSNBC5}) can be partially motivated by considering an isolated system which, globally, is described by an asymptotically flat spacetime. Therefore, if the outer boundary is placed far away enough from the strong field region, one may linearize the field equations on a Minkowski background to a first approximation. In this case, one is in the same situation as in Example~\ref{Example:Weyl}, where the Weyl scalar $\Psi_0$ is an outgoing characteristic field when constructed from the adapted null tetrad. Furthermore, one can also appeal to the peeling behavior of the Weyl tensor \cite{rP65} in which $\Psi_0$ is the fastest decaying component along an outgoing null geodesics and describes the incoming radiation at past null infinity. While $\Psi_0$ can only be defined in an unambiguous way at null infinity, where a preferred null tetrad exists, the boundary condition~(\ref{Eq:BSSNBC5}) has been successfully numerically implemented and tested for truncated domains with artificial boundaries in the context of the harmonic formulation, see for example~\cite{oRlLmS07}. Estimates on the amount of spurious reflection introduced by this condition have also been derived in~\cite{lBoS06,lBoS07}, see also~\cite{eDjS11}.
\end{itemize}

%===================================================================
\subsection{Geometric existence and uniqueness}
\label{SubSec:GeometricUniqueness}
%===================================================================

The results mentioned so far concerning the well posed IBVP for Einstein's field equations in the tetrad formulation of \cite{hFgN99}, in the metric formulation with harmonic coordinates described in Section~\ref{SubSec:HarmonicBC}, or in the linearized BSSN formulation described in Section~\ref{SubSec:BSSNBC} allow, from the PDE point of view, to construct unique solutions on a manifold of the form $M = [0,T] \times \Sigma$ given appropriate initial and boundary data. However, since general relativity is a diffeomorphism invariant theory, one needs to pose the IBVP from a geometric perspective. In particular, the following questions arise, which, for simplicity, we only formulate for the vacuum case:
\begin{itemize}
\item \textbf{Geometric existence}. Let $(M,g)$ be any smooth solution of Einstein's vacuum field equations on the manifold $M = [0,T]\times\Sigma$ corresponding to initial data $(h,k)$ on $\Sigma_0$ and boundary data $\psi$ on ${\cal T}$, where $h$ and $k$ represent, respectively, the first and second fundamental forms of the initial surface $\Sigma_0$ as embedded in $(M,g)$. Is it possible to reproduce this solution with any of the well posed IBVP mentioned so far, at least on a submanifold $M' = [0,T']\times\Sigma$ with $0 < T' \leq T$? That is, does there exist initial data $f$ and boundary data $q$ for this IBVP and a diffeomorphism $\phi: M' \to \phi(M') \subset M$ which leaves $\Sigma_0$ and ${\cal T}'$ invariant, such that the metric constructed from this IBVP is equal to $\phi^* g$ on $M'$?
\item \textbf{Geometric uniqueness}. Is the solution $(M,g)$ uniquely determined by the data $(h,k,\psi)$? Given a well posed IBVP for which geometric existence holds, the question about geometric uniqueness can be reduced to the analysis of this particular IBVP in the following way: let $u_1$ and $u_2$ be two solutions of the IBVP on the manifold $M = [0,T]\times\Sigma$ with corresponding data $(f_1,q_1)$ and $(f_2,q_2)$. Suppose the two solutions induce the same data $(h,k)$ on $\Sigma_0$ and $\psi$ on ${\cal T}$. Does there exist a diffeomorphism $\phi: M' = [0,T']\times\Sigma \to \phi(M') \subset M$ which leaves $\Sigma_0$ and ${\cal T}'$ invariant, such that the metrics $g_1$ and $g_2$ corresponding to $u_1$ and $u_2$ are related to each other by $g_2 = \phi^* g_1$ on $M'$?
\end{itemize}

These geometric existence and uniqueness problems have been solved in the context of the Cauchy problem without boundaries, see~\cite{cBrG69} and Section~\ref{SubSubSec:ivpGeometric}. When boundaries are present, however, several new difficulties appear as pointed out in~\cite{hF09}, see also Refs.~\cite{hFgN99,hF05}:
\begin{enumerate}
\item[(i)] It is a priori not clear what the boundary data $\psi$ should represent geometrically. Unlike the case of the initial surface, where the data represents the first and second fundamental forms of $\Sigma_0$ as a spatial surface embedded in the constructed spacetime $(M,g)$, it is less clear what the geometric meaning of $\psi$ should be since it is restricted by the characteristic structure of the evolution equations, as discussed in Section~\ref{section:ibvp}.
\item[(ii)] The boundary data $(q_K,q_L,q_Q,q_{QQ})$ in the boundary conditions~(\ref{Eq:KKK}, \ref{Eq:KKL}, \ref{Eq:KKQ}, \ref{Eq:KQQ-QQK}) for the harmonic formulation and the boundary data $G_{ij}$ in the boundary condition~(\ref {Eq:BSSNBC5}) for the BSSN formulation ultimately depend on a specific choice of a future-directed time-like vector field $T$ at the boundary surface ${\cal T}$. Together with the unit outward normal $N$ to ${\cal T}$, this vector defines the preferred null directions $K = T + N$ and $L = T - N$ which are used to construct the boundary adapted null tetrad in the harmonic case and the projection operators $\Pi^\mu{}_\nu = \delta^\mu{}_\nu + T^\mu T_\nu - N^\mu N_\nu$ and $P^{\mu\nu}{}_{\alpha\beta} = \Pi^\mu{}_\alpha\Pi^\nu{}_\beta - \frac{1}{2}\Pi^{\mu\nu}\Pi_{\alpha\beta}$ in the BSSN one. Although it is tempting to define $T$ as the unit, future-directed time-like vector tangent to ${\cal T}$ which is orthogonal to the cross sections $\partial\Sigma_t$, this definition would depend on the particular foliation $\Sigma_t$ the formulation is based on, and so the resulting vector $T$ would be gauge-dependent. A similar issue arises in the tetrad formulation of Ref.~\cite{hFgN99}.
\item[(iii)] When addressing the geometric uniqueness issue, an interesting question is whether or not it is possible to determine \emph{from the data sets $(f_1,q_1)$ and $(f_2,q_2)$ alone} if they are equivalent in the sense that their solutions $u_1$ and $u_2$ induce the same geometric data $(h,k,\psi)$. Therefore, the question is whether or not one can identify equivalent data sets by considering only transformations on the initial and boundary surfaces $\Sigma_0$ and ${\cal T}$, without knowing the solutions $u_1$ and $u_2$.
\end{enumerate}

Although a complete answer to these questions remains a difficult task, there has been some recent progress towards their understanding. In~\cite{hF09} a method was proposed to geometrically single out a preferred time direction $T$ at the boundary surface ${\cal T}$. This is done by considering the trace-free part of the second fundamental form, and proving that under certain conditions which are stable under perturbations, the corresponding linear map on the tangent space possesses a unique time-like eigenvector. Together with the unit outward normal vector $N$, the vector field $T$ defines a distinguished adapted null tetrad at the boundary, from which geometrical meaningful boundary data could be defined. For instance, the complex Weyl scalar $\Psi_0$ can then be defined as the contraction $\Psi_0 = C_{\alpha\beta\gamma\delta} K^\alpha Q^\beta K^\gamma Q^\delta$ of the Weyl tensor $C_{\alpha\beta\gamma\delta}$ associated to the metric $g_{\mu\nu}$ along the null vectors $K$ and $Q$, and the definition is unique up to the usual spin rotational freedom $Q\mapsto e^{i\varphi} Q$, and therefore, the Weyl scalar $\Psi_0$ is a good candidate for forming part of the boundary data $\psi$.

In~\cite{oRoS10} it was suggested that the unique specification of a vector field $T$ may not be a fundamental problem, but rather the manifestation of the inability to specify a non-incoming radiation condition correctly. In the linearized case, for example, setting to zero the Weyl scalar $\Psi_0$ computed from the boundary adapted tetrad is transparent to gravitational plane waves traveling along the specific null direction $K = T + N$, see Example~\ref{Example:Weyl}, but it induces spurious reflections for outgoing plane waves traveling in other null direction. Therefore, a genuine non-incoming radiation condition should be, in fact, independent of any specific null or time-like direction at the boundary, and can only depend on the normal vector $N$. This is indeed the case for much simpler systems like the scalar wave equation on a Minkowski background~\cite{bEaM77}, where perfectly absorbing boundary conditions are formulated as a nonlocal condition which is independent of a preferred time direction at the boundary.

Aside from controlling the incoming gravitational degrees of freedom, the boundary data $\psi$ should also comprise information related to the geometric evolution of the boundary surface. In~\cite{hFgN99} this was achieved by specifying the mean curvature of ${\cal T}$ as part of the boundary data. In the harmonic formulation described in Section~\ref{SubSec:HarmonicBC} this information is presumably contained in the functions $q_K$, $q_L$ and $q_Q$, but their geometric interpretation is not clear.

In order to illustrate some of the issues related to the geometric existence and uniqueness problem in a simpler context, in what follows we analyze the IBVP for linearized gravitational waves propagating on a Minkowski background. Before analyzing this case, however, we make two remarks. First, it should be noted~\cite{hF09} that the geometric uniqueness problem, especially an understanding of point (iii), also has practical interest, since in long term evolutions it is possible that the gauge threatens to break down at some point, requiring a redefinition. The second remark concerns the formulation of the Einstein IBVP in generalized harmonic coordinates, described in Sections~\ref{SubSec:Harmonic} and \ref{SubSec:HarmonicBC}, where general covariance was maintained by introducing a background metric $\gz_{\mu\nu}$ on the manifold $M$. IBVPs based on this approach have been formulated in~\cite{mRoRoS07} and \cite{hKoRoSjW09} and further developed in~\cite{jW09a} and \cite{jW09b}. However, one has to emphasize that this approach does not automatically solve the geometric existence and uniqueness problems described here: although it is true that the IBVP is invariant with respect to any diffeomorphism $\phi: M \to M$ which acts on the dynamical \emph{and} the background metric at the same time, the question on the dependency of the solution on the background metric remains.

\subsubsection{Geometric existence and uniqueness in the linearized case}

Here we analyze some of the geometric existence and uniqueness issues of the IBVP for Einstein's field equations in the much simpler setting of linearized gravity on Minkowski space, where the vacuum field equations reduce to
\begin{equation}
-\nabla^\mu\nabla_\mu h_{\alpha\beta} - \nabla_\alpha\nabla_\beta h
 + 2\nabla^\mu\nabla_{(\alpha} h_{\beta)\mu} = 0,
\label{Eq:LinEinstein}
\end{equation}
where $h_{\alpha\beta}$ denotes the first variation of the metric, $h := \eta^{\alpha\beta} h_{\alpha\beta}$ its trace with respect to the Minkowski background metric $\eta_{\alpha\beta}$, and $\nabla_\mu$ is the covariant derivative with respect to $\eta_{\alpha\beta}$. An infinitesimal coordinate transformation parametrized by a vector field $\xi^\mu$ induces the transformation
\begin{equation}
h_{\alpha\beta}\mapsto 
\tilde{h}_{\alpha\beta} = h_{\alpha\beta} + 2\nabla_{(\alpha}\xi_{\beta)},
\label{Eq:LinEinsteinGT}
\end{equation}
where $\xi_\alpha := \eta_{\alpha\beta}\xi^\beta$.

Let us consider the linearized Cauchy problem without boundaries first, where initial data is specified at the initial surface $\Sigma_0 = \{ 0 \} \times \Real^3$. The initial data is specified geometrically by the first and second fundamental forms of $\Sigma_0$, which in the linearized case are represented by a pair $(h^{(0)}_{ij},k^{(0)}_{ij})$ of covariant symmetric tensor fields on $\Sigma_0$. We assume $(h^{(0)}_{ij},k^{(0)}_{ij})$ to be smooth and to satisfy the linearized Hamiltonian and momentum constraints
\begin{equation}
G^{ijrs}\partial_i\partial_j h^{(0)}_{rs} = 0,\qquad
G^{ijrs}\partial_j k^{(0)}_{rs} = 0,
\end{equation}
where $G^{ijrs} := \delta^{i(r}\delta^{s)j} - \delta^{ij}\delta^{rs}$. A solution $h_{\alpha\beta}$ of Equation~(\ref{Eq:LinEinstein}) with the induced data corresponding to $(h^{(0)}_{ij},k^{(0)}_{ij})$ up to a gauge transformation (\ref{Eq:LinEinsteinGT}) satisfies
\begin{equation}
 \left. h_{ij} \right|_{\Sigma_0} = h^{(0)}_{ij} + 2\partial_{(i} X_{j)},\qquad
\left. \partial_t h_{ij} - 2\partial_{(i} h_{j)0} \right|_{\Sigma_0} 
 = -2(k^{(0)}_{ij} + \partial_i \partial_j f),
\label{Eq:LinEinsteinID}
\end{equation}
where $X_j = \xi_j$ and $f = \xi_0$ are smooth and represent the initial gauge freedom. Then, one has:

\begin{theorem}
The initial-value problem~(\ref{Eq:LinEinstein}, \ref{Eq:LinEinsteinID}) possesses a smooth solution $h_{\alpha\beta}$ which is unique up to an infinitesimal coordinate transformation $\tilde{h}_{\alpha\beta} = h_{\alpha\beta} + 2\nabla_{(\alpha}\xi_{\beta)}$ generated by a vector field $\xi^\alpha$.
\end{theorem}

\proof We first show the existence of a solution in the linearized harmonic gauge $C_\beta = \nabla^\mu h_{\beta\mu} - \frac{1}{2}\nabla_\beta h = 0$, for which Equation~(\ref{Eq:LinEinstein}) reduces to the system of wave equations $\nabla^\mu\nabla_\mu h_{\alpha\beta} = 0$. The initial data, $(\left. h_{\alpha\beta} \right|_{\Sigma_0},\left. \partial_t h_{\alpha\beta} \right|_{\Sigma_0})$, for this system is chosen such that $\left. h_{ij} \right|_{\Sigma_0} = h^{(0)}_{ij}$, $\left. \partial_t h_{ij} \right|_{\Sigma_0} = \left. 2\partial_{(i} h_{j)0} \right|_{\Sigma_0} - 2 k^{(0)}_{ij}$ and $\left. \partial_t h_{00} \right|_{\Sigma_0} = 2\delta^{ij} k^{(0)}_{ij}$, $\left. \partial_t h_{0j} \right|_{\Sigma_0} = \partial^i(h^{(0)}_{ij} - \frac{1}{2}\delta_{ij}\delta^{kl} h^{(0)}_{kl}) + \frac{1}{2}\left. \partial_j h_{00} \right|_{\Sigma_0}$, where the initial data for $h_{00}$ and $h_{0j}$ is chosen smooth but otherwise arbitrary. This choice implies the satisfaction of Equation~(\ref{Eq:LinEinsteinID}) with $X_j=0$ and $f=0$ and the initial conditions $\left. C_\beta \right|_{\Sigma_0} = 0$ and $\left. \partial_t C_\beta \right|_{\Sigma_0} = 0$ on the constraint fields $C_\beta$. Therefore, solving the wave equation $\nabla^\mu\nabla_\mu h_{\alpha\beta} = 0$ with such data, we obtain a solution of the linearized Einstein equations~(\ref{Eq:LinEinstein}) in the harmonic gauge with initial data satisfying~(\ref{Eq:LinEinsteinID}) with $X_j=0$ and $f=0$. This shows geometric existence for the linearized harmonic formulation.

As for uniqueness, suppose we had two smooth solutions of Equations~(\ref{Eq:LinEinstein}, \ref{Eq:LinEinsteinID}). Then, since the equations are linear, the difference $h_{\alpha\beta}$ between these two solutions also satisfies the Equations~(\ref{Eq:LinEinstein}, \ref{Eq:LinEinsteinID}) with trivial data $h^{(0)}_{ij} = 0$, $k^{(0)}_{ij} = 0$. We show that $h_{\alpha\beta}$ can be transformed away by means of an infinitesimal gauge transformation~(\ref{Eq:LinEinsteinGT}). For this, define $\tilde{h}_{\alpha\beta} := h_{\alpha\beta} + 2\nabla_{(\alpha}\xi_{\beta)}$ where $\xi_\beta$ is required to satisfy the inhomogeneous wave equation
\begin{displaymath}
0 = \nabla^\alpha\tilde{h}_{\alpha\beta} - \frac{1}{2}\nabla_\beta\tilde{h}
 = \nabla^\alpha h_{\alpha\beta} - \frac{1}{2}\nabla_\beta h 
 + \nabla^\alpha\nabla_\alpha\xi_\beta
\end{displaymath}
with initial data for $\xi_\beta$ defined by $\left. \xi_0 \right|_{\Sigma_0} = f$, $\left. \xi_i \right|_{\Sigma_0} = -X_i$, $\left. \partial_t\xi_0 \right|_{\Sigma_0} = -h_{00}/2$, $\left. \partial_t\xi_i \right|_{\Sigma_0} = -h_{0i} + \partial_i f$. Then, by construction, $\tilde{h}_{\alpha\beta}$ satisfies the harmonic gauge, and it can be verified that $\left. \tilde{h}_{\alpha\beta} \right|_{\Sigma_0} = \left. \partial_t\tilde{h}_{\alpha\beta} \right|_{\Sigma_0} = 0$. Therefore, $\tilde{h}_{\alpha\beta}$ is a solution of the wave equation $\nabla^\mu\nabla_\mu\tilde{h}_{\alpha\beta} = 0$ with trivial initial data, 
and it follows that $\tilde{h}_{\alpha\beta} = 0$ and that $h_{\alpha\beta} = -2\nabla_{(\alpha}\xi_{\beta)}$ is a pure gauge mode.
\qed

It follows from the existence part of the proof that the quantities $\left. h_{00} \right|_{\Sigma_0}$ and $\left. h_{0j} \right|_{\Sigma_0}$, corresponding to linearized lapse and shift, parametrize pure gauge modes in the linearized harmonic formulation. 

Next, we turn to the IBVP on the manifold $M = [0,T]\times \Sigma$. Let us first look at the boundary conditions~(\ref{Eq:KKK}--\ref{Eq:KQQ-QQK}), which, in the linearized case reduce to
\begin{displaymath}
\left. \nabla_K h_{KK} \right|_{\cal T} = q_K,\qquad
\left. \nabla_K h_{KL} \right|_{\cal T} = q_L,\qquad
\left. \nabla_K h_{KQ} \right|_{\cal T} = q_Q,\qquad
\left. \nabla_K h_{QQ} - \nabla_Q h_{QK} \right|_{\cal T} = q_{QQ}.
\end{displaymath}
There is no problem in repeating the geometric existence part of the proof on $M$ imposing this boundary condition, and using the IBVP described in Section~\ref{SubSec:HarmonicBC}. However, there is a problem when trying to prove the uniqueness part. This is because a gauge transformation~(\ref{Eq:LinEinsteinGT}) induces the following transformations on the boundary data,
\begin{eqnarray}
&& \tilde{q}_K = q_K + 2\nabla_K^2\xi_K,\qquad
\tilde{q}_L = q_L + \nabla_K^2\xi_L + \nabla_K\nabla_L\xi_K,\qquad
\tilde{q}_Q = q_Q + \nabla_K^2\xi_Q + \nabla_K\nabla_Q\xi_K,
\nonumber\\
&& \tilde{q}_{QQ} = q_{QQ} + \nabla_Q(\nabla_K\xi_Q - \nabla_Q\xi_K),
\nonumber
\end{eqnarray}
which overdetermines the vector field $\xi_\beta$ at the boundary. On the other hand, replacing the boundary condition~(\ref{Eq:KQQ-QQK}) by the specification of the Weyl scalar $\Psi_0$, leads to~\cite{lLmSlKrOoR06,mRoRoS07}
\begin{equation}
\left. \nabla_K^2 h_{QQ} + \nabla_Q(\nabla_Q h_{KK} - 2\nabla_K h_{KQ})
 \right|_{\cal T} = \Psi_0.
\label{Eq:LinEinsteinBC}
\end{equation}
Since the left-hand side is gauge-invariant, there is no overdetermination of $\xi_\beta$ at the boundary any more, and the transformation properties of the remaining boundary data $q_K$, $q_L$ and $q_Q$ provides a complete set of boundary data for $\xi_K$, $\xi_L$ and $\xi_Q$ which may be used in conjunction with the wave equation $\nabla^\mu\nabla_\mu\xi_\beta = 0$ in order to formulate a well-posed IBVP~\cite{mRoRoS07}. Provided $\Psi_0$ is smooth and the compatibility conditions are satisfied at the edge $S = \Sigma_0 \cap {\cal T}$, it follows:

\begin{theorem}\cite{oRoS10}
The IBVP~(\ref{Eq:LinEinstein}, \ref{Eq:LinEinsteinID}, \ref{Eq:LinEinsteinBC}) possesses a smooth solution $h_{\alpha\beta}$ which is unique up to an infinitesimal coordinate transformation $\tilde{h}_{\alpha\beta} = h_{\alpha\beta} + 2\nabla_{(\alpha}\xi_{\beta)}$ generated by a vector field $\xi^\alpha$.
\end{theorem}

In conclusion, we can say that in the simple case of linear gravitational waves propagating a Minkowksi background we have resolved the issues (i--iii). Correct boundary data is given to the linearized Weyl scalar $\Psi_0$ computed from the boundary adapted tetrad. To linear order, $\Psi_0$ is invariant with respect to coordinate transformations, and the time-like vector field $T$ appearing in its definition can be defined geometrically by taking the future-directed unit normal to the initial surface $\Sigma_0$ and parallel transport it along the geodesics orthogonal to $\Sigma_0$.

Whether or not this result can be generalized to the full nonlinear case it not immediately clear. In our linearized analysis we have imposed no restrictions on the normal component $\xi_N$ of the vector field generating the infinitesimal coordinate transformation. However, such a restriction is necessary in order to keep the boundary surface fixed under a diffeomorphism. Unfortunately, it does not seem possible to restrict $\xi_N$ in a natural way with the current boundary conditions constructed so far.

%===================================================================
\subsection{Alternative approaches}
\label{SubSec:OutToInfinity}
%===================================================================
 
Although the formulation of Einstein's equations on a finite space domain with an artificial time-like boundary is currently the most used approach in numerical simulations, there are a number of difficulties associated with it. First, as discussed above, spurious reflections from the boundary surface may contaminate the solution unless the boundary conditions are chosen with great care. Second, in principle there is a problem with wave extraction, since gravitational waves can only be defined in an unambiguous (gauge-invariant) way at future null infinity. Third, there is an efficiency problem, since in the far zone the waves propagate along outgoing null geodesics so that hyperboloidal surfaces which are asymptotically null should be better adapted to the problem. These issues have become more apparent as numerical simulations have achieved higher accuracy to the point that boundary and wave extraction artifacts are noticeable, and have driven a number of other approaches.

One of them is that one of compactification schemes which include spacelike or null infinity into the computational domain.  For schemes compactifying spacelike infinity, see Refs.~\cite{fP05,fP06}. Conformal compactifications are reviewed in Refs.~\cite{jF04,Friedrich:2002xz}, and a partial list of references to date includes~\cite{rP65,hF81,hF83,hF86a,hF86b,jF98,pH99,jF04,sHcStVaZ05,gCcGdH05,aZ08a,aZ08b,Ohme:2009gn,Buchman:2009ew,aZdNsH09a, aZmT09b,aZ10a, Zenginoglu:2010cq, aZlK10b,vMoR09,oR10a,jBoSlB11,Bardeen:2011ip}. 

Another approach is Cauchy-characteristic matching (CCM)~\cite{gC06,Sperhake:2001si,Szilagyi:2000xu,dInverno:2000ae,Dubal:1998ei,nBrGlLbSjWrI98}, which combines a Cauchy approach in the strong field regime (thereby avoiding the problems that the presence of caustics would cause on characteristic evolutions) with a characteristic one in the wave zone. Data from the Cauchy evolution is used as inner boundary conditions for the characteristic one and, viceversa, the latter provides outer boundary conditions for the Cauchy IBVP. An understanding of the Cauchy IBVP is still a requisite. CCM is reviewed in~\cite{Winicour:2008tr,Winicour:2008tr}. A related idea is Cauchy-perturbative matching~\cite{bZePpDmT06,lRaArMmRsS99,Abrahams:1997ut,mRaAlR98,Abrahams:1997ut}, where the Cauchy code is instead coupled to one solving gauge-invariant perturbations of Schwarzschild black holes or flat spacetime. The multipole decomposition in the Regge--Wheeler--Zerilli equations~\cite{tRjW57,fZ70a,oSmT01,Martel:2005ir,Nagar:2005ea} implies that the resulting equations are 1+1 dimensional and can therefore extend the region of integration to very large distances from the source. As in CCM, an understanding of the IBVP for the Cauchy sector is still a requisite.  

One way of dealing with the ambiguity of \textit{extracting}  gravitational waves from Cauchy evolutions at finite radii is by extrapolating procedures, see for example~\cite{Boyle:2009vi, Pollney:2009ut} for some approaches and quantification of their accuracies. Another approach is Cauchy characteristic extraction (CCE)~\cite{cRnBdPbS10,Babiuc:2010ze, Reisswig:2009us, Babiuc:2008qy, Babiuc:2005pg, Bishop:1996gt}. In CCE a Cauchy IBVP is solved, and the numerical data on a world tube is used to provide inner boundary conditions for a characteristic evolution that ``transports'' the data to null infinity. The difference with CCM is that in CCE there is no ``feedback'' from the characteristic evolution to the Cauchy one, and the extraction is done as a post-processing step.

\newpage
%===================================================================
%===================================================================
\section{Numerical Stability}
\label{sec:num_stability}
%===================================================================
%===================================================================

In the previous sections we have discussed continuum initial and initial-boundary value problems. In this section we start with the study of the discretization of such problems. In the same way that a PDE can have a unique solution yet be ill posed\epubtkFootnote{See, for instance Example~\ref {Example:BackwardsHeat} with initial data $f\in {\cal S}^\omega$.}, a numerical scheme can be consistent yet not convergent due to the unbounded growth of small perturbations as resolution is increased. The definition of numerical stability is the discrete version of well posedness. One wants to ensure that small initial perturbations in the numerical solution, which naturally appear due to discretization errors and finite precision, remain bounded for all resolutions at any given time $t>0$. Due to the classical Lax--Richtmyer theorem~\cite{LaxRichtmyer}, this property, combined with consistency of the scheme,  is equivalent in the linear case to convergence of the numerical solution, and the latter approaches the continuum one as resolution is increased (at least within exact arithmetic). Convergence of a scheme is in general difficult to prove directly, especially because the exact solution is in general not known. Instead, one shows stability. 

The different definitions of numerical stability follow those of well posedness, with the $L^2$ norm in space replaced by a discrete one, which is usually motivated by the spatial approximation. For example, discrete norms under which the Summation By Parts property holds are natural in the context of some finite difference approximations and collocation spectral methods (see Sections~\ref{sec:fd} and \ref{sec:spec}). 

We start with a general discussion of some aspects of stability, and explicit analyses of simple, low order schemes for test models. There follows a discussion of different variations of the von Neumann condition, including an eigenvalue version which can be used to analyze in practice necessary conditions for initial-boundary value problems. Next, we discuss a rather general stability approach for the method of lines, the notion of time-stability, Runge--Kutta methods, and we close the section with some references to other approaches not covered here, as well as some discussion in the context of numerical relativity.   

%=============================================
\subsection{Definitions and examples}
%=============================================

Consider a well posed linear initial value problem (see Definition~\ref{Def:WPVC}) 
\begin{eqnarray}
u_t(t,x) = P(t,x,\partial/\partial x) u(t,x), 
&& x\in \Real^n, \quad t \geq 0, \label{eq:ibv1_num}
\label{eq:cauchy1}\\
u(0,x) = f(x),
&& x\in\Real^n \, . 
\label{eq:cauchy2}
\end{eqnarray}

\begin{definition}
\label{def:num_stability}
An approximation-discretization to the Cauchy problem (\ref{eq:cauchy1}, \ref{eq:cauchy2})  is numerically stable if there is some {\tt d}iscrete norm in space $\| \cdot \|_{\tt d}$ and constants $K_{\tt d}, \alpha_{\tt d}$ such that the corresponding approximation $v$ satisfies  
\be
\| v (t,\cdot) \|_{\tt d} \leq K_{\tt d} e^{\alpha_{\tt d} t} \| f \|_{\tt d} \, , \label{eq:num_stability_ivp}
\ee
for high enough resolutions, smooth initial data $f$, and $t \geq 0$. 
\end{definition} 

Note:
\begin{itemize}
\item The previous definition applies both to the semidiscrete case (where space but not time is discretized) as well as the fully discrete one. In the latter case, Equation~(\ref{eq:num_stability_ivp}) is to be interpreted at \emph{fixed} time. For example, if the timestep discretization is constant, 
$$
t_k = k \Delta t,\qquad k= 0,1,2 \ldots 
$$
then Equation~(\ref{eq:num_stability_ivp})  needs to hold for fixed $t_k$ and arbitrarily large $k$. In other words, the solution is allowed to grow with time, but not with the number of timesteps at fixed time when resolution is increased. 
\item The norm $\| \cdot \|_{\tt d}$ in general depends on the spatial approximation, and in Sections~\ref{sec:fd} and \ref{sec:spec} we discuss some definitions for the finite difference and spectral cases.
\item From Definition~\ref{def:num_stability}, one can see that an ill posed problem cannot have a stable discretization, since otherwise one could take the continuum limit in (\ref{eq:num_stability_ivp}) and reach the contradiction that the original system was well posed.
\item As in the continuum, Equation~(\ref{eq:num_stability_ivp}) implies uniqueness of the numerical solution $v$. 
\item In Section \ref{section:ivp} we discussed that if in a well posed homogeneous Cauchy problem a forcing term is added to Equation~(\ref{eq:ibv1_num}), 
\begin{equation}
u_t(t,x) = P(t,x,\partial/\partial x) u(t,x) 
\qquad\mapsto\qquad
u_t(t,x) = P(t,x,\partial/\partial x) u(t,x) + F(t,x), 
\label{eq:forcing_term}
\end{equation}
then the new problem admits another estimate, related to the original one via Duhamel's formula, Equation~(\ref{Eq:DuhamelFormula}). A similar concept holds at the semidiscrete level, and  the discrete estimates change accordingly (in the fully discrete case the integral in time is replaced by a discrete sum),  
\begin{equation}
\| v (t,\cdot) \|_{\tt d} \leq K_{\tt d} e^{\alpha_{\tt d} t} \| f \|_{\tt d} 
\qquad\mapsto\qquad
\| v (t,\cdot) \|_{\tt d} \leq K_{\tt d} e^{\alpha_{\tt d} t} \left( \| f \|_{\tt d} + \int\limits_0^t \| F(s,\cdot) \|_{\tt d} ds \right).
\label{eq:discrete_estimate_forcing_term} 
\end{equation}

In other words, the addition of a lower order term does not affect numerical stability, and without loss of generality one can restrict stability analyses to the homogeneous case.
\item The difference $w:=u-v$ between the exact solution and its numerical approximation satisfies an equation analogous to (\ref{eq:forcing_term}), where $F$ is related to the truncation error of the approximation. If the scheme is numerically stable, then in the linear and semidiscrete cases Equation~(\ref{eq:discrete_estimate_forcing_term}) implies
\be
\| w (t,\cdot) \|_{\tt d} \leq K_{\tt d} e^{\alpha_{\tt d} t} \int\limits_0^t \| F(s,\cdot) \|_{\tt d} ds \, . \label{eq:lax_estimate}
\ee
If the approximation is consistent, the truncation error converges to zero as resolution is increased, and Equation (\ref{eq:lax_estimate}) implies that so does the norm of the error $\| w (t,\cdot) \|_{\tt d}$. That is, stability implies convergence. The inverse is also true and this equivalence between convergence and stability is the celebrated Lax--Richtmyer theorem. The equivalence also holds in the fully discrete case.  
\item In the quasi-linear case one follows the principle of linearization, as described in Section \ref{SubSec:QLP}. One  linearizes the problem, and constructs a stable numerical scheme for the linearization. The expectation, then, is that the scheme also converges for the nonlinear scheme. For particular problems and discretizations this expectation can be rigorously proved (see, for example, \cite{KL89}).

\end{itemize}
From hereon $\{ x_j, t_k \}$ denotes some discretization of space and time. This includes both finite difference and spectral collocation methods, which are the ones discussed in Sections~\ref{sec:fd} and \ref{sec:spec}, respectively. In addition, we use the shorthand notation
$$
v^k_j := v(t_k,x_j) \,.
$$

In order to gain some intuition into the general problem of numerical stability we start with some examples of simple, low order approximations for a test problem. Consider uniform grids both in space and time
$$
t_k = k \Delta t,\quad x_j = j \Delta x, \qquad
k=0,1,2,\ldots,\quad j=0,1,2,\ldots N,
$$
and the advection equation,
\begin{equation}
u_t = a u_x \, , \qquad x\in [0,2\pi],\quad t\geq 0,
\label{eq:adv_periodic}
\end{equation}
on a periodic domain with $2\pi = N\Delta x$, and smooth periodic initial data.  Then the solution $u$ can be represented by a Fourier series:
\begin{equation}
u(t,x) 
 = \frac{1}{\sqrt{2\pi}}\sum_{\omega \in \Integer }\hat{u}(t,\omega) e^{i \omega x}, \label{eq:interp_fourier}
\end{equation}
where 
$$
\hat{u}(t,\omega) =  \frac{1}{\sqrt{2\pi}} \int_0^{2\pi} e^{-i \omega x} u(t,x) dx \, , 
$$
and the stability of the following schemes can be analyzed in Fourier space.

\begin{example} 
\label{ex:one_sided_euler}
The one-sided Euler scheme.\\
Equation~(\ref{eq:adv_periodic}) is discretized with a \textit{ one-sided} finite difference approximation for the spatial derivative and evolved in time with the \textit{forward Euler} scheme,
$$
\frac{v^{k+1}_j - v^{k}_j}{\Delta t} = a \frac{v^{k}_{j+1} - v^{k}_j}{\Delta x}. 
$$
In Fourier space the approximation becomes 
\be
\hat{v}^{k+1}(\omega ) = \hat{q}(\omega) \hat{v}^{k}(\omega ) = \left[ \hat{q}(\omega)\right]^{k+1}\hat{v}^{0}(\omega )  \, , \label{eq:fourier_test}
\ee
where  
$$
\hat{q}(\omega) = 1+ a \lambda \left( e^{i \omega \Delta x} -1 \right) 
$$
is called the \textbf{amplification factor} and  
$$
\lambda = \frac{\Delta t}{\Delta x} \, 
$$
the \textbf{Courant--Friedrich--Levy (CFL) factor}. 

Using Parseval's identity, we find
$$
\| v (t_k,\cdot) \| ^2 
 = \sum_{\omega\in\Integer} | \hat{q}(\omega) |^{2k} | \hat{v}^0(\omega) |^2, 
$$
and therefore, we see that the inequality~(\ref{eq:num_stability_ivp}) can only hold for all $k$ if
\be
| \hat{q} (\omega) | \leq 1\quad \hbox{for all $\omega\in\Integer$}.
\label{eq:vn_example}
\ee
For $a > 0$, this is the case if and only if the CFL factor satisfies
\be
0 < \lambda \leq \frac{1}{a},
\label{eq:cfl_example} 
\ee
and in this case the well posedness estimate~(\ref{eq:num_stability_ivp}) holds with $K_{\tt d}=1$ and $\alpha_{\tt d}=0$. The upper bound in condition~(\ref{eq:cfl_example}) for this example is known as the \textbf{CFL limit}, and~(\ref{eq:vn_example}) as the von~Neumann condition. If $a=0$, $\hat{q}(\omega)=1$,  while for $a<0$ the scheme is \textbf{unconditionally unstable} even though the underlying continuum problem is well posed.
\end{example}

Next we consider a scheme very similar to the previous one, but which turns out to be unconditionally unstable for $a\neq 0$, regardless of the direction of propagation.

\begin{example}
\label{ex:centered_euler}
A centered Euler scheme.\\
Consider first the semidiscrete approximation to Equation~(\ref{eq:adv_periodic}),
\be
\frac{d}{dt}v_j = a \frac{v_{j+1} - v_{j-1}}{2 \Delta x} \, ; \label{eq:centered_semidiscrete}
\ee
it is easy to check that it is stable for all values of $\Delta x$. Next discretize time through an Euler scheme, leading to 
\be
\frac{v^{k+1}_j - v^{k}_j}{\Delta t} = a \frac{v^{k}_{j+1} - v^{k}_{j-1}}{2 \Delta x} \, . \label{eq:euler_centered}
\ee
The solution again has the form given by Equation~(\ref{eq:fourier_test}), now with 
$$
|\hat{q}(\omega) | = | 1 + i a\lambda\sin(\omega\Delta x) | \geq 1.
$$
At fixed time $t_k$, the norm of the solution to the fully discrete approximation (\ref{eq:euler_centered}) for arbitrary small initial data with $\omega\Delta x\notin \pi\Integer$ grows without bound as the timestep decreases.

The semidiscrete centered approximation (\ref{eq:centered_semidiscrete})  and the fully discrete centered Euler scheme (\ref{eq:euler_centered}) constitute the simplest example of an approximation which is not fully discrete stable even though its semidiscrete version is. This is related to the fact that the Euler time integration is not \textit{locally stable}, as discussed in Section~\ref{sec:fully_discrete}. 
\end{example}

The previous two examples were \textbf{one-step methods}, where $v^{k+1}$ can be computed in terms of $v^k$. The following is an example of a two-step method.

\begin{example} 
\label{ex:leap_frog}
Leap-frog.\\
A way to stabilize the centered Euler scheme is by approximating the time derivative by a centered difference instead of a forward, one-sided operator: 
$$
v_j^{k+1} = v_j^{k-1}  + a\lambda\left( v_{j+1}^k - v_{j-1}^k  \right) \,.  
$$
Enlarging the system by introducing
$$
w^k_j:= \left( \begin{array}{l} v^k_j \\ v^{k-1}_j \end{array} \right) 
$$
it can be cast into the one-step method
$$
\hat{w}^{k+1} = \hat{Q}(\omega) \hat{w}^k = \hat{Q}(\omega)^{k+1} \hat{w}^0,
\qquad \mbox{with }
\hat{Q}(\omega) = \left( \begin{array}{cc}
 2i a\lambda\sin(\omega\Delta x) & 1 \\ 1 & 0 \end{array} \right).
$$
By a similar procedure, a general multi-step method can always be reduced to a one-step one. Therefore, in the stability results below we can assume without loss of generality that the schemes are one-step.

In the above example the amplification matrix $\hat {Q}(\omega)$ can be diagonalized through a transformation that is uniformly bounded: 
$$
\hat{Q}(\omega) = \hat{T}(\omega )\left( \begin{array}{cc}
 \mu_+ & 0 \\ 0 & \mu_-  \end{array} \right) \hat{T}^{-1}(\omega ),\qquad
\hat{Q}(\omega)^k = \hat{T}(\omega )\left( \begin{array}{cc}
 \mu_+^k & 0 \\ 0 & \mu_-^k \end{array} \right) \hat{T}^{-1}(\omega ),
$$
with $ \mu_{\pm} = z\pm \left( 1 + z^2 \right)^{1/2}$, $z:=ia\lambda\sin{\left(\omega \Delta x \right)}$, and 
$$
\hat{T}(\omega ) = \left( \begin{array}{cc}
 \mu_+ & \mu_- \\ 1 & 1 \end{array} \right) \, . 
$$
The eigenvalues $\mu_{\pm}$ are of unit modulus, $|\mu_{\pm}|=1$. In addition, the norms of $\hat{T}(\omega )$ and its inverse are
$$
| \hat{T}(\omega ) | = \sqrt{2\left( 1 + |z| \right) } \quad \, , \quad | \hat{T}^{-1}(\omega ) | = \frac{1}{\sqrt{2\left( 1 - |z| \right) }} \, . 
$$
Therefore the condition number of $\hat{T} (\omega )$ can be bounded for all $\omega$:
$$
| \hat{T}(\omega ) | \cdot | \hat{T}^{-1}(\omega ) |= \left(\frac{1+|z|}{1-|z|} \right)^{1/2} < \left(\frac{1+|a|\lambda }{1-|a|\lambda} \right)^{1/2} < \infty \, , 
$$
provided that
\begin{equation}
\lambda < \frac{1}{|a|} \, ,
\label{eq:leap_frog_stability}
\end{equation}
and it follows that the Leap-frog scheme is stable under the condition~(\ref{eq:leap_frog_stability}).
\end{example}

The previous examples were \textbf{explicit methods}, where the solution $v^{k+1}_j$ (or $w^{k+1}_j$) can be explicitly computed from the one at the previous timestep, without inverting any matrices.

\begin{example}
\label{ex:crank_nicholson}
Crank--Nicholson.\\
Approximating Equation (\ref{eq:adv_periodic}) by 
$$
\left( 1 - a\frac{\Delta t}{2} D_0 \right) v^{k+1}_j 
 = \left( 1 + a\frac{\Delta t}{2} D_0 \right) v^{k}_j,
$$
with
\begin{equation}
D_0 v_j:= \frac{1}{2\Delta x}\left(v_{j+1}  - v_{j-1} \right),
\label{eq:D0}
\end{equation}
defines an \textbf{implicit method}. Fourier transform leads to
$$
\left[ 1 - i a\frac{\lambda}{2}\sin(\omega x) \right]\hat{v}^{k+1}_j
 = \left[ 1 + i a\frac{\lambda}{2}\sin(\omega x) \right]\hat{v}^k_j.
$$
The expressions inside the square brackets on both sides are different from zero and have equal magnitude. As a consequence, the amplification factor in this case satisfies 
\begin{equation}
|\hat{q}(\omega)| = 1\quad\mbox{ for all $\omega\in\Integer$ and $\lambda > 0$},
\label{eq:amp_factor_crank_nicholson}
\end{equation}
and the scheme is \textbf{unconditionally stable} at the expense of having to invert a matrix to advance the solution in time. 
\end{example}

\begin{example}
\label{ex:ICN}
Iterated Crank--Nicholson.\\
Approximating the Crank-Nicholson scheme through an iterative scheme with a  {\em fixed} number of iterations is usually referred to as the Iterated Crank-Nicholson (ICN) method. For  Equation (\ref{eq:adv_periodic})  it proceeds as follows \cite{Teukolsky:1999rm}:
\begin{itemize}
\item First iteration: an intermediate variable $^{(1)}\tilde{v}$ is calculated using a second order in space centered difference (\ref{eq:D0}) and an Euler, first order forward time approximation, 
$$
\frac{1}{\Delta t} \left( ^{(1)}\tilde{v}_j^{n+1} - v_j^n \right) = D_0 \, v_j^{n} \, . 
$$
Next, a second intermediate variable is computed through averaging, 
$$
^{(1)}\bar{v}_j^{n + 1/2} = \frac{1}{2} \left( ^{(1)}\tilde{v}_j^{n + 1} + v_j^n  \right) \, .
$$
The full time step for this first iteration is 
$$
\frac{1}{\Delta t }\left( v_j^{n+1}  - v_j ^{n}\right) = D_0  \, ^{(1)}\bar{v}_j^{n+1/2}
$$
\item  Second iteration: it follows the same steps. Namely, the intermediate variables
\begin{eqnarray*}
\frac{1}{\Delta t} \left( ^{(2)}\tilde{v}_j^{n+1} - v_j^n \right) &=& D_0\,  ^{(1)}\bar{v}_j^{n+1/2}  \\
^{(2)}\bar{v}_j^{n + 1/2} &=& \frac{1}{2} \left( ^{(2)}\tilde{v}_j^{n + 1} + v_j^n  \right) \, . 
\end{eqnarray*}
are computed, and the full step is obtained from
$$
\frac{1}{\Delta t }\left( v_j^{n+1}  - v_j ^{n}\right) = D_0   \, ^{(2)}\bar{v}_j^{n+1/2} \, . 
$$
\item  Further iterations proceed in the same way. 
\end{itemize}
The resulting discretization is numerically stable for $\lambda \leq 2/a$ and $p=2,3,6,7,10,11,\ldots$ iterations, and unconditionally unstable otherwise. In the limit $p \rightarrow \infty$ the ICN scheme becomes the implicit, unconditionally stable Crank--Nicholson scheme of the previous example. For any fixed number of iterations, though, the method is explicit and stability is contingent on the CFL condition $\lambda \leq 2/a$. The method is unconditionally unstable for $p=4,5,8,9,12,13,\ldots$ because the limit of the amplification factor approaching one in absolute value [cf.\ Equation~(\ref{eq:amp_factor_crank_nicholson})] as $p$ increases is not monotonic. See \cite{Teukolsky:1999rm} for details and \cite{Schanze200515} for a similar analysis for ``theta'' schemes. 

\end{example}

Similar definitions to that one of Definition~\ref{def:num_stability} are introduced for the initial-boundary value problem. For simplicity we explicitly discuss the semi-discrete case. In analogy with the definition of a strongly well posed IBVP (Definition~\ref{Def:WPIBVP}) one has 

\begin{definition}
\label{def:strong_stability}
A semi-discrete approximation to the linearized version of the IBVP (\ref{Eq:QLIBVP1}, \ref{Eq:QLIBVP2}, \ref{Eq:QLIBVP3})  is numerically stable if there are {\tt d}iscrete norms $\| \cdot \|_{\tt d}$ at $\Sigma$ and $\| \cdot \|_{\partial,{\tt d}}$ at $\partial\Sigma$ and constants  $K_{\tt d} = K_{\tt d}(T)$ and $\varepsilon_{\tt d} = \varepsilon_{\tt d}(T) \geq 0$  such that for high enough resolution the corresponding approximation $v$ satisfies  
\begin{equation}
\| v(t,\cdot) \|_{\tt d}^2 
 + \varepsilon_{\tt d} \int\limits_0^t \| v(s,\cdot) \|_{\partial,{\tt d}}^2\, ds
 \leq K_{\tt d}^2\left[ \| f \|_{\tt d}^2
 + \int\limits_0^t\left( \| F(s,\cdot) \|_{\tt d}^2 
 + \| g(s,\cdot) \|_{\partial,{\tt d}}^2 \right) ds\right],
\label{eq:num_strongly_stable}
\end{equation}
for all $t\in [0,T]$. If the constant $\varepsilon_{\tt d}$ can be chosen strictly positive, the problem is called \textbf{strongly stable}.
\end{definition}

In addition, the semi-discrete version of Definitions~\ref{Def:SWPGS} and \ref{Def:BoundaryStable} lead to the concepts of strong stability in the generalized sense and boundary stability, respectively, which we do not write down explicitly here. The definitions for the fully discrete case are similar, with time integrals such as those in Equation~(\ref{eq:num_strongly_stable}) replaced by discrete sums.

%=============================================
\subsection{The von~Neumann condition}
%=============================================

Consider a discretization for a linear system with variable in space but time-independent coefficients such that 
\begin{eqnarray}
\mathbf{v}^{k +1 } &=& \mathbf{Q} \mathbf{v}^k \, , \label{eq:amp_matrixa} \\
\mathbf{v}^{0} &=& \mathbf{f} \, , \label{eq:amp_matrixb} 
\end{eqnarray}
where $\mathbf{v}^k$ denotes the gridfunction $\mathbf{ v}^k = \{ v^k_j : j= 0,1,\ldots,N \}$ and 
$\mathbf{Q}$ is called the \textbf{amplification matrix}. We assume that $\mathbf{Q}$ is also time-independent. Then 
$$
\mathbf{v}^{k } = \mathbf{Q}^{k} \mathbf{f}  
$$
and the approximation (\ref{eq:amp_matrixa}, \ref{eq:amp_matrixb}) is stable if and only if there are constants $K_{\tt d}$ and $\alpha_{\tt d}$ such that 
\be
\| \mathbf{Q}^{k} \|_{\tt d} \leq  K_{\tt d} e^{\alpha_{\tt d} t_k} \label{eq:power_bounded}
\ee
for all $k=0,1,2,\ldots$ and high enough resolutions. 

In practice, condition (\ref{eq:power_bounded}) is not very manageable as a way of determining if a given scheme is stable since it involves computing the norm of the power of a matrix. A simpler condition based on the eigenvalues $\{ q_i \}$ of $\mathbf{Q}$ as opposed to the norm of $\mathbf{Q}^{k}$ is von~Neumann's one: 
\be
| q_i | \leq e^{\alpha_{\tt d} \Delta t}\quad
\mbox{for all eigenvalues $q_i$ of $\mathbf{Q}$ and all $\Delta t > 0$.}
\label{eq:von_neumann}
\ee
This condition is necessary for numerical stability: if $q_i$ is an eigenvalue of $\mathbf{Q}$, $q_i^k$ is an eigenvalue of $\mathbf{Q}^k$ and 
$$
| q_i ^k | \leq  \| \mathbf{Q}^k\| \leq K_{\tt d} e^{\alpha_{\tt d} t_k} = K_{\tt d}  e^{\alpha_{\tt d} k\Delta t}. 
$$
That is, 
$$
| q_i | \leq K_{\tt d}^{1/k} e^{\alpha_{\tt d} \Delta t}, 
$$
which, in order to be valid for all $k$, implies Equation~(\ref{eq:von_neumann}). 

As already mentioned, in order to analyze numerical stability one can drop low order terms. Doing so typically leads to $\mathbf{Q}$ depending on $\Delta t$ and $\Delta x$ only through a quotient (the CFL factor) of the form (with $p=1$ for hyperbolic equations)
$$
\lambda = \frac{\Delta t}{(\Delta x)^p}\, , 
$$
\be
\mathbf{Q}(\Delta t, \Delta x) = \mathbf{Q}(\lambda ) \, . \label{eq:Q_cfl}
\ee
Then for Equation~(\ref{eq:von_neumann}) to hold for all $\Delta t > 0$ while keeping the CFL factor fixed (in particular, for small $\Delta t > 0$), the following condition has to be satisfied: 
\be
| q_i | \leq 1\quad \mbox{for all eigenvalues $q_i$ of $\mathbf{Q}$},
\label{eq:von_neumann_strong}
\ee
and one has a stronger version of the von Neumann condition, which is the one encountered in Example~\ref{ex:one_sided_euler}, see Equation~(\ref{eq:vn_example}).

%%%%%%%%%%%%%%%%%%%%%%%%%%%%%%
\subsubsection{The periodic, scalar case}
%%%%%%%%%%%%%%%%%%%%%%%%%%%%%%

We return to the periodic scalar case, such as the schemes discussed in Examples~\ref{ex:one_sided_euler}, \ref{ex:centered_euler}, \ref{ex:leap_frog}, and \ref{ex:crank_nicholson} with some more generality. Suppose then, in addition to the linearity and time-independent assumptions of the previous subsection, that the continuum problem, initial data and discretization~(\ref{eq:amp_matrixa}, \ref{eq:amp_matrixb}) are periodic on the interval $[0,2\pi]$. Through a Fourier expansion we can write the grid function $\mathbf{f} = (f(x_0),f(x_1),\ldots,f(x_N))$ corresponding to the initial data as
$$
\mathbf{f} = \frac{1}{\sqrt{2\pi}}\sum_{\omega\in\Integer}\hat{f}(\omega) \mathbf{e}^{i \omega}, 
$$
where $\mathbf{ e}^{i \omega} = ( e^{i \omega x_0},e^{i\omega x_1},\ldots,e^{i\omega x_N} )$. The approximation becomes 
\begin{eqnarray}
\mathbf{v}^{k} =  \frac{1}{\sqrt{2\pi}} \sum_{\omega\in\Integer}\hat{f}(\omega) \mathbf{Q}^{k} \mathbf{e}^{i \omega} . 
\end{eqnarray}
Assuming that $\mathbf{Q}$ is diagonal in the basis $\mathbf{e}^{i\omega}$, such that
\be
\mathbf{Q} \mathbf{e}^{i \omega} = \hat{q}(\omega ) \mathbf{e}^{i \omega} \label{eq:qdiag}
\ee
as is many times the case, we obtain, using Parseval's identity,
$$
\| \mathbf{v}^{k}  \| = \left( \sum_{\omega\in\Integer}  | \hat{f}(\omega) |^2 | \hat{q}(\omega)^k |^2 \right)^{1/2} .
$$
If 
\be
 | \hat{q}(\omega) | \leq e^{\alpha_{\tt d} \Delta t}
 \quad \mbox{for all eigenvalues $\omega\in\Integer$ and $\Delta t > 0$},
\label{eq:von_neumann_fourier}
\ee
for some constant $\alpha_{\tt d}$ then 
$$
\| \mathbf{v}^{k }  \|\leq e^{\alpha_{\tt d} k \Delta t} \left( \sum_{\omega}  | \hat{f}(\omega) |^2 \right)^{1/2}= e^{\alpha_{\tt d} k \Delta t} \| \mathbf{f} \|  = e^{\alpha_{\tt d} t_k}  \| \mathbf{f}  \|
$$
and stability follows. Conversely, if the scheme is stable and (\ref{eq:qdiag}) holds, (\ref{eq:von_neumann_fourier}) has to be satisfied: take 
$$
\mathbf{f} = \mathbf{e}^{i \omega} \,, 
$$
then 
$$
| \hat{q}^{k}(\omega) | \| \mathbf{f} \| 
 = \| \mathbf{v}^{k}  \| \leq   K_{\tt d} e^{\alpha_{\tt d} t_k} \| \mathbf{f} \|,
$$
or 
$$
| \hat{q} (\omega) | \leq K_{\tt d}^{1/k}  e^{\alpha_{\tt d} \Delta t}
$$
for arbitrary $k$, which implies (\ref{eq:von_neumann_fourier}). Therefore, provided the condition~(\ref{eq:qdiag}) holds, stability is equivalent to the requirement~(\ref{eq:von_neumann_fourier}) on the eigenvalues of $\mathbf{Q}$.

%%%%%%%%%%%%%%%%%%%%%%%%%%%%%%%%%%%%%%%%%%%%%%%%%%%
\subsubsection{The general, linear, time-independent case}
\label{sec:von_neumann_linear}
%%%%%%%%%%%%%%%%%%%%%%%%%%%%%%%%%%%%%%%%%%%%%%%%%%%%

However, as mentioned, the von~Neumann condition is not sufficient for stability, neither in its original form~(\ref{eq:von_neumann}) nor in its strong one~(\ref{eq:von_neumann_strong}), unless, for example, $\mathbf{Q}$ can be uniformly diagonalized. This means that there exists a matrix $\mathbf{T}$ such that 
$$
\mathbf{\Lambda} = \mathbf{T}^{-1} \mathbf{Q} \mathbf{T} = \diag(q_0, \ldots, q_N)
$$
is diagonal and the condition number of $\mathbf{T}$ with respect to the same norm,
$$
\kappa_{\tt d} (\mathbf{T}):= \| \mathbf{T} \|_{\tt d} \| \mathbf{T}^{-1} \|_{\tt d}
$$
is bounded
$$
\kappa_{\tt d} (\mathbf{T}) \leq K_{\tt d}
$$ 
for some constant $K_{\tt d}$ independent of resolution (an example is
that one of $\mathbf{Q}$ being normal, $\mathbf{Q} \mathbf{Q}^{*} = \mathbf{Q}^{*} \mathbf{Q}$).  In that case
$$
\mathbf{v}^{k} = \mathbf{T} \mathbf{\Lambda}^{k} \mathbf{T}^{-1} \mathbf{f}
$$
and
$$
\| \mathbf{v}^{k} \|_{\tt d} \leq \kappa (\mathbf{T}) \max_i | q _i |^{k} \|  \mathbf{f} \|_{\tt d} \leq K_{\tt d} e^{\alpha_{\tt d} k \Delta t }\|  \mathbf{f} \|_{\tt d}  = 
K_{\tt d} e^{\alpha_{\tt d} t_{k}}\|  \mathbf{f} \|_{\tt d} \, . 
$$

Next, we discuss two examples where the von~Neumann condition is satisfied but the resulting scheme is unconditionally unstable. The first one is for a well posed underlying continuum problem and the second one for an ill posed one. 

\begin{example} An unstable discretization which satisfies the von~Neumann condition for a trivially well posed problem~\cite{GKO95}.

Consider the following system on a periodic domain with periodic initial data 
$$
u_t = 0\, , \qquad u= \left( \begin{array}{l} u_1 \\ u_2 \end{array} \right) \, , 
$$
discretized as
\be
\frac{\mathbf{v}^{k+1}-\mathbf{v}^k}{\Delta t} = - \Delta x \left( \begin{array}{ll} 0 & 1 \\ 0 & 0 \end{array} \right) D_0^2\mathbf{v}^k \label{eq:vN_counterexample}
\ee
with $D_0$ given by Equation~(\ref{eq:D0}). The Fourier transform of the amplification matrix and its $k$-th power are 
$$
\hat\mathbf{Q} = \left( \begin{array}{cc} 1 & \lambda \sin^2(\omega \Delta x) \\ 0 & 1 \end{array} \right) \, , \qquad 
\hat\mathbf{Q}^k = \left( \begin{array}{cc} 1 & k \lambda \sin^2(\omega \Delta x) \\ 0 & 1 \end{array} \right) \, . 
$$
The von~Neumann condition is satisfied, since the eigenvalues are $1$. However, the discretization is unstable for any value of $\lambda > 0$. For the unit vector $\mathbf{e} =  \left( 0, 1 \right)^T$, for instance, we have 
$$
| \hat\mathbf{Q}^k \mathbf{e} | = \sqrt{1+ \left( k\lambda \right)^2\sin^4{\left( \omega \Delta x \right)}} \, ,  
$$
which grows without bound as $k$ is increased for $\sin{(\omega \Delta x)} \neq 0$. 

The von-Neumann condition is clearly not sufficient for stability in this example because the amplification matrix not only cannot be uniformly diagonalized, but it cannot be diagonalized at all because of the Jordan block structure in (\ref{eq:vN_counterexample}).
\end{example}

\begin{example} Ill posed problems are unconditionally unstable, even if they satisfy the von~Neumann condition. The following example is drawn from~\cite{gCjPoSmT02a}.

Consider the periodic Cauchy problem 
\be
u_t =  A u_x\, , \label{eq:wh_example}
\ee
where $u=(u_1,u_2)^{\mathrm{T}}$, $A$ is a $2\times 2$ constant matrix, and the following discretization. The right hand side of the equation is approximated by a second order centered derivative plus higher (third) order numerical dissipation (see Section~\ref{sec:dissipation})
$$
A u_x \rightarrow AD_0 v - \epsilon I (\Delta x)^3D_+^2D_-^2 v \, ,  
$$
where $I$ is the $2\times 2$ identity matrix, $\epsilon \geq 0$ an arbitrary parameter regulating the strength of the numerical dissipation and $D_+,D_-$ are first order forward and backward approximations of $d/dx$, 
\be
D_+ v_j := \frac{v_{j+1} - v_j}{\Delta x}, \quad 
D_- v_j := \frac{v_{j} - v_{j-1}}{\Delta x}. \label{eq:one_sided}
\ee
The resulting system of ordinary differential equations is marched in time (method of lines, discussed in the next section) through an explicit method: the iterated Crank--Nicholson (ICN) one with an arbitrary but fixed number of iterations $p$ (see Example \ref{ex:ICN}).

If the matrix $A$ is diagonalizable, as in the scalar case of Example \ref{ex:ICN}, the resulting discretization is numerically stable for $\lambda \leq 2/a$ and $p=2,3,6,7,10,11,\ldots$, even without dissipation. On the other hand, if the system (\ref{eq:wh_example}) is weakly hyperbolic,   as when the principal part has a Jordan block, 
$$
A= \left( \begin{array}{cc} a & 1 \\ 0 & a \end{array} \right)\, , 
$$
one can expect on general grounds that any discretization will be unconditionally unstable. As an illustration, this was explicitly shown in Ref.~\cite{gCjPoSmT02a} for the above scheme and variations of it. In Fourier space the amplification matrix and its $k$-th power take the form
$$
\hat{Q}=
\left(
\begin{array}{cc}
c & b \\
0 & c
\end{array}
\right) \,, \qquad 
\hat{Q}^k = \left(\begin{array}{cc}
c^k & k c^{k-1} b \\
0 & c^k 
\end{array}\right),
$$
with coefficients $c,b$ depending on $\{ a, \lambda, \omega \Delta x, \epsilon \}$ such that for an arbitrary small initial perturbation at just one gridpoint, 
$$
v^0_0 = (0,2\pi \epsilon)^{\mathrm{T}} \, , \qquad v^0_j = (0,0)^{\mathrm{T}} \quad \mbox{otherwise}, 
$$
the solution satisfies
$$
\|\mathbf{v}^{k} \|_{\tt d}^2 \geq C k^{5/4}\|\mathbf{v}^{0} \|_{\tt d}^2 \quad \mbox{for some constant $C$},  
$$ 
and is therefore unstable regardless of the value of $\lambda $ and $\epsilon$. On the other hand, the von Neumann condition $|a| \leq 1$ is satisfied if and only if
\be
0 \leq \epsilon \lambda \leq 1/8.
\label{eq:von_neumann_example} 
\ee
Notice that, as expected, the addition of numerical dissipation cannot stabilize the scheme independently of its amount. Furthermore, adding dissipation with a strength parameter $\epsilon >1/(8\lambda)$ violates the von Neumann condition (\ref{eq:von_neumann_example}) and the growth rate of the numerical instability becomes worse. 

\end{example}

%===================================================================
\subsection{The method of lines}
\label{sec:mol}
%===================================================================

A convenient approach both from an implementation point of view as well as for analyzing numerical stability or constructing numerically stable schemes is to decouple spatial and time discretizations. That is, one first analyzes stability under some spatial approximation assuming time to be continuous (\textit{semi-discrete stability}) and then finds conditions for time integrators to preserve stability in the \textit{fully discrete} case. 

In general, this method provides only a subclass of numerically stable approximations. However, it is a very practical one, since spatial and time stability are analyzed separately and stable semi-discrete approximations and appropriate time integrators can then be combined at will, leading to modularity in implementations. 

%===================================================================
\subsubsection{Semi discrete stability}

Consider the approximation 
\begin{eqnarray}
\mathbf{v}_t (t) &=& \mathbf{L} \mathbf{v} \,, \quad t>0  \label{eq:semi_discretea} \\
\mathbf{v}(0) &=& \mathbf{f}   \label{eq:semi_discreteb}
\end{eqnarray}
for the initial value problem (\ref{eq:ibv1_num}, \ref{eq:cauchy2}).
The scheme is semi-discrete stable if the solution to Equations~(\ref{eq:semi_discretea}, \ref{eq:semi_discreteb}) satisfies the estimate~(\ref{eq:num_stability_ivp}). 

In the time-independent case, the solution to (\ref{eq:semi_discretea}, \ref{eq:semi_discreteb}) is 
$$
\mathbf{v} (t) = e^{\mathbf{L} t} \mathbf{f}
$$
and stability holds if and only if there are constants $K_{\tt d}$ and $\alpha_{\tt d}$ such that
$$
\| e^{\mathbf{L} t} \|_{\tt d} \leq K_{\tt d} e^{\alpha_{\tt d} t}\quad
\mbox{for all $t\geq 0$}.
$$
The von Neumann condition now states that there exists a constant $\alpha_{\tt d}$, independent of spatial resolution (i.e. the size of the matrix $\mathbf{L}$), such that the eigenvalues ${\ell}_i$ of $\mathbf{L}$ satisfy
\be
\max_i \re ({\ell}_i)  \leq \alpha_{\tt d}.
\label{eq:von_neumann_semidiscrete}
\ee
This is the discrete in space version of the Petrovskii condition, see Lemma~\ref{Lem:Petrovskii}. As already pointed out, it is not always a sufficient condition for stability, unless $\mathbf{L}$ can be uniformly diagonalized. Also, if the lower order terms are dropped from the analysis then 
$$
\mathbf{L} = \frac{1}{(\Delta x)^p}\tilde\mathbf{L}
$$
with $\tilde\mathbf{L}$ independent of $\Delta x$, and in order for (\ref{eq:von_neumann_semidiscrete}) to hold for all $\Delta x$ (in particular small $\Delta x$), 
\be
\max_i \re ({\ell}_i)  \leq 0 \, ,  \label{eq:strong_von_neumann_semidiscrete}
\ee
which is a stronger version of the semidiscrete von Neumann condition. 

Semi-discrete stability also follows if $\mathbf{L}$ is semi-bounded, that is, there is a constant $\alpha_{\tt d}$ independent of resolution such that (cf. Equation~(\ref{Eq:Symmetrizer}) in Theorem~\ref{Thm:MatrixTheorem})
\begin{equation}
\langle \mathbf{v}, \mathbf{L} \mathbf{v} \rangle_{\tt d} + \langle \mathbf{L} \mathbf{v},  \mathbf{v} \rangle_{\tt d}  \leq 2 \alpha_{\tt d} \| \mathbf{v}\|^2_{\tt d}\quad  \mbox{ for all $\mathbf{v}$ }.
\label{eq:semi-bounded}
\end{equation}
In that case the semi-discrete approximation (\ref{eq:semi_discretea}, \ref{eq:semi_discreteb})  is numerically stable, as follows immediately from the following energy estimate arguments,  
$$
\frac{d}{dt}\| \mathbf{v} \|^2_{\tt d} = \frac{d}{dt} \langle \mathbf{v} , \mathbf{v} \rangle_{\tt d} = \langle \mathbf{L v} ,\mathbf{v} \rangle_{\tt d} + \langle \mathbf{v} , \mathbf{L v} \rangle_{\tt d} \leq 2\alpha_{\tt d} \| \mathbf{v} \|^2_{\tt d}. 
$$ 

For a large class of problems which can be shown to be well posed using the energy estimate, one can construct semi-bounded operators $\mathbf{L}$ by satisfying the   discrete counterpart of the properties of the differential operator $P$ in Equation~(\ref{eq:cauchy1}) that were used to show well posedness. This leads to the construction of spatial differential approximations satisfying the \textit{Summation By Parts} property, discussed in Sections~\ref{sec:sbp} and \ref{sec:gauss_and_sbp}. 

%==============================================================================
\subsubsection{Fully discrete stability}
\label{sec:fully_discrete}

Now we consider explicit time integration for systems of the form (\ref{eq:semi_discretea}, \ref{eq:semi_discreteb}) with time-independent coefficients. That is, if there are $N$ points in space we consider the system of ordinary differential equations (ODEs)
\be
\mathbf{v}_t = \mathbf{L} \mathbf{v} \, , \label{eq:MoL}
\ee
where $\mathbf{L }$ is an $N \times N$ matrix. 

In the previous subsection we derived necessary conditions for semidiscrete stability of such systems. Namely, the von~Neumann one in its weak (\ref{eq:von_neumann_semidiscrete}) and strong (\ref{eq:strong_von_neumann_semidiscrete}) forms.  Below we shall derive necessary conditions for \emph{fully discrete} stability for a large class of time integration methods, including Runge--Kutta ones. Upon time discretization, stability analyses of (\ref{eq:MoL}) require the introduction of the notion of the region of \textit{absolute stability} of ODE solvers. Part of the subtlety in the stability analysis of fully discrete systems is that the size $N$ of the system of ODEs is not fixed; instead, it depends on the spatial resolution.  However,  the obtained necessary conditions for fully discrete stability will also turn out to be sufficient when combined with additional assumptions. We will also discuss sufficient conditions for fully discrete stability using the energy method.  

\paragraph*{Necessary conditions.}

Recall the von~Neumann condition for the semidiscrete system (\ref{eq:MoL}): if ${\ell }_i$ is an eigenvalue of $\mathbf{L}$ a necessary condition for semidiscrete stability is [cf.\ Equation~(\ref{eq:von_neumann_semidiscrete})]
\be
\max_i \re(\ell_i) \leq \alpha_{\tt d} \, . 
\label{eq:spectrum_neg}
\ee 
for some $\alpha_{\tt d}$ independent of $N$. 

Suppose now that the system of ODEs (\ref{eq:MoL}) is evolved in time using a one-step explicit scheme, 
\begin{equation}
\mathbf{v}^{k+1} = \mathbf{Q}  \mathbf{v}^{k}\, ,  \label{eq:Q_fullydiscrete}
\end{equation}
and recall (cf.\ Equation~(\ref{eq:von_neumann_strong})) that a necessary condition for the stability of the fully discrete system (\ref{eq:Q_fullydiscrete}) under the assumption (\ref{eq:Q_cfl}) that $\mathbf{Q}$ depends only on the CFL factor is that its eigenvalues $\{q_i \}$ satisfy 
\begin{equation}
\max_i | q_i | \leq 1
\label{eq:abs_stability}
\end{equation}
for all spatial resolutions $N$. Next, assume that the ODE solver is such that  
\begin{equation}
\mathbf{Q} = {\tt R}(\Delta t \mathbf{L}) \,, \mbox{ where } {\tt R} \mbox{ is a polynomial in } \Delta t\mathbf{L},\epubtkFootnote{As we will discuss later, this includes explicit, one-step Runge--Kutta methods.}
\label{eq:mol_rk} 
\end{equation}
and  notice that if $\ell _i$ is an eigenvalue of $\mathbf{L}$, then 
$$
r_i:= {\tt R}(\Delta t \ell_i)
$$
is an eigenvalue of $\mathbf{Q}$,  
$$
\{ r_i \} \subset \{ q_i \} \,, 
$$
and (\ref{eq:abs_stability}) implies that a necessary condition for fully discrete stability is  
\begin{equation}
\max_i | r_i | \leq 1 \label{eq:abs_stability_r} \, . 
\end{equation}

\begin{definition}
The region of absolute stability of a one-step time integration scheme $\tt R$ is defined as
$$
{\cal S_{\tt R}}:= \{ z \in  \Complex : |{\tt R}(z)| \leq 1 \} \,. 
$$ 
\end{definition}
The necessary condition (\ref{eq:abs_stability_r}) can then be restated as:
\begin{lemma}[Fully discrete von~Neumann condition for the method of lines.]  \label{lemma:vn_MoLa} 
Consider the semidiscrete system (\ref{eq:MoL}) and a one-step explicit time discretization (\ref{eq:Q_fullydiscrete}) satisfying the assumptions (\ref{eq:mol_rk}).  Then, a necessary condition for fully discrete stability is that the spectrum of the scaled spatial approximation $\Delta t \mathbf{L}$  is contained in the region of absolute stability of the ODE solver ${\tt R}$, 
$$
\sigma(\Delta t\mathbf{L})\subset {\cal S_{\tt R}} \, .
$$
for all spatial resolutions $N$.
\end{lemma}

In the absence of lower order terms and under the already assumed conditions (\ref{eq:mol_rk}) the strong von Neumann condition (\ref{eq:strong_von_neumann_semidiscrete}) then implies that ${\cal S_{\tt R}}$ must overlap the half complex plane $\{ z\in\Complex : \re(z)\leq 0\}$. In particular, this is guaranteed by locally stable schemes, defined as follows. 
\begin{definition}
An ODE solver ${\tt R}$ is said to be locally stable if its region of absolute stability ${\cal S_{\tt R}}$ contains an open half disc $D_-(r) := \{ \mu\in\Complex : |\mu| < r, \re(\mu) < 0 \}$ for some $r > 0$ such that 
$$
D_-(r)\subset {\cal S_{\tt R}}  \, . 
$$ 
\end{definition}

As usual, the von~Neumann condition is not sufficient for numerical stability and we now discuss an example, drawn from \cite{Kreiss1993213}, showing that the particular version of Lemma \ref{lemma:vn_MoLa} is not either. 

\begin{example}
Consider the following advection problem with boundaries: 
\begin{eqnarray}
u_t = - u_x, && x\in [0,1],\quad t\geq 0,\nonumber\\
u(t,0) = 0, && x=0,\quad t\geq 0,\nonumber\\
u(0,x) = f(x), && x\in [0,1],\quad t=0,
\nonumber  
\end{eqnarray}
where $f$ is smooth and compactly supported on $[0,1]$, and the one-sided forward Euler discretization, cf.\ Example~\ref{ex:one_sided_euler}, with \textit{injection} of the boundary condition (see Section~\ref{sec:num_boundary})
\begin{eqnarray}
\frac{v^{k+1}_j - v^{k}_j}{\Delta t} = - \frac{v^k_j - v^k_{j-1}}{\Delta x} \,, && \mbox{ for } j=1,2,\ldots N,\nonumber\\
v^k_0 = 0, && \mbox{ for } j=0.\nonumber
\end{eqnarray}

The corresponding semidiscrete scheme can be written in the form
$$
\frac{d}{dt}\mathbf{v } = \mathbf{L} \mathbf{v}
$$
for (notice that in the following expression the boundary point is excluded and not evolved) 
$$
\mathbf{v} = \left( v_1,v_2,\ldots, v_N \right)^{T}\, , 
$$
with $\mathbf{L}$ the banded matrix 
$$
\mathbf{L} = \frac{1}{\Delta x}\left( \begin{array}{ccccc}
-1 & 0 & 0 & \cdots & 0 \\
1  & -1 & 0 & \cdots &  0 \\
0  & 1 & -1 & \cdots& 0 \\
\vdots  & \vdots  & \ddots  & \ddots & 0 \\
0 & 0  & \cdots & 1 & -1  
\end{array}
\right ) \, , 
$$
followed by integration in time through the Euler method, 
\be
\mathbf{v}^{k+1} = \left( \mathbf{1} + \Delta t \mathbf{L} \right)\mathbf{v}^k \, .  \label{eq:euler_example_mol}
\ee

Since $\mathbf{L}$ is triangular, its eigenvalues are the elements of the diagonal; namely, $\{ {\ell}_i\} = \{ -1/\Delta x \}$, i.e., there is a single, degenerate eigenvalue $q=-1/\Delta x$. 

The region of local stability of the Euler method is
$$
S_{E} = \{ z \in \Complex : |1 + z| \leq 1  \}, 
$$
which is a closed disk of radius $1$ in the complex plane centered at $z_0 = -1$, see Figure~\ref{fig:RK}. The von~Neumann condition then is 
$$
| 1 + q \Delta t | = \left| 1 - \frac{\Delta t}{\Delta x} \right|\leq 1 
$$
or 
$$
\frac{\Delta t}{\Delta x}\leq 2.
$$
On the other hand, if the initial data has compact support, the numerical solution at early times is that one of the periodic problem discussed in Example~\ref{ex:one_sided_euler}, for which the Fourier analysis gave a stability condition of 
$$
\frac{\Delta t}{\Delta x} \leq 1.
$$
\end{example}

\paragraph*{Sufficient conditions.}

Under additional assumptions, fully discrete stability does follow from semi-discrete stability if the time integration is locally stable:
\begin{theorem}[Kreiss--Wu~\cite{Kreiss1993213}]
\label{theorem:kreiss_wu}
Assume that 
\begin{enumerate}
\item[(i)] A consistent semidiscrete approximation to a constant-coefficient, first order initial-boundary value problem is stable in the generalized sense (see Definition~\ref{def:strong_stability} and the following remarks).
\item[(ii)] The resulting system of ODEs is integrated with a locally stable method of the form (\ref{eq:mol_rk}), with stability radius $r>0$.
\item[(iii)] If $\alpha \in \Real$ is such that $|\alpha | <  r$ and  
$$
{\tt R}(i \alpha ) = e^{i \phi} \, , \quad \phi \in \Real \, , 
$$
then there is no $\beta \in \Real $ such that $| \beta | < r $, ${\tt R}(i \beta ) = e^{i \phi}$,  and $\beta \neq \alpha$.  
\end{enumerate}
Then the fully discrete system is numerically stable, also in the generalized sense, under the CFL condition 
$$
\Delta t \|  \mathbf{L} \|_{\tt d} \leq \lambda \, \;\;\; \mbox{ for any } \lambda < r \, .
$$
\end{theorem}

\textbf{Remarks}
\begin{itemize}
\item Condition (iii) can be shown to hold for any consistent approximation, if $r$ is sufficiently small~\cite{Kreiss1993213}.

\item Explicit, one-step Runge--Kutta (RK) methods,  which will be discussed in Section~\ref{sec:time_integration}, are in particular of the form~(\ref{eq:mol_rk}) when applied to linear, time-independent problems. In fact, consider an arbitrary, consistent, one-step, explicit ODE solver (\ref{eq:Q_fullydiscrete}) of the form given in Equation~(\ref{eq:mol_rk}), 
\begin{equation}
\mathbf{Q} = {\tt R}(\Delta t \mathbf{L}) = \sum_{j=0}^{s}  \alpha_j \frac{\left( \Delta t \mathbf{L} \right)^j}{j!} \, \mbox{ with } \alpha_s \neq 0;  \label{eq:stages_mol}
\end{equation}
the integer $s$ is referred to as the \textit{number of stages}. 

Since the exact solution to Equation~(\ref{eq:MoL}) is $\mathbf{v}(t) = e^{t \mathbf{L}}\mathbf{v}(0)$ and, in particular, 
$$
\mathbf{v}(t_{k+1}) = e^{\Delta t \mathbf{L}}\mathbf{v}(t_k) 
 = \sum_{j=0}^\infty\frac{\left( \Delta t \mathbf{L} \right)^j}{j!} \mathbf{v}(t_k) \, , 
$$ 
Equation~(\ref{eq:stages_mol}) must agree with the first $n$ terms of the Taylor expansion of $e^{\Delta t \mathbf{L}}$, where $n$ is the order of the \emph{global} \epubtkFootnote{This refers to the truncation error at a fixed final time, as opposed to the local one after an iteration. } truncation error of the ODE solver, defined through  
$$
{\tt R}(\Delta t \mathbf{L}) - e^{\Delta t\mathbf{L}} = {\cal O}\left( \Delta t \mathbf{L} \right)^{n+1} \, . 
$$
Therefore, we must have
\begin{equation}
\alpha_j= 1 \, \mbox{ for } 0\leq j \leq n \, . 
\label{eq:neqs_mol}  
\end{equation}
We then see that a scheme of order $n$ needs at least $n$ stages, $s\geq n$, and
$$
{\tt R}(\Delta t \mathbf{L}) = \sum_{j=0}^{n} \frac{\left( \Delta t \mathbf{L} \right)^j}{j!} + \sum_{j=n+1}^s \alpha_j \frac{\left( \Delta t \mathbf{L} \right)^j}{j!} \, . 
$$
The above expression in particular shows that when $s=n$ (i.e. when the second sum on the right hand side is zero) the scheme is unique, with coefficients given by Equation~(\ref{eq:neqs_mol}). In particular, for $n=1=s$, such scheme corresponds to the Euler one discussed in the example above, see Equation~(\ref{eq:euler_example_mol}).

\item As we will discuss in Section~\ref{sec:time_integration}, in the nonlinear case it is possible to choose RK methods with $s=n$ if and only if $n<5$.
 
\item When $s=n$, first and second order RK are not locally stable, while third and fourth order are. The fifth order Dormand--Prince scheme (also introduced in Section~\ref{sec:time_integration}) is also locally stable.

\epubtkImage{}{
\begin{figure}[h]
\centerline{\includegraphics[width=15cm]{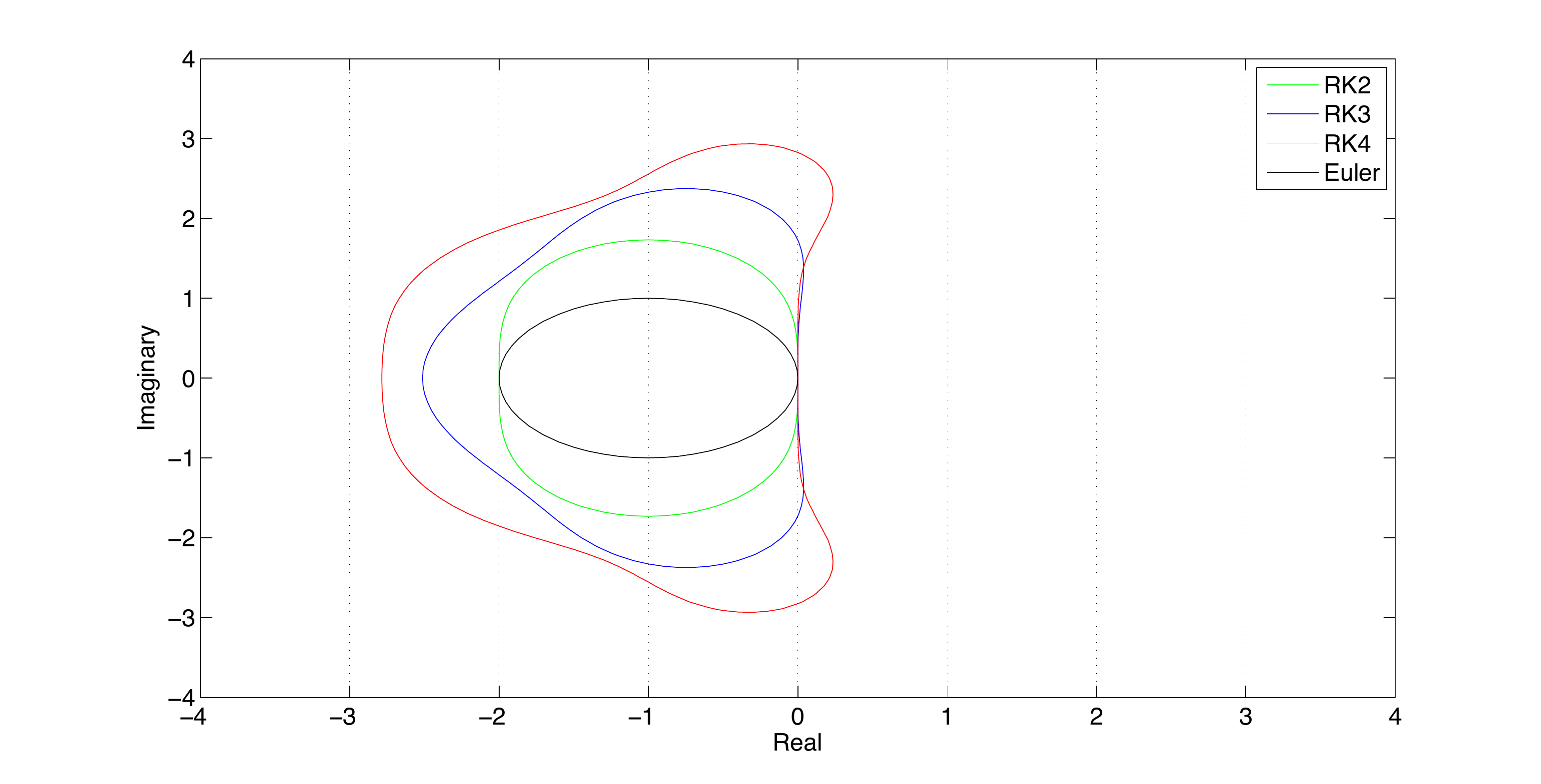}}
\caption{Regions of absolute stability of RK time integrators with $s=n$. Second order RK (RK2) and the Euler method are not locally stable.}
\label{fig:RK}
\end{figure}}

\end{itemize}

Using the energy method, fully discrete stability can be shown (resulting in a more restrictive CFL limit) for third order Runge--Kutta integration and arbitrary dissipative operators $\mathbf{L}$~\cite{Levy98fromthe, Tadmor2002}:

\begin{theorem}[Levermore]
\label{Thm:Levermore}
Suppose $\mathbf{L}$ is dissipative, that is Equation~(\ref{eq:semi-bounded}) with $\alpha_{\tt d}=0$ holds. Then, the third order Runge-Kutta approximation $\mathbf{v}(t_{k+1}) = {\tt R}_3(\Delta t\mathbf{L})\mathbf{v}(t_k)$, where
$$
{\tt R}_3(z) = 1 +  z + \frac{z^2}{2} + \frac{z^3}{6},
$$
for the semi-discrete problem~(\ref{eq:MoL}) is strongly stable,
$$
\| {\tt R}_3(\Delta t\mathbf{L}) \|_{\tt d} \leq 1
$$
under the CFL timestep restriction $\Delta t\| \mathbf{L} \|_{\tt d} \leq 1$.
\end{theorem}

Notice that the restriction $\alpha_{\tt d} = 0$ is not so severe, since one can always achieve it by replacing $\mathbf{L}$ with $\mathbf{L} - \alpha_{\tt d} I$. A generalization of Theorem~\ref{Thm:Levermore} to higher order Runge-Kutta methods does not seem to be known.

%===================================================================
\subsection{Strict or time-stability}
\label{sec:time_stability}

Consider a well posed problem. For the sake of simplicity, we assume it is a linear initial value one; a similar discussion holds for linear initial-boundary value problems. According to Definition~\ref{Def:WPVC} there are constants $K$ and $\alpha$ such that smooth solutions satisfy
\begin{equation}
\| u(t,\cdot) \| \leq K e^{\alpha t} \| u(0,\cdot) \| \qquad \mbox{ for all } t \geq 0.
\label{eq:estimate-cont}
\end{equation}

\begin{definition} \label{def:strict-stability}
For a numerically stable semidiscrete approximation, there are resolution independent constants $K_{\tt d},\alpha_{\tt d}$ such that for all initial data
\begin{equation}
\| v(t,\cdot) \|_{\tt d} \leq K_{\tt d} e^{\alpha_{\tt d} t} \| v(0,\cdot) \|_{\tt d}
\qquad \mbox{ for all } t \geq 0.
\label{eq:estimate-num} 
\end{equation}
The approximation is called strict or time stable if, for finite differences,  
\begin{equation}
\alpha_{\tt d} \leq \alpha + {\cal O}(\Delta x).
\label{eq:strict-stability}
\end{equation}
\end{definition}
Similar definitions hold in the fully discrete case and/or when the spatial approximation is not a finite difference one. Essentially, (\ref{eq:strict-stability}) attempts to capture the notion that the numerical solution should not have, at a fixed resolution, growth in time which is not present at the continuum. The problem with the definition, however, is that it is not useful if the estimate (\ref{eq:estimate-cont}) is not sharp, since neither will be (\ref{eq:estimate-num}), and the numerical solution can still exhibit artificial growth. 

\begin{example}
Consider the problem (drawn from Ref.~\cite{Lehner:2004cf})
\begin{eqnarray}
u_t = u' + \frac{u}{x}\,, && a\leq x\leq b,\quad t\geq 0,\label{eq:strict-ex1}\\
u(t,b) = \frac{1}{b}\,, && x=b,\quad t\geq 0,\label{eq:strict-ex2}\\
u(0,x) = \frac{1}{x}\,, && a\leq x\leq b,\quad t=0,\label{eq:strict-ex3}
\end{eqnarray}
where $b > a > 0$, for which the solution is the stationary one 
$$
u(t,x) = \frac{1}{x} \, . 
$$ 

Defining 
$$
E(t)= \langle u(t,\cdot), u(t,\cdot) \rangle \mbox{ with } 
\langle u, v \rangle = \int\limits_a^b u(x) v(x) dx, 
$$
it follows that 
\begin{equation}
\frac{d}{dt}E(t) = \frac{1}{b^2} - u(a)^2 + 2\langle u, u/x \rangle 
 \leq \frac{1}{b^2} + 2\langle u, u/x \rangle 
 \leq \frac{1}{b^2} + \frac{2}{a}\langle u, u \rangle = \frac{1}{b^2} + \frac{2}{a}2 E(t).
\label{eq:non-sharp-estimate}
\end{equation}
\end{example}
This energy estimate implies that the spatial norm of $u(t)$ cannot grow faster than $e^{t/a}$. However, in principle this need not be a sharp bound. In fact, Equation~(\ref{eq:strict-ex1}) is, in disguise, the advection equation $(x u)_t = (x u)_x$ for which the general solution $(x u)(t,x) = F(t+x)$ does not grow in time. 
Therefore, a numerical scheme for the problem~(\ref{eq:strict-ex1}, \ref{eq:strict-ex2}, \ref{eq:strict-ex3}) whose solutions are allowed to grow exponentially at the rate $e^{t/a}$ is strictly stable according to Definition~\ref{def:strict-stability} but the classification of the scheme as such is of not much use if the growth does take place. In order to illustrate this, we show the results for two schemes. In the first one, spurious growth takes place although the scheme is  strictly stable, and in the second case one obtains a strictly stable scheme with respect to a sharp energy estimate which does not exhibit such growth.

If the system (\ref{eq:strict-ex1}, \ref{eq:strict-ex2}, \ref{eq:strict-ex3}) is approximated by 
\begin{equation}
v_t = D v + \frac{v}{x} \, ,  \label{eq:nosharp}
\end{equation}
where $D$ is a finite difference  operator satisfying Summation By Parts (SBP) (Section~\ref{sec:sbp}) and the boundary condition is imposed through a projection method (Section~\ref{sec:num_boundary}), it can be shown --as discussed in the next section-- that the following semi-discrete estimate holds, at least for analytic boundary conditions, with the discrete SBP scalar product $\langle u,v\rangle_\mathbf{\Sigma}$ defined by Equation~(\ref{eq:prod}) in the next section,
\begin{equation}
\frac{d}{dt}E_{\tt d}(t) \leq \frac{1}{b^2} + \frac{2}{a} E_{\tt d}(t)\, , \qquad E_{\tt d}:= \langle v , v \rangle_{\tt d} \, . 
\label{eq:non-sharp-estimate-discrete}
\end{equation}
Technically, the semi-discrete approximation~(\ref{eq:nosharp}) is strictly stable, since the continuum~(\ref{eq:non-sharp-estimate}) and semi-discrete~(\ref{eq:non-sharp-estimate-discrete}) estimates agree. However, this does not preclude spurious growth,  as the bounds are not sharp. The left panel of Figure~\ref{fig:strict-stability} shows results for the $D_{2-1}$ SBP operator (see the definition in Example~\ref{ex:D21} below), boundary conditions through an orthogonal projection, and third order Runge--Kutta time integration. At any time, the errors do converge to zero with increasing resolution as expected since the scheme is numerically stable. However, at any fixed resolution there is spurious growth in time. 

\epubtkImage{}{
\begin{figure}[ht]
\centerline{
\includegraphics[width=0.5\textwidth]{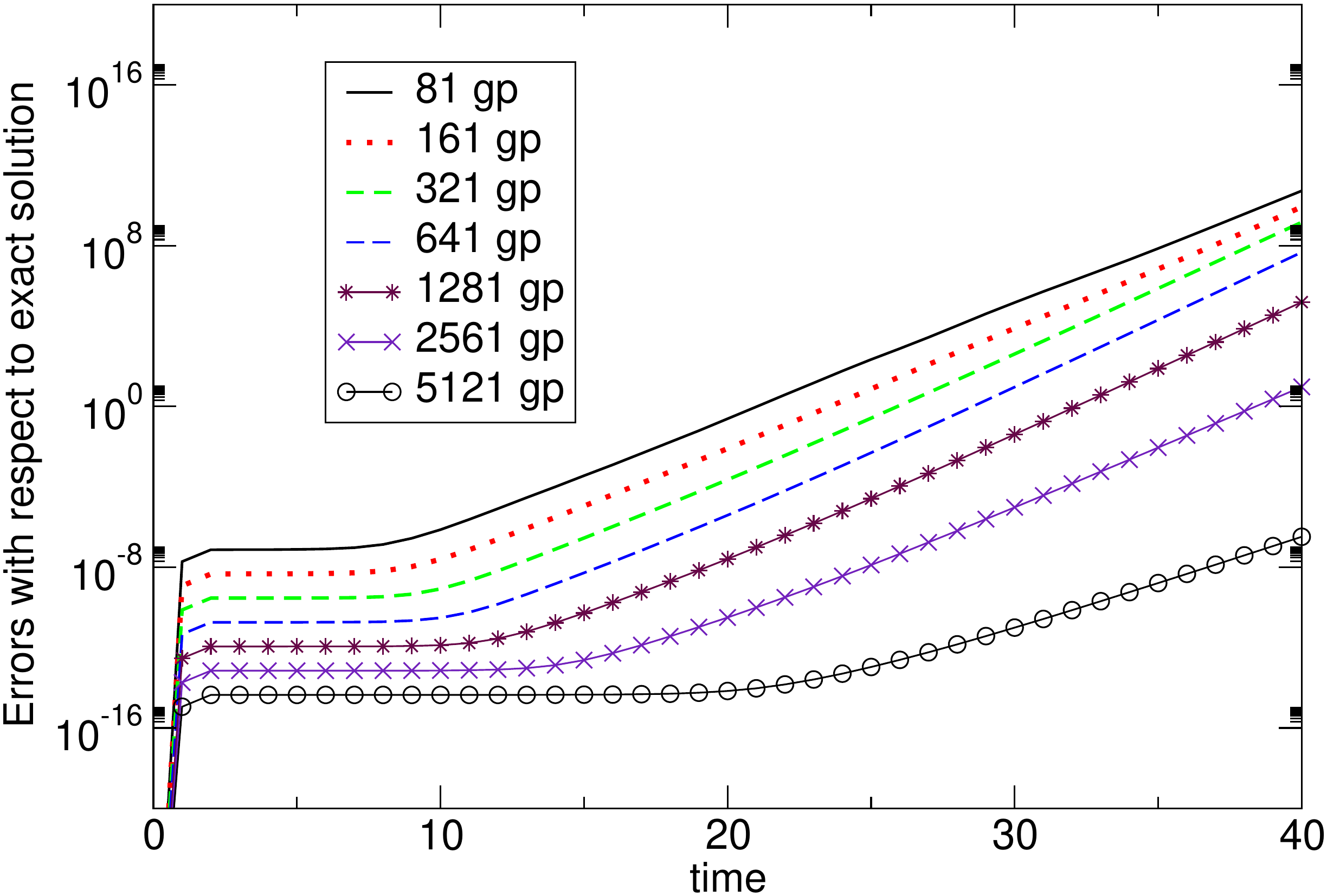} 
\includegraphics[width=0.5\textwidth]{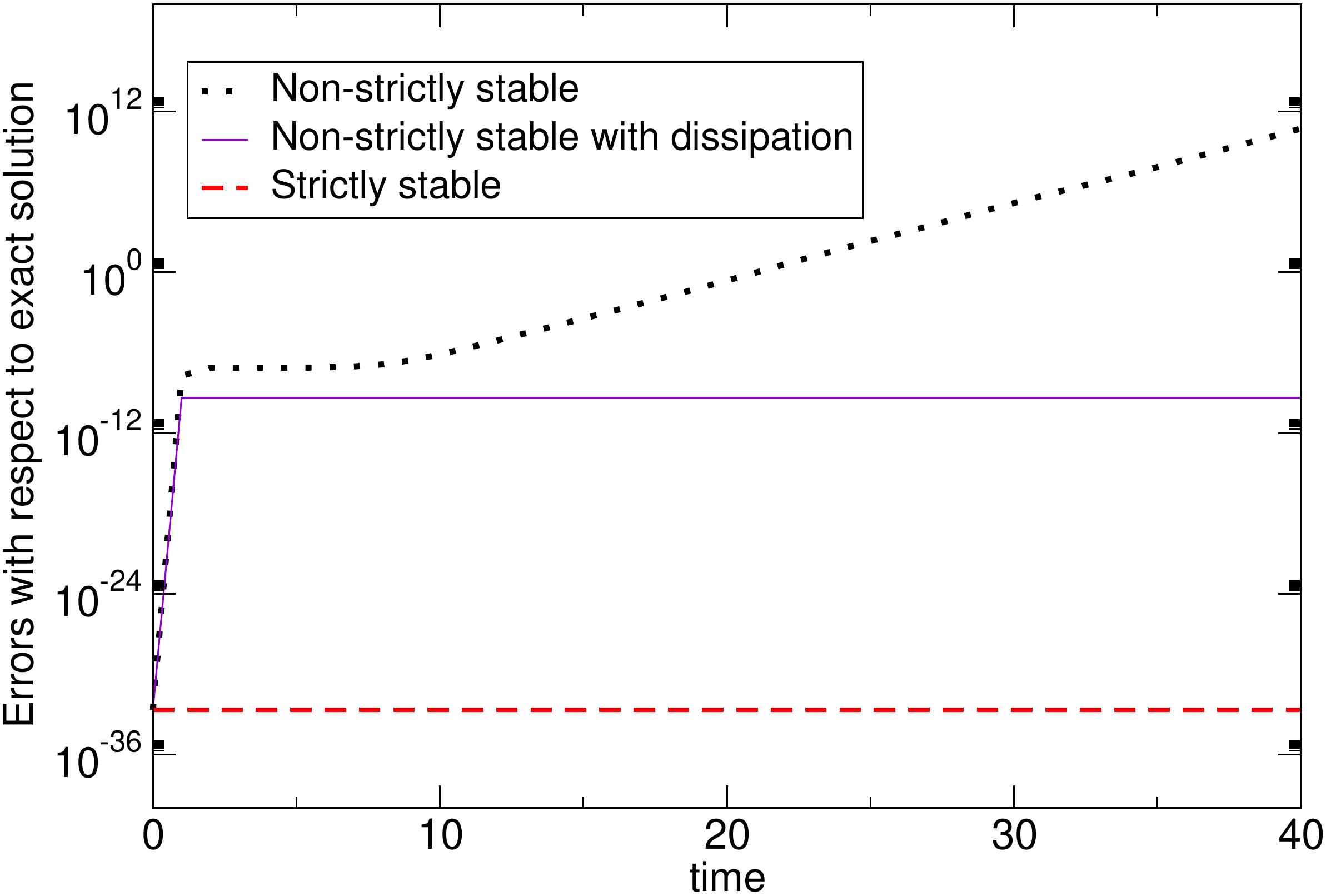}
}
\caption{Comparison~\cite{Lehner:2004cf} between numerical solutions
  to (\ref{eq:nosharp}) (left, and right labeled as \textit{Non-strictly
    stable}) and (\ref{eq:sharp}) (right, labeled as \textit{Strictly
    stable}). The left panel shows that the scheme is numerically
  stable but not time-stable. The right panel shows that a
  conservative scheme is time-stable and considerably more accurate.}
\label{fig:strict-stability}
\end{figure}}

On the other hand, discretizing the system as 
\begin{equation}
v_t = \frac{1}{x}D\left(xv \right) 
\label{eq:sharp}
\end{equation}
is in general not equivalent to (\ref{eq:nosharp}) because difference operators do not satisfy the Leibnitz rule exactly. In fact, defining 
$$
\tilde{E}_{\tt d} := \langle xv, xv \rangle_\mathbf{\Sigma}
$$ 
it follows that 
\begin{equation}
 \frac{d}{dt}\tilde{E_{\tt d}}(t) =  -\frac{a^2v(a)^2}{2}+ \frac{b^2v(b)^{2}}{2}  \, , 
\end{equation}
which, being an equality, is as sharp as an estimate can be. 

The right panel of Figure~\ref{fig:strict-stability} shows a comparison between discretizations (\ref{eq:nosharp}) and (\ref{eq:sharp}), as well as  (\ref{eq:nosharp}) with the addition of numerical dissipation (see Section~\ref{sec:dissipation}), in all cases at the same fixed resolution. Even though numerical dissipation does stabilize the spurious growth in time, the strictly stable discretization (\ref{eq:sharp}) is considerably more accurate. Technically, according to Definition~\ref{def:strict-stability} the approximation (\ref{eq:nosharp}) is also strictly stable, but it is more useful to reserve the term to the cases in which the estimate is sharp. The approximation (\ref{eq:sharp}), on the other hand, is (modulo the flux at boundaries, discussed in Section~\ref{sec:num_boundary}) \textit{energy preserving} or \textit{conservative}. 

In order to construct conservative or time stable semidiscrete schemes one essentially needs to write the approximation by grouping terms in such a way that when deriving at the semidiscrete level what would be the conservation law at the continuum, the need of using the Leibnitz rule is avoided. In addition, the numerical imposition of boundary conditions also plays a role (see Section~\ref{sec:num_boundary}). 

In many application areas, conservation or time-stability play an important role in the design of numerical schemes. That is not so much (at least so far) the case for numerical solutions of Einstein's equations because in General Relativity there is no gauge-invariant local notion of conserved energy unlike many other nonlinear hyperbolic systems (most notably, in Newtonian or special relativistic Computational Fluid Dynamics), see however \cite{Szabados:2009vb}. In addition, there are no generic sharp estimates for the growth of the solution that can guide the design of numerical schemes. 
However, in simpler settings such as fields propagating on some stationary fixed background geometry, there is a notion of conserved local energy and accurate conservative schemes are possible. Interestingly, in several cases such as Klein--Gordon or Maxwell fields in stationary background spacetimes the resulting conservation of the semidiscrete approximations follows regardless of the constraints being satisfied (see, for example, \cite{Lehner:2004cf}). 
A local conservation law in stationary spacetimes can also guide the construction of schemes to guarantee stability in the presence of coordinate singularities \cite{Calabrese:2003vy, Sarbach:2003dz, Gundlach:2010et, Neilsen:2004aj}, as discussed in the Section \ref{sec:remarks_stability}.  

In addition, there has been work done on variational, symplectic or mimetic integration techniques for Einstein's equations, which aim at exactly or approximately preserving the discrete constraints while solving the discrete evolutions equations. See, for example, \cite{Meier:2003bn, DiBartolo:2004dn, Gambini:2005jm, Gambini:2005pg, Brown:2005jc, Campiglia:2006vy, Richter:2006qa, Frauendiener:2006dx, Richter:2008ur, Frauendiener:2008bt, Richter:2008pr, Richter:2009ff}.

%==================================================================
\subsection{Runge--Kutta methods}
\label{sec:time_integration}

In the method of lines approach one ends up effectively integrating a system of ordinary differential equations, which we now generically denote  by 
$$
\frac{d}{dt}y(t) = f(t,y) \, . 
$$
The majority of approaches in numerical relativity use one-step, multi-stage, explicit Runge--Kutta (RK) methods, which take the following form. 
\begin{definition} \label{def:runge_kutta}
An explicit, s-stage, Runge--Kutta method is of the form
\begin{eqnarray*}
k_1 &=& f\left( t_n,y_n \right) \, ,  \\
k_2 &=& f\left( t_n + c_2 \Delta t, y_n + \Delta t a_{21} k_1 \right) \, ,  \\
k_3 &=& f\left( t_n + c_3 \Delta t, y_{n} + \Delta t \left( a_{31}k_1 + a_{32}k_2 \right) \right) \, , \\
\vdots && \\
k_s&=& f\left( t_n + c_s \Delta t, y_n + \Delta t \left( a_{s1}k_1 + \ldots + a_{s,s-1} k_{s-1} \right) \right) \, ,  \\
y_{n+1} &= & y_n + \Delta t (b_1 k_1 + \ldots + b_s k_s) \, . 
\end{eqnarray*}
with $b_l \, (1\leq l\leq s)$,  $c_j \,  (2 \leq j \leq s)$, and $a_{ji} (1 \leq i \leq j-1)  $ real numbers. 
\end{definition}
Next we present a few examples, in increasing order of accuracy. The simplest one is a forward finite difference scheme [cf. Equation~\ref{eq:one_sided}]. 
\begin{example} \label{ex:rk1} The Euler method (one stage, first order): 
\begin{eqnarray*} 
k_1 &=& f\left( t_n,y_n \right) \\
y_{n+1} &=& y_n + \Delta t k_1
\end{eqnarray*}
It corresponds to all the coefficients set to zero except $b_1=1$. 
\end{example}

\begin{example} \label{ex:rk2} Second order, two-stages RK :
\begin{eqnarray*}
k_1 &=& f\left( t_n,y_n \right) \\
k_2 &=& f\left( t_n + \frac{\Delta t}{2}, \frac{\Delta t }{2} k_1 \right) \, ,  \\
y_{n+1} &=&  y_{n} + \Delta t k_2 \, , 
\end{eqnarray*}
The non-vanishing coefficients are $c_{2}=1/2=a_{21},b_2=1$. 
\end{example}

\begin{example} \label{ex:rk3}Third order, four-stages RK:
\begin{eqnarray*}
k_1 &=& f\left( t_n,y_n \right) \\
k_2 &=& f\left( t_n + \frac{\Delta t}{2}, y_n + \frac{\Delta t}{2}  k_1 \right) \\
k_3 &=& f\left( t_n + \Delta t, y_{n} + \Delta t k_2 \right)\\
k_4 &=& f\left( t_n + \Delta t, y_n + \Delta t k_3 \right) \\
y_{n+1} &= & y_n + \Delta t \left( \frac{k_1}{6} + \frac{2k_2}{3} + \frac{k_4}{6} \right)
\end{eqnarray*}
\end{example}
In the previous case the number of stages is larger than the order of the scheme. It turns out that it is possible to have a third order RK scheme with three stages. 
\begin{example} \label{ex:heun} Third order, three-stages Heun--RK method 
\begin{eqnarray*}
k_1 &=& f\left( t_n,y_n \right) \\
k_2 &=& f\left( t_n + \frac{\Delta t}{3}, y_n + \frac{\Delta t }{3}k_1 \right) \\
k_3 &=& f\left( t_n + \frac{2\Delta t}{3}, y_{n} + \frac{2\Delta t }{3}k_2\right)\\
y_{n+1} &= & y_n + \Delta t \left( \frac{k_1}{4} + \frac{3k_3}{4} \right)
\end{eqnarray*}
It is also possible to find a fourth order, four-stages RK scheme. Actually, there are multiple ones, with one and two-dimensional free parameters. A popular choice is
\begin{example} \label{ex:rk4}The standard fourth order, four-stages RK method:
\begin{eqnarray*}
k_1 &=& f\left( t_n,y_n \right) \\
k_2 &=& f\left( t_n + \frac{\Delta t}{2}, y_n + \frac{ \Delta t}{2}k_1 \right) \\
k_3 &=& f\left( t_n + \frac{\Delta t}{2}, y_{n} + \frac{\Delta t }{2}k_2 \right)\\
k_4 &=& f\left( t_n + \Delta t, y_{n} + \Delta t k_3 \right)\\
y_{n+1} &= & y_n + \Delta t \left( \frac{k_1}{6}+ \frac{2k_2}{6} + \frac{2k_3}{6} + \frac{k_4}{6} \right)
\end{eqnarray*}
\end{example}
\end{example}
At this point it is clear that since the RK methods have the same structure, it is not very efficient to explicitly write down all the stages and the final step. Butcher's tables are a common way to represent them, they have the following structure:

\begin{table}[h]
\caption{The structure of a Butcher table.}
\label{tab:butcher}
\centering
\begin{tabular}{c | c c c c c }
0 &   &  &  & & \\
$c_2$ & $a_{21}$ &  &  & &  \\
$c_3$ &  $a_{31}$  & $a_{32}$  &  & &   \\
\vdots &  \vdots  & \vdots &  $\ddots$ & &  \\
$c_s$ &   $a_{s1}$ & $a_{s2}$ & \dots & $a_{s,s-1}$ &   \\
\hline 
& $b_1$ & $b_2$ & \dots & $b_{s-1}$ & $b_s$ 
\end{tabular}
\end{table}

Table~\ref{tab:butcher-low-order} shows representations of
Examples~\ref{ex:rk1}, \ref{ex:rk2}, \ref{ex:rk3}, \ref{ex:heun},
while Table~\ref{tab:butcher-rk4} shows the classical fourth order
Runge--Kutta method of Example~\ref{ex:rk4}. 

\begin{table}[h]
\caption[The Euler method, second and third order RK, third order Heun.]{From left to right: The Euler method, second and third order RK, third order Heun.} 
\label{tab:butcher-low-order}
\centering
\begin{tabular}{c | c   }
0 &      \\
\hline 
& 1  
\end{tabular}
\hspace{0.5cm}
\begin{tabular}{c | c c  }
0 &   & \\
0 & 1/2 &     \\
\hline 
& 0 & 1 
\end{tabular}
\hspace{0.5cm}
\begin{tabular}{c | c c c c  }
0 &   & & & \\
1/2 & 1/2 & & &     \\
1 & 0 & 1 &  &   \\
1 & 0 &0  & 1 &    \\
\hline 
& 1/6 & 2/3 & 0 & 1/6 
\end{tabular}
\hspace{0.5cm}
\begin{tabular}{c | c c c  }
0 &   & &  \\
1/3 & 1/3 &  &     \\
2/3 & 0 & 2/3 &     \\
\hline 
& 1/4 & 0 & 3/4 
\end{tabular}
\end{table}

\begin{table}
\caption{The standard fourth-order Runge--Kutta method.} 
\label{tab:butcher-rk4}
\centering
\begin{tabular}{c | c c c c }
0 &   & &  & \\
1/2 & 1/2 &  & &     \\
1/2 & 0 & 1/2 & &    \\
1 & 0 & 0 & 1 &    \\
\hline 
& 1/6 & 1/3 & 1/3 & 1/6
\end{tabular}
\end{table}

The above examples explicitly show that up to, and including, fourth order accuracy there are Runge--Kutta methods of order $p$ and $s$ stages with $s=p$. It is interesting that even though the first RK methods date back to the end of the nineteenth century, the question of whether there are higher order (than four) RK methods remained open until the following result was shown by Butcher in 1963~\cite{Butcher1964}: $s=p$ cannot be achieved anymore starting with fifth order accurate schemes, and there are a number of \textit{barriers}. 

\begin{theorem}
For $p\geq 5$ there are no Runge--Kutta methods with $s=p$ stages. 
\end{theorem}

There are however, fifth and sixth order RK methods with six and seven stages, respectively. Butcher in 1965~\cite{Butcher1965} and 1985~\cite{Butcher1985} respectively showed the next following barriers.
\begin{theorem} For $p\geq 7$ there are no Runge--Kutta methods with $s=p+1$ stages. 
\end{theorem}
\begin{theorem} For $p\geq 8$ there are no Runge--Kutta methods with $s=p+2$ stages. 
\end{theorem}
Seventh and eighth order methods with $s=9$ and $s=11$ stages, respectively, have been constructed, as  well as a tenth order one with $s=17$ stages.

\subsubsection{Embedded methods}

In practice many approaches in numerical relativity use an adaptive timestep method. One way of doing so is to evolve the system of equations two steps with timestep $\Delta t$ and one with $2\Delta t$. The difference in both solutions at $t+2\Delta t$ can be used, along with Richardson extrapolation, to estimate the new timestep needed to achieve any given tolerance error. 

In more detail: if we call $y_2$ the solution at $t+2\Delta t$ evolved from time $t$ in two steps of size $\Delta t$, and $\tilde{y}_1$ the solution at the same time advanced from time $t$ in one step of size $2\Delta t$, then the following holds 
$$
y(t + 2\Delta t) - y_2 = \frac{y_2-\tilde{y_1}}{2^p-1} + {\cal O}\left( \left( \Delta t \right) ^{p+1} \right)
$$
where $y$ denotes the exact solution. Therefore, the term
$$
\frac{y_2-\tilde{y_1}}{2^p-1}
$$
can be used as an estimate of the error and choose the next timestep.  
  
\textit{Embedded methods} also compute two solutions and use their
difference to estimate the error and adapt the timestep. However, this
is done by reusing the stages. Two Runge--Kutta methods, of order $p$
and $p'$ (in most cases -- but not always -- $p'=p+1$ or $p=p'+1$) are
constructed, which share the intermediate function values so that
there is no overhead cost. Therefore their Butcher table looks as
follows:

\begin{table}[ht]
\caption{The structure of embedded methods.} 
\label{tab:embedded-rk}
\centering
\begin{tabular}{c | c c c c c }
0 &   &  &  & & \\
$c_2$ & $a_{21}$ &  &  & &  \\
$c_3$ &  $a_{31}$  & $a_{32}$  &  $\ddots$ & &   \\
\vdots &  \vdots  & \vdots & & &  \\
$c_s$ &   $a_{s1}$ & $a_{s2}$ & \dots & $a_{s,s-1}$ &   \\
\hline 
& $b_1$ & $b_2$ & \dots & $b_{s-1}$ & $b_s$ \\
\hline
& $ b'_1$ & $b'_2$ & \dots & $b'_{s-1}$ & $b'_s$ 
\end{tabular}
\end{table}

Embedded methods are denoted by $p(p')$, where $p$ is the order of the scheme which advances the solution. For example, a $5(4)$ method would be of fifth order, with a fourth order scheme which shares its function calls used to estimate the error.

Table~\ref{tab:dormandprince54} shows the coefficients for a popular
embedded method, the seven stages Dormand--Prince $5(4)$
\cite{Dormand198019}. Dormand--Prince methods are embedded methods
which minimize a quantification of the truncation error for the
highest order component, which is the one used for the evolution. 

\begin{table}[ht]
\caption{The $5(4)$ Dormand--Prince method.} 
\label{tab:dormandprince54}
\centering
\begin{tabular}{c | c c c c c c c }
0 &   &  &  & & \\
$\ds \frac{1}{5}$ & $\ds \frac{1}{5}$ &  &  & & & &  \\
&   &  &  & &  & & \\
$\ds \frac{3}{10}$ &  $\ds \frac{3}{40}$  & $\ds \frac{9}{40}$  &  & &  & &   \\
&   &  &  & &  & & \\
$\ds \frac{4}{5}$ &  $\ds \frac{44}{45}$  & $\ds -\frac{56}{15}$ & $\ds \frac{32}{9}$ & &  & & \\
&   &  &  & &  & & \\
$\ds \frac{8}{9}$ &   $\ds \frac{19372}{6561}$ & $\ds -\frac{25360}{2187}$ & $\ds \frac{64448}{6561}$ & $\ds - \frac{212}{729}$ &    & & \\
&   &  &  & &  & & \\
$\ds 1$ &   $\ds \frac{9017}{3168}$ & $\ds -\frac{355}{33}$ & $\ds \frac{46732}{5247}$ & $\ds \frac{49}{176}$ & $\ds - \frac{5103}{18656}$ & &  \\
&   &  &  & &  & & \\
$\ds 1$ &   $\ds \frac{35}{384}$ & $\ds 0$ & $\ds \frac{500}{1113}$ & $\ds \frac{125}{192}$ & $\ds - \frac{2187}{6784}$  & $\ds  \frac{11}{84}$ &  \\
&   &  &  & &  & & \\
\hline 
&   &  &  & &   & &\\
$\ds y_1$& $\ds \frac{35}{384}$ & $\ds 0$ & $\ds \frac{500}{1113}$ & $\ds \frac{125}{192}$ & $\ds - \frac{2187}{6784}$  & $\ds  \frac{11}{84}$ & $\ds  0$\\
&   &  &  & &   & &\\
\hline
&   &  &  & &  & & \\
$\ds y_1'$ & $\ds \frac{5179}{57600}$ & $\ds 0$ & $\ds \frac{7571}{16695}$ & $\ds \frac{393}{640}$ & $\ds - \frac{92097}{339200}$  & $\ds  \frac{187}{2100}$ & $\ds\frac{1}{40}$ 
\end{tabular}
\end{table}
  
%=====================================================
\subsection{Remarks} \label{sec:remarks_stability}
%=====================================================

The classical reference for the stability theory of finite-difference for time dependent problems is \cite{RichtmyerMorton}. A modern account of stability theory for initial-boundary value discretizations is~\cite{GKO95}. Ref.~\cite{GustafssonBook} includes a discussion of some of the main stability definitions and results, with emphasis on multiple aspects of high order methods, and~\cite{ThomasVol1,ThomasVol2} many examples at a more introductory level. We have omitted discussing the discrete version of the Laplace theory for IBVP, developed by Gustafsson, Kreiss and Sundstr\"{o}m (known as GKS theory or GKS stability) \cite{GKS} since it has been used very little (if at all) in numerical relativity, where most stability analyses instead rely on the energy method.  

The simplest stability analysis is that one of a periodic, constant coefficient test problem. An eigenvalue analysis can include boundary conditions and is typically used as a rule of thumb for CFL limits or to remove some instabilities. The eigenvalues are usually numerically computed for a number of different resolutions. See~\cite{Frauendiener:2002iv, Frauendiener:2002ix} for some examples within numerical relativity. 

Our discussion of Runge--Kutta methods follows \cite{Butcher} and \cite{HairerVol1}, which we refer to, along with \cite{HairerVol2}, for the rich area of methods for solving ordinary differential equations, in particular Runge--Kutta ones. We have only mentioned (one-step) explicit methods, which are the ones used the most in numerical relativity, but they are certainly not the only ones. For example, stiff problems in general require implicit integrations. Ref.~\cite{Lau:2008fb,Palenzuela:2008sf,Lau:2011we} explored implicit-explicit (IMEX) time integration schemes in numerical relativity. Among many of the topics that we have not included is that one of dense output. This refers to methods which allow the evaluation of an approximation to the numerical solution at \emph{any} time between two consecutive timesteps, at an order comparable or close to that one of the integration scheme, and at low computational cost.

%===================================================================
%===================================================================
\newpage
\section{Spatial Approximations: Finite Differences}
\label{sec:fd}
%===================================================================
%===================================================================

As mentioned in the previous section, a general stability theory (referred to as GKS) for initial-boundary value problems was developed by Gustafsson, Kreiss and Sundstr\"{o}m \cite{GKS}, and a simpler approach, when applicable, is the energy method. The latter is particularly and considerably simpler than a GKS analysis for complicated systems such as Einstein's field equations, high order schemes, spectral methods, and/or complex geometries. The Einstein vacuum equations can be written in linearly degenerate form and are therefore expected to be free of physical shocks and ideally suited for methods which exploit the smoothness of the solution to achieve fast convergence, such as high order finite difference and spectral methods. In addition, an increasing number of approaches in numerical relativity use some kind of multi-domain or grid structure approach (see Section~\ref{sec:complex}). If the problem at the continuum admits an energy estimate, which as discussed in Sections~\ref{section:IVFEinstein} and \ref{section:ibvpEinstein} is usually the case in General Relativity, then at least in the linearized case there are multi-domain schemes for which numerical stability can  relatively easily be established for a large class of linear symmetric hyperbolic problems and maximal dissipative boundary conditions through the energy method. In this section we discuss spatial finite difference (FD) approximations of arbitrary high order for which the energy method can be applied, and in Section~\ref{sec:spec} boundary closures for them. We start by reviewing polynomial interpolation, followed by the systematic construction of finite difference  approximations of arbitrary high order and stencils through interpolation. Next, we introduce the concept of operators satisfying Summation By Parts, present a semidiscrete stability analysis, and the construction of high order operators optimized in terms of minimizing their boundary truncation error and their associated timestep (CFL) limits (more specifically, their spectral radius). Finally, we discuss numerical dissipation, with emphasis on the region near boundaries or grid interfaces. 

%===================================================================
\subsection{Polynomial interpolation}
\label{sec:interpolation}
%===================================================================

Although interpolation is not strictly a finite differencing topic, we briefly present it here because it is used below and in Section \ref{sec:spec}, when discussing spectral methods. 

Given a set of $(N+1)$ distinct points $\{x_j\}_{j=0}^N$ (sometimes referred to as \textit{nodal points} or \textit{nodes}) and arbitrary associated function values $f(x_j)$, the interpolation problem amounts to finding (in this case) a polynomial ${\cal I}_N [f](x)$ of degree less than or equal to $N$ such that ${\cal I}_N [f] (x_j) = f(x_j)$ for $j=0,1,2,\ldots,N$. 

It can be shown that there is one and only one such polynomial. Existence can be shown by explicit construction: suppose one had 
\be
{\cal I}_N[f](x) = \sum_{j=0}^N f(x_j) l_j^{(N)}(x),
\label{eq:Lag_int}
\ee
where, for each $j=0,1,\ldots, N$, $l_j^{(N)}(x)$ is a polynomial of degree less than or equal to $N$ such that
\be
l_j^{(N)}(x_i) = \delta_{ij} \qquad \mbox{for } i=0,1,\ldots, N.
\label{def:Lag_pol}
\ee
Then ${\cal I}_N[f](x)$ as given by Equation~(\ref{eq:Lag_int}) would interpolate $f(x)$ at the $(N+1)$ nodal points $\left\{ x_i \right\}$. The Lagrange polynomials, defined as 
\be
l_j^{(N)}(x) = \left( \prod_{k=0,k \neq j}^N (x-x_k) \right)
\left(  \prod_{k=0,k \neq j}^N (x_j-x_k) \right)^{-1},
\label{eq:lagrange}
\ee
indeed do satisfy Equation~(\ref{def:Lag_pol}). Uniqueness of the interpolant can be shown by using the property that polynomials of order $N$ can have at most $N$ roots, applied to the difference between any two interpolants.

Defining the interpolation error by 
$$
E_N(x) = | f(x)- {\cal I}_N[f](x) |
$$
and assuming that $f$ is differentiable enough, it can be seen that $E_N$ satisfies
\be
E_N (x) = \frac{1}{(N+1)!} | f^{(N+1)}(\xi _x) \omega _{N+1}(x) | \, , 
\label{error_interp}
\ee
where $\omega _{N+1}(x) := \prod_{j=0}^N (x-x_j)$ is called the nodal polynomial of degree $(N+1)$,  and $\xi_x$ is in the smallest interval ${\cal I}_x$ containing $\{ x_0,x_1,\ldots, x_N \}$ \emph{and} $x$. In other words, if we assume the ordering $x_0 < x_1 < \ldots < x_N$, then $x$ can actually be outside $[x_0,x_N]$. For example, if $x< x_0$, then ${\cal I}_x = [x,x_N]$. Sometimes, approximating $f(x)$ by ${\cal I}_N[f](x)$ when $x \notin [x_0,x_N]$ is called \textit{extrapolation}, and interpolation only if $x\in [x_0,x_N]$, even though an interpolating polynomial is used as approximation. 

%===============================================
\subsection{Finite differences through interpolation}
\label{sec:fd_interpolation}
%===============================================

Finite difference (FD) approximations of a $p$-th derivative are local linear combinations of function values at node points,
\begin{eqnarray}
D^{(p)}f(x) &:= &\frac{1}{( \Delta x)^p}\sum_{i=0}^N f(x_i) a_i(x)  \label{eq:fd_defn} \\
 &=& \frac{d^p}{dx^p}f(x) + {\cal O }\left( \left( \Delta x\right)^r\right) \, ,  \nonumber
\end{eqnarray}
where the nodes $\{ x_i \}$ do not need to include $x$ (that is, the grid can be staggered). In the finite difference case the number of nodes is usually kept fixed as resolution is changed, resulting in a fixed convergence order $r$. This is in contrast to discrete spectral collocation methods, discussed in Section \ref{sec:spec}, which can be seen as approximation through global polynomial interpolation at special nodal points. 

A way of systematically constructing FD operators with an arbitrary distribution of nodes, any desired convergence order, and which are centered, one-sided or partially off-centered in any way is through interpolation. A local polynomial interpolant is used to approximate the function $f$, and the FD approximation is defined as the exact derivative of the interpolant. That is, 
$$
f(x) \approx {\cal I}[f](x) := \sum_{i=0}^{N} f(x_i) \ell_i^{(N)} (x)
$$
and, for instance for a first derivative, 
\be
Df(x) := \frac{d}{dx} {\cal I}[f](x)= \sum_{i=0}^N f(x_i) \frac{d}{dx} \ell_i^{(N)} (x) \, . \label{eq:diff_interpolation}
\ee
Notice that the expression~(\ref{eq:diff_interpolation}) does have the form of Equation~(\ref{eq:fd_defn}), where $\{ x_ i\}$ are the nodal points of the interpolant and 
$$
a_i (x) =( \Delta x) \frac{d}{dx} \ell_i^{(N)}(x) . 
$$

The truncation error for the FD approximation (\ref{eq:diff_interpolation}) to the first derivative can be estimated by differentiating the error formula for the interpolant, Equation~(\ref{error_interp}), 
\begin{eqnarray}
E_{f'}(x) &:= & \frac{d}{dx}f(x) - \frac{d}{dx}{\cal I}[f](x) =  \frac{d}{dx} \left(  \frac{1}{(N+1)!} f^{(N+1)}(\xi _x ) \omega (x) \right)  \nonumber \\
& = &  \frac{1}{(N+1)!} \left( \frac{d}{dx} f^{(N+1)}(\xi _x) \right) \omega (x) +  \frac{1}{(N+1)!} f^{(N+1)}(\xi_x ) \frac{d}{dx} \omega (x).
\label{eq:error_df_interp} 
\end{eqnarray}
The derivative of the first term in Equation~(\ref{eq:error_df_interp}) is more complicated to estimate than the second one without analyzing the details of the dependence of $\xi$ on $x$. But if we restrict $x$ to be a nodal point $x=x_k$, then $\omega (x_k) =0 $ and the previous equation simplifies to
\begin{equation}
E_{f'}(x_k) =  \frac{1}{(N+1)!} f^{(N+1)}(\xi_{x_k} ) \prod_{i=0,i\neq k}^N(x_k-x_i). 
\label{eq:error_diff_interpolation}
\end{equation}
Notice that the Equation~(\ref{eq:error_diff_interpolation}) implies that the resulting finite difference approximation has design convergence order $r=N$. For example, assume that the nodes are equally spaced, as is many times the case (the fact that the convergence rate is $N$ does not rely on such assumption). Then,  
\be
E_{f'}(x_k) =  \frac{(\Delta x)^N}{(N+1)!} f^{(N+1)}(\xi_{x_k} ) \prod_{i=0,i\neq k}^N(k-i). \label{eq:conv_diff_interpolation}
\ee

\begin{example}
A first order, one sided, finite difference approximation for $d/dx$.\\
We construct a first degree interpolant using two nodal points $\{x_0,x_1\}$, 
$$
{\cal I}[f](x)= f(x_0) \ell_0^{(1)}(x) + f(x_1) \ell_1^{(1)}(x),
$$ 
with 
$$
\ell_0^{(1)} (x) = \frac{x-x_1}{x_0-x_1}\, ,\qquad
\ell_1^{(1)} (x) = \frac{x-x_0}{x_1-x_0} \, . 
$$
Then 
\be
Df(x):= \frac{d}{dx} {\cal I}[f](x) = \frac{f(x_1) - f(x_0)}{x_1-x_0} ,.
\label{eq:fd_example1}
\ee
If we evaluate $x$ at $x_0$ or $x_1$ we obtain the standard first order forward and backward finite difference approximations $D_+$ and $D_-$, respectively [cf. Equations~(\ref{eq:one_sided})]. From Equation~(\ref{eq:conv_diff_interpolation}) we recover the known first order convergence for these approximations, $r=1$, which can also be obtained through a Taylor expansion in $\Delta x$ of Equation~(\ref{eq:fd_example1}).
\end{example}

\begin{example}
A second order, centered, finite difference approximation for $d/dx$.\\
Now we construct a second degree interpolant using three nodal points $\{x_0,x_1,x_2\}$, 
$$
{\cal I}[f](x)= f(x_0) \ell_0^{(2)}(x) + f(x_1) \ell_1 ^{(2)} (x) + f(x_2) \ell_2 ^{(2)} (x) \, ,
$$ 
with 
$$
\ell_0^{(2)} (x) = \frac{(x-x_1)(x-x_2)}{(x_0-x_1)(x_0-x_2)} \, , \qquad
\ell_1^{(2)} (x) = \frac{(x-x_0)(x-x_2)}{(x_1-x_0)(x_1-x_2)} \, , \qquad
\ell_2^{(2)} (x) = \frac{(x-x_0)(x-x_1)}{(x_2-x_0)(x_2-x_1)} \, . 
$$
If we assume the points to be equally spaced, $x_2-x_1=\Delta x = x_1-x_0$, and evaluate the derivative at the center one,  $x=x_1$,  we obtain 
$$
Df(x_1) := \frac{d}{dx} {\cal I}[f](x_1)  = \frac{f(x_2) - f(x_0)}{2 \Delta x} 
$$
the standard, second order, centered finite difference operator $D_0$  [cf. Equation~(\ref{eq:D0})]. 
\end{example}

One can proceed in this way to systematically construct any finite difference approximation to any derivative with any  desired convergence order and distribution of nodal points. The result for centered differences approximations to $d/dx$ with even accuracy order $r$ at equally spaced nodes can be written in terms of $D_0,D_+,D_-$ as follows \cite{GKO95},
\be
D_r = D_0 \sum_{\nu =0 }^{r/2-1}(-1)^{\nu} \alpha_{\nu}\left( (\Delta x)^2 D_+D_- \right)^{\nu} \, ,  \label{eq:high_order_centered_fd}
\ee
with  
\begin{eqnarray*}
\alpha_0 &=& 1 \, ,  \\
\alpha_{\nu } &=& \frac{\nu }{4\nu +2 }\alpha_{\nu -1}  \qquad \mbox{ for } \nu =1,2,\ldots,(r/2-1) \, . 
\end{eqnarray*}
\begin{example} The fourth, sixth, eighth and tenth order centered FD approximations to $d/dx$ are:
\begin{eqnarray*}
D_4 & =& D_0 - D_0 \frac{(\Delta x)^2}{6}D_+D_- \,,\\
D_6 & =& D_4 + D_0 \frac{(\Delta x)^4}{30}D_+^2D_-^2\,, \\
D_8 & =& D_6 - D_0\frac{(\Delta x)^6}{140}D_+^3D_-^3\,,  \\
D_{10} & =& D_8 + D_0 \frac{(\Delta x)^8}{630}D_+^4D_-^4\, . 
\end{eqnarray*}
\end{example}

%===============================================
\subsection{Summation by parts}
\label{sec:sbp} 
%===============================================

Since numerical stability is by definition the discrete counterpart of well posedness, one way to come up with schemes which are by construction numerically stable is by designing them so that they satisfy the same properties used at the continuum when showing well posedness through an energy estimate. As discussed in Section~\ref{section:ivp} one such property is integration by parts, which leads to its numerical counterpart: \textit{Summation by Parts} \cite{Kreiss1974a, Kreiss1977a}.
 
Consider a discrete grid consisting of points $\{ x_i \}_{i=0}^N$ and uniform spacing $\Delta x$ on some, possibly unbounded, domain $[a,b]$.

\begin{definition}
\label{def:sbp}
A difference operator $D$ approximating $\partial /\partial x$ is said to satisfy Summation By Parts (SBP) on a given domain with respect to a positive
definite scalar product $\mathbf{\Sigma}$, 
$$
\mathbf{\Sigma}= \Delta x \left( \begin{array}{ccc} 
&  &  \\ 
 & \sigma_{ij} &   \\
 &  & 
\end{array} \right) \, , 
$$
\begin{equation}
\langle u,v\rangle_\mathbf{\Sigma} := \Delta x \sum_{i,j=0}^n u_i v_j \sigma_{ij}\, ,
\label{eq:prod}
\end{equation}
if the property
\begin{equation}
\langle u,Dv \rangle_\mathbf{\Sigma} + \langle Du,v \rangle_\mathbf{\Sigma} 
 = u(b)v(b) -u(a) v(a) \label{eq:sbp}
\end{equation}
holds for all grid functions $u$ and $v$. 
\end{definition}
This is the discrete counterpart of integration by parts for the $\frac{d}{dx}$ operator, 
$$
\langle f, \frac{d}{dx} g \rangle + \langle \frac{d}{dx}f, g \rangle = f(b)g(b) - f(a)g(a), 
$$
for all $f,g$ and 
$$
\langle f, g \rangle := \int\limits_a^b f(x) g(x) dx. 
$$
Similar definitions for SBP can be introduced for
 higher dimensional domains. 

If the interval is infinite, say $(-\infty, b)$ or $(-\infty,\infty)$, certain fall-off conditions are required and Equation~(\ref{eq:sbp}) replaced by dropping the corresponding boundary term(s).

\begin{example}
Standard centered differences as defined by Equation~(\ref{eq:high_order_centered_fd}) in the domain $(-\infty, \infty)$ or for periodic domains and functions satisfy SBP with respect to the trivial scalar product ($\sigma _{ij} = \delta_{ij}$), 
$$
\langle u,D_0v \rangle_\mathbf{\Sigma} + \langle D_0u,v \rangle_\mathbf{\Sigma} = 0,\qquad
\langle u,v\rangle_\mathbf{\Sigma} := \Delta x \sum_{i\in\Integer} u_i v_i \, . 
$$
\end{example}

The scalar product or associated norm are said to
be \emph{diagonal} if
\begin{equation}
\sigma_{ij} = \sigma_{ii} \delta_{ij} \;\textrm{,}
\end{equation}
that is, if $\mathbf{\Sigma}$ is diagonal.  It is called \emph{restricted
  full} if
\begin{equation}
\sigma_{i_bj} = \sigma_{i_bi_b} \delta_{i_bj} \;\textrm{,}
\end{equation}
where $i_b$ denote boundary point indices, $i_b=0$ or $i_b=N$:
$$
\mathbf{\Sigma}= \Delta x \left( \begin{array}{ccccc}
\sigma_{00}& 0  & \cdots &  & 0  \\
 0 &   &  &  &   \\
\vdots &  & \mathbf{\Sigma_{\tt interior}} &  & \vdots  \\
& & & & 0 \\
 0 & & \cdots & 0  & \sigma_{NN} 
\end{array} 
\right)  \, . 
$$
In the case of bounded, non-periodic domains one possibility is to use centered differences and the trivial scalar product in the interior and modify both of them at and near boundaries. That is, the scalar product has the form
\be
\mathbf{\Sigma}= \Delta x \left( \begin{array}{ccc} 
\mathbf{\Sigma_l }&  &  \\ 
 & I &   \\
 &  & \mathbf{\Sigma_r} 
\end{array} \right) \, , \label{eq:sbp_block}
\ee
with $\mathbf{\Sigma_l }$ and $\mathbf{\Sigma_r }$ blocks of size \emph{independent} of $\Delta x$.

\paragraph*{Accuracy and Efficiency.}

As mentioned, in the absence of boundaries standard centered finite differences (which have even order of accuracy $2p$) satisfy SBP with respect to the trivial ($\mathbf{\Sigma} = \Delta x \mathbf{I}$) scalar product. In their presence the operators can be modified at and near boundaries so as to satisfy SBP, examples are given below. It can be seen that the accuracy at those points drops to $p$ in the diagonal case and to $2p-1$ in the restricted full one. Therefore the latter is more desirable from an accuracy perspective, but less so from a stability one. Depending on the system, numerical dissipation might be enough to stabilize the discretization in the restricted full case. This is discussed below in Section~\ref{sec:dissipation}. 

When constructing SBP operators the discrete scalar product cannot be arbitrarily fixed and afterward the difference operator solved for so that it satisfies the SBP property~(\ref{eq:sbp}) --in general this leads to no solutions. The coefficients of $\mathbf{\Sigma}$ \emph{and} those of $D$ have to be simultaneously solved for. The resulting systems of equations lead to SBP operators being in general not unique, with increasing freedom with the  accuracy order. In the diagonal case the resulting norm is automatically positive definite but not so in the full-restricted case. 

We label the operators by their order of accuracy in the interior and near boundary points. For diagonal norms and restricted full ones this would be $D_{2p-p}$ and $D_{2p-(p-1)}$, respectively.

\begin{example}
\label{ex:D21}{$\mathbf{D_{2-1}}$:}
For the simplest case, $p=1$, the SBP operator and scalar product are unique: 
\begin{displaymath}
D_{2-1} u_i = \frac{1}{\Delta x} \left\{ 
\begin{array}{lcl} 
\displaystyle
\left( - u_{0} + u_{1} \right) & & \mbox{ for } i =0 \, , \\
\displaystyle
\frac{\left( u_{i+1} - u_{i-1} \right)}{2} & &  \mbox{ for } i=1 \ldots (N-1)\, ,   \\ 
\displaystyle
\left( u_{n} - u_{n-1} \right)  & & \mbox{ for } i=N \, . 
\end{array} \right.
\end{displaymath}
with $\sigma_{00}=\sigma_{NN}=1/2$ and $\sigma_{ij} = \delta_{ij}$ otherwise. That is, the SBP scalar product (\ref{eq:prod}) is the simple trapezoidal rule for integration.
\end{example}

The operator $D_{4-2}$ and its associated scalar product are also unique in the diagonal norm case:
\begin{example} 
\label{ex:D42}{$\mathbf{D_{4-2}}$:}
\begin{displaymath}
D_{4-2} u_i =  \frac{1}{\Delta x} \left\{ 
\begin{array}{lcl} 
\displaystyle
\left(-\frac{24}{17}u_{0} + \frac{59}{34}u_{1}-\frac{4}{17} u_{2} -\frac{3}{34} u_{3} \right) & & \mbox{ for } i =0 \, , \\
\displaystyle
\left(-\frac{1}{2} u_{0} +\frac{1}{2} u_{2} \right) & & \mbox{ for } i =1 \, , \\
\displaystyle
\left( \frac{4}{43} u_{0} - \frac{59}{86} u_{1} + \frac{59}{86} u_{3} - \frac{4}{43}  u_{4} \right) & & \mbox{ for } i =2 \, , \\
\displaystyle
\left( \frac{3}{98} u_{0} - \frac{59}{98} u_{2} + \frac{32}{49} u_{4} - \frac{4}{49}  u_{5} \right) & & \mbox{ for } i =3 \, , \\
& & \\
\displaystyle
\left( \frac{1}{12}u_{i-2} - \frac{2}{3} u_{i-1} + \frac{2}{3}u_{i+1} -\frac{1}{12}u_{i+2} \right) & &  \mbox{ for } i=4 \ldots (N-4)\, ,   \\ 
& & \\
\displaystyle
- \left( \frac{3}{98} u_{N} - \frac{59}{98} u_{N-2} + \frac{32}{49} u_{N-4} - \frac{4}{49}  u_{N-5}  \right)  & & \mbox{ for } i=N-3 \\ 
\displaystyle
- \left( \frac{4}{43} u_{N} - \frac{59}{86} u_{N-1} + \frac{59}{86} u_{N-3} - \frac{4}{43}  u_{N-4}  \right)  & & \mbox{ for } i=N-2 \\ 
\displaystyle
- \left( -\frac{1}{2} u_{N} +\frac{1}{2} u_{N-2} \right)  & & \mbox{ for } i=N-1 \\ 
\displaystyle
- \left(-\frac{24}{17}u_{N} + \frac{59}{34}u_{N-1}-\frac{4}{17} u_{N-2} -\frac{3}{34} u_{N-3} \right)  & & \mbox{ for } i=N \, .
\end{array} \right.
\end{displaymath}
with scalar product
$$
\mathbf{\Sigma} = \Delta x \mbox{ diag } \left\{\frac{17}{48} , \frac{59}{48}, \frac{43}{48}, \frac{49}{48}, 1,1,\ldots,1,1, \frac{49}{48} , \frac{43}{48}, \frac{59}{48}, \frac{17}{48}  \right \} \, . 
$$
\end{example}

On the other hand, the operators $D_{6-3},D_{8-4},D_{10-5}$ have one, three and ten free parameters, respectively. Up to $D_{10-5}$ their associated scalar products are unique, while for $D_{10-5}$ one of the free parameters enters in $\mathbf{\Sigma}$. For the full-restricted case, $D_{4-3}, D_{6-5}, D_{8-7}$ have three, four and five free parameters, respectively, all of which appear in the corresponding scalar products. 

A possibility \cite{Strand199447} is to use the non-uniqueness of SBP operators to minimize the boundary stencil {\tt s}ize $s$. If the difference operator in the interior is a standard centered difference with accuracy order $2p$ then there are $b$ points at and near each boundary where the accuracy is of order $q$ (with $q=p$ in the diagonal case and $q=2p-1$ in the full restricted one). The integer $b$ can be referred to as the {\tt b}oundary width. 
 The boundary stencil size $s$ is the number of gridpoints that the difference operator uses to evaluate its approximation at those $b$ boundary points. 

However, minimizing such size, as well as any naive or arbitrary choice of the free parameters, easily leads to a large spectral radius and as a consequence restrictive CFL (Courant--Friedrich--Lewy, see Section~\ref{sec:num_stability}) limit in the case of explicit evolutions. Sometimes it also leads to rather large boundary truncation errors. Thus, an alternative is to numerically compute the spectral radius for these multi-parameter families of SBP operators and find in each case the parameter choice that leads to a minimum~\cite{Svaerd:2005,Lehner:2005bz}. It turns out that in this way the order of accuracy can be increased from the very low one of $D_{2-1}$ to higher order ones such as $D_{10-5}$ or $D_{8-7}$ with a very small change in the CFL limit. It involves some work, but since the SBP property~(\ref{eq:sbp}) is independent of the system of equations one wants to solve, it only needs to be done once. In the full-restricted case, when marching through parameter space and minimizing the spectral radius, this minimization has to be constrained with the condition that the resulting norm is actually positive definite. 

The non-uniqueness of high order SBP operators can be further used to minimize a combination of the average of the boundary truncation error (ABTE), defined below, without a significant increase in the spectral radius. For definiteness consider a left boundary. If a Taylor expansion of the finite difference operator is written as
$$
Du|_{x_i} = \frac{du}{dx}|_{x_i} + c_i(\Delta x)^q\frac{d^{q+1}u}{dx^{q+1}}|_{x_i} \mbox{ for } i=0\ldots b
$$
then 
$$
\mbox{ABTE}:= \left(\frac{1}{b}\sum_{i=0}^b c_i^2 \right)^{1/2} \, . 
$$
Table~\ref{tab:10-5-comparison} illustrates the results of this optimization procedure for the $D_{10-5}$ operator, see \cite{Diener:2005tn} for more details. 

The coefficients for the SBP operators 
$$
D_{2-1},D_{4-2},D_{4-3},D_{6-3},D_{6-5},D_{8-4},D_{8-7},D_{10-5} \, , 
$$ 
and in particular for their optimized versions are available, along with their associated dissipation operators described below in Section \ref{sec:dissipation}, from the arXiv in \cite{Diener:2005tn} and also as complete source code from the Einstein Toolkit \cite{einstein-toolkit}.

\begin{table}
\caption[Comparison, for the $D_{10-5}$ operator, of both the spectral
  radius and average boundary truncation error (ABTE) when minimizing
  the bandwidth or a combination of the spectral radius and
  ABTE.]{Comparison, for the $D_{10-5}$ operator, of both the spectral
  radius and average boundary truncation error (ABTE) when minimizing
  the bandwidth or a combination of the spectral radius and ABTE. For
  comparison, the spectral radius and ABTE for the lowest accuracy
  operator, $D_{1-2}$, (which is unique) are 1.414 and 0.25,
  respectively. Note: the ABTE, as defined, is larger for this
  operator but its convergence rate is faster.} 
\label{tab:10-5-comparison}
\centering
\begin{tabular}{l || c |  r}
\hline \hline
Operator & Min. bandwidth &Min. ABTE and spectral radius \\ \hline
Spectral radius & 995.9 & 2.240 \\
ABTE & 20.534 & 0.7661 \\
\hline \hline
\end{tabular}
\end{table}

\textbf{Remarks:}
\begin{itemize}
\item The requirement of uniform spacing is not an actual restriction, since a coordinate transformation can always be used so that the computational grid is uniformly spaced even though the physical distance between gridpoints varies. In fact, this is routinely done in the context of multiple-domains or curvilinear coordinates (see Section~\ref{sec:complex}). 
\item In that case, though, stability needs to be guaranteed for systems with variable coefficients, since they appear due to the coordinate transformation(s) even if the original system had constant coefficients. This has relevance in terms of the distinction between diagonal and block-diagonal SBP norms, as mentioned below. 
\item A similar concept of SBP holds for discrete expansions into Legendre polynomials using Gauss-type quadratures, as discussed in Section~\ref{sec:gauss_and_sbp}.
\item The definition of SBP depends only on the computational domain, not on the system of equations being solved. This allows the construction of optimized SBP operators once and for all. 
\item Difference operators satisfying SBP, which are genuinely multi-dimensional, can be explicitly constructed (see, for example, Refs.~\cite{Calabrese:2003vx,Calabrese:2003yd}). However, they become rather complicated even for simple geometries as higher order accuracy is sought. An easier approach, for the case in which the domain is the cross product of one dimensional ones (say, topologically a cube in three dimensions), which is usually the case in many domain-decompositions for complex geometries (Section~\ref{sec:complex}) is to simply apply a one-dimensional operator satisfying SBP in each direction, and this is the approach that we will discuss from hereon. The question then is whether SBP holds in several dimensions; the answer is affirmative in the case of diagonal norms but not necessarily otherwise.

\end{itemize}

\subsection{Stability}

As usual, the addition of lower order, undifferentiated terms to the right-hand side of Equation~(\ref{eq:system_sbp}) still gives an energy estimate at the semidiscrete level, modulo boundary terms.

If a simple symmetric hyperbolic system with constant coefficients in one dimension of the form
\be
u_t = A u_x
\label{eq:system_sbp}
\ee
with $A = A^T$ symmetric, is discretized in space by approximating the space derivative with an operator satisfying SBP, a semidiscrete energy estimate can be derived, modulo boundary conditions (discussed in Section~\ref{sec:num_boundary}). For this, we define
$$
E_{\tt d}(t):=  \langle u, u \rangle_\mathbf{\Sigma} \, .
$$ 
Taking a time derivative, using the symmetry of $A$, the fact that $D$ and $A$ commute because the latter has constant coefficients, and the SBP property,
\begin{equation}
\frac{dE_{\tt d}}{dt} = \langle \frac{d}{dt}u, u \rangle_\mathbf{\Sigma} + \langle u , \frac{d}{dt}u \rangle_\mathbf{\Sigma} = \langle D Au, u \rangle_\mathbf{\Sigma} + \langle  Au , Du \rangle_\mathbf{\Sigma} = \left[ (Au)^T u\right]_a^b \,.
\label{eq:semidiscrete_sbp}
\end{equation}
Therefore, modulo the boundary terms, an energy estimate and semidiscrete stability follow. 

When considering variable coefficients, Equation~(\ref{eq:semidiscrete_sbp}) becomes
$$
\frac{dE}{dt} = \left[ (Au)^T u\right]_a^b - \langle u , [D,A] u \rangle_\mathbf{\Sigma}
$$
and the commutator between $A$ and $D$ needs to be uniformly bounded for all resolutions in order to obtain an energy estimate. 

Estimates for the term involving the commutator $[D,A]$ have been
given in Refs.~\cite{Olsson:1995a, OlssonSupplement, Olsson:1995b, Tadmor87}.  For this, we first notice that the SBP property of $D$ with respect to $\mathbf{\Sigma}$ and the symmetry of $A$ imply that the operator $B:=[D,A]$ is symmetric with respect to $\mathbf{\Sigma}$,
$$
\langle u, [D,A] v \rangle_\mathbf{\Sigma} = \langle [D,A] u, v \rangle_\mathbf{\Sigma}
$$
for all grid functions $u$ and $v$. Therefore, its norm is equal to its spectral radius and we have the estimate
\begin{equation}
\langle u , [D,A] u \rangle_\mathbf{\Sigma} \leq \rho([D,A]) \| u \|_\mathbf{\Sigma}^2
\end{equation}
for all grid functions $u$. Hence, the problem is reduced to finding an upper bound for the spectral radius of $[D,A]$, which is independent of the scalar product $\mathbf{\Sigma}$.

Next, in order to find such an upper bound, we write the finite difference operator as
\begin{equation}
(D u)_j = \frac{1}{\Delta x}\sum\limits_{k=0}^N d_{jk} u_k,\qquad j=0,1,\ldots N,
\label{Eq:FDCoeffs}
\end{equation}
where the $d_{jk}$'s are the coefficient of a banded matrix, that is, there exists $b > 0$ which is independent of $N$ such that $d_{jk} = 0$ for $|k-j| > b$. Then, we have 
$$
B u_j = [D,A] u_j = \sum\limits_{k=0}^N d_{jk} \frac{A(x_k) - A(x_j)}{\Delta x} u_k,\qquad
j=0,1,\ldots N,
$$
from which it follows, under the assumption that $A$ is continuously differentiable and that its derivative is bounded, $| B u_j | \leq |A_x|_\infty \sum\limits_{k=0}^N |k-j| |d_{jk}| |u_k|$, $j=0,1,\ldots N$, where $|A_x|_\infty := \sup\limits_{a\leq x\leq b} |A_x(x)|$. Now we can easily estimate the spectral radius of $B$, based on the simple observation that
$$
\rho(B) \leq \| B \|,\qquad
\| B \| := \sup\limits_{u\neq 0} \frac{\|B u\|}{\|u\|}.
$$
for \emph{any} norm $\|\cdot\|$ on the space of grid functions $u$. Choosing the $1$-norm $\| u \| := \sum\limits_{j=0}^N |u_j|$ we find
$$
\| B u \| = \sum\limits_{j=0}^N |Bu_j| \leq  
 |A_x|_\infty \sum\limits_{j,k=0}^N |k-j| |d_{jk}| |u_k|
 \leq  |A_x|_\infty \left( \max\limits_{k=0,\ldots,N}\sum\limits_{j=0}^N |k-j| |d_{jk}| \right) 
 \| u \|,
$$
from which it follows that
$$
\rho([D,A]) \leq C_1 |A_x|_\infty,\qquad
C_1 :=  \max\limits_{k=0,\ldots,N}\sum\limits_{j=0}^N |k-j| |d_{jk}|.
$$
The important point to notice here is that for each fixed $k$, the sum in the expression for $C_1$ involves at most $2b+1$ non-vanishing terms, since $d_{jk} = 0$ for $|k-j|>b$. For the SBP operators and scalar products used in practice, namely those for which the latter has the structure given by Equation~(\ref{eq:sbp_block}), $C_1$ can be bounded by a constant which is independent of resolution. Since the spectral radius of $B$ is equal to the one of its transposed, we may interchange the roles of $j$ and $k$ in the definition of the constant $C_1$, and with these observations we arrive at the following result:

\begin{lemma}[Discrete commutator estimate]
Consider a finite difference operator $D$ of the form~(\ref{Eq:FDCoeffs}) which satisfies the SBP property with respect to a scalar product $\mathbf{\Sigma}$, and let $A = A^T$ be symmetric. Then, the following commutator estimate holds:
\begin{equation}
| \langle u , [D,A] u \rangle_\mathbf{\Sigma} | \leq C |A_x|_\infty  \| u \|_\mathbf{\Sigma}^2
\label{Eq:DiscreteCommEstimate}
\end{equation}
for all grid functions $u$ and resolutions $N$, where $C := \min\{ C_1,C_2 \}$ with
\be
C_1 :=  \max\limits_{k=0,\ldots,N}\sum\limits_{j=0}^N |k-j| |d_{jk}|,\quad
C_2 :=  \max\limits_{j=0,\ldots,N}\sum\limits_{k=0}^N |k-j| |d_{jk}|. \label{eq:c1c2}
\ee
\end{lemma}

\textbf{Remarks}
\begin{itemize}
\item A key ingredient used above to uniformly bound the norm of $[D,A]$ is that the SBP scalar products used in practice have the form (\ref{eq:sbp_block}). In those cases, both the boundary width and the boundary stencil size (defined below Example~\ref{ex:D42}) associated with the corresponding difference operators are  independent of $N$. Therefore, the constants $C_1$ and $C_2$ can also be bounded independently of $N$.
\item For the $D_{2-1}$ operator defined in Example~\ref{ex:D21}, for instance, Equation~(\ref{eq:c1c2}) gives $C_1 = 3/2$, $C_2 = 1$, and we obtain the optimal estimate corresponding to the one in the continuum limit,
$$
\left| \ \langle u, [\frac{d}{dx},A] u \rangle \right| \leq |A_x|_\infty \| u \|^2 \, .
$$
\item For the $D_{4-2}$ operator defined in Example~\ref{ex:D42}, in turn, Equation~(\ref{eq:c1c2}) gives $C_1 = 1770/731 \approx 2.421$ and $C_2 = 62/27 \approx 2.296$
\item For spectral methods the constants $C_1$ and $C_2$ typically grow with $N$ as the coefficients $d_{jk}$ do not form a banded matrix anymore. This leads to difficulties when estimating the commutator, see Ref.~\cite{Tadmor87} for a discussion on this point.
\item 
A straightforward energy estimate shows that the skew-symmetric discretization  
$$
u_t = \frac{1}{2}\left( AD + DA \right)u - \frac{1}{2}A_x u, 
$$ 
leads to (with appropriate imposition of boundary conditions, discussed in Section~\ref{sec:num_boundary}) strict stability.

\item Summation by Parts by itself is not enough to obtain an energy estimate since the boundary conditions yet still need to be imposed, and in a way such that the boundary terms in the estimate after SBP are under control. This is the topic of Section~\ref{sec:num_boundary}.
\end{itemize}

%%%%%%%%%%%%%%%%%%%%%%%%%%%%%%%%%%
\subsection{Numerical dissipation}
\label{sec:dissipation}
%%%%%%%%%%%%%%%%%%%%%%%%%%%%%%%%%%

The use of numerical dissipation consistently with the underlying system of equations is a standard way of filtering unresolved modes, stabilizing a scheme, or both, without spoiling the convergence order of the scheme. As an example of unresolved modes, for centered differences, the mode with highest frequency for any given resolution does not propagate at all, while the semidiscrete group velocity with highest frequency is exactly in the \emph{opposite} direction to the one at the continuum. In addition, the speed increases with the order of the scheme. See, for example, Ref.~\cite{Lehner:2005bz} for more details.

Some schemes, such as those with upwind finite differences, are intrinsically dissipative, with a fixed ``amount'' of dissipation for a given resolution. Another approach is to add to the discretization a dissipative operator $Q_d$ with a tunable strength factor $\epsilon \geq 0$,
$$
\dot{u} = (\ldots ) \rightarrow \dot{u} = (\ldots )  + \epsilon Q_d u \, .
$$
The operator $Q_d$ is a higher order derivative compared to the principal part of the equation, mimicking dissipative physical systems and/or parabolic equations but in such a way that $\| Q_d \| \rightarrow 0$ as $\Delta x \rightarrow 0$. Furthermore,   $Q_d$ is usually chosen so that $Q_d$ scales with the gridspacing as the highest finite difference approximation (so that the amplification factor depends only on $\Delta t /\Delta x$). For example, for first order in space systems finite differences scale as 
\begin{equation}
D \sim \frac{1}{\Delta x} \label{eq:scaling_fd} \,  
\end{equation} 
and $Q_d$ is usually chosen to scale in the same way.  More precisely, in the absence of boundaries the standard way to add numerical dissipation to a first order in space system is (Kreiss--Oliger)  
\begin{equation}
Q_d = (-1)^{r-1}(\Delta x)^{2r-1}D_+^{(r)}D_-^{(r)} \, ,  \label{eq:KO}
\end{equation}
where $D^{(r)}$ denotes the application of $D$ $r$ times and $D_+,D_-$ denote forward and backward one-sided finite differences, respectively. Thus $Q_d$ scales with the gridspacing as $(\Delta x)^{-1}$, like Equation~(\ref{eq:scaling_fd}). If the accuracy order of the scheme is not higher than $(2r-1)$ in the absence of dissipation, it is not decreased by the addition of numerical dissipation of the form~(\ref{eq:KO}).  

The main property that sometimes allows numerical dissipation to stabilize otherwise unstable schemes is when they strictly carry away energy (as in the energy definitions involved in well posedness or numerical stability analysis) from the system. For example, the operators (\ref{eq:KO}) are semi-negative definite
\begin{equation}
\langle u, Q_d u \rangle_\mathbf{\Sigma} \leq 0. \label{eq:dis_neg}
\end{equation}
with respect to the trivial scalar product $\mathbf{\Sigma} = \mathbb{I}$, under which centered differences satisfy SBP.

In the presence of boundaries it is standard to simply set the operators (\ref{eq:KO}) to zero near them. The result is in general not semi-negative definite as in (\ref{eq:dis_neg}), which can not only not help resolve instabilities but also trigger them. Many times this is not the case in practice if the boundary is an outer one where the solution is weak, but not for inter-domain boundaries (see Section~\ref{sec:num_boundary} ). For example, for a discretization of the standard wave equation on a multi-domain, curvilinear grid setting, using the $D_{6-5}$ SBP operator with Kreiss--Oliger dissipation set to zero near interpatch boundaries does not lead to stability while the more elaborate construction below does \cite{Diener:2005tn}. 

For SBP-based schemes adding artificial dissipation
may lead to an unstable scheme unless the dissipation operator is semi-negative under the SBP scalar product. In addition, the dissipation operator should ideally be non-vanishing all the way up to the boundary and preserve the accuracy of the scheme everywhere (which is more difficult in the SBP case, as it is non-uniform). In Ref.~\cite{Mattsson:2004},
a prescription for operators satisfying both conditions for arbitrary high order SBP scalar products, is presented. A compatible dissipation operators is constructed as
\begin{equation}
Q_{d} = - (\Delta x)^{2p}\; \Sigma^{-1} D_p^T B_p D_p, \label{dismat}
\end{equation}
where $\Sigma$ is the SBP scalar product, $D_p$
is a consistent approximation of $d^p/dx^p$ with minimal
bandwidth (other choices are presumably possible), and $B_p$ is the so-called \emph{boundary
  operator}.  The latter has to be positive semi-definite and its role
is to allow boundary points to be treated differently from interior
points.  $B_p$ cannot be chosen freely, but has to follow certain 
restrictions (which become somewhat involved in the non-diagonal SBP case) based on preserving the accuracy of the schemes near and at boundaries, see \cite{Mattsson:2004} for more details.

%==============================================================================
\subsection{Going further}
\label{sec:fd_going_further}

Besides the applications already mentioned, high order finite difference operators satisfying SBP have been used, for example,  in simulations of binary black holes  immersed in an external magnetic field in the force free approximation \cite{Palenzuela:2010nf}, orbiting binary black holes in vacuum \cite{Pazos:2009vb}, and for the metric sector in binary black hole-neutron star evolutions \cite{Chawla:2010sw} and binary neutron star ones which include magneto hydrodynamics \cite{Anderson:2008zp}. Other works are referred to in combination to multi-domain interface numerical methods in Section~\ref{sec:num_boundary}.

In Ref.~\cite{Svaerd:2004} the authors present a numerical spectrum stability analysis for block-diagonal based SBP operators in the presence of curvilinear coordinates. However, the case of non-diagonal SBP norms and the full Einstein equations in multi-domain scenarios for orders higher than four in the interior needs further development and analysis. 

Efficient algorithms for efficiently computing the weights for generic finite differences operators (though not necessarily satisfying SBP or with proven stability) are given in \cite{2011arXiv1103.5182H}.

Discretizing second order time dependent problems without reducing them to first order leads to a similar concept of Summation By Parts for operators approximating second derivatives. There is steady progress in an effort to combine SBP with penalty interface and outer boundary conditions for high order multi-domain simulations of second order in space systems. At present though these tools have not yet reached the state of those for first order systems, and they have not been used within numerical relativity except for the test case of a `shifted advection equation' \cite{MattssonParisi2010}. The difficulties appear in the variable coefficient case. We discuss some of these difficulties and the state of the art in Section~\ref{sec:num_boundary}. In short, unlike the first order case, SBP by itself does not imply an energy estimate in the variable coefficient case, even if using diagonal norms, unless the operators are built taking into account the PDE as well. In Ref.~\cite{Mattsson2004503} the authors explicitly constructed minimal width diagonal norms SBP difference operators approximating $d^2/dx^2$ up to eighth order in the interior, and in Ref.~\cite{Carpenter:1999} non minimal width operators up to sixth order using full norms are given. 

Ref.~\cite{Witek:2010es} presents a stability analysis around flat space-time for a family of generalized BSSN-type formulations, along with numerical experiments which include binary black hole inspirals. 

SBP operators have also been constructed to deal with coordinate singularities in specific systems of equations~\cite{Calabrese:2003vy,Sarbach:2003dz,Gundlach:2010et}.  Since a sharp semi-discrete energy estimate is explicitly derived in these references, (strict) stability is guaranteed. In particular, in Ref.~\cite{Gundlach:2010et} schemes which converge \emph{pointwise} everywhere --including the origin-- are derived for wave equations on arbitrary space dimensions decomposed in spherical harmonics.  
 Interestingly enough, popular schemes \cite{Evans86} to deal with the singularity at the origin, which had not been explicitly designed to satisfy SBP, were found a posteriori to do so at the origin and closed at the outer boundary, see \cite{Gundlach:2010et} for more details. In these cases the SBP operators are tailored to deal with specific equations and coordinate singularities, therefore they \emph{are} problem dependent.  For this reason their explicit construction has so far been restricted to second and fourth order operators (with diagonal scalar products), though the procedure conceptually extends to arbitrary orders. For higher order operators, optimization of at least the spectral radius might become necessary to address.  
 
In Ref.~\cite{2011arXiv1103.5182H} the authors use SBP operators to design high order quadratures. The reference also includes a detailed description of many properties of SBP operators. 

Superconvergence of some estimates in the case of diagonal SBP operators is discussed in \cite{Hicken:2011:SFE:2078801.2078822}.

\newpage
%===================================================================
%===================================================================
\section{Spatial Approximations: Spectral Methods}
\label{sec:spec}
%===================================================================
%===================================================================

In this section we review some of the theory for spectral spatial approximations, and their applications in numerical relativity. These are global representations, which display very fast convergence for smooth enough functions. They are therefore very well suited for Einstein's vacuum equations, where physical shocks are not expected since they can be written in linearly degenerate form. 

We start in Section~\ref{sec:spectral_convergence} discussing expansions onto orthogonal polynomials which are solutions to Sturm--Liouville problems. In those cases it is easy to see that for smooth functions the decay of the error in truncated expansions with respect to the number of polynomials is in general faster than any power law, which is usually referred to as \textit{spectral convergence}. Next, in Section~\ref{sec:properties_orthogonal_polynomials} we discuss a few properties of general orthogonal polynomials; most important that they can be generated through a three-term recurrence formula. Follows Section~\ref{sec:legendre_and_chebyshev} with a discussion of the most used families of polynomials in bounded domains; namely, Legendre and Chebyshev ones, including the minmax property of Chebyshev points. Approximating integrals through a global interpolation with a careful choice of nodal points makes it possible to maximize the degree with respect to which they are exact for polynomials (Gauss quadratures). When applied to compute \emph{discrete} truncated expansions, they lead to two remarkable features. One of them is Summation by Parts (SBP) for Legendre polynomials, in analogy with the finite difference version discussed in Section~\ref{sec:sbp}. As in that case, SBP can also be sometimes used to show semidiscrete stability when solving time-dependent partial differential equations (PDEs). The second one is an exact equivalence, for general Jacobi polynomials, between the resulting discrete expansions and interpolation at the Gauss points, a very useful property for collocation methods. Gauss quadratures and SBP are discussed in Section~\ref{sec:gauss_and_sbp}, followed by interpolation at Gauss points in Section~\ref{sec:expansion_interpolation}. In Sections~\ref{sec:spectral_differentiation}, \ref{sec:collocation} and \ref{sec:spectral_numrel} we discuss spectral differentiation, 
the collocation method for time-dependent PDEs, and applications to numerical relativity. 

The results for orthogonal polynomials to be discussed are classical ones, but we present them because spectral methods are less widespread in the relativity community, at least compared to finite differences. The proofs and a detailed discussion of many other properties can be found in, for example, \cite{Funaro} and references therein. Ref.~\cite{HGG} is a modern approach to the use of spectral methods in time-dependent problems with the latest developments, while \cite{Boyd} discusses many issues which appear in applications, and \cite{Fornberg} presents a very clear practical guide to spectral methods, in particular to the collocation approach. A good fraction of our presentation of this section follows \cite{Funaro} and \cite{HGG}, to which we refer when we do not provide any other references, or for further material.

%==============================================================================
\subsection{Spectral convergence}
\label{sec:spectral_convergence}

\subsubsection{Periodic functions} 
An intuition about expansion of smooth functions into orthogonal polynomials and spectral convergence can be obtained by first considering the periodic case in $[0,2\pi]$ and expansion in Fourier modes, 
\be
p_j(x):= \frac{1}{\sqrt{2\pi}}e^{ijx} \mbox{ for integer } j \in\Integer.
\label{eq:fourier_modes}
\ee
These are orthonormal under the standard complex scalar product in $L^2([0,2\pi])$, 
\be
\langle f, g\rangle := \int\limits_0^{2\pi} \overline{f(x)} g(x) dx,\qquad f,g\in L^2([0,2\pi]),
\label{eq:fourier_sp}
\ee
\be
\langle p_j , p_{j'}\rangle = \delta_{jj'} \, .
\label{eq:fourier_orthogonality}
\ee
Furthermore, they form a complete basis under the norm  induced by the above scalar product. More explicitly, the expansion of a continuous, periodic function in these modes, 
\be
f(x) = \sum_{j=-\infty}^{\infty} \hat{f}_j p_j(x) \, , \label{eq:fourier} 
\ee
converges to $f$ in the $L^2$ norm if 
$$
\sum_{j=-\infty}^{\infty} | \hat{f}_j | ^2 < \infty \,. 
$$

The Fourier coefficients $\hat{f}_j$ can be computed from the orthonormality condition~(\ref{eq:fourier_orthogonality}) of the basis elements defined in Equation~(\ref{eq:fourier_modes}),
\be
\hat{f}_j = \langle  p_j, f \rangle 
 =  \frac{1}{\sqrt{2\pi}}\int\limits_{0}^{2\pi} f(x) e^{-i jx} dx
\, . 
\label{eq:fourier_coeff}
\ee
The \textbf{truncated expansion} of $f$ is (assuming $N$ to be even)
\be
{\cal P}_N [f](x) = \sum_{j=-N/2}^{N/2} \hat{f}_j p_j(x) \, ,
\label{eq:fourier_truncated} 
\ee
where the notation is motivated by the fact that the ${\cal P}_N$ operator can also be seen as the orthogonal projection under the above scalar product to the space spanned by $\{ p_j: j=-N/2 \ldots N/2 \}$ {see also Section~\ref{sec:properties_orthogonal_polynomials}). The error of the truncated expansion, using the orthonormality of the basis functions (Parseval's property) is 
\be
\| f - {\cal P}_N[f]  \|^2 = \sum_{|j|>N/2} | \hat{f}_j | ^2,
\label{eq:error_fourier}
\ee
from which it can be seen that a fast decay in the error relies on a fast decay of the high frequency Fourier coefficients $|\hat{f}_j|$ as $j\rightarrow \infty$. Using the explicit definition of the basis elements $p_j$ in Equation~(\ref{eq:fourier_modes}) and the scalar product in~(\ref{eq:fourier_sp}), 
$$
| \hat{f}_j  | = | \langle p_j , f \rangle | 
 = \frac{1}{\sqrt{2\pi}} \left | \int\limits_{0}^{2\pi} f(x) e^{-i jx} dx \right | \, . 
$$
Integrating by parts multiple times, 
\begin{eqnarray*}
| \hat{f}_j | & = & \frac{1}{\sqrt{2\pi} |j|} \left | \int\limits_{0}^{2\pi} f'(x) e^{-i jx} dx \right | 
 = \ldots 
 =  \frac{1}{\sqrt{2\pi} |j|^s} \left | \int\limits_{0}^{2\pi} f^{(s)}(x) e^{-i jx} dx \right |
 = \frac{1}{|j| ^s} \left | \langle f^{(s)},  p_j \rangle \right | \\
& \leq & \frac{1}{|j|^s}\left\| f^{(s)} \right \| \left\| p_j \right \|  =  \frac{1}{|j|^s}\left\| f^{(s)} \right \|  ,
\end{eqnarray*}
and the process can be repeated for increasing $s$ as long as the s-derivative $f^{(s)}$ remains bounded in the $L^2$ norm. In particular, if $f \in C^{\infty}$, then the Fourier coefficients decay to zero
$$
|\hat{f}_j| \rightarrow 0 \mbox{ as } j \rightarrow \infty  
$$
faster than any power law, which is usually referred to as \textbf{spectral convergence}. The \textit{spectral} denomination comes from the property that the decay rate of the error is dominated by the spectrum of an associated Sturm--Liouville problem, as discussed below. The convergence rate for each Fourier mode in the remainder can be extended to the whole sum (\ref{eq:error_fourier}). More precisely, the following result can be shown (see, for example, \cite{HGG}):

\begin{theorem}
\label{theorem:estimate_fourier}
For any $f \in H^s_p[0,2\pi]$ (p standing for periodic) there exists a constant $C>0$ independent of $N$ such that for all $N\geq 1$ 
\begin{equation}
\| f - {\cal P}_{N} [f] \| \leq CN^{-s} \left \| \frac{d^s f }{dx^s} \right \|\, . 
\label{eq:estimate_fourier}
\end{equation}
\end{theorem}
In fact, an estimate for the difference between $u$ and its projection similar to (\ref{eq:estimate_fourier}) but on the infinity norm can also be obtained \cite{HGG}.  

In preparation for the discussion below for non-periodic functions, we rephrase and re-derive the previous results in the following way. Integrating by parts twice, the differential operator 
$$
{\cal D} = -\partial_x^2
$$
is self-adjoint under the standard scalar product~(\ref{eq:fourier_sp}),
$$
\la f, {\cal D} g  \ra = \la {\cal D} f ,  g \ra
$$
for periodic, twice continuously differentiable functions $f$ and $g$. 
Therefore the eigenfunctions $p_j$ of the problem 
\be
{\cal D} p_j (x) = \lambda _j p_j (x)
\label{eq:sturm_liouvile}
\ee
are orthogonal (and can be chosen orthonormal) -- they turn out to be the Fourier modes (\ref{eq:fourier_modes}) -- represent an orthonormal complete set for periodic functions in $L^2$, the expansion (\ref{eq:fourier}) converges, and the error in the truncated expansion (\ref{eq:fourier_truncated}) is given by the decay of high order coefficients, see Equation~(\ref{eq:error_fourier}). Assuming $f$ is smooth enough, the fast decay of such modes is a consequence of ${\cal D}$ being self-adjoint, the basis elements $p_j$ being solutions to the problem (\ref{eq:sturm_liouvile}), 
$$
p_j = \frac{1}{\lambda_j} {\cal D}p_j \, , 
$$
and the eigenvalues satisfying $\lambda_j \simeq j^2$ for large $j$ (in the Fourier case, $\lambda_j = j^2$ holds exactly). Combining these properties, 
\begin{eqnarray}
| \hat{f}_j | &=& \left| \langle f , p_j \rangle \right| =  \frac{1}{| \lambda_j |} \left|  \langle f , {\cal D}p_j  \rangle \right|  = \frac{1}{| \lambda_j |} \left| \langle {\cal D}  f , p_j 
\rangle \right| = 
\frac{1}{| \lambda_j |^2} \left| \langle {\cal D}^{(2)} f, p_j \rangle \right| = \ldots = \frac{1}{| \lambda_j |^s} \left| \langle   {\cal D}^{(s)} f , p_j \rangle \right|  \nonumber \\
& \leq & \frac{1}{| \lambda_j |^s} \left\|  {\cal D}^{(s)} f \ \right\|,
\label{eq:spectral_fourier}
\end{eqnarray}
where $ {\cal D}^{(s)}$ denotes the application of ${\cal D}$ $s$ times (in this case, ${\cal D}^{(s)}$ is equal to $(-1)^s\partial_x^{2s}$).

The main property that leads to spectral convergence is then the fast decay of the Fourier coefficients, see Equation~(\ref{eq:spectral_fourier}), provided the norm of $ {\cal D}^{(s)}$ remains bounded for large $s$. 

Before moving to the non-periodic case we notice that in either the full or truncated expansions the integrals~(\ref{eq:fourier_coeff}) need to be computed. Numerically approximating the latter leads to \textbf{discrete} expansions and an appropriate choice of quadratures for doing so leads to a powerful connection between the discrete expansion and interpolation. We discuss this in Section~\ref{sec:gauss_and_sbp}, directly for the non-periodic case. 

%==============================================================================
\subsubsection{Singular Sturm--Liouville problems}

Next, consider non-periodic domains $(a,b)$ (which can actually be unbounded; for example, $(0,\infty)$ as in the case of Laguerre polynomials) in the real axis. We discuss how bases of orthogonal polynomials with spectral convergence properties arise as solutions to singular Sturm--Liouville problems. 

For this we need to consider more general scalar products. For a continuous, strictly positive weight function $\omega$ on the open interval $(a,b)$ we define
\be
\la h, g \ra _{\omega } = \int\limits_a^b h(x)g(x) \omega (x) dx,
\label{eq:sp} 
\ee
and its induced norm, $\| h \|_{\omega}:= \sqrt{\langle h, h \rangle_{\omega}}$, on the Hilbert space $L^2_\omega(a,b)$ of all real-valued, measurable functions $h,g$ on the interval $(a,b)$ for which $\| h \|_\omega$ and $\| g \|_\omega$ are finite.

Consider now the Sturm--Liouville problem
\be
{\cal D} p_j (x)  = \omega(x) \lambda_j p_j(x)
\label{eq:sturm_liouville}
\ee
where ${\cal D}$ is a second order linear differential operator on $(a,b)$ along with appropriate boundary conditions so that it is self-adjoint under the \emph{non-weighted} scalar product, 
\be
\langle f, {\cal D} g  \rangle_{\omega =1 } = \langle {\cal D} f ,  g  \rangle _{\omega =1} \, ,
\label{eq:selfadjoint}
\ee
for all twice continuously differentiable functions $f,g$ on $(a,b)$ which are subject to the boundary conditions. Then the set of eigenfunctions is also complete, and orthonormal under the \emph{weighted} scalar product, and there is again a full and truncated expansions as in the Fourier case,   
$$
f(x) = \sum_{j=0}^{\infty } \hat{f}_j p_j(x) \, , 
$$
$$
{\cal P}_N[f](x) = \sum_{j=0}^{N} \hat{f}_j p_j(x) \, , 
$$
with coefficients 
$$
\hat{f}_j = \langle p_j, f \rangle _{\omega} \, . 
$$
The truncation error is similarly given by 
$$
\| f - {\cal P}_N[f] \|^2_{\omega } = \sum_{j> N} \hat{f}_j ^2 
$$
and spectral convergence is again obtained if the coefficients $\hat{f}_j$ decay to zero as $j\rightarrow \infty$ faster than any power law. Consider then, the singular Sturm--Liouville problem
\begin{equation}
{\cal D} p_j (x)  = -\partial_x\left( m(x) \partial_x p_j(x) \right) + n(x) p_j(x) 
 = \omega(x) \lambda_j p_j(x)
\label{eq:sing_sturm_liouville}
\end{equation}
with the functions $m,n: (a,b)\to \Real$ being continuous and bounded and such that $n(x)\geq 0$ and $m(x) > 0$ for all $x\in (a,b)$ and -- thus the \emph{singular}  part of the problem --,
\be
m(a) = m(b) = 0.
\label{eq:singular_sl}
\ee
For twice continuously differentiable functions with bounded derivatives, the boundary terms arising from integration by parts of the expression $\langle f, {\cal D} g  \rangle_{\omega=1}$ cancel due to Equation~(\ref{eq:singular_sl}), and it follows that the operator ${\cal D}$ is symmetric, see Equation~(\ref{eq:selfadjoint}). Therefore one can proceed as in the Fourier case and arrive to 
$$
| \hat{f}_j | \leq  \frac{1}{|  \lambda_j |^s}\left\| \left( \frac{{\cal D}}{\omega}\right)^{(s)} f \right \| _{\omega}\, 
$$
with spectral convergence if $f \in C^{\infty}$ and, for example, 
\be
| \lambda _j | \simeq j^2 \, ,   \mbox{ for large enough }  j > j_{\tt min} \, . \label{eq:min_eigen_jacobi}
\ee

\begin{theorem}
\label{th:jacobi}
The solutions to the singular Sturm--Liouville problem~(\ref{eq:sing_sturm_liouville}) with $(a,b) = (-1,1)$ and
\begin{eqnarray}
m(x) &=& (1-x)^{\alpha +1}(1+x)^{1+\beta}, \label{eq:jacobi1} \\
\omega(x) &=& (1-x)^{\alpha}(1+x)^{\beta}, \label{eq:jacobi2} \\
n(x) &=& 0, \label{eq:jacobi3}
\end{eqnarray}
where $\alpha,\beta > -1$, are the Jacobi polynomials $P_j^{(\alpha,\beta)}(x)$. Here, $P_j^{(\alpha,\beta)}(x)$ has degree $j$, and it corresponds to the eigenvalue
$$
\lambda _j = j (j+\alpha + \beta +1)  \, . 
$$
\end{theorem}

Notice that the eigenvalues satisfy the asymptotic
condition~(\ref{eq:min_eigen_jacobi}) and, roughly speaking,
guarantees spectral convergence. More precisely, the following holds
(see, for example, \cite{Funaro}) -- in analogy with
theorem~\ref{theorem:estimate_fourier} for the Fourier case -- for the
expansion ${\cal P}_{N} [f]$ of a function $f$ in Jacobi polynomials: 
\begin{theorem} \label{theorem:estimate_jacobi}
For any $f(x) \in H^s_{\omega}[0,2\pi]$ ($\omega $ refers to the weight function) there exists a constant $C>0$ independent of $N$ such that  
\be
\| f - {\cal P}_{N} [f] \| _{\omega} \leq C N^{-s} \left \|  (1-x^2)^{s/2} \frac{d^s f}{dx^s} \right\| _{\omega} \qquad \forall N > s .  \label{eq:estimate_jacobi}
\ee
\end{theorem}
Sturm-Liouville problems are discussed in, for example,  \cite{Weidmann80}. Below we discuss some properties of general orthogonal polynomials.

%==============================================================================
\subsection{Some properties of orthogonal polynomials}
\label{sec:properties_orthogonal_polynomials}

Given any weighted scalar product $\langle \cdot, \cdot \rangle_{\omega}$ as in Equation~(\ref{eq:sp}), the expansion of a function onto polynomials is optimal in the associated norm if orthogonally projected to the space of polynomials $\mathbf{P}_N$ of degree at most $N$. More precisely, given a function $f\in L^2_\omega(a,b)$ on some interval $(a,b)$ and an orthonormal basis $\mathbf{P}_N$ of polynomials $\{ p_j \}_{j=0}^N$, where
$p_j$ has degree $j$, its orthogonal projection, 
\be
{\cal P}_N [f] = \sum_{j=0}^N \langle p_j,f \rangle_{\omega } p_j
\label{eq:orthogonalp}
\ee
onto the subspace spanned by $\mathbf{P}_N$} satisfies
\be
{\cal P}_N [f] = \min_{f_N\in \mathbf{P}_N}\| f - f_N \|_{\omega }^2  \label{min_norm}\,, 
\ee
or 
\be
\| {\cal P}_N [f] - f \|_\omega \leq \| f_N - f \|_\omega
\ee
for all $f_N$ in the span of $\mathbf{P}_N$, that is, it minimizes the error for such $f_N$.

The operator~(\ref{eq:orthogonalp}) is a projection in the sense that 
$$
{\cal P}_N ^2 = {\cal P}_N \, , 
$$
and it is orthogonal with respect to $\langle \cdot, \cdot \rangle_{\omega}$: the residual $r:= f - {\cal P}_N [f]$ satisfies
$$
\langle r, {\cal P}_N [f] \rangle_{\omega} = 0 \,.
$$
Notice that unlike the interpolation problem (discussed below) here the solution of the above least squares problem (\ref{min_norm}) is not required to agree with $f$ at any prescribed set of points. 

In order to obtain an orthonormal basis of $\mathbf{P}_N$ a Gram-Schmidt procedure could be applied to the standard basis $\{ x^j \}_{j=0}^n$. However, exploiting properties of polynomials, a more efficient approach can be used, where the first two polynomials $p_0, p_1$ are constructed and then a three-term recurrence formula is used. 

In the following construction each orthonormal polynomials is chosen to be monic, meaning that its leading coefficient is one.

\begin{itemize}
\item \textbf{The zero-th order polynomial:} 

The conditions that $p_0$ has degree zero and that it is monic only leaves the choice
\be
p_0(x) = 1
\label{eq:0term} \,.
\ee
\item \textbf{The first order one:}

Writing $p_1(x) = x + b_1$ the condition $\la p_0,p_1\ra _{\omega}=0$ yields
\be
p_1(x) = x  - \frac{\la 1,x\ra _{\omega}}{\la 1,1\ra _{\omega}}\, .
\label{eq:1term}
\ee

\item \textbf{The higher order polynomials:}

\begin{theorem} (Three-term recurrence formula for orthogonal polynomials)
 
For \emph{monic} polynomials $\{ p_k\}_{k=0}^N$ which are orthogonal with respect to the scalar product $\langle \cdot , \cdot \rangle_{\omega}$, where each $p_k$ is of degree $k$, the following relation holds
\be
p_{k+1} = x p_k - \frac{\la x p_k,p_k\ra _{\omega }}{\la p_k,p_k\ra _{\omega }} p_k - \frac{\la x p_k,p_{k-1}\ra _{\omega }}{\la p_{k-1},p_{k-1}\ra _{\omega }} p_{k-1}
\label{eq:kterm} \,.
\ee
for $k=1,2,3,\ldots,N-1$.
\end{theorem}

\proof
Let $1\leq k \leq N-1$. Since $x p_k$ is a polynomial of degree $k+1$, it can be expanded as
$$
x p_k(x) = \sum\limits_{j=0}^{k+1} a_j p_j(x),
$$
where the orthogonality of the polynomials $\{ p_k\}_{k=0}^N$ implies that $a_j = \la x p_k, p_j \ra_\omega/\la p_j,p_j \ra_\omega$ for $j=0,1,2,\ldots,k+1$. However, since $\la x p_k,p_j \ra_\omega = \la p_k, x p_j \ra_\omega$ and $x p_j$ can be expanded in terms of the polynomials $p_0,p_1,\ldots,p_{j+1}$, it follows again by the orthogonality of $\{ p_k\}_{k=0}^n$ that $a_j=0$ for $j\leq k-2$. Finally, $a_{k+1}=1$ since $p_k$ and $p_{k+1}$ are both monic. This proves Equation~(\ref{eq:kterm}).
\qed

Notice that $p_{k+1}$ as defined in Equation~(\ref{eq:kterm}) remains monic and can therefore be automatically used for constructing $p_{k+2}$, without any rescaling.
\end{itemize}
Equations~(\ref{eq:0term}, \ref{eq:1term}, \ref{eq:kterm}) allow to
compute orthogonal polynomials for any weight function $\omega$,
without the expense of a Gram-Schmidt procedure. For specific weight
cases, there are even more explicit recurrence formulae, such as those
in Equations~(\ref{eq:leg_first_two}, \ref{eq:leg_three_term}) and
(\ref{eq:cheb_first}, \ref{eq:cheb_second}, \ref{def:cheb_3termrec})
for Legendre and Chebyshev polynomials, respectively.

%%%%%%%%%%%%%%%%%%%%%%%%%
\subsection{Legendre and Chebyshev polynomials}
\label{sec:legendre_and_chebyshev}
%%%%%%%%%%%%%%%%%%%%%%%%%

For finite intervals, Legendre and Chebyshev polynomials are the ones most typically used. In the Chebyshev case, the polynomials themselves, their roots and  quadrature points in general can be computed in closed form. They also satisfy a minmax property and lend themselves to using a Fast Fourier Transform (FFT).

\subsubsection{Legendre}

Legendre polynomials correspond to the trivial weight function $\omega \equiv 1$ and the choice $\alpha=\beta=0$ in Equations~(\ref{eq:jacobi1}, \ref{eq:jacobi2}, \ref{eq:jacobi3}).  The eigenvalues are 
$$
\lambda_j = j(j+1)\,, 
$$
and the first two polynomials 
\be
P_0(x) = 1\, , \qquad P_1(x) = x \, . \label{eq:leg_first_two}
\ee
A variation (in that it leads to non-monic polynomials) of the three term recurrence formula~(\ref{eq:kterm}) is 
\be
P_{j+1}(x) = \left( \frac{2j+1}{N+1}\right) xP_j(x) - \left(\frac{j}{j+1} \right)P_{j-1}(x),
\qquad j=1,2,3,\ldots,  \label{eq:leg_three_term}
\ee
leading to the normalization
$$
\langle P_i, P_j \rangle_{\omega} 
 = \int\limits_{-1}^1 P_i(x) P_j(x) dx = \frac{2}{2i+1}\delta_{ij}\, . 
$$

\subsubsection{Chebyshev}

Chebyshev polynomials correspond to the choice $\omega (x) = \left( 1- x^2\right)^{-1/2}$, $x\in (-1,1)$ and $\alpha=\beta=-1/2$ in Equations~(\ref{eq:jacobi1}, \ref{eq:jacobi2}, \ref{eq:jacobi3}). In particular, the eigenvalues are
$$
\lambda_j = j^2 \, . 
$$
A closed-form expression for Chebyshev polynomials (which lends itself to the use of FFT) is 
\be
T_j(x) = \cos\left( j \cos^{-1} (x) \right),\qquad j = 0,1,2,\ldots.
\label{def:cheb}
\ee
At first sight it might appear confusing that, given the above definition through trigonometric functions, $T_n(x)$ are actually polynomials in $x$ (of degree $n$, in fact). To get an idea of why this is so, we can compute the first few:
\begin{eqnarray*}
T_0(x) &=& \cos(0) \\
& = & 1 ,\\
T_1(x) &=& \cos (\cos^{-1}(x)) \\
 & = & x ,\\
T_2(x) &=& \cos (2 \cos^{-1}(x)) = -1 +2  \cos ^2  (\cos^{-1}(x)) \\
&=& -1 +2x^2.
\end{eqnarray*}
Notice from the above expressions that $T_2(x)  = 2 x T_1(x) - T_0$. In fact, a slight variation of the three-term recurrence formula~(\ref{eq:kterm}) for Chebyshev polynomials becomes
\begin{eqnarray}
T_0(x) &=& 1 \label{eq:cheb_first}\\
T_1(x) &=& x \label{eq:cheb_second}\\
T_{j+1}(x) &=&  2 x T_j(x) - T_{j-1},\qquad j = 1,2,3,\ldots.
\label{def:cheb_3termrec}
\end{eqnarray}
As in the Legendre case this is a variation of Equations~(\ref{eq:0term}, \ref{eq:1term}, \ref{eq:kterm}) in that the resulting polynomials are not monic: the leading coefficient is given by 
$$
T_{j}(x) =  2^{j-1} x^j + \ldots,
$$
That is, the polynomial $2^{1-j}T_j(x)$ is monic. 

Both Equation~(\ref{def:cheb}) and Equation~(\ref{def:cheb_3termrec}) lead to the normalization
$$
\langle T_i, T_j \rangle_{\omega} 
 = \int\limits_{-1}^1 T_i(x) T_j(x)\left( 1-x^2 \right)^{-1/2}dx = \delta_{ij}
\left\{ 
\begin{array}{lcl} 
\displaystyle 
\pi & & \mbox{ for } i =0  \\ 
& & \\
\displaystyle 
\frac{\pi}{2} & &  \mbox{ for } i > 0 
\end{array} \right.
$$

From the expression (\ref{def:cheb}), it can be noticed that the roots $\{ x_j \}$ of $T_{N+1}$ are 
\be
x_j = - \cos \left( \frac{2j+1}{2N+2} \pi \right),\qquad j=0,1,2,\ldots,N.
\label{eq:cheb_pts}
\ee
These points play an important role below in Gauss quadratures and collocation methods (Section~\ref{sec:gauss_and_sbp}). 

%===================================================================
\subsubsection{The minmax property of Chebyshev points}
\label{sec:minmax}
%===================================================================

Chebyshev polynomials satisfy a rather remarkable property in the context of interpolation. In Section~\ref{sec:interpolation} we pointed out that the error in polynomial interpolation of a function $f$ at $(N+1)$  nodal points $x_j$ satisfies [cf.\ Equation~(\ref{error_interp})]
$$
E_N (x) = \frac{1}{(N+1)!} \left|  f^{(N+1)}(\xi _x) \omega _{N+1}(x) \right| \, , 
$$
where $\omega _{N+1}(x) := \prod_{j=0}^N (x-x_j)$. 

When doing global interpolation, that is, keeping the endpoints $\{ x_0,x_N \}$ fixed and increasing $N$, it is not true that the error converges to zero even if the function is $C^{\infty}$. For example, for each $N$, $f^{(N+1)}(x)$ could remain bounded as a function of $x$, 
$$
| f^{(N+1)}(x) | \leq C_N \qquad \forall x\in [x_0,x_N]\, , 
$$
but $C_n$ could grow with $N$.  A classical example of non-convergence of polynomial interpolation on uniform grids is the following (see, for example, \cite{Epperson}):

\begin{example} Runge phenomenon.\\
Consider the function 
$$
f(x) = \frac{1}{1+x^2} \, , \qquad x\in[-5,5] 
$$
and its interpolating polynomial ${\cal I}_N[f](x)$ [cf. Equation~(\ref{eq:Lag_int})]  at equally spaced points 
$$
x_i = -5 + \frac{i}{N} 10\, , \qquad i=0,\ldots, N \, . 
$$
Then 
$$
|f(x) - {\cal I}_N[f](x) | \rightarrow \infty \qquad \mbox{ as } N\rightarrow \infty \qquad \forall x \mbox{ such that } x_c < |x| < 5 \, , 
$$
where $x_c \approx 3.63$. 
\end{example}

The error (\ref{error_interp}) can be decomposed into two terms, one related to the behavior of the derivatives of $f$, $f^{(N+1)}(\xi_x)/(N+1)! $, 
and the another one related to the distribution of the nodal points, $\omega _{N+1}(x)$. We assume in what follows that $x\in [x_0,x_N]$ and that $[x_0,x_N] =[-1,1]$. The analogue results for an arbitrary interval can easily be obtained by a shifting and rescaling of coordinates. It can then be shown  that for all choices of nodal points
\be
\max _{x\in[-1,1]} | \omega _{N+1} (x) | \geq 2^{-N}. 
\ee
Furthermore, the nodes which minimize this maximum (thus the \emph{minmax} term) are the roots of the Chebyshev polynomial of order $(n+1)$, for which the equality is achieved: 
$$
\max _{x\in[-1,1]} | \omega _{N+1} (x) |= 2^{-N} 
$$
when $\{ x_0,x_1,\ldots, x_N \}$ are given by Equation~(\ref{eq:cheb_pts}); see, for example, \cite{DeuflhardHohmann}, for the proof. 

In other words, using Chebyshev points, that is, the roots of the Chebyshev polynomials, as interpolating nodes minimizes the maximum error associated with the nodal polynomial term. Notice that in this case, the nodal polynomial is given by $T_{N+1}(x)/2^N$.

%==============================================================================
\subsection{Gauss quadratures and Summation By Parts} 
\label{sec:gauss_and_sbp} 
When computing a discrete expansion in terms of orthogonal polynomials 
$$
f_N (x) = \sum_{j=0}^N \hat{f}_j p_j(x)
$$
one question is how to efficiently numerically approximate the coefficients $\{ \hat{f}_j \}$ as given by Equation~(\ref{eq:orthogonalp}). This involves computing weighted integrals of the form  
\be
\int\limits_a^b q(x) \omega(x) dx.
\label{eq:weighted_int}
\ee
If approximating the weighted integral (\ref{eq:weighted_int}) by a quadrature rule, 
\be
\int\limits_a^b q(x) \omega(x) dx \approx \sum_{i=0}^N A_i q(x_i),
\label{eq:quad2}
\ee
where the points $\{ x_i \}$ are given but having the freedom to choose the coefficients $\{ A_i \}$,  by a counting argument one would expect to be able to choose the latter in such a way that Equation~(\ref{eq:quad2}) is \emph{exact} for all polynomials of degree at most $N$. That is indeed the case, and the answer  is obtained by approximating $q(x)$ by its polynomial interpolant (\ref{eq:Lag_int}) and integrating the latter, 
$$
\int\limits_a^b q(x) \omega(x) dx 
 \approx \int\limits _a^b \sum_{i=0}^N q(x_i) \ell_i^{(N)} (x) \omega(x) dx 
  =  \sum_{i=0}^N A_i^{(N)} q(x_i)
$$
where the $\{ {\ell}_i ^{N}\}_{i=0}^N$ are the Lagrange polynomials (\ref{eq:lagrange}) and the coefficients 
\be
A^{(N)}_i = \int\limits_a^b  \ell_i^{(N)} (x) \omega(x) dx \qquad
i=0,1,\ldots,N, \label{eq:quad}
\ee
are independent of the integrand $q(x)$. If the weight function is non-trivial they might not be known in closed form, but since they are independent of the function being integrated they need to be computed only once for each set of nodal points $\{ x_i \}$.

Suppose now that, in addition to having the freedom to choose the coefficients  $\{ A_i \}$ we can choose the nodal points $\{ x_i \}$. Then we have $(N+1)$ points and $(N+1)$ $\{A_i \}$, i.e.,  $(2N+2)$ degrees of freedom. We therefore expect that we can make the quadrature exact for all polynomials of degree at most $(2N+1)$. This is indeed true and is referred to as \textit{Gauss quadratures}. Furthermore, the optimal choice of $A_i$ remains the same as in Equation~(\ref{eq:quad}), and only the nodal points need to be adjusted.

\begin{theorem}[Gauss quadratures] \label{theorem:gauss}
Let $\omega$ be a weight function on the interval $(a,b)$, as introduced in Equation~(\ref{eq:sp}), and let be $p_{n+1}$ be the associated orthogonal polynomial of degree $N+1$. Then, the quadrature rule~(\ref{eq:quad2}) with the choice~(\ref{eq:quad}) for the discrete weights, and as nodal points $\{ x_j \}$ the roots of $p_{N+1}$ is exact for all polynomials of degree at most $(2N+1)$.
\end{theorem}

The following remarks are in order:
\begin{itemize}
\item The roots of $p_{nN1}(x)$ are referred to as Gauss points or nodes.
\item Suppose that $\omega (x)  = (1-x^2)^{-1/2}$. Then the $(N+1)$ Gauss points, i.e., the roots of the Chebyshev polynomial $T_{N+1}(x)$ [see Equation~(\ref{eq:cheb_gauss})], are exactly the points that minimize the infinity norm of the nodal polynomial in the interpolation problem, as discussed in Section~\ref{sec:minmax}. 
\end{itemize}

One can see that the Gauss points actually lie inside the interval $(a,b)$, and do not contain the endpoints $a$ or $b$. Now suppose that for some reason we want the nodes to include the end points of integration,
$$
x_0 = a,\qquad x_N = b.
$$
One reason for including the end points of the interval in the set of nodes is when applying boundary conditions in the collocation approach, as discussed in Section~\ref{sec:num_boundary}. Then we are left with two less degrees of freedom compared to Gauss quadratures and therefore expect to be able to make the 
quadrature exact for polynomials of order up to $(2N+1) - 2 = (2N-1)$. This leads to:

\begin{theorem}[Gauss--Lobatto quadratures] \label{theorem:gauss_lobatto}
If we choose the discrete weights according to Equation~(\ref{eq:quad}) as before but as nodal points the so called Gauss-Lobatto ones, i.e., the roots of the polynomial 
\be
m_{N+1}(x) = p_{N+1}(x) + \alpha p_{N}(x) + \beta p_{N-1}(x),
\label{eq:mpol}
\ee
with $\alpha$ and $\beta$ chosen so that $m_{N+1}(a) = 0 = m_{n+1}(b)$, then the quadrature rule~(\ref{eq:quad2}) is exact for all polynomials of degree at most $(2N-1)$. 
\end{theorem}
Note that the coefficients $\alpha$ and $\beta$ in the previous equations are obtained by solving the simple system
\begin{eqnarray*}
m_{N+1}(a) = & 0   &= p_{N+1}(a) + \alpha p_{N}(a) + \beta p_{N-1}(a), \\
m_{N+1}(b) = & 0   &= p_{N+1}(b) + \alpha p_{N}(b) + \beta p_{N-1}(b).
\end{eqnarray*}

One can similarly enforce that only one of the end points coincides
with a quadrature one, leading to Gauss--Radau quadratures. The proofs
of theorems~\ref{theorem:gauss} and \ref{theorem:gauss_lobatto} can be
found in most numerical analysis books, in
particular~\cite{Hildebrand}. 

For Chebyshev polynomials there are closed form expressions for the
nodes and weights in Equations~(\ref{eq:quad2} and \ref{eq:quad}):

\paragraph*{Chebyshev--Gauss quadratures.}

For $j=0,1,\ldots, N$,
\begin{eqnarray}
x_j &=& -\cos \left( \frac{2j+1}{2N+2} \pi \right) \label{eq:cheb_gauss} \\
A^{(N)}_j &=& \frac{\pi}{N+1} 
\end{eqnarray}

\paragraph*{Chebyshev--Gauss--Lobatto quadratures.}

\be
x_j = - \cos \left( \frac{\pi j}{N} \right), \qquad \mbox{ for } j =0,1,\ldots ,N,
\label{eq:cheb_gauss_lobatto}
\ee
\begin{displaymath}
A^{(N)}_j = \left\{ 
\begin{array}{lcl} 
\displaystyle 
\frac{\pi}{N} & & \mbox{ for } j =1,2,\ldots,(N-1)  \\ 
& & \\
\displaystyle 
\frac{\pi}{2N} & &  \mbox{ for } j = 0 \mbox{ and } N
\end{array} \right.
\end{displaymath}

\paragraph*{Summation By Parts.}

For any two polynomials $p(x),q(x)$ of degree $N$, in the Legendre case Summation By Parts follows for Gauss, Gauss--Lobatto or Gauss--Radau quadratures, in analogy with the finite difference case (Section~\ref{sec:sbp}).

Since both products $q(x) dp(x)/dx$ and $p(x)dq(x)/dx$ are polynomials of degree $(2n-1)$, their quadratures are exact (in fact, the equality holds for each term separately):
$$
\left \langle \frac{d}{dx}p , q \right \rangle_{\omega}^{\tt d} + \left \langle p , \frac{d}{dx}q \right \rangle_{\omega}^ {\tt d} = \left \langle \frac{d}{dx}p , q \right \rangle_{\omega} + \left \langle p , \frac{d}{dx}q \right \rangle_{\omega}  \, , 
$$
where we have introduced the discrete counterpart of the weighted scalar product (\ref{eq:sp}),
$$
\left \langle h, g \right \rangle_{\omega}^{\tt d} := \sum_{i=0}^N A_i h(x_i) g(x_i) \, ,
$$
with the nodes $\{ x_i \}$ and discrete weights $\{ A_i\}$ those of the corresponding quadrature. 

On the other hand, in the Legendre case, 
$$
\left \langle \frac{d}{dx}p , q \right \rangle_{\omega =1} + \left \langle p , \frac{d}{dx}q \right \rangle_{\omega =1}  = p(b)q(b) - p(a)q(a) \, , 
$$
and therefore
\be
\left \langle \frac{d}{dx}p , q \right \rangle_{\omega =1}^{\tt d} + \left \langle p , \frac{d}{dx}q \right \rangle_{\omega =1}^{\tt d}  = p(b)q(b) - p(a)q(a) \, . \label{eq:leg_sbp}
\ee
Property~(\ref{eq:leg_sbp}) will be used in Section~\ref{sec:collocation} when discussing stability through the energy method, much as in the finite difference case. 

%==============================================================================
\subsection{Discrete expansions and interpolation}
\label{sec:expansion_interpolation}

Suppose one approximates a function $f$ through its \emph{discrete} truncated expansion,  
\be
f(x) \approx {\cal P}^{\tt d}_N[f] (x) = \sum_{j=0}^N \hat{f}^{\tt d}_j p_j(x).
\label{eq:spectral}
\ee
That is, instead of considering the exact projection coefficients $\{ \hat{f}_j \}$, these are  approximated by discretizing the corresponding integrals using Gauss, Gauss--Lobatto or Gauss--Radau quadratures, 
$$
\hat{f}_j = \langle p_j,f \rangle_\omega 
 = \int\limits_{a}^b f(x) p_j(x) \omega(x) dx 
 \approx \hat{f}^{\tt d}_j := \sum_{i=0}^N f(x_i) p_j(x_i) A_i
$$
with $A_i$ given by Equation~(\ref{eq:quad}) and $\{ x_i \}$ any of the Gauss-type points. 
Putting the pieces together, 
$$
{\cal P}^{\tt d}_N [f] (x) = \sum_{i=0}^N  f(x_i) \left( A_i \sum_{j=0}^N p_j(x_i)  p_j(x)  \right) \, . 
$$
Since if $f$ is a polynomial of degree smaller than or equal to $N$, the discrete truncated expansion ${\cal P}^{\tt d}_N[f] = f$ is exact for Gauss or Gauss--Radau quadratures according to the results in Section~\ref{sec:gauss_and_sbp}, it follows that the above term inside the square parenthesis is exactly the $i$-th Lagrange interpolating polynomial, 
$$
A_i \sum_{j=0}^N p_j(x_i)  p_j(x) = \ell^{N}_i(x).
$$
Therefore, we arrive at the remarkable result:

\begin{theorem}
Let $\omega$ be a weight function on the interval $(a,b)$, as introduced in Equation~(\ref{eq:sp}), and denote by ${\cal P}^{\tt d}_n [f]$ the discrete truncated expansion of $f$ corresponding to Gauss, Gauss--Lobatto or Gauss--Radau quadratures. Then,
\be
{\cal P}^{\tt d}_N [f] (x) = {\cal I}[f](x) = \sum_{i=0}^N f(x_i) \ell^N_i(x) \, . 
\label{eq:expansion_interpolation}
\ee
That is, the discrete truncated expansion in orthogonal polynomials of $f$ is exactly equivalent to the interpolation of $f$ at the Gauss, Gauss--Lobatto or Gauss--Radau points.
\end{theorem}
The above simple proof did not assume any special properties of the polynomial basis, but does not hold for the Gauss-Lobatto case (for which the associated quadrature is exact for polynomials of degree $2N-1$). The result, however still holds (at least for Jacobi polynomials), see, for example, \cite{HGG}. 

Examples of Gauss-type nodal points $\{ x_i \}$ are those given in Equation~(\ref{eq:cheb_gauss}) or Equation~(\ref{eq:cheb_gauss_lobatto}). As we will see below, the identity (\ref{eq:expansion_interpolation}) is very useful for spectral differentiation and collocation methods, among other things, since one can equivalently operate with the interpolant, which only requires knowledge of the function at the nodes.

%==============================================================================
\subsection{Spectral collocation differentiation}
\label{sec:spectral_differentiation}

The equivalence (\ref{eq:expansion_interpolation}) between the discrete truncated expansion and interpolation at Gauss-type points allows the approximation of the derivative of a function in a very simple way, 
$$
\frac{d}{dx} f(x) \approx \frac{d}{dx} {\cal P}^{\tt d}_N [f] (x) = \frac{d}{dx}{\cal I}[f](x) =  \sum_{i}^N f(x_i)  \frac{d}{dx} \ell_j(x). 
$$
Therefore, knowing the values of the function $f$ at the collocation points, i.e., the Gauss-type points, we can construct its interpolant ${\cal P}^{\tt d}_N$, take an exact derivative thereof, and evaluate the result at the collocation points to obtain the values of the discrete derivative of $f$ at these points. This leads to a matrix-vector multiplication, where the corresponding matrix elements $D_{ij}$ can be computed once and for all:
$$
\frac{d}{dx} f(x_i) \approx D_{ij} f(x_i) \, \qquad i=0,1,\ldots, N \,,
$$
with 
$$
D_{ij}:= \frac{d}{dx}\ell_i^N (x_j) \, . 
$$
We give the explicit expressions for this differentiation matrix for Chebyshev polynomials both at Gauss and Gauss--Lobatto points (see, for example, \cite{Fornberg, HGG}). 

\paragraph*{Chebyshev--Gauss.}

\begin{displaymath}
D_{ij} = \left\{ 
\begin{array}{lcl} 
\displaystyle
\frac{x_i}{2(1-x_i^2)}  & & \mbox{ for } i=j,\\ 
& & \\
\displaystyle 
\frac{T'_{N+1}(x_i)}{(x_i-x_j) T'_{N+1}(x_j)}  & &  \mbox{ for } i\neq j,
\end{array} \right.
\end{displaymath}
with a prime denoting differentiation. 
 
\paragraph*{Chebyshev--Gauss--Lobatto.}

\begin{displaymath}
D_{ij} = \left\{
\begin{array}{lcl} 
\displaystyle
-\frac{2N^2+1}{6} & &  \mbox{ for } i=j=0, \\
& & \\
\displaystyle
\frac{c_i}{c_j} \frac{(-1)^{i+j}}{x_i-x_j} & &  \mbox{ for } i \neq j, \\
& & \\
\displaystyle
- \frac{x_i}{2(1-x_i^2)}  & &  \mbox{ for } i = j = 1,\ldots, (N-1), \\
& & \\
\displaystyle
\frac{2N^2+1}{6} & &  \mbox{ for } i=j=N,
\end{array} \right.
\end{displaymath}
where $c_i=1$ for $i=1,\ldots (N-1)$ and $c_i=2$ for $i=0,N$.

%==============================================================================
\subsection{The collocation approach}
\label{sec:collocation}

When solving a quasilinear evolution equation 
\be
u_t = P(t,x,u,\partial/\partial x) u + F(t,x,u)
\label{eq:pde}
\ee
using spectral expansions in space and some time evolution scheme, one could proceed in the following way: Work with the truncated expansion of $u(t,x)$, 
$$
{\cal P}_N[u] (t,x) = \sum_{i=0}^N b_i(t) p_i(x)
$$
and write the PDE (\ref{eq:pde}) as a system of $(n+1)$ coupled evolution ordinary differential equations for the $b_i(t)$ coefficients, subject to the initial condition
$$
b_i(t=0) = \langle u (t=0, \cdot), p_i \rangle_{\omega} \,, 
$$
where the quadratures can be approximated using, say, Gauss-type points. One problem with this approach is that if the equation is nonlinear or already at the linear level with variable coefficients, the right hand side of Equation~(\ref{eq:pde}) needs to be re-expressed  in terms of truncated expansions at each timestep. Besides the complexity of doing so, accuracy is lost because higher order modes due to nonlinearities or coupling with variable coefficients are not represented. Or, worse, inaccuracies from those absent modes move to lower frequency ones. This is one of the reasons why collocation methods are instead usually preferred for this class of problems. 

In the collocation approach the differential equation is exactly solved, \emph{in physical space}, at the collocation points $\{ x_i \}_{i=0}^n$, which are those  appearing in Gauss quadratures (Section~\ref{sec:gauss_and_sbp}). Assume for definiteness that we are dealing with a symmetric hyperbolic system in three dimensions, 
$$
u_t (t,x) = \sum_{j=1}^3A^j(t,x,u) \frac{\partial}{\partial x^j} u + F(t,x,u),
$$
see Section~\ref{SubSec:QLP}. We approximate $u$ by its discrete truncated expansion 
\be
u_N := {\cal P}^{\tt d}_N [u].
\label{eq:projection_collocation}
\ee
Then the system is solved at the collocation points, 
\be
\frac{d}{dt}u_N(t,x_i)  = \sum_{j=1}^3A^j(t,x_i, u_N)  \frac{\partial}{\partial x^j} u_N(t,x_i) + F(t,x_i, u_N),
\label{eq:pde_collocation}
\ee
where the spatial derivatives are approximated using spectral differentiation as described in Section~\ref{sec:spectral_differentiation}. The system can then be evolved in time using the preferred time integration method, see Section~\ref{sec:mol}.

From an implementation perspective, there is actually very little difference between a spectral collocation method and a finite difference one: the only two being that the grid points need to be Gauss-type ones and that the derivative is computed using global interpolation at those points. In fact, the actual projection (\ref{eq:projection_collocation}) never needs to be computed for actually solving the system (\ref{eq:pde_collocation}): given initial data $u(t=0,x)$, by construction the interpolant coincides with it at the nodal points, 
\be
u_N(t=0,x_i) = u_N(t=0,x_i)  \label{eq:idata_collocation} \, ,  
\ee
and the system (\ref{eq:pde_collocation}) for $u_n(t,x_i)$ is directly numerically evolved subject to the initial condition (\ref{eq:idata_collocation}). 
 
\paragraph*{Stability.}

As discussed in Section~\ref{sec:gauss_and_sbp}, in the Legendre case the discrete truncated expansion using Gauss-type quadratures leads to Summation by Parts (SBP). In analogy with the finite difference case (Section~\ref{sec:sbp}), when the continuum system can be shown to be well posed through the energy method, a semi-discrete energy estimate can be shown by using the SBP property, at least for constant coefficient systems, and modulo boundary conditions (discussed in the following section).  Consider the same case discussed in Section~\ref{sec:gauss_and_sbp}, a constant coefficient symmetric hyperbolic system in one dimension, 
$$
u_t = A u_x, 
$$
and a collocation approach
\be
\frac{d}{dt}u_N(t,x_i)  = A \frac{\partial}{\partial x} u_N(t,x_i), 
\ee
at Legendre--Gauss-type nodes. Then defining 
\be
E^{\tt d} = \langle u_N, u_N \rangle^{\tt d}_{\omega =1}  \label{eq:energy_spectral}
\ee
taking a time derivative, as in, Section~\ref{sec:gauss_and_sbp}, 
\begin{equation}
\frac{dE^{\tt d} }{dt} = \left \langle \dot{u}_N, u_N \right \rangle^{\tt d}_\mathbf{\omega =1} + \left \langle u_N , \dot{u}_N \right \rangle^{\tt d}_\mathbf{\omega =1} =  \left \langle \frac{\partial}{\partial x} Au_N, u_N \right \rangle^{\tt d}_\mathbf{\omega =1} + \left \langle  Au_N , \frac{\partial }{\partial x}u_n \right \rangle^{\tt d}_\mathbf{\omega =1} \label{eq:spectral_sbp} \, . 
\end{equation}
Now, $u_N \partial u_N /\partial_x$ is a polynomial of degree $(2N-1)$ so the above discrete scalar product is exact for Gauss, Gauss--Lobatto and Gauss--Radau collocation points [cf. Equation~(\ref{eq:leg_sbp})], and if the system (such as symmetric hyperbolic ones) admits an energy estimate at the continuum under the $L^2$ norm, then semidiscrete numerical stability follows (modulo boundary conditions). 

The weighted norm case $\omega\neq 1$ is more involved. In fact, already the advection equation is not well posed under the Chebyshev norm, see for example~\cite{HGG}.

\paragraph*{Spectral viscosity.}

In analogy with numerical dissipation (Section~\ref{sec:dissipation}), spectral viscosity (SV) adds a resolution dependent dissipation term to the evolution equations without sacrificing spectral convergence. SV was introduced by Tadmor in \cite{Tadmor:1989}.  For simplicity consider the Fourier case. Then spectral viscosity involves adding to the evolution equations a dissipative term of the form
$$
\frac{d}{dt}u_N = \left( \ldots \right) - \epsilon_N(-1)^{s}\frac{\partial^s}{\partial x^s}\left[ Q_m(t,x)\frac{\partial^s u_N}{\partial x^s} \right], \quad s \geq 1,
$$
where $s$ is the (fixed) order of viscosity, the viscosity amplitude scales as
$$
\epsilon_N = \frac{C_s}{N^{2s-1}} \, , \qquad
C_s > 0,
$$
and the smoothing functions $Q_m$ effectively apply the viscosity to only the upper portion of the spectrum. In more detail, if $\hat{Q}_j(t)$ are the Fourier coefficients [cf.\ Equation~(\ref{eq:fourier_modes})] of $Q_m(t,x)$ then they are only applied to frequencies $j> m_N$ in such a way that they satisfy
$$
1- \left( \frac{m_N}{|j|} \right)^{(2s-1)/\theta} \leq \hat{Q}_{j}(t) \leq 1
$$
with 
$$
m_N \sim N^{\theta } \, \qquad \theta < \frac{2s-1}{2s} \, . 
$$
The case $s=1$ corresponds to a dissipation term involving a second derivative and $s>1$ is referred to as \textit{super} (or \textit{hyper}) viscosity. Higher values of $s$ (up to $s \sim \sqrt{N}$) dissipate `less aggressively'. 

The Legendre and Chebyshev cases are similar and are discussed
in~\cite{Ma:1998a, Ma:1998b}. The webpage~\cite{SV} keeps a selected
list of publications on spectral viscosity.

%===================================================================
\subsection{Going further, applications in numerical relativity}
\label{sec:spectral_numrel}
%===================================================================

Based on the minimum gridspacing between spectral collocation points, one would naively expect the CFL limit to scale as $1/N^2$, where $N$ is the number of points. The expectation indeed holds, but the reason is related to the ${\cal O} \left( N^2 \right)$ scaling of the eigenvalues of Jacobi polynomials as solutions to Sturm-Liouville problems (in fact, the result holds for non-collocation spectral methods as well) \cite{Gottlieb91thecfl}.

There are relatively few rigorous results on convergence and stability of Chebyshev collocation methods for initial-boundary value problems; some of them are \cite{GottliebLustmanTadmor1987} and \cite{GottliebLustmanTadmor1987b}. 

Even though this review is concerned with time-dependent problems, we note in passing that there is a significant number of efforts in relativity using spectral methods for the constraint equations. The use of spectral methods in relativistic \emph{evolutions} can be traced back to pioneering work in the mid eighties \cite{bonazzola-86} (see also~\cite{bonazzola-90,bonazzola-91,gourgoulhon-92}). Over the last decade they have gained popularity, with applications in scenarios as diverse as relativistic hydrodynamics~\cite{Novak:1997hw, Villain:2002hs, Villain:2004kw}, characteristic evolutions~\cite{Bartnik2000}, absorbing and/or constraint-preserving boundary conditions~\cite{jNsB04,mRoRoS07,oRlBmShP09,oR06}, constraint projection~\cite{mHlLrOhPmSlK04}, late time ``tail'' behavior of black hole perturbations \cite{Scheel:2003vs, Tiglio:2007jp},  cosmological studies \cite{Amorim:2008ff, Beyer:2008gr, Beyer:2008gq}, extreme mass ratio inspirals within perturbation theory and self-forces \cite{Canizares:2008dp, Field:2009kk, Canizares:2009ay, Vega:2009qb, Canizares:2010yx, Canizares:2011kw, Chakraborty:2011gx} and, prominently, binary black hole simulations (see, for example, \cite{mShPlLlKoRsT06,Pfeiffer:2007yz, Boyle:2007ft, Scheel:2008rj, Cohen:2008wa, Lovelace:2008hd, Szilagyi:2009qz, Chu:2009md, Buonanno:2010yk, Lovelace:2010ne}) and black hole-neutron star ones \cite{Duez:2008rb, Foucart:2010eq}. The method of lines (Section~\ref{sec:mol}) is typically used with a small enough timestep so that the time integration error is smaller than the one due to the spatial approximation and spectral convergence is observed. Spectral collocation methods were first used in spherically symmetric black hole evolutions of the Einstein equations in~\cite{Kidder:2000yq} and in three dimensions in~\cite{lKmSsT01}. The latter work showed that some constraint violations in the Einstein-Christoffel \cite{aAjY99} type of formulations do not go away with resolution but are a feature of the continuum evolution equations (though the point --namely, that time instabilities are in some cases not a product of lack of resolution, applies to many other scenarios). 

Most of these references use explicit symmetric hyperbolic first order formulations.  More recently, progress has been made towards using spectral methods for the BSSN formulation of the Einstein equations directly in second order form in space~\cite{Tichy:2009zr, Field:2010mn}, and, generally, on multi-domain interpatch boundary conditions for second order systems~\cite{Taylor:2010ki} (numerical boundary conditions are discussed in the next section). A spectral spacetime approach (as opposed to spectral approximation in space and marching in time) for the 1+1 wave equation in compactified Minkowski was proposed in~\cite{Hennig:2008af}; in higher dimensions and dynamical spacetimes the cost of such approach might be prohibitive though. 

Reference \cite{Bruegmann:2011zj} presents an implementation of the harmonic formulation of the Einstein equations on a spherical domain using a double Fourier expansion and, in addition, significant speed-ups using Graphics Processing Units (GPUs). 

Ref.~\cite{GrandclementNovakLR} presents a detailed review of spectral methods in numerical relativity. 

A promising approach   which, until recently, has been largely unexplored within numerical relativity is the use of discontinuous Galerkin methods~\cite{dGHesthaven, Zumbusch:2009fe, Field:2009kk, Field:2010mn, Radice:2011qr}.

\newpage
%===================================================================
%===================================================================
\section{Numerical Boundary Conditions} 
\label{sec:num_boundary}
%===================================================================
%===================================================================

In practical computations one usually inevitably deals with an initial-boundary value problem and numerical boundary conditions in general have to be imposed (but see the discussion in Section~\ref{SubSec:OutToInfinity}). In this section we discuss some approaches for doing so, with emphasis on sufficient conditions for stability based on the energy method, simplicity, and applicability to high order and spectral methods. In addition to outer boundaries, we also discuss interface ones appearing when there are multiple grids.

General stability results through the energy method are available for symmetric hyperbolic first order linear systems with maximal dissipative boundary conditions. Unfortunately, in many cases of physical interest the boundary conditions are often neither in maximal dissipative form nor is the system linear. In particular, this is true for Einstein's field equations, which are nonlinear, and, as we have seen in Section~\ref{section:ibvpEinstein}, require constraint-preserving absorbing boundary conditions which do not always result in algebraic conditions on the fields at the boundary. Therefore, in many cases one does ``the best that one can'', implementing the outer boundary conditions using discretizations which are known to be stable at least in the linearized, maximal dissipative case. Fortunately, since the outer boundaries are usually placed in the weak field, wave zone, more often than not this approach works well in practice. At the same time, it should be noted that the initial-boundary value problems (IBVPs) for General Relativity formulated in~\cite{hFgN99} and \cite{hKoRoSjW09} (discussed in Section~\ref{section:ibvp}) are actually based on a symmetric hyperbolic first order reduction of Einstein's field equations with maximal dissipative boundary conditions (including constraint-preserving ones). Therefore, it should be possible to construct numerical schemes which can be provably stable, at least in the linearized regime, using the techniques described in the last two sections and in Section~\ref{SubSec:OuterBC}. A numerical implementation of the formulations of Refs.~\cite{hFgN99} and \cite{hKoRoSjW09} has not yet been pursued. 

The situation at interface boundaries between grids, which are at least partially contained in the strong field region, is more subtle. Fortunately, only the characteristic structure of the equations is in principle needed at such boundaries, and not constraint-preserving boundary conditions. Methods for dealing with interfaces are discussed in Section~\ref{SubSec:InterfaceBC}.

Finally, in Section~\ref{sec:boundary_numrel} we give an overview of some applications to numerical relativity of the boundary treatments discussed in Sections~\ref{SubSec:OuterBC} and \ref{SubSec:InterfaceBC}. As mentioned above, most of the techniques that we discuss have been mainly developed for first order symmetric hyperbolic systems with maximal dissipative boundary conditions. In Section~\ref{sec:boundary_numrel} we also point out ongoing and prospective work for second order systems, as well as the important topic of absorbing boundary conditions in General Relativity. 

Most of the methods reviewed below involve decomposition of the principal part, its time derivative, or both, into characteristic variables, imposing the boundary conditions and changing back to the original variables. This can be done a priori, analytically, and the actual online numerical computational cost of these operations is negligible.

\subsection{Outer boundary conditions}
\label{SubSec:OuterBC}

\subsubsection{Injection}

\textit{Injecting} boundary conditions is presumably the simplest way to numerically impose them. It implies simply overwriting, at every or some number of timesteps,  the numerical solution  for each  incoming characteristic variable or its time derivative with the conditions that they should satisfy. 

Stability of the injection approach can be analyzed through GKS theory \cite{GKS}, since energy estimates are in general not available for it (the reason for this should become more clear below when  discussing the projection and penalty methods). Stability analyses in general not only depend on the spatial approximation (and time integration in the fully discrete case) but are in general also equation-dependent. Stability is therefore in general difficult to establish, especially for high order schemes and non-trivial systems. For this reason a semidiscrete eigenvalue analysis is many times performed. Even though this only provides necessary conditions (namely, the von~Neumann condition~(\ref{eq:von_neumann_semidiscrete})) for stability, it serves as a rule of thumb and to discard obviously unstable schemes. 

As an example, we discuss the advection equation with ``freezing'' boundary condition, 
\begin{eqnarray}
u_t = u_x,    && x\in [-1,1],\quad t\geq 0, \label{eq:adv1}\\
u(0,x) = f(x), && x\in [-1,1],\label{eq:adv2}\\
u(t,1)  = g(t), && t \geq 0, \label{eq:adv3} 
\end{eqnarray}
where the boundary data is chosen to be constant, $g(t) = f(1)$. As space approximation we use a Chebyshev collocation method at Gauss--Lobatto nodes. The approximation then takes the form
$$
\frac{d}{dt}u_N (t, x_i) = \left( Q u_N \right)(t,x_i), \qquad i=0,1,\ldots,N. 
$$
where $Q$ is the corresponding Chebyshev differentiation matrix. The boundary condition is then imposed by replacing the $N-th$ equation by 
$$
\frac{d}{dt}u_N (t, x_n) = \frac{d}{dt} g(t) = 0. 
$$
For this problem and approximation an energy estimate can be
derived~\cite{Gottlieb81}, but we present an eigenvalue analysis as a
typical example of those done for more complicated systems and
injection boundary conditions. Figure~\ref{fig:spectral_injection}
shows the semi-discrete spectrum using different number of collocation
points $N$. As needed by the strong version of the von~Neumann
condition, no positive real component is present
[cf. Equation~(\ref{eq:strong_von_neumann_semidiscrete})]. We also note
that, as discussed at the beginning of Section \ref{sec:spectral_numrel}, the spectral radius scales as $\sim N^2$. 

\epubtkImage{}{
\begin{figure}[ht]
\centerline{
\includegraphics[width=0.3\textwidth,angle=0]{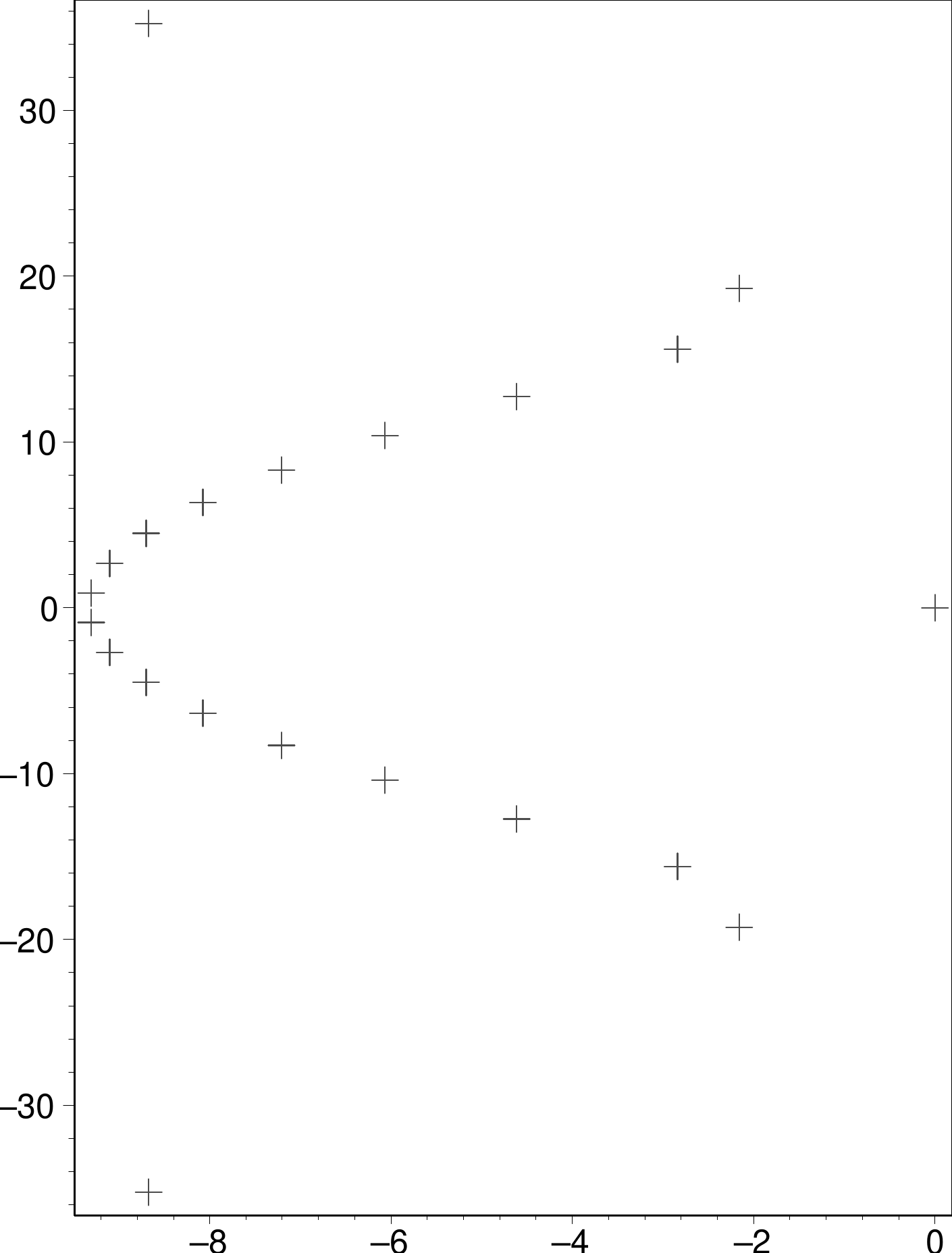}\quad
\includegraphics[width=0.3\textwidth,angle=0]{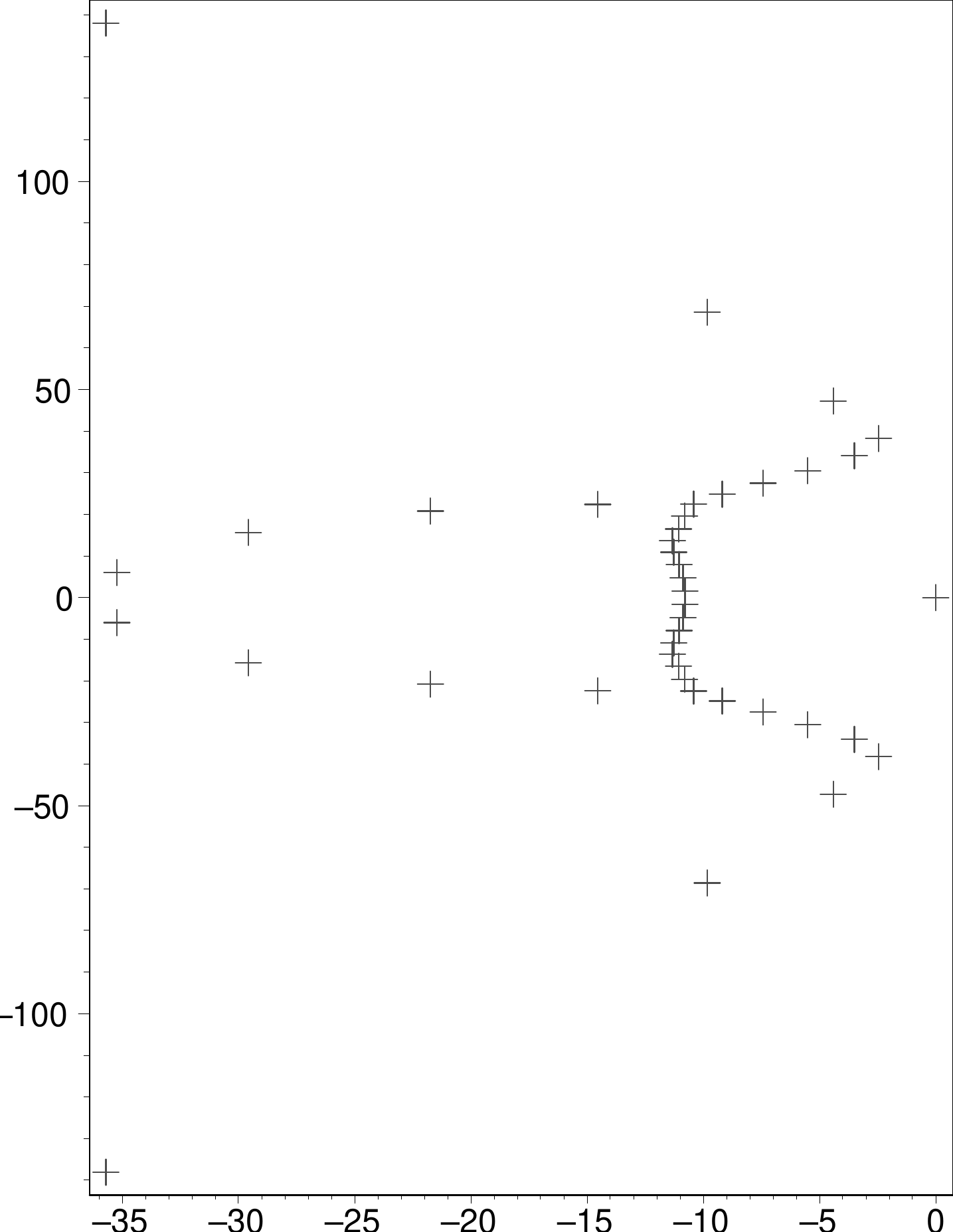}\quad
\includegraphics[width=0.3\textwidth,angle=0]{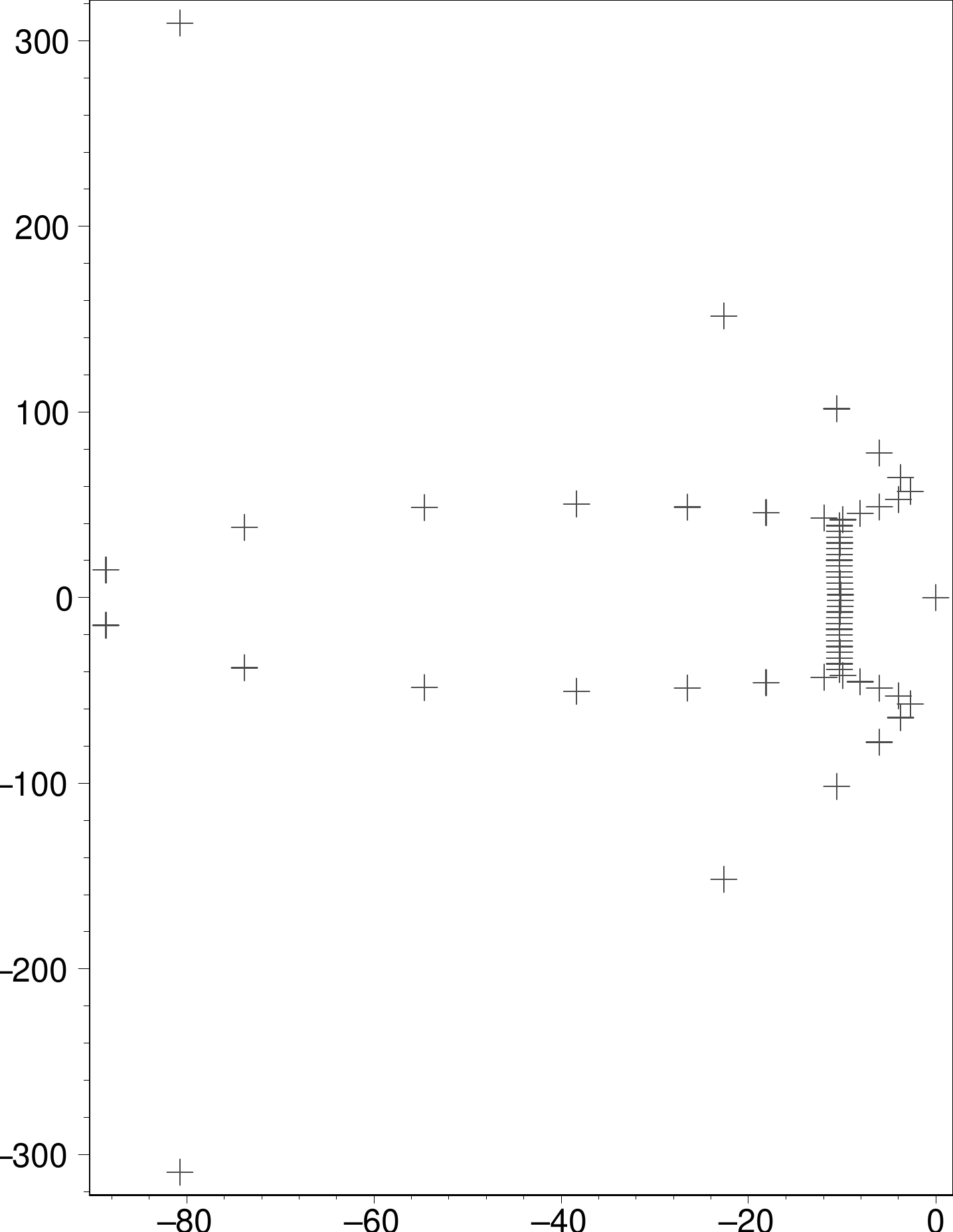}
}
\caption{Semidiscrete spectrum for the advection equation and freezing
  boundary conditions [Eqs.~(\ref{eq:adv1}, \ref{eq:adv2},
  \ref{eq:adv3}], a Chebyshev collocation method, and numerical
  injection of the boundary condition. The number of collocation
  points is  (from left to right) $N = 20, 40, 60$. The scheme passes the
  von~Neumann condition (no positive real component in the
  spectrum). In fact, as discussed in the body of the text, the method can be
  shown to actually be numerically stable.}
\label{fig:spectral_injection}
\end{figure}}

There are other difficulties with the injection method, besides the
fact that stability results are usually partial and incomplete for
realistic systems and/or high order or spectral methods. One of them
is that it sometimes happens that a full GKS analysis can actually be
carried out for a simple problem and scheme, and the result turns out to be stable but not time-stable (see
Section~\ref{sec:time_stability} for a discussion on
time-stability). Or the scheme is not time-stable when applied to a
more complicated system (see, for example, \cite{Carpenter:1993,
  Abarbanel:2000}). 

Seeking stable numerical boundary conditions for realistic systems
which preserve the accuracy of high order finite difference and
spectral methods has been a recurring theme in numerical methods for time-dependent partial differential equations for a long time, especially for non-trivial domains, with
substantial progress over the last decade -- especially with the
penalty method discussed below. Before doing so we review another
method, which improves on the injection one in that stability can be
shown for rather general systems and arbitrary high order finite difference schemes. 

\subsubsection{Projections}

Assume that a given IBVP is well posed and admits an energy estimate, as discussed in Section~\ref{section:MaxDiss}. Furthermore, assume that, up to the control of boundary terms, a semi-discrete approximation to it also admits an energy estimate. The key idea of the projection method \cite{Olsson:1995a,OlssonSupplement, Olsson:1995b} is to impose the boundary conditions by projecting at each time the numerical solution to the space of gridfunctions satisfying those conditions, the central aspect being that the projection is chosen to be orthogonal with respect to the scalar product under which a semi-discrete energy estimate in the absence of boundaries can be shown. The orthogonality of the projection then guarantees that the estimate including the control of the boundary term holds.

In more detail, if the spatial approximation prior to the imposition of the boundary conditions is written as 
\begin{equation}
u_t = Q u \label{eq:prior_bc}
\end{equation} 
and [for example, as is many times the case when Summation By Parts (SBP) holds] there is a semi-discrete energy, 
$$
E_{\tt d} = \langle u, u \rangle \, , 
$$
with respect to some scalar product, for which an estimate holds up to boundary terms: 
$$
\frac{d}{dt}E_{\tt d} = \langle Qu, u \rangle + \langle u, Qu \rangle \,,
$$
with 
$$
\langle Qu, u \rangle + \langle u Qu \rangle \leq F(t)  + \mbox{ boundary terms } 
$$
for some $F(t)$ independent of the initial data and resolution. Due to the SBP property, the boundary terms are exactly those present in the continuum energy estimate, except that so far they cannot be bounded because the numerical solution is not yet required to satisfy the boundary conditions. 

The latter are then imposed by changing the semidiscrete equation~(\ref{eq:prior_bc}) to
$$
u_t = {\cal P} Q u
$$
where ${\cal P}$ projects $u$ to the space of gridfunctions satisfying the desired boundary conditions, is time-independent (which in particular requires the assumption that the boundary conditions are time-independent) and symmetric  
$$
\langle u, {\cal P} v \rangle = \langle {\cal P}u, v\rangle  
$$
(and, being a projection, $ {\cal P} ^2 = {\cal P} $). Since the projection is assumed to be time-independent, projecting both sides of Equation~(\ref{eq:prior_bc}) implies that
$$
({\cal P} u)_t = {\cal P} Q u
$$
and for any solution of (\ref{eq:prior_bc}) satisfying the boundary conditions, ${\cal P} u=u$, 
$$
u_t = {\cal P} Q u \, . 
$$
Then 
\begin{eqnarray*}
\frac{d}{dt}E_{\tt d} &=& \langle {\cal P} Qu, u \rangle + \langle u, {\cal P} Q u \rangle = \langle Qu, {\cal P} u \rangle + \langle {\cal P} u, Q u \rangle = 
\langle Qu, u \rangle + \langle u, Q u \rangle \\
& \leq & F(t)  + \mbox{ boundary terms } \, . 
\end{eqnarray*}
Since the solution now satisfies the boundary conditions, the boundary terms can be bounded as in the continuum and stability of the semi-discrete IBVP follows. In principle, this method only guarantees stability for time-independent, homogeneous boundary conditions. Homogeneity can be assumed by a redefinition of the variables, but time-independent boundary conditions is a more severe restriction.

Details on how to explicitly construct the projection can be found in~\cite{GustafssonBook}. The orthogonal projection method guarantees stability for a large class of problems admitting a continuum energy estimate. However, its implementation is somewhat involved.

%%%%%%%%%%%%%%%%%%%%%%%%%%
\subsubsection{Penalty conditions}
\label{SubSubSec:PenaltyOuterBC}
%%%%%%%%%%%%%%%%%%%%%%%%%%

A simple and robust method for imposing numerical boundary conditions, either at outer or interpatch boundaries, such as those appearing in domain decomposition approaches, is through penalty terms. The boundary conditions are not imposed strongly but weakly, preserving the convergence order of the spatial approximation and leading to numerical stability for a large class of problems. It can be applied both to the case of finite differences and spectral approximations. In fact, the spirit of the method can be traced back to Finite Elements-discontinuous Galerkin methods (see \cite{DouglasDupont1976} and~\cite{Arnold:2001} for more recent results). Terms are added to the evolution equations at the boundaries  to consistently penalize the mismatch between the numerical solution and the boundary conditions the exact solution is subject to.

\paragraph*{Finite differences.}

In the finite difference context the method is known as the Simultaneous Approximation Technique (SAT)~\cite{Carpenter1994220}. For semidiscrete approximations of initial-boundary value problems of arbitrary high order with an energy estimate, both the order of accuracy and the presence of energy estimates are preserved when imposing the boundary conditions through it. 

As an example, consider the half-space IBVP for the advection equation, 
\begin{eqnarray}
u_t = \lambda u_x, && x\leq 0,\quad t\geq 0, \label{eq:penalty_advection1}\\ 
u(0,x)  =  f(x), && x\leq 0,\label{eq:penalty_advection2} \\
u(t,0)  =  g(t), && t \geq 0,\label{eq:penalty_advection3}
\end{eqnarray}
where $\lambda\geq 0$ is the characteristic speed in the $-x$ direction.

We first consider a semi-discrete approximation using some  finite difference  operator $D$ satisfying SBP with respect to a scalar product $\mathbf{\Sigma}$, which we assume to be either diagonal or restricted full (see Section~\ref{sec:sbp}). As usual, $\Delta x$ denotes the spacing between gridpoints,
$$
x_i = i\Delta x,\qquad i=0,-1,-2,\ldots.
$$
In the SAT approach the boundary conditions are imposed through a penalty term which is applied at the boundary point $x_0$, 
\begin{equation}
\frac{d}{dt}u_i = \lambda D u_i + \frac{\delta_{i,0}S}{\sigma_{00} \Delta x} (g - u_0), \label{eq:adv_outer}
\end{equation}
where $\delta_{i,0}$ is the Kronecker delta, $S$ a real parameter to be restricted below, and $\sigma_{00}$ is the $00$-component of the SBP scalar product $\mathbf{\Sigma}$.
Introducing the semi-discrete SBP energy 
$$
E_{\tt} = \langle u,u \rangle_\mathbf{\Sigma} \,,
$$
its time derivative is 
\begin{equation}
\frac{d}{dt}E = (\lambda - 2S)u_0^2 + 2 g u_0 S  \label{eq:SAT1} \, .
\end{equation}
Next, define $\delta$ through $S=\lambda + \delta$. For homogeneous boundary conditions ($g=0$) both numerical and time-stability follow if $\delta \geq -\lambda/2$: 
$$
\frac{d}{dt}E = - ( \lambda + 2 \delta )u_0^2 \leq 0 \,.
$$
Strong stability follows if $\delta > -\lambda /2$, including the case of inhomogeneous boundary conditions: for any $\varepsilon $ such that 
$$
0< \varepsilon ^2 < 2S - \lambda \, , 
$$
Equation~(\ref{eq:SAT1}) implies
\be
\frac{d}{dt}E = (\lambda - 2S )u_0^2 + 2 (\varepsilon u_0) \left( \frac{gS}{\varepsilon} \right)  \leq (\lambda - 2S )u_0^2 + (\varepsilon ^2 u_0 ^2) + \left( \frac{gS}{\varepsilon }\right)^2  = -\tilde{\varepsilon } u_0^2 + \left( \frac{S}{\varepsilon }\right)^2 g^2\, , \label{eq:penalty_tmp}
\ee
where 
$$
\tilde{\varepsilon}:= 2S - \lambda - \varepsilon ^2 > 0 \, . 
$$
Integrating Equation~(\ref{eq:penalty_tmp}) in time, 
$$
\| u (t)  \|_\mathbf{\Sigma}^2 + \tilde{\varepsilon} \int_0^t u_0(\tau )^2 d\tau \leq \left( \frac{S}{\varepsilon }\right)^2 \int_0^t g(\tau )^2 d\tau\, , 
$$
thereby proving strong stability.

In the case of diagonal SBP norms it is straightforward to derive similar energy estimates for general linear symmetric hyperbolic systems of equations in several dimensions,  simply by working with each characteristic variable at a time, at each boundary.  A penalty term is applied to the evolution term of each incoming  characteristic variable at a time as in Equation~(\ref{eq:adv_outer}), where $\lambda $ is replaced by the corresponding characteristic speed. In particular, edges and corners are dealt by simply imposing the boundary conditions with respect to the normal to each boundary, and an energy estimate follows.

One can show what the global semidiscrete convergence rate is as follows. Define the error gridfunction as the difference between the numerical solution $u$ and the exact one $u^{(e)}$ evaluated at the gridpoints, 
\be
e_i(t) = e(t,x_i) := u(t,x_i) - u^{(e)}(t,x_i), \quad i=0,-1,-2,\ldots,
\label{eq:error_sbp_penalty}
\ee
It satisfies the equation
$$
\frac{d}{dt} e_i = \lambda D e_i + \frac{\delta_{i,0}S}{\sigma_{00} \Delta x} e_0 - F_i \, , 
$$
where $F_i$ denotes the truncation error, here solely depending on the differentiation approximation:
\begin{eqnarray}
F_i(t) =  F(t,x_i) &=& \lambda \left( u^{(e)}_x(t,x_i) - Du^{(e)}(t,x_i) \right) \, .\\
& =  &   
\left \{
\begin{array}{l }
{\cal O} (\Delta x )^{r} \quad \mbox{ at and close to boundaries} \, ,  \label{eq:penalty_rate} \\
{\cal O} (\Delta x )^{2p} \quad   \mbox{in the interior}\, , 
\end{array}
\right .
\end{eqnarray}
with $r<2p$ in general. For example, in the diagonal case $r=p$, and in the restricted full one $r=2p-1$ [see the discussion below Equation~(\ref{eq:sbp_block})]. 

Using Equation~(\ref{eq:error_sbp_penalty}) and the SBP property the norm of the error satisfies
\be
\frac{d}{dt} \| e \|_\mathbf{{\Sigma }}^2  =  (\lambda -2S )e_0^2  - 2 \langle e, F \rangle_\mathbf{\Sigma } \leq  2\| e\|_\mathbf{\Sigma} \| F \|_\mathbf{\Sigma}\, ,  \label{eq:strong_penalty}
\ee
where we have used $S\leq \lambda/2$ and the Cauchy-Schwarz inequality in the second step. Dividing both sides of the inequality by $2\| e \|_\mathbf{{\Sigma }}$ and integrating we obtain
$$
 \| e(t) \|_\mathbf{{\Sigma }} \leq \int\limits_0^t \| F(\tau) \|_\mathbf{\Sigma} d\tau,\qquad
 t\geq 0.
$$
Since by Equation~(\ref{eq:penalty_rate}) the norm of the truncation error is of the order $(r+1/2)$, it follows that the $\mathbf{\Sigma}$-norm of the error converges to zero with rate $(r+1/2)$. This also implies that the error converges \emph{pointwise} to zero with rate,
$$
e_i(t) = {\cal O} \left( (\Delta x )^r\right),\qquad t > 0,\quad i=0,-1,-2,\ldots
$$
In particular, this proves that the error at the boundary point $x_0 = 0$ converges to zero, and therefore, the penalty term in Equation~(\ref{eq:adv_outer}) is finite. Two remarks are in order:
\begin{itemize}
\item In special cases it is possible to improve the error estimate exploiting the strong stability of the problem. Consider for instance the case of the $D_{2-1}$ operator defined in Example~\ref{ex:D21} with the associated diagonal scalar product $(\sigma_{ij}) = \diag(1/2,1,1,1,...)$. Then, Equation~(\ref{eq:strong_penalty}) gives
\begin{eqnarray}
\frac{d}{dt}\| e \|_\mathbf{{\Sigma }}^2  
 &=&  (\lambda -2S )e_0^2  - 2 \langle e, F \rangle_\mathbf{\Sigma } 
 = (\lambda - 2S) e_0^2 - \Delta x e_0 F_0 - 2\Delta x\sum\limits_{i=-\infty}^{-1} e_i F_i
\nonumber\\
 &\leq& \left[ (\lambda - 2S) + \frac{\varepsilon^2}{2} \right] e_0^2 
 + \frac{\Delta x^2}{2\varepsilon^2} F_0^2
 +  \Delta x\sum\limits_{i=-\infty}^{-1} \left( e_i^2 + F_i^2 \right)
\nonumber\\
 &\leq& \| e \|_\mathbf{{\Sigma }}^2  +  {\cal O} \left( (\Delta x )^4\right),
\nonumber
\end{eqnarray}
where we have chosen $0 < \varepsilon^2/2\leq 2S - \lambda$. This implies that the 
error converges to zero in the $\mathbf{\Sigma}$-norm with second order and pointwise with order $3/2$.
\item At any fixed resolution, the error \emph{typically}\epubtkFootnote{There are cases in which this is not true, at least for not too large penalty strengths.} decreases with larger values of the penalty parameter $\delta$, but the spectral radius of the discretization grows quickly in the process by effectively introducing dissipative eigenvalues (on the left half plane) in the spectrum, leading to demanding CFL limits (see, for example \cite{Lehner:2005bz}). Because the method is usually applied along with high order methods, decreasing the error for a fixed resolution at the expense of increasing the CFL limit does not seem worthwhile. In practice values of $\delta$ in the range $-\lambda/2 < \delta < 0$ give reasonable CFL limits. 
\end{itemize}

\paragraph*{Spectral methods.}

The penalty method for imposing boundary conditions was actually introduced for spectral methods prior to finite differences in Refs.~\cite{Funaro88,Funaro91}. In fact, as we will see below, the finite difference and spectral cases follow very similar paths. Here we only discuss its application to the collocation method. Furthermore, as discussed in Section~\ref{sec:gauss_and_sbp}, when solving an initial-boundary value problem, Gauss--Lobatto collocation points are natural among Gauss-type nodes, because they include the end points of the interval. We restrict our review to them, but the penalty method applies equally well to the other nodes. We refer to~\cite{Hesthaven200023} for a thorough analysis of spectral penalty methods. 

Like the finite difference case we summarize the method through the example of the advection problem~(\ref{eq:penalty_advection1}, \ref{eq:penalty_advection1}, \ref{eq:penalty_advection2}), except that now we consider the bounded domain $x\in[-1,1]$ and apply the boundary condition at $x=1$. Furthermore, we first consider a truncated expansion in Legendre polynomials. A penalty term with strength $\tau $ is added to the evolution equation at the last collocation point: 
\begin{equation}
\frac{d}{dt}u_N (t,x_i) = \lambda u_N'(t,x_i) +  \tau \lambda \left( g(t) - u_N(t,x_n) \right)\delta_{iN}  \qquad i=0,1,\ldots, N,
\label{eq:adv_outer_spectral}
\end{equation}
where $\{ x_i \}_{i=0}^N$ are now the Gauss--Lobatto nodes; in particular, $x_0=-1$ and $x_N=1$.
  
Using the discrete energy given by Equation~(\ref{eq:energy_spectral}) and Summation by Parts property associated with Gauss quadratures discussed in Section~\ref{sec:gauss_and_sbp} a discrete energy estimate follows exactly as in the finite difference case if 
\be
\tau  \geq \frac{1}{2A_N^N} \geq \frac{N(N+1)}{4} \, , 
\label{eq:tau_limit_legendre} 
\ee
where $A_N^N$ is the Legendre--Gauss--Lobatto quadrature weight at the
right end point $x_N=1$  [cf.\ Equations~(\ref{eq:quad2}, \ref{eq:quad})]. Notice that this scaling of the penalty with the weight is the analog of the $(\sigma_{00} \Delta x)$ SBP weight term for the finite difference case [Equation~\ref{eq:adv_outer}]. The similitude is not accidental: both weights arrive from the discrete integration formulae for the energy. Notice: 
\begin{itemize}
\item The approach for linear symmetric hyperbolic systems is the same one that we discussed for finite differences: the evolution equation for each characteristic variable is penalized as in Equation~\ref{eq:adv_outer_spectral}, where $\lambda$ is replaced by the corresponding characteristic speed. 
\end{itemize}

Devising a Chebyshev penalty scheme that guarantees stability is more convoluted. In particular, the advection equation is already not well posed in the Chebyshev norm, for example (see \cite{Gottlieb81}).  Stability in the $L_2$ norm can be established using the \textit{Chebyshev--Legendre} method~\cite{DonGottlieb94}, where the Chebyshev--Gauss--Lobatto nodes are used for the approximation but the Legendre ones for satisfying the equation. In this approach, the penalty method is global, because it adds terms to the right-hand side of the equations not only at the endpoint, but at all other collocation points as well.

A simpler approach, where the penalty term is only applied to the boundary collocation point, as in the Legendre and finite differences case, is to show stability in a different norm. For example, in Ref.~\cite{DettoriYang96} it is shown that a penalty term as in Equation~(\ref{eq:adv_outer_spectral}) is stable for the Chebyshev--Gauss--Lobatto case in the norm defined by the weight, cf. Equation~(\ref{eq:sp}),
$$
\tilde{\omega}(x) = (1+x) \omega_{\tt Chebyshev}(x) = \sqrt{\frac{1+x}{1-x}}\, ,
\qquad -1 < x < 1,
$$
if 
\be
\tau \geq \frac{N^2}{2} \, .  \label{eq:tau_limit_chebyshev}
\ee
In Figure~\ref{fig:openbc_pen03} we show the spectrum of an
approximation of the advection equation (with speed $\lambda =1$ and
$g(t)=0$) using the Chebyshev penalty method, $N=20,40,60$ collocation
points and $S:=\tau N^2=0.3$. There are no eigenvalues with positive
real component. Even though this is only a necessary condition for
stability, it suggests that it might be possible to lower the bound
(\ref{eq:tau_limit_chebyshev}) on the minimum penalty needed for
stability. On the other hand, for $S=0.2$ (and lower values) real
positive values do appear in the spectrum. Compared to the injection
case (Figure~\ref{fig:spectral_injection}), the spectral radius for
the cases shown in Figure~\ref{fig:openbc_pen03} are about a factor of
two larger. However, in most applications using the method of lines
the timestep is dictated by the condition of keeping the time
integration error smaller than the spatial one, so as to maintain
spectral convergence, and not the CFL limit.

\epubtkImage{}{
\begin{figure}[htbp]
\centerline{
\includegraphics[width=0.3\textwidth,angle=0]{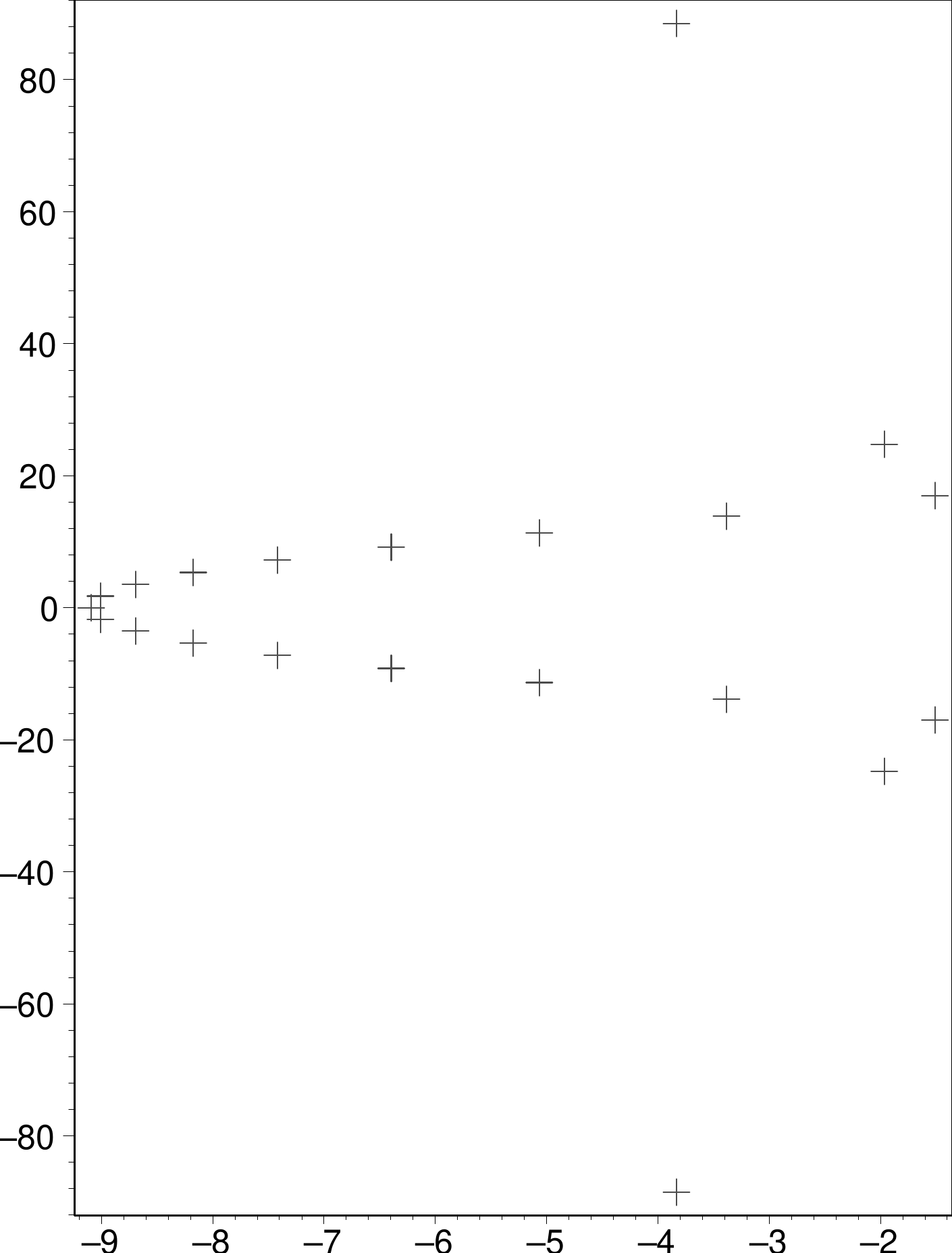}\quad
\includegraphics[width=0.3\textwidth,angle=0]{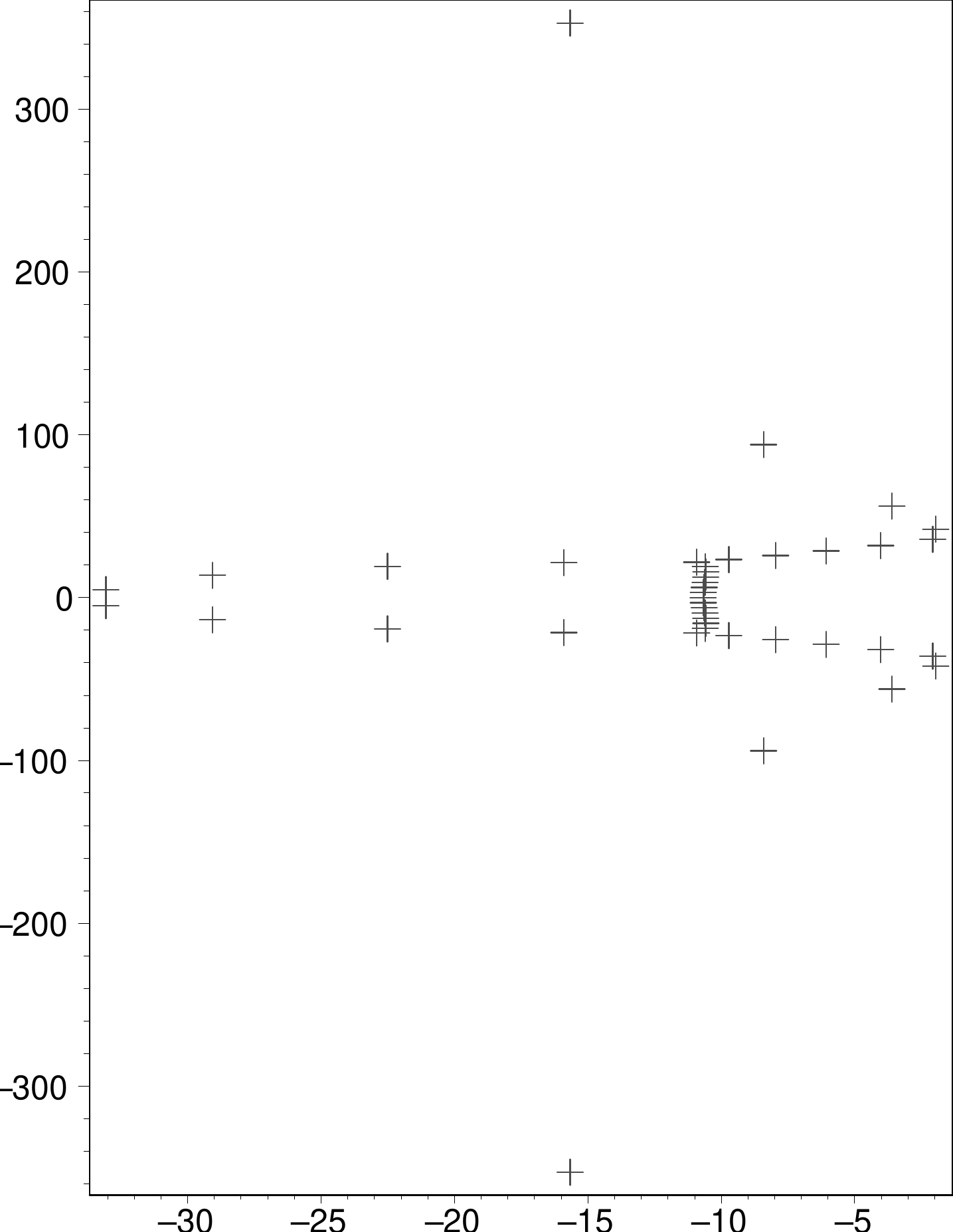}\quad
\includegraphics[width=0.3\textwidth,angle=0]{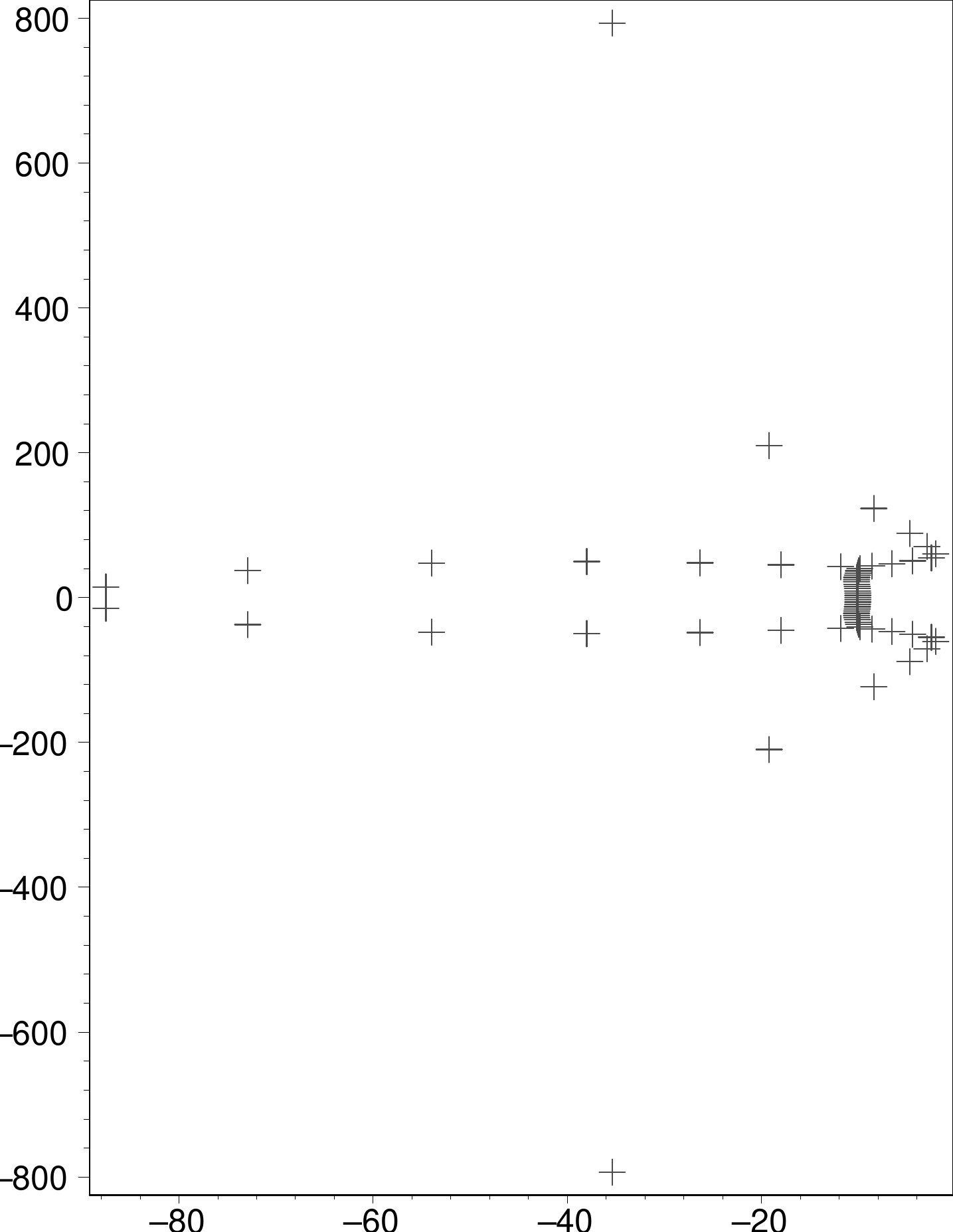}
}
\caption{Spectrum of the Chebyshev penalty method for the advection
  equation and (from left to right) $N=20,40,60$ collocation points
  and $S=0.3$.}
\label{fig:openbc_pen03}
\end{figure}}

\subsection{Interface boundary conditions}
\label{SubSec:InterfaceBC}

Interface boundary conditions are needed when there are multiple grids in the computational domain, as discussed in Section~\ref{sec:complex} below. This applies to complex geometries, when multiple patches are used for computational efficiency, to adapt the domain decomposition to the wave zone, mesh refinement, or a combination of these.

A simple approach for exchanging the information between two or more grids is a combination of interpolation and extrapolation of the whole state vector at the intersection of the grids. This is the method of choice in mesh refinement, for example, and works well in practice. In the case of curvilinear grids and very high order finite difference or spectral methods, though, it is in general not only difficult to prove numerical stability, but even to find a scheme that exhibits stability from a practical point of view.

\subsubsection{Penalty conditions}

The penalty method discussed above for outer boundary conditions can
also be used for multi-domain interface ones, including those present
in complex geometries~\cite{Carpenter:1996, Hesthaven:1997,
  Carpenter:1999, Nordstrom:1999, Nordstrom:2001, Diamessis:2005}. It
is simple to implement, robust, leads to stability for a very large
class of problems and preserves the accuracy of arbitrary high order
difference and spectral methods. 

\paragraph*{Finite differences.}

As an example consider again an advection equation but now on the whole real line
\begin{eqnarray*}
u_t = \lambda u_x, && -\infty < x < \infty, \quad t \geq 0,\\
u(0,x) = f(x), && -\infty < x < \infty, 
\end{eqnarray*}
where we assume that the initial data $f$ is $C^\infty$ smooth and has compact support. The domain of the real line is chosen just for simplicity, to focus on the interface procedure at $x=0$. In the realistic case of compact domains,
outer boundaries are also present and these can be treated by any of the methods discussed in Section~\ref{SubSec:OuterBC} above.

Consider two grids, \emph{left} and \emph{right}, covering the intervals $(-\infty,0]$, and
 $[0,+\infty)$, respectively:
 \begin{eqnarray*}
 x_i^l &=& i\Delta x^l,\qquad i=0,-1,-2,\ldots, \\
 x_i^r &=& i\Delta x^r,\qquad i=0,1,2,\ldots,
 \end{eqnarray*}
where the gridspacings need not agree, $\Delta x^l \neq \Delta x^r$, and the difference operators $D^l$ and $D^r$, which are not necessarily equal to each other either, and do not even need to be of the same order of accuracy, satisfy SBP with respect to scalar products given by the weights $\sigma^l, \sigma^r$ on their individual grids:
 \begin{equation}
\langle v^l,v^l \rangle_{\mathbf{\Sigma}_l}  =  \Delta x^l\sum_{i,j=-\infty}^{0} \sigma^l_{ij} v^l_i v^l_j  \;\;\; , \;\;\;
\langle v^r,v^r \rangle_{\mathbf{\Sigma}_r}  =  \Delta x^r\sum_{i,j=0}^{+\infty} \sigma^r_{ij}  v^r_i v^r_j \, . 
\end{equation}

For the time being assume that both SBP scalar products are diagonal, though. The SAT semi-discrete approximation to the problem then is 
\begin{eqnarray}
\frac{d}{dt}v^l_i &=& \lambda D^l v^l_i + \frac{\delta_{i,0} S^l}{\Delta x^l\sigma^l_{00}}(v^r_0-v^l_0),  \qquad i=0,-1,-2,\ldots, \label{eq:sat_left}\\
\frac{d}{dt}v^r_i &=& \lambda D^r v^r_i + \frac{\delta_{i,0} S^r}{\Delta x^r\sigma^r_{00}}(v^l_0-v^r_0), \qquad i=0,1,2,\ldots. \label{eq:sat_right}
\end{eqnarray}
Notice that in this approach the numerical solution at any fixed resolution is bi-valued at the interface boundary $x=0$, and in the same spirit as the penalty approach for outer boundary conditions, continuity of the fields at the interface is not enforced strongly but weakly.  

Defining the energy 
$$
E:= \langle v^l, v^l\rangle_{\mathbf{\Sigma}_l} + \langle v^r, v^r\rangle_{\mathbf{\Sigma}_r}\, , 
$$
and using the approximation (\ref{eq:sat_left}, \ref{eq:sat_right}) and the SBP property of the difference operators, its time derivative is
$$
\frac{d}{dt}E = (\lambda - 2S^l)(u_0^l)^2 + (-\lambda - 2S^r)(u_0^r)^2 + 2(S^l+S^r)u_0^lu_0^r \,. 
$$
Then an estimate follows if two
conditions are satisfied. One of them is 
$\lambda + S_r - S_l = 0$. The other one imposes an additional
constraint on the values of $S_l$
and $S_r$:
\begin{itemize}
\item \emph{Positive $\lambda$:}
$$
S_l = \lambda + \delta, \;\;\; S_r = \delta, \;\;\; \mbox{ with } \delta \geq
- \frac{\lambda}{2} \label{pos_lam}
$$
The estimate is 
$$
\frac{d}{dt}E= -(u_0^l-u_0^r)^2(\lambda + 2 \delta ) \leq 0 \, . 
$$
\item \emph{Negative $\lambda$:} this is obtained from the previous case after the transformation $\lambda\mapsto -\lambda$
$$
S_r = -\lambda + \delta, \;\;\; S_l = \delta, \;\;\; \mbox{ with } \delta \geq
 \frac{\lambda}{2} \label{neg_lam}
$$
and
$$
\frac{d}{dt}E= (u_0^l-u_0^r)^2(\lambda - 2 \delta ) \leq 0 \, . 
$$
\item \emph{Vanishing $\lambda$:} this can be seen as the limiting case of any of the above two, with 
$$
\frac{d}{dt}E= -(u_0^l-u_0^r)^22 \delta  \leq 0 \, . 
$$
\end{itemize}
The following remarks are in order:
\begin{itemize}
\item For the minimum values of $\delta $ allowed by the above inequalities the energy estimate is the same as for the single grid case with outer boundary conditions, see Section~\ref{SubSubSec:PenaltyOuterBC}, and the discretization is time-stable (see Section~\ref{sec:time_stability}), while for larger values
of $\delta$ there is damping in the energy which is proportional to the mismatch at the interface. 

\item Except for the case of the most natural choice $\delta =0$, the evolution equations for outgoing modes also need to be penalized in a consistent way in order to derive an energy estimate. However, as is always the case, the lack of an energy-type estimate does not mean that the scheme is unstable, since the energy method provides sufficient but not always necessary conditions for stability. 

\item The general case of symmetric hyperbolic systems follows along the same lines: a decomposition into characteristic variables is performed and the evolution equation for each of them is penalized as in the advection equation example. At least for diagonal norms, stability also follows for general linear symmetric hyperbolic systems in several dimensions. With the standard caveats for non-diagonal norms, the procedure follows in a similar way except that penalty terms are not only added to the evolution equations at the interface on each grid but also \emph{near} them. In practice, though, applying penalties just at the interfaces appears to work well in practice in many situations.
\end{itemize}

\paragraph*{Spectral methods.}

The standard procedure for interface spectral methods is to penalize each incoming characteristic variable, exactly as in the outer boundary condition case. Namely, as in Equation~(\ref{eq:adv_outer_spectral}) with lower bounds for the penalty strengths given by Equations~(\ref{eq:tau_limit_legendre}) and (\ref{eq:tau_limit_chebyshev}) for Legendre and Chebyshev polynomials, respectively. We know from the finite difference analysis above, though, that in general this does not imply an energy estimate and in order to achieve one outgoing modes also need to be penalized, with strengths that are coupled to the penalty for incoming modes. 
However, the procedure of penalizing just incoming modes at interfaces appears to works well in practice, so we analyze this in some detail. 

Figure~\ref{fig:penalty_chebyshev} shows the spectrum of the Chebyshev penalty method as described, for an advection equation, 
$$
u_t = u_x \, , \qquad -1\leq x\leq 1,\quad t\geq 0,
$$
where there is an interface boundary at $x=0$. In more detail: the figure shows the maximum real component in the spectrum for $N=20$ collocation points as a function of the penalty strength $S$. There are no real positive values, and this remains true for different values of $N$, in agreement with the fact that penalizing just the incoming mode works well in practice. The figure also supports the possibility that $S\gtrsim 0.3$ might actually be enough for stability, as in the outer boundary case. 

The figure also shows the spectral radius as a function of the penalty
strength. Beyond $S=1$ it grows very quickly, and even though as
mentioned the timestep is usually determined by keeping the time
integration error below that one due to spatial discretization, that
might not be the case if the spectral radius is too large. Thus, it is
probably good to keep $S \lesssim 1$. 

\epubtkImage{}{
\begin{figure}[htbp]
\centerline{
\includegraphics[width=0.3\textwidth,angle=0]{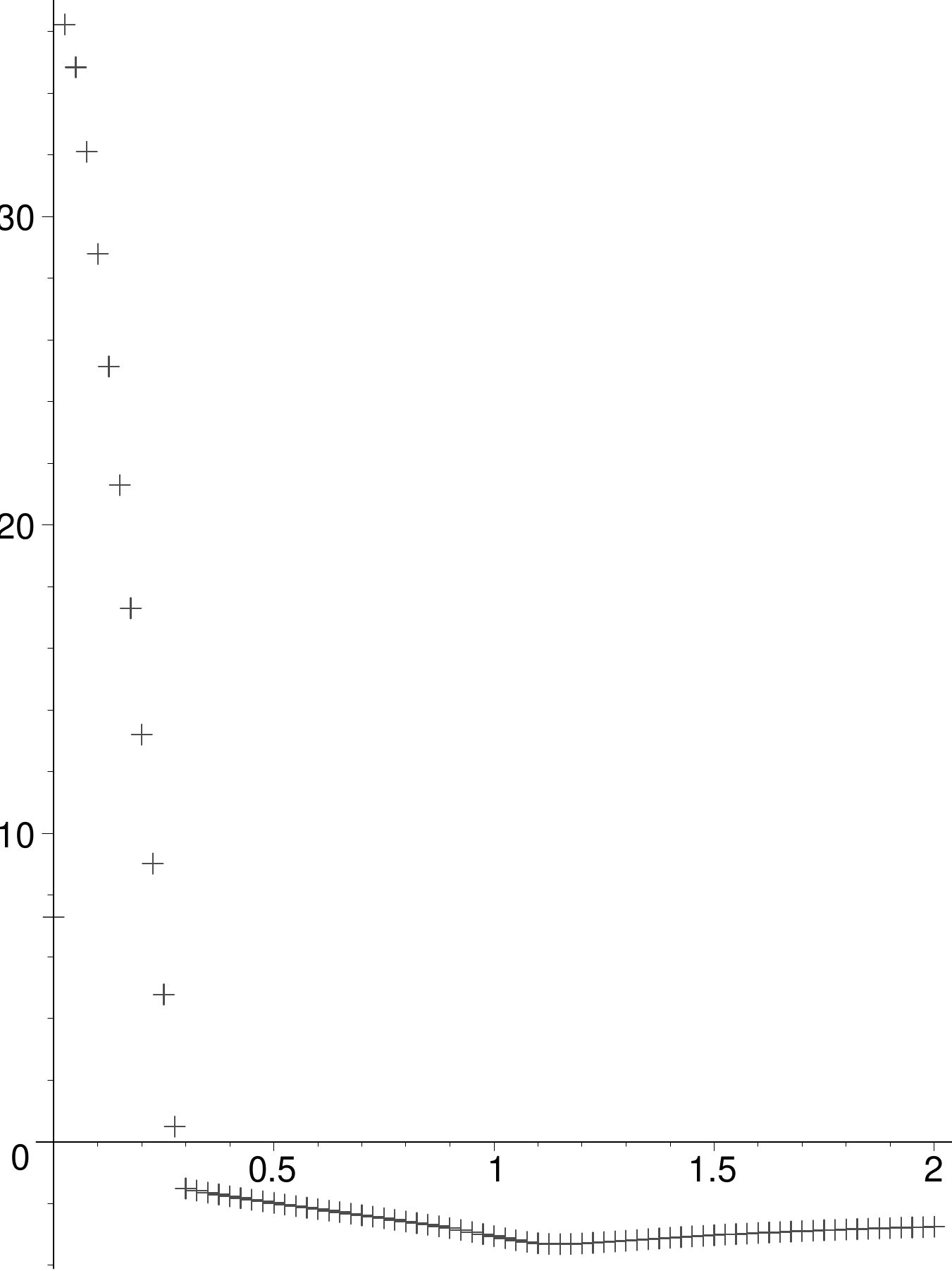}\quad
\includegraphics[width=0.3\textwidth,angle=0]{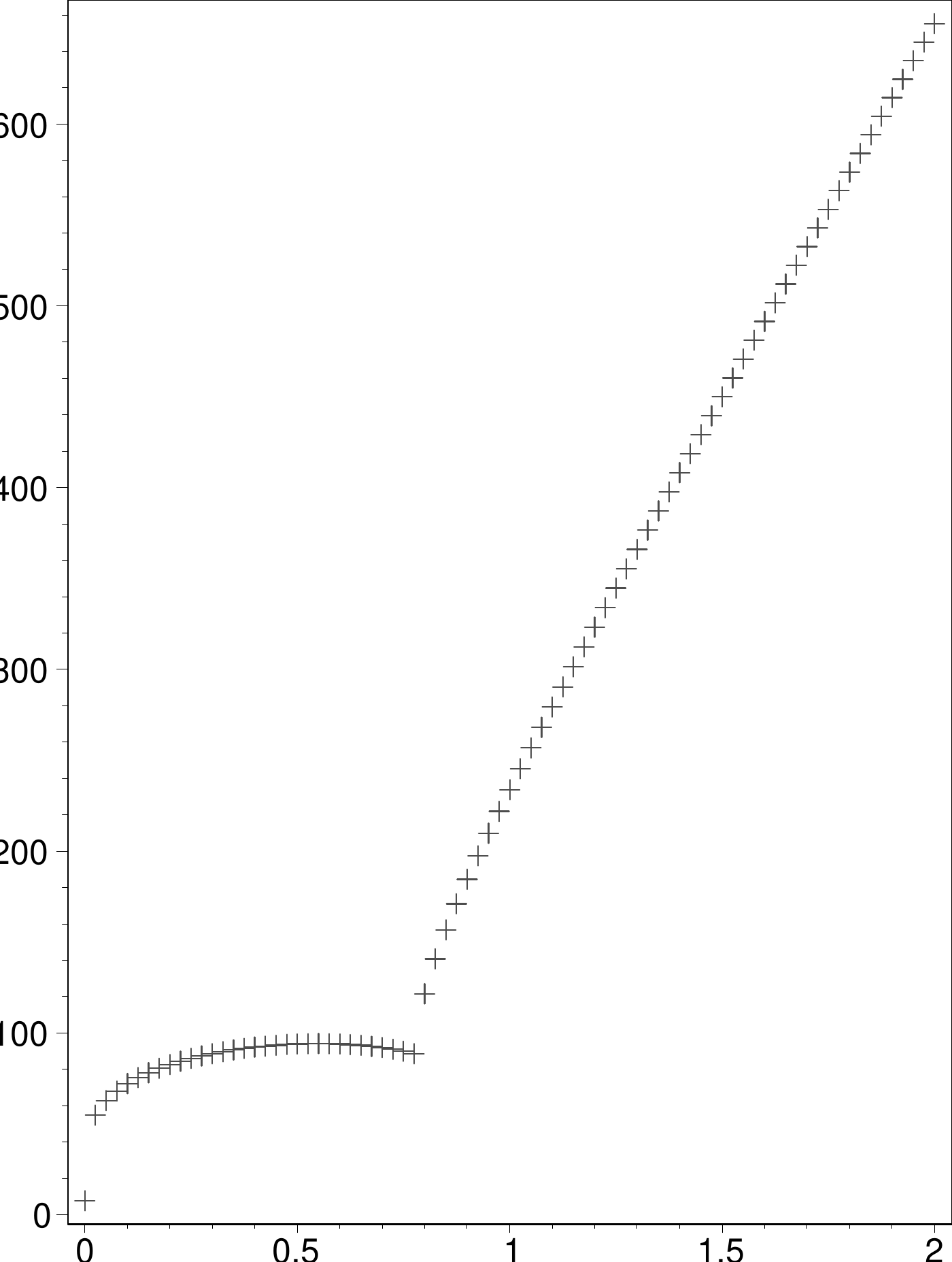}
}
\caption{Maximum real component (left) and spectral radii (right)
  versus penalty strength $S$, for the Chebyshev penalty method for
  the advection equation with two domains ($N=20$ collocation
  points).}
\label{fig:penalty_chebyshev}
\end{figure}}

%%%%%%%%%%%%%%%%%%%%%%%%%%%%%%%%%%%%%%%
\subsection{Going further, applications in numerical relativity}
\label{sec:boundary_numrel}
%%%%%%%%%%%%%%%%%%%%%%%%%%%%%%%%%%%%%%%

In numerical relativity, the projection method for outer boundary conditions has been used in references~\cite{Calabrese:2003yd, Calabrese:2003vx, Tiglio:2003xm, Lehner:2004cf, Gundlach:2010et}, the penalty finite difference one for multi-domain boundary conditions in~\cite{Lehner:2005bz, Diener:2005tn, Schnetter:2006pg, Dorband:2006gg, Pazos:2006kz, Pazos:2009vb, Vega:2009qb, Korobkin:2010qh, Zink:2010bq, Vega:2011wf}, and for spectral methods in -- among many other references -- \cite{Chu:2010yu, Lovelace:2010ne, Buonanno:2010yk, Foucart:2010eq, Mroue:2010re,aZlK10b, Szilagyi:2009qz, Chu:2009md, Duez:2009yy, Lovelace:2009dg, Aylott:2009tn, Buonanno:2009qa, Aylott:2009ya, Scheel:2008rj, Duez:2008rb,mShPlLlKoRsT06}. 

Ref.~\cite{Mattsson:2003} presents a comparison, for the finite difference case, of numerical boundary conditions through injection, orthogonal  projections, and penalty terms. One additional advantage of the penalty approach is that for advection-diffusion problems it damps away in time violations of the compatibility conditions between the initial and boundary data, see Section~\ref{section:ibvp}. Figure~\ref{fig:bc_comparison} shows the error as a function of time for an advection-diffusion equation where the initial data is perturbed so that an inconsistency with the boundary condition is present, and the error is defined as the difference between the unperturbed and perturbed solutions. See Ref.~\cite{Mattsson:2003} for more details. One expects this not to be the case for hyperbolic problems, though. For example, at the continuum an incompatibility between the initial data and initial-boundary condition for the advection would propagate, not dissipate in time, and a consistent and convergent numerical scheme should reflect this behavior. 

\epubtkImage{}{
\begin{figure}[h]
\centerline{\includegraphics[width=0.5\textwidth,angle=90]{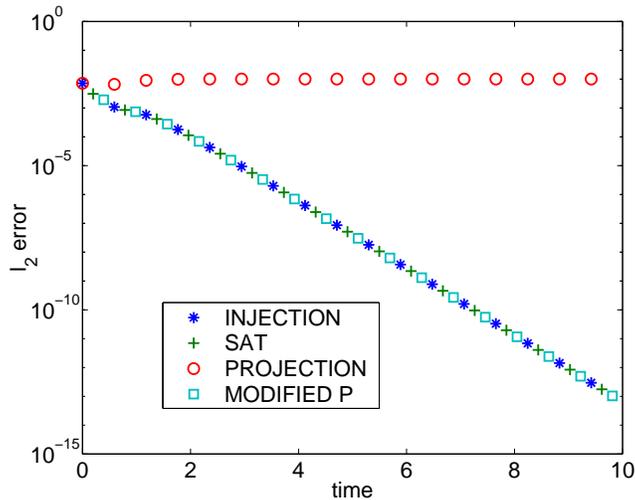}}
\caption{Comparison of different numerical boundary approaches for an
  advection-diffusion equation where the initial data is perturbed,
  introducing an inconsistency with the boundary condition
  \cite{Mattsson:2003}. The SAT approach,
  besides guaranteeing time stability for general systems, ``washes
  out'' this inconsistency in time. ``MODIFIED P'' corresponds to a
  modified projection \cite{Gustafsson:1998} for which
  this is solved at the expense of losing an energy
  estimate. Courtesy: Ken Mattsson.}
\label{fig:bc_comparison}
\end{figure}}

Many systems in numerical relativity, starting with Einstein's equations themselves, are numerically solved by reducing them to first order systems because there is a large pool of advanced and mature analytical and numerical techniques for them. However, this is at the expense of enlarging the system and, in particular, introducing extra constraints (though this seems to be less of a concern; see, for example, \cite{lLmSlKrOoR06,dBetal12}). It seems more natural to solve such equations directly in second order (at least in space) form. It turns out, though, that it is considerably more complicated to ensure stability for such systems. Trying to ``integrate back'' an algorithm for a first order reduction to its original second order form is in general not possible; see, for example, Refs.~\cite{Calabrese:2005ft,Taylor:2010ki} for a discussion of this apparent paradox and the difficulties associated with constructing boundary closures for second order systems such that an energy estimate follows.  
 
The SAT approach was generalized to a class of wave equations in
second order form in space in Refs.~\cite{Mattsson2004503,
  MattssonParisi2010, Mattsson:2009,Cecere:2011aa}. Part of the difficulty in
obtaining an energy estimate for second order variable coefficient systems in these approaches is related to the property that the finite
difference operators now depend on the equations being solved. This
also complicates their generalization to arbitrary systems. For example, in Ref.~\cite{Mattsson:2009} deals with systems of the form
$$
au_{tt} = (bu_x)_x \, ,\qquad 0\leq x \leq 1,
$$
where $a,b: [0,1]\to\Real$ are strictly positive, smooth functions, and 
 difference and dissipative operators of the desired order satisfying SBP approximating $(bu_x)_x$ are constructed. See also Ref.~\cite{MattssonParisi2010} for generalizations which include shift-type terms.

In~\cite{Mattsson2006249} the projection method and high order SBP operators approximating second derivatives were used to provide interface boundary conditions for a wave equation (directly in second order in space form) in discontinuous media while guaranteeing an energy estimate.  The domain in this work is  a rectangular multi-block domain with piecewise constant coefficients, where the interfaces coincide with the location of the discontinuities. This approach was generalized, using the SAT for the jump discontinuities instead of projection, to variable coefficients and complex geometries in~\cite{Mattsson20088753}. 

The difficulty is not related to finite differences: in~\cite{Taylor:2010ki} a penalty multi-domain method was derived for second order systems. For the Legendre and constant coefficient case the method guarantees an energy estimate but difficulties are reported in guaranteeing one in the variable coefficient case. Nevertheless, it appears to work well in practice in the variable coefficient case as well. An interesting aspect of the approach of~\cite{Taylor:2010ki} is that an energy estimate is obtained by applying the penalty in the whole domain (as an analogy we recall the above discussion about the Legendre--Chebyshev penalty method in Section~\ref{SubSubSec:PenaltyOuterBC}).   

A recent generalization, valid both for finite differences and -- at
least Legendre -- collocation methods (as discussed, the underlying
tool is the same: Summation by Parts), to more general penalty
couplings,  where the penalty terms are not scalar but matrices
(i.e., there is coupling between the penalty for different
characteristic variables) can be found
in~\cite{Carpenter:2010}. 

Energy estimates are in general lost when the different grids are not \textit{conforming} (different types of domain decompositions are discussed in the next section), and interpolation is needed. This is the case when using overlapping patches with curvilinear coordinates but also mesh refinement with Cartesian, nested boxes (see, for example, Ref.~\cite{Lehner:2005vc} and references therein). A recent promising development has been the introduction of a procedure for systematically constructing interpolation operators preserving the Summation by Parts property for arbitrary high order cases, see Ref.~\cite{MattssonCarpenter2010}. Numerical tests are presented with a 2:1 refinement ratio, where the design convergence rate is observed. It is not clear whether reflections at refinement boundaries such as those reported in Ref.~\cite{Baker:2005xe} would still need to be addressed or they would be taken care of by the high order accuracy.

\subsubsection{Absorbing boundary conditions}
\label{SubSubSec:AbsorbingNum}

Finally, we mention some results in numerical relativity concerning absorbing artificial boundaries. In Ref.~\cite{jNsB04}, boundary conditions based on the work of Ref.~\cite{aBeT80}, which are perfectly absorbing for quadrupolar solutions of the flat wave equation, were numerically implemented via spectral methods, and proposed to be used in a constrained evolution scheme of Einstein's field equations~\cite{sBeGpG04}. For a different method which provides exact, nonlocal outer boundary conditions for linearized gravitational waves propagating on a Schwarzschild background, see~\cite{sL04a,sL04b,sL05}. A numerical implementation of the well-posed harmonic IBVP with Sommerfeld-type boundary conditions given in Ref.~\cite{hKjW06} was worked out in~\cite{mBhKjW06}, where the accuracy of the boundary conditions was also tested.

In Ref.~\cite{oRlLmS07}, various boundary treatments for the Einstein equations were compared to each other using the test problem of a  Schwarzschild black hole perturbed by an outgoing gravitational wave. The solutions from different boundary algorithms were compared to a reference numerical solution obtained by placing the outer boundary at a  distance large enough to be causally disconnected from the interior spacetime region where the comparison was performed. The numerical implementation in~\cite{oRlLmS07} was based on the harmonic formulation described in~\cite{lLmSlKrOoR06}.

In Figure~\ref{fig:absorbing}, a comparison is shown between (a) simple boundary conditions which just freeze the incoming characteristic fields to their initial value, and (b) constraint-preserving boundary conditions controlling the complex Weyl scalar $\Psi_0$ at the boundary. The boundary surface is an approximate metric sphere of areal radius $R = 41.9\,M$, with $M$ the mass of the black hole. The left side of the figure demonstrates that case (a) leads to significantly larger reflections than case (b). The difference with the reference solution after the first reflection at the boundary is not only large in case (a), but also it does not decrease with increasing resolution. Furthermore, the violations of the constraints shown in the right side of the figure do not converge away in case (a), indicating that one does not obtain a solution to Einstein's field equations in the continuum limit. In contrast to this, the difference with the reference solution and the constraint violations both decrease with increasing resolution in case (b).

\epubtkImage{}{
\begin{figure}[ht]
\centerline{
\includegraphics[width=0.48\textwidth]{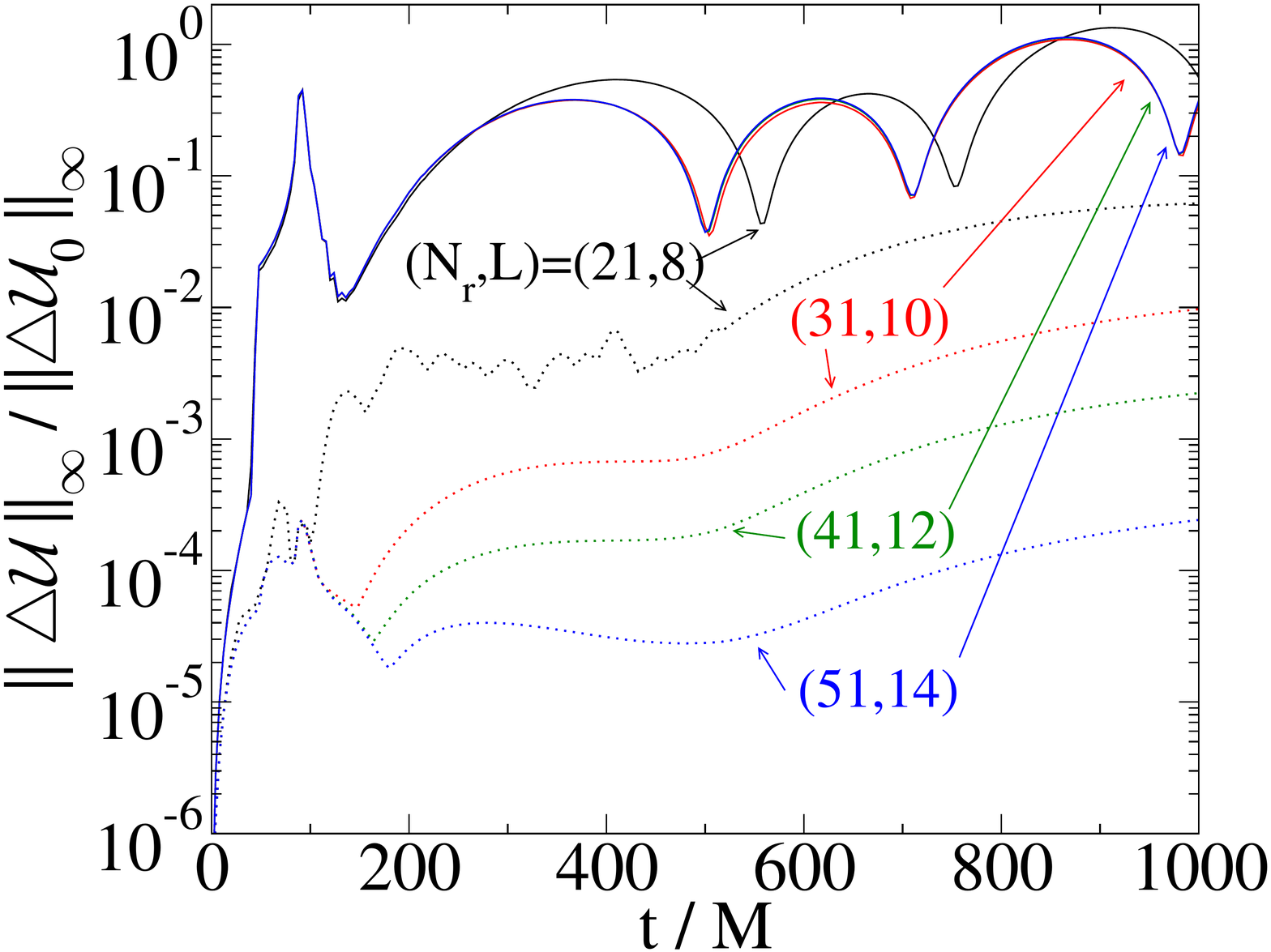}
\includegraphics[width=0.48\textwidth]{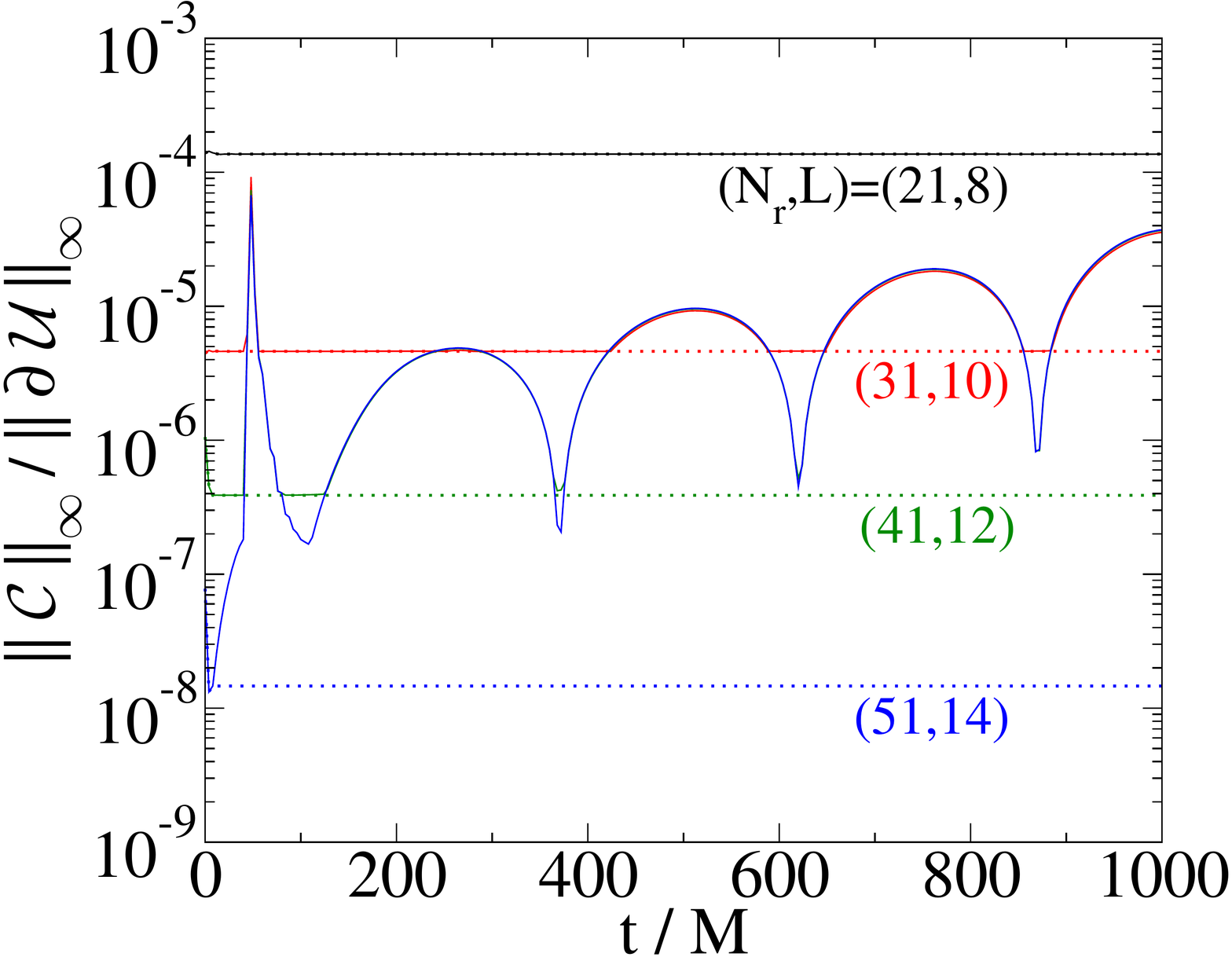}
}
\caption{Comparison between boundary conditions in case (a) (solid)
  and case (b) (dotted); see the body of the text for more details. Four different resolutions are shown:
  $(N_r,L) = (21,8)$, $(31,10)$, $(41,12)$ and $(51,14)$, where $N_r$
  and $L$ refer to the number of collocation points in the radial and
  angular directions, where Chebyshev and spherical harmonics are
  used, respectively. Left panel: the difference $\Delta{\cal U}$
  between the solution with outer boundary at $R=41.9\,M$ and the
  reference solution. Right panel: the constraint violation ${\cal C}$
  (see Ref.~\cite{oRlLmS07} for precise definitions of these
  quantities and further details). Courtesy: Oliver Rinne.}
\label{fig:absorbing}
\end{figure}}

A similar comparison was performed for alternative boundary conditions, including spatial compactifications and sponge layers. The errors in the gravitational waves were also estimated in~\cite{oRlLmS07}, by computing the complex Weyl scalar $\Psi_4$ for the different boundary treatments, see Figure~10 in that reference.

Based on the construction of exact in- and outgoing solutions to the linearized Bianchi identities on a Minkowksi background (cf. Example~\ref{Example:Weyl}), the reflection coefficient $\gamma$ for spurious reflections at a spherical boundary of areal radius $R$ which sets the Weyl scalar $\Psi_0$ to zero was estimated to be~\cite{lBoS06}
\begin{equation}
|\gamma(kR)| \approx \frac{3}{2} (k R)^{-4},
\end{equation}
for outgoing quadrupolar gravitational radiation with wave number $k\gg R^{-1}$. Figure~\ref{fig:reflection} shows the Weyl scalars $\Psi_0$ and $\Psi_4$ computed for the boundary conditions in case (b) and extracted at $1.9\,M$ inside the outer boundary. By computing their Fourier transform in time, the overall dependence of the ratio agrees very well with the predicted reflection coefficient.

For higher-order absorbing boundary conditions, which involve
derivatives of the Weyl scalar $\Psi_0$, see Ref.~\cite{mRoRoS07}, and
Ref.~\cite{oRlBmShP09} for their numerical implementation.

\epubtkImage{}{
\begin{figure}[ht]
\centerline{
\includegraphics[width=0.48\textwidth]{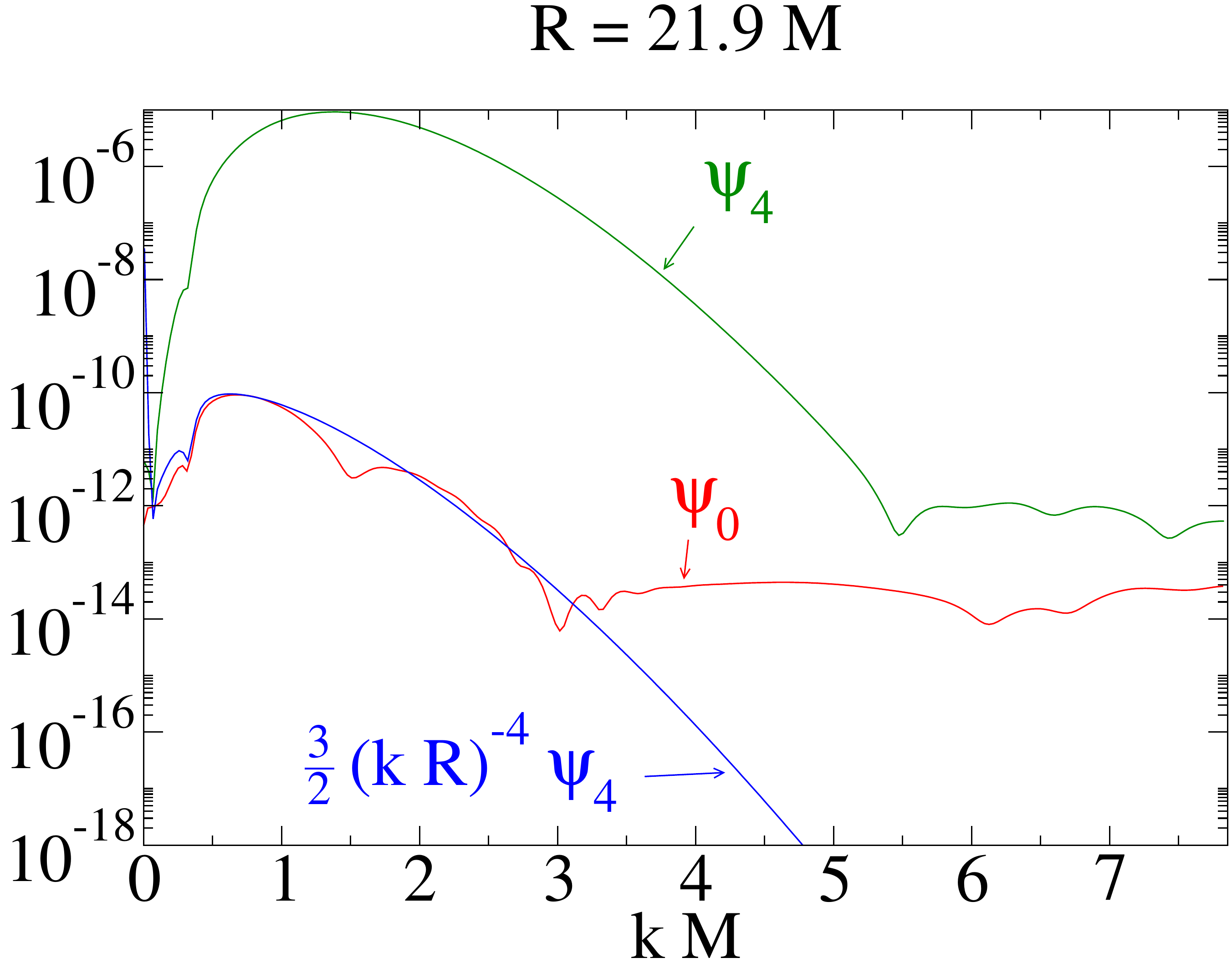}
\includegraphics[width=0.48\textwidth]{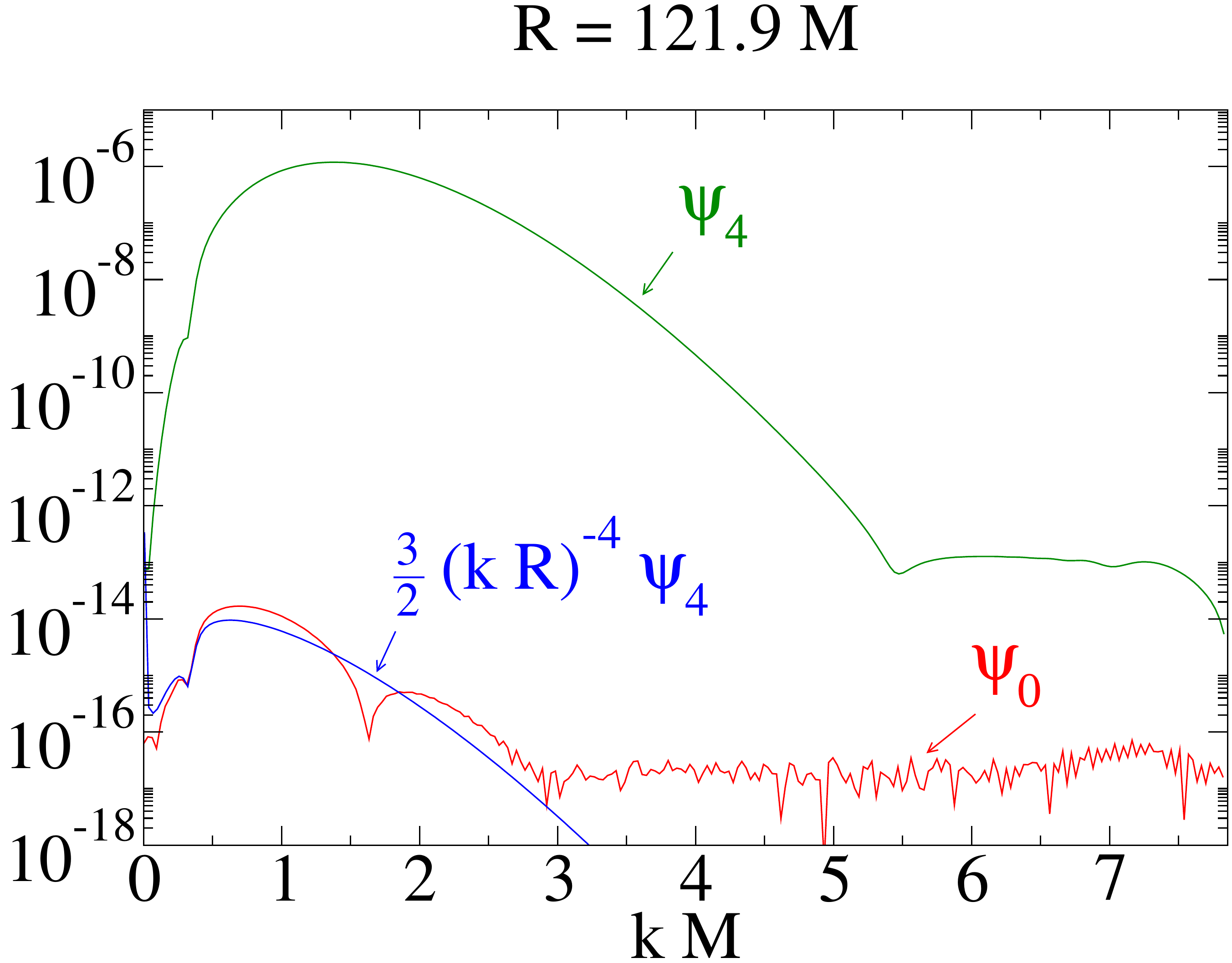}
}
\caption{Comparison of the time Fourier transform of $\Psi_0$ and
  $\Psi_4$ for two different radii of the outer boundary. The leveling
  off of $\Psi_0$ for $k M\gtrsim 3$ is due to numerical roundoff
  effects (note the magnitude of $\Psi_0$ at those
  frequencies). Courtesy: Oliver Rinne.}
\label{fig:reflection}
\end{figure}}

\clearpage
%===================================================================
%===================================================================
\section{Domain Decomposition}
\label{sec:complex}
%===================================================================
%===================================================================

Most three-dimensional codes solving the Einstein equations currently use several non-uniform grids/numerical domains. Adaptive mesh refinement (AMR) a la Berger-Oliger \cite{Berger1984484}, where the computational domain is covered with a set of nested grids, usually taken to be Cartesian ones, is used by many efforts. See, for instance, Refs.~\cite{Schnetter:2003rb, Pretorius:2005ua, Sperhake:2005uf, Lehner:2005vc, Evans:2005mt, Anderson:2006ay, Sperhake:2006cy, Baker:2006yw, Bruegmann:2006at, Campanelli:2007ew, Washik:2008jr, Yamamoto:2008js, Witek:2010zz, Etienne:2010ui, Palenzuela:2010nf}). Other approaches use multiple patches with curvilinear coordinates, or a combination of both. Typical simulations of Einstein's equations do not fall into the category of complex geometries and usually require a fairly ``simple'' domain decomposition (in comparison to fully unstructured approaches in other fields).

Below we give a brief overview of some domain decomposition approaches. Our discussion is far from exhaustive, and only a few representative samples from the rich variety of efforts are mentioned.  In the context of Cauchy evolutions, the use of multiple patches in numerical relativity was first advocated and pursued by Thornburg~\cite{Thornburg:2000cb, Thornburg:2004dv}.

\subsection{The power and need of adaptivity}

Physical problems governed by nonlinear equations can develop small scale structures which are rarely easy to predict. The canonical example in hydrodynamics is turbulence, which results in short wavelength features up to the viscous scale. An example in General Relativity of arbitrarily small scales is that one of critical phenomena \cite{Choptuik:1992jv,cGjG07}, where the solution develops a self-similar behavior revealing a universal approach to a singular one describing a naked singularity. Uncovering this phenomena crucially required to dynamically adjust the grid structure to respond to the (exponentially) ever-shrinking features of the solution. Recently, such need was also demonstrated quite clearly by the resolution of the final fate of unstable black strings \cite{lLfP10,lLfP11}. This work followed the dynamics, in five dimensional spacetimes, of an unstable \emph{black string}: a black hole with topology $S^2 \times S^1$, with the asymptotic length of $S^1$ over the black hole mass per unit length above the critical value for linearized stability \cite{Gregory:1993vy,Gregory:1994bj}. As the evolution unfolds, pieces of the string shrink and elongate (so that the area increases), yielding another unstable stage, see Figure~\ref{fig:LP}. This behavior repeats in a self-similar manner, the black string developing a fractal structure of thin black strings joining spherical black holes. This behavior was followed through four generations and the numerical grid refined in some regions up to a factor of $2^{17}$ compared to the initial one. This allowed the authors to extrapolate the observed behavior and conclude that the spacetime will develop naked singularities in finite time from generic conditions, thereby providing a counterexample to the cosmic censorship conjecture in five dimensions.

\epubtkImage{}{
\begin{figure}[htbp]
\centerline{
\includegraphics[width=0.3\textwidth]{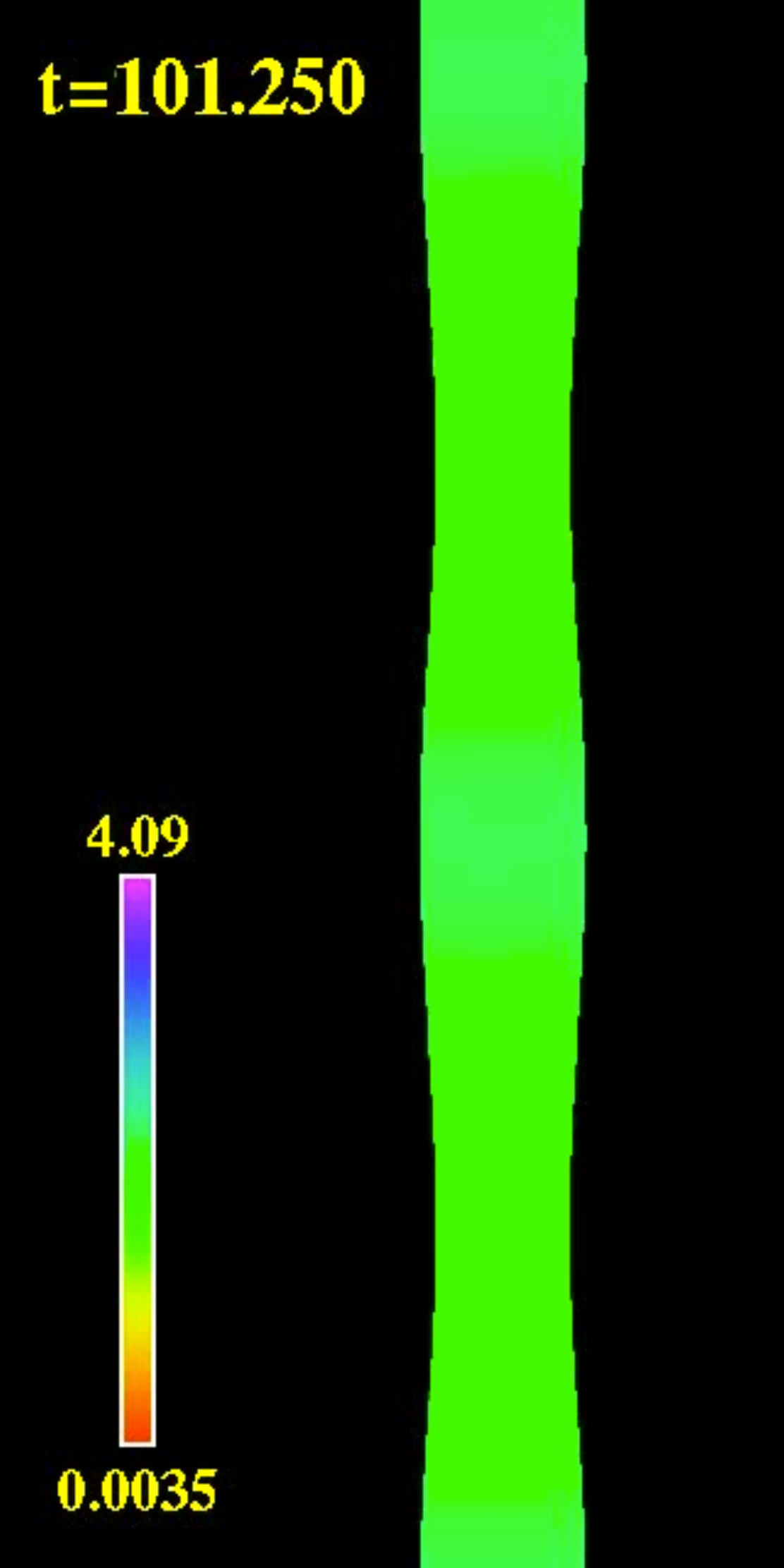}
\includegraphics[width=0.3\textwidth]{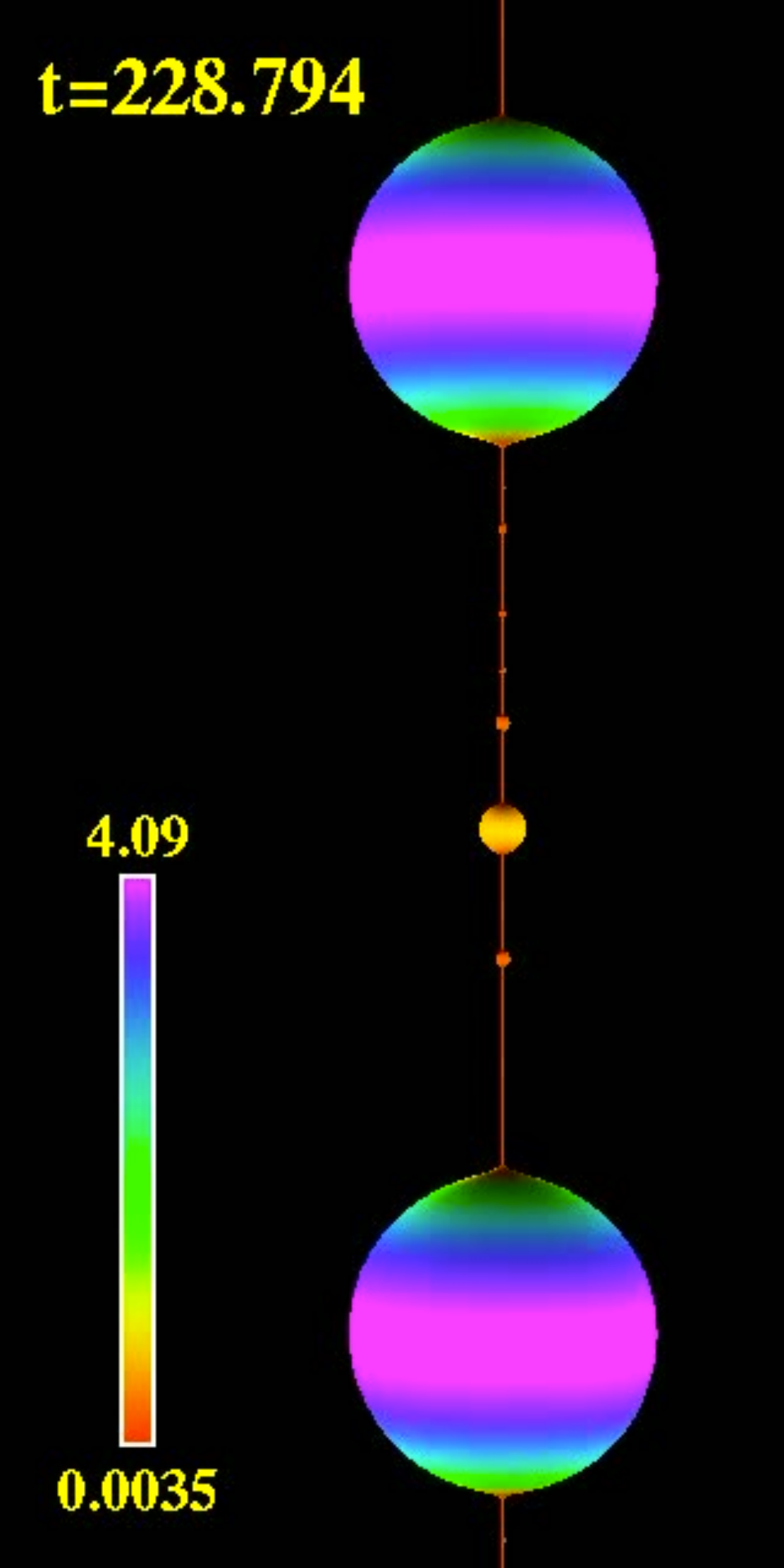}
}
\caption{Early (left) and late (right) stages of the apparent horizon describing the evolution of an unstable black string.  Courtesy: Luis Lehner and Frans Pretorius.}
\label{fig:LP}
\end{figure}}

\subsection{Adaptive mesh refinement for BBH in higher dimensional gravity}

In~\cite{Witek:2010zz} the authors numerically evolved binary black holes ``in a box'' by imposing reflecting outer boundary conditions which mimic the Anti-de~Sitter spacetime. These conditions are imposed on all the fields and the outer boundary is taken to have spherical shape, which is approximated by a ``Lego'' sphere. Figure~\ref{fig:amr}  on the left illustrates a Lego-sphere around a black hole binary, mimicking  an asymptotically Anti-de Sitter spacetime. A computational domain is schematically displayed using four refinement levels with one or two  components each. The individual components are labeled $G^i_m$, where  the indices $i$ and $m$ denote the refinement level and component number, respectively. At the spherical boundary, marked by $X$, reflecting boundary conditions were imposed.
The right figure shows mesh refinement around two black holes, with the apparent horizons represented by the white grid. Further extensions of this research program to numerically study black holes in higher dimensional gravity can be found in~\cite{Witek:2010xi,Witek:2010az}. 

\epubtkImage{}{
\begin{figure}[htbp]
\centerline{
\includegraphics[width=0.4\textwidth]{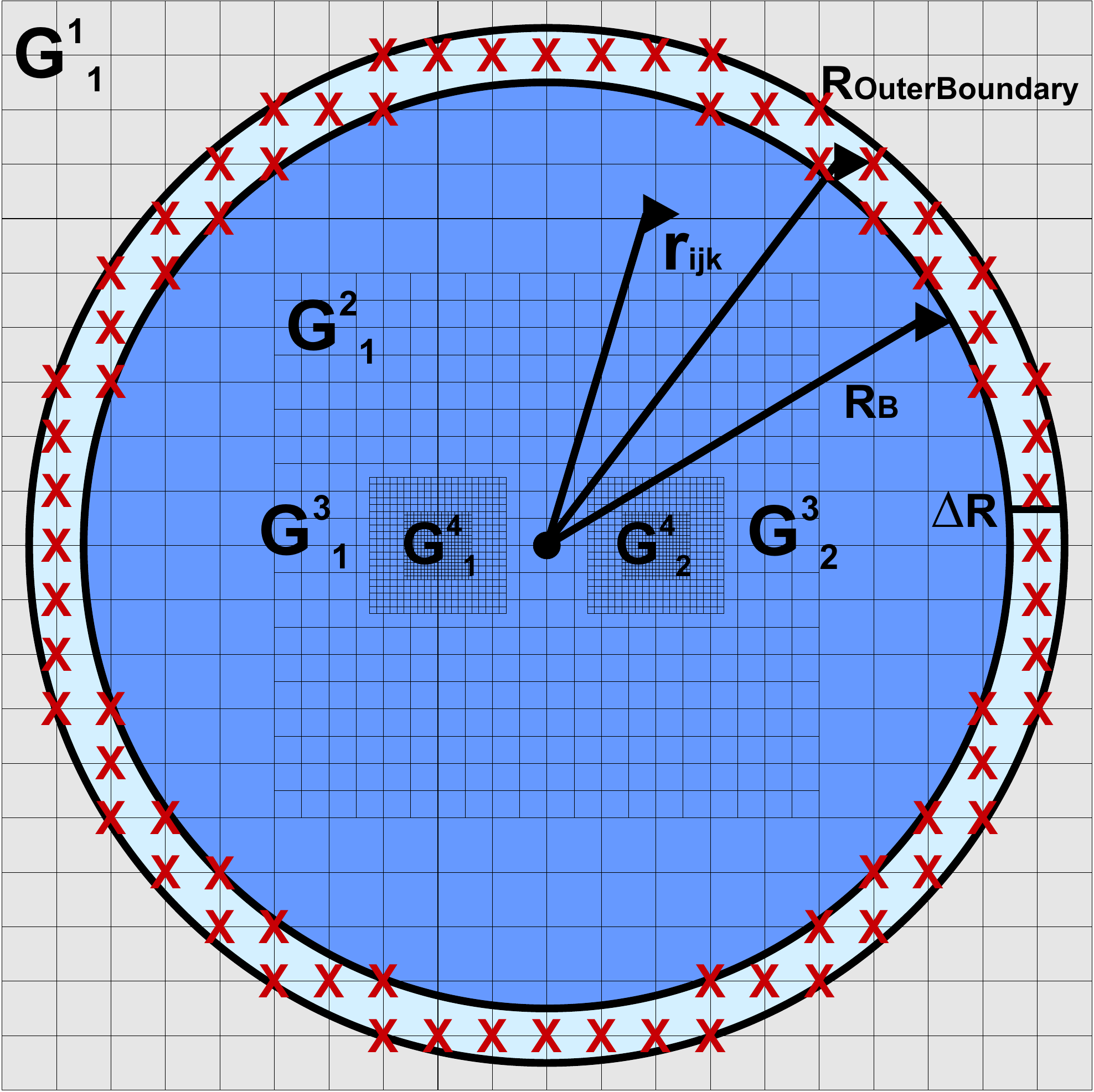}
\includegraphics[width=0.55\textwidth]{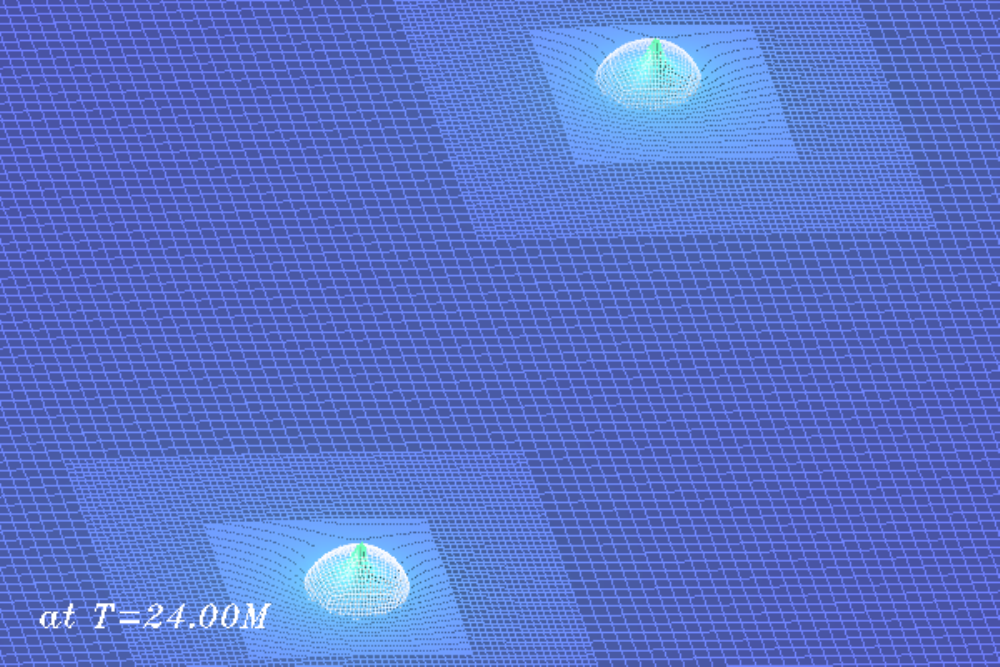}
}
\caption{Left: Lego-sphere around a black hole binary. Right: Mesh refinement  
around two black holes. Courtesy: Vitor Cardoso, Ulrich Sperhake and Helvi Witek.}
\label{fig:amr}
\end{figure}}

\subsection{Adaptive mesh refinement and curvilinear grids}

In Refs.~\cite{Pollney:2009ut,Pollney:2009yz} the authors introduced an approach which combines the advantages of adaptive mesh refinement near the ``sources'' (say, black holes) with curvilinear coordinates adapted to the wave zone, see Figure~\ref{fig:amr_mp}. The patches are communicated using polynomial interpolation in its Lagrange form, as explained in Section~\ref{sec:interpolation}, and centered stencils are used, both for finite differencing and interpolation. Up to eighth order finite differencing is used, with an observed convergence rate between six and eight in the $(\ell =2=m)$ modes of the computed gravitational waves (parts of the scheme have of lower order convergence rate, but they do not appear to dominate). Presently, the BSSN formulation of the Einstein's equations as described in Section~\ref{SubSec:BSSN} is used directly in its second order in space form, with outgoing boundary conditions for all the fields.  The implementation is generic and flexible enough to allow for other systems of equations, though.  As in most approaches using curvilinear coordinates in numerical relativity, the field variables are expressed in a global coordinate frame. This might sound unnatural and against the idea of using local patches and coordinates. However, it simplifies dramatically any implementation. It is also particularly important when taking into account that most formulations of Einstein's equations and coordinate conditions used are not covariant. 

This hybrid approach has been used in several applications. In particular, for validating extrapolation procedures of gravitational waves extracted from numerical simulations at finite radii to large distances from the ``sources''~\cite{Pollney:2009ut}. Since the outermost grid structure is well adapted to the wave zone the outer boundary can be located at large distances with only linear cost on its location.  Other applications include Cauchy-Characteristic extraction (CCE) of gravitational waves~\cite{cRnBdPbS10, Bishop:2011iu}, a waveform hybrid development~\cite{Santamaria:2010yb}, and studies of memory effect in gravitational waves~\cite{Pollney:2010hs}. The accuracy necessary to study small memory effects is enabled both by the grid structure -- being able to locate the outer boundary far away -- and CCE. 

\epubtkImage{}{
\begin{figure}[htbp]
\centerline{\includegraphics[width=0.6\textwidth]{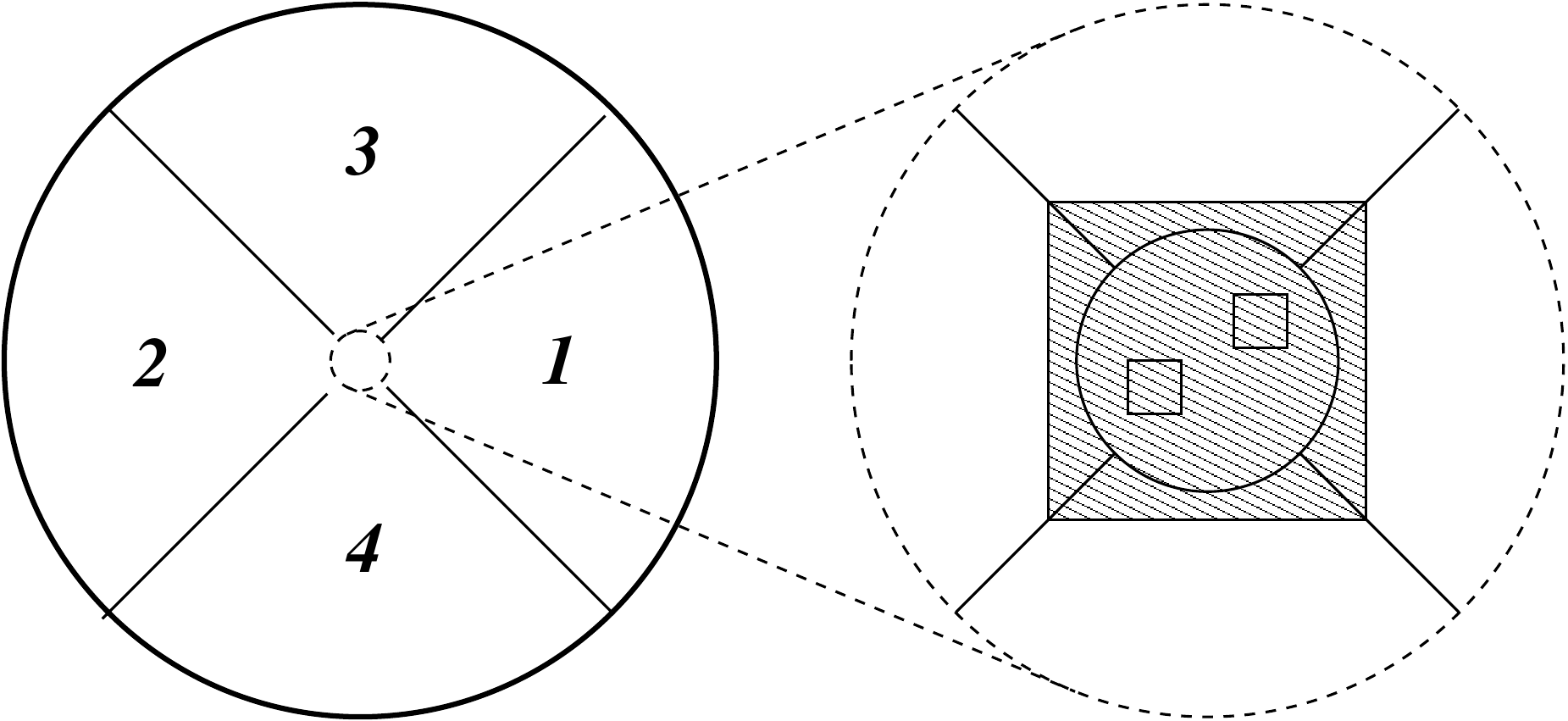}}
\centerline{\includegraphics[width=0.5\textwidth]{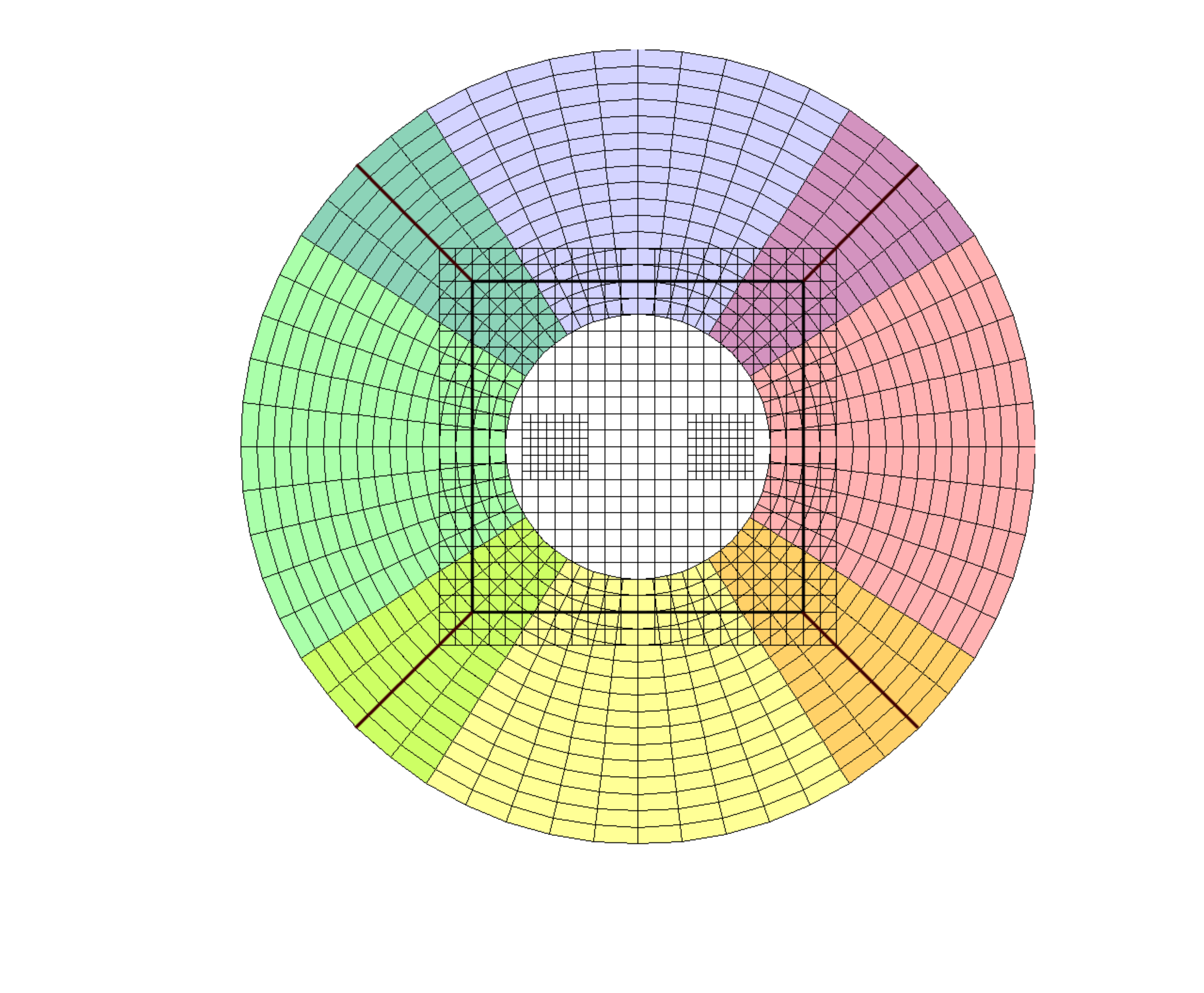}}
\caption{Combining adaptive mesh refinement with curvilinear grids
  adapted to the wave zone. Courtesy: Denis Pollney.}
\label{fig:amr_mp}
\end{figure}}

\subsection{Spectral multi-domain binary black hole evolutions}

In current binary black hole evolutions using spectral collocation methods there are typically three sets of spherical shells, one around each black hole and one in the wave zone. These three shells are connected by subdomains of various shapes and sizes. Figure~\ref{fig:spec} shows the global structure of one such grid-structure, emphasizing the spherical patches in the wave zone and how the different domains are connected. Inter-domain boundary conditions are set by the spectral penalty method described in Section~\ref{sec:num_boundary}. The adaptivity provided by domain decomposition in addition to the spectral convergence rate has lead to the highest accuracy binary black hole simulations to date. Currently these evolutions use a first order symmetry hyperbolic reduction of the harmonic system with constraint damping as derived in~\cite{lLmSlKrOoR06} and summarized in Section~\ref{SubSec:Harmonic}, with constraint-preserving boundary conditions as designed in~\cite{lLmSlKrOoR06,oRlLmS07,oR06}. The field variables are expressed in an ``inertial'' Cartesian coordinate system which is related to one fixed to the computational domain through a dynamically defined coordinate transformation tracking the black holes (the ``dual frame'' method)~\cite{mShPlLlKoRsT06}. 

\epubtkImage{}{
\begin{figure}[htbp]
\centerline{
\includegraphics[width=0.4\textwidth]{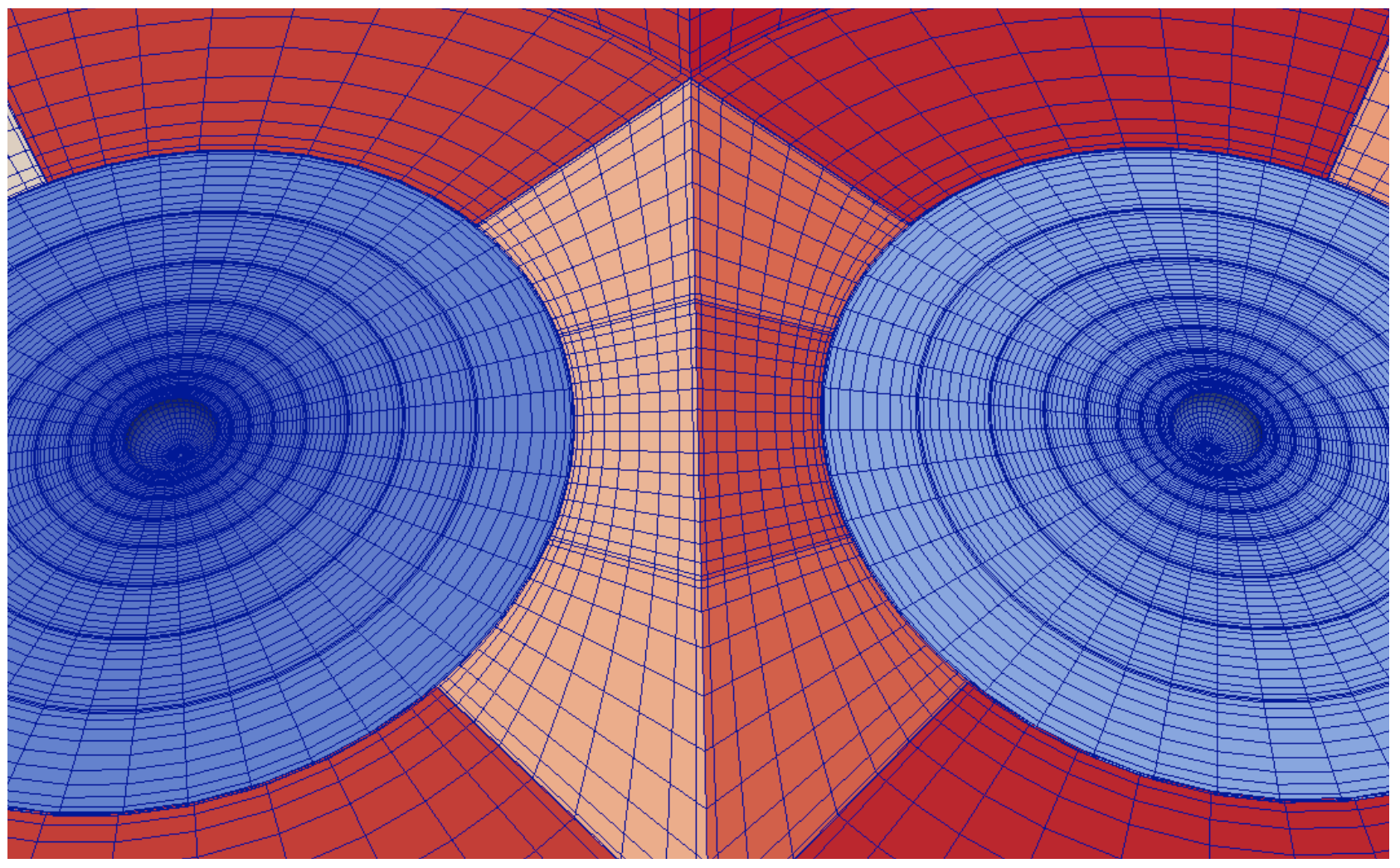}\quad
\includegraphics[width=0.4\textwidth]{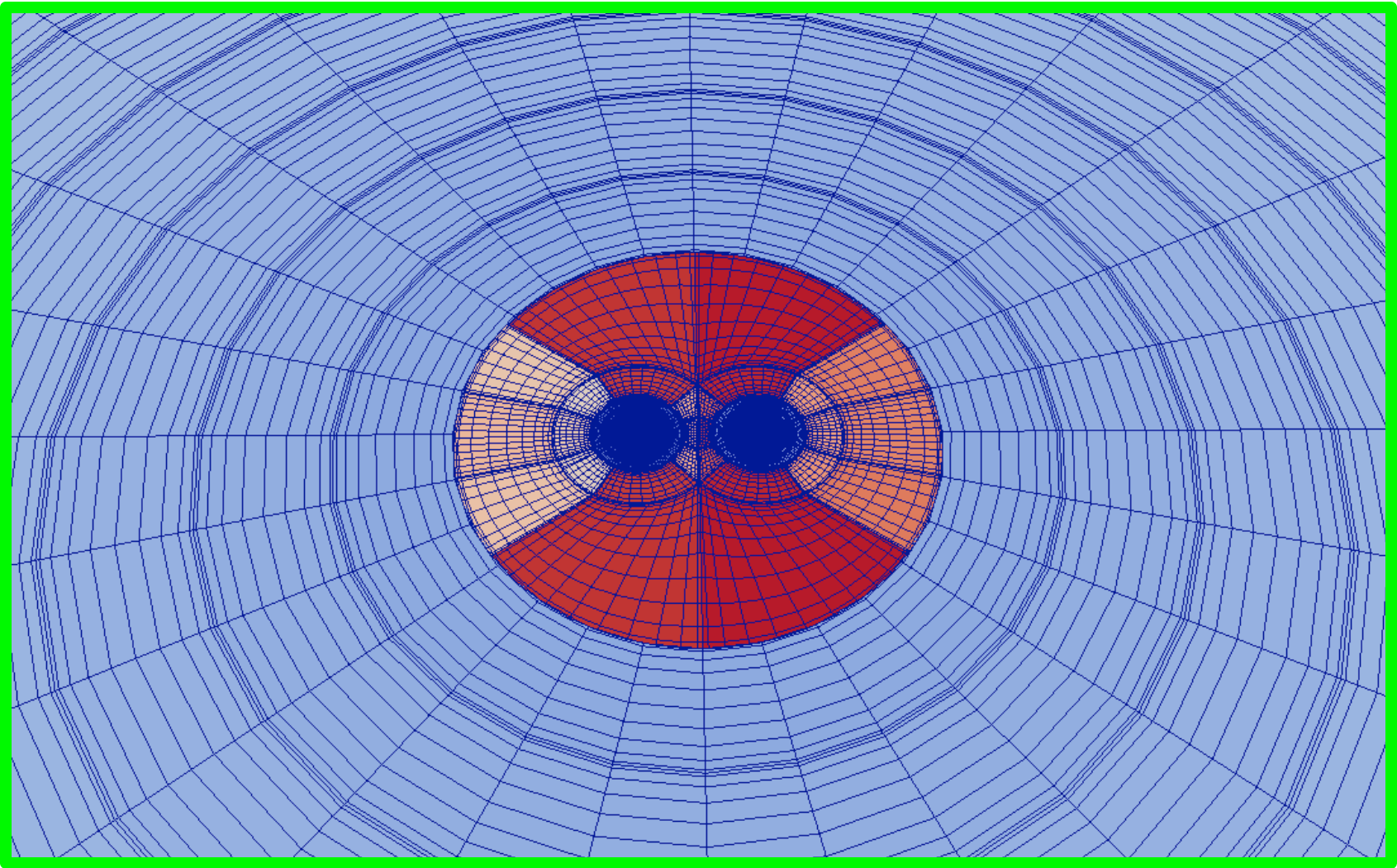}
}
\centerline{
\includegraphics[width=0.3\textwidth]{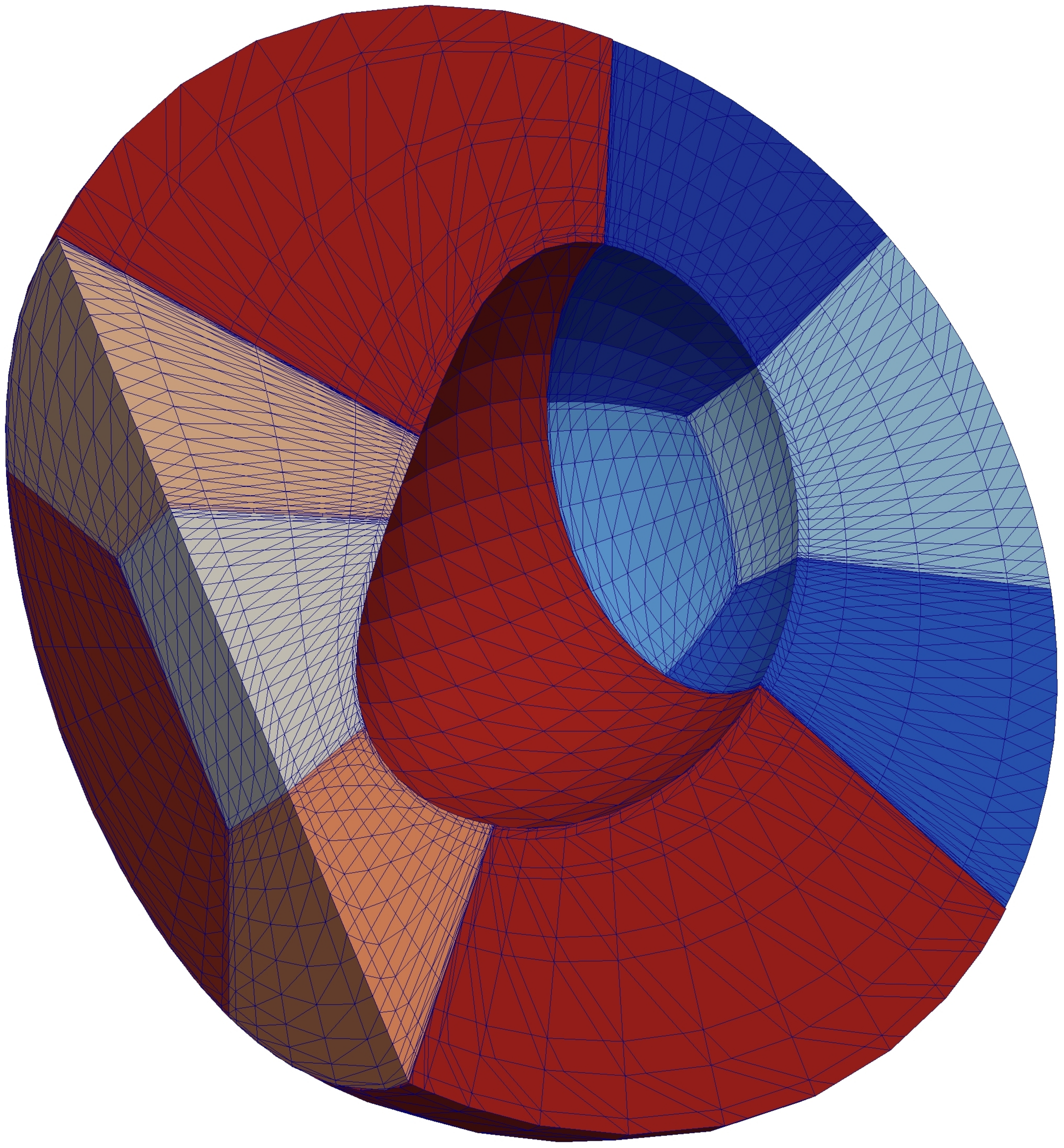}
}
\caption{Sample domain decomposition used in spectral evolutions of
  binary black holes. The bottom plot illustrates how the coordinate
  shape of the excision domain is kept proportional to the coordinate
  shape of the black holes themselves. Courtesy: Bela Szil{\'{a}}gyi.}
\label{fig:spec}
\end{figure}}

Simulating non-vacuum systems such as relativistic hydrodynamical ones using spectral methods can be problematic, particularly when surfaces, shocks, or other non-smooth behavior appears in the fluid.  Without further processing the fast convergence is lost, and Gibbs' oscillations can destabilize the simulation. A method that has been successfully used to overcome this in general relativistic hydrodynamics is evolving the spacetime metric and the fluid on two different grids, each  using different numerical techniques. The spacetime is evolved spectrally, while the fluid is evolved using standard finite difference/finite volume shock-capturing techniques on a separate uniform grid. The first code adopting this approach was described in Ref.~\cite{Dimmelmeier:2004me}, which is a stellar collapse code assuming a conformally flat three-metric, with the resulting elliptic equations being solved spectrally. The two-grid approach was adopted for full numerical relativity simulations of black hole-neutron star binaries in Refs.~\cite{Duez:2008rb, Duez:2009yy, Foucart:2010eq}. The main advantage of this method when applied to binary systems is that at any given time the fluid evolution grid only needs to extend as far as the neutron star matter.  During the pre-merger phase, then, this grid can be chosen to be a small box around the neutron star, achieving very high resolution for the fluid evolution at low computational cost. More recently, in Ref.~\cite{Foucart:2010eq} an automated re-gridder was added, so that the fluid grid automatically adjusts itself at discrete times to accommodate expansion or contraction of the nuclear matter. The main disadvantage of the two-grid method is the large amount of interpolation required for the two grids to communicate with each other. Straightforward spectral interpolation would be prohibitively expensive, but a combination of spectral refinement and polynomial interpolation~\cite{Boyd92} reduces the cost to about 20\,--\,30 percent of the simulation time.

\subsection{Multi-domain studies of accretion disks around black holes}

The freedom to choose arbitrary (smooth) coordinate transformations allows the design of  sophisticated problem-fitted meshes to address a number of practical issues. In~\cite{Korobkin:2010qh} the authors used a hybrid multiblock approach for a general relativistic hydrodynamics code developed in~\cite{Zink:2007xn} to study instabilities in accretion disks around black holes in the context of gamma-ray burst central engines. They evolved the spacetime metric using the first order form of the generalized harmonic formulation of the Einstein equations (see Section~\ref{SubSec:Harmonic}) on conforming grids, while using a high resolution shock capturing scheme for relativistic fluids on the same grid but with additional overlapping boundary zones (see Ref.~\cite{Zink:2007xn} for details on the method). The metric differentiation was performed using the optimized $D_{8-4}$ finite difference operators satisfying the Summation by Parts property, as described in Section~\ref{sec:sbp}. The authors made extensive use of adapted curvilinear coordinates in order to achieve desired resolutions in different parts of the domain and to make the coordinate lines conform to the shape of the solution. Maximal dissipative boundary conditions as defined in Section~\ref{section:MaxDiss} were applied to the incoming fields, and inter-domain boundary conditions for the metric were implemented using the finite-difference version of the penalty method described in Section~\ref{sec:num_boundary}.

Figure~\ref{fig:oleg} shows examples of the type of mesh adaptation used. The top left panel shows the meridional cut of an accretion disk on a uniform multiblock mesh (model C in~\cite{Korobkin:2010qh}). The top right panel gives an example of a mesh with adapted radial coordinate lines which resolves the disk more accurately than the mesh with uniform grid resolution (see~\cite{Korobkin:2010qh} for details on the particular coordinate transformations used to obtain such grid).  A 3D view of such multi-domain mesh at large radii is shown on the bottom left panel of Figure~\ref{fig:oleg}. In the area near the inner radius where the central black hole is located, the resolution is high enough to accurately resolve the shape and the dynamics of the black hole horizon. 
 Finally, near the disk the resolution across the disk in both radial and angular directions is made approximately
 equal and sufficiently high to resolve the transverse disk
 dynamics. The bottom right panel of Figure~\ref{fig:oleg} shows the 3D view of the adapted mesh in the vicinity of the disk.

\epubtkImage{}{
\begin{figure}[htbp]
\centerline{
\includegraphics[width=0.4\textwidth]{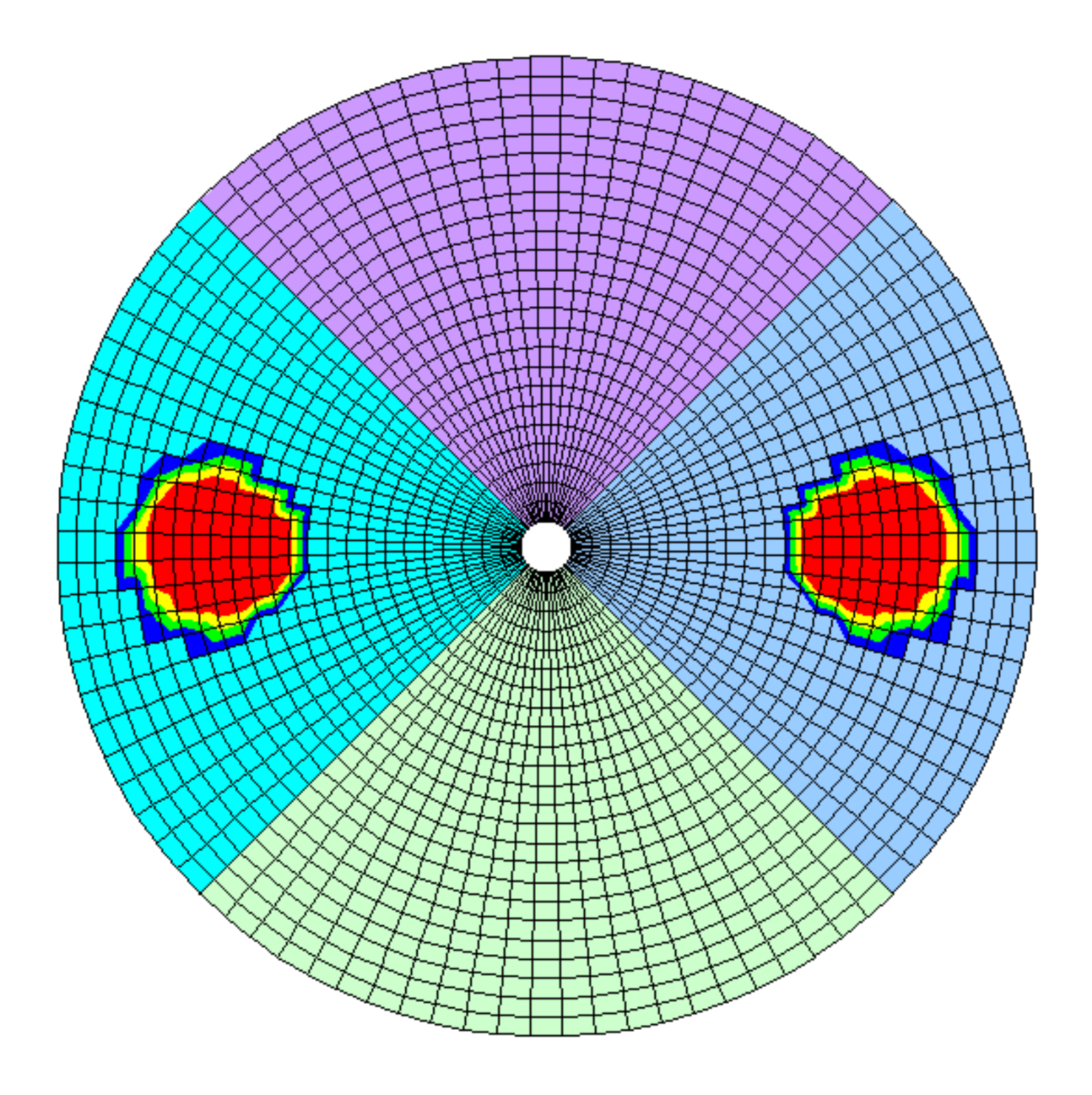}
\includegraphics[width=0.4\textwidth]{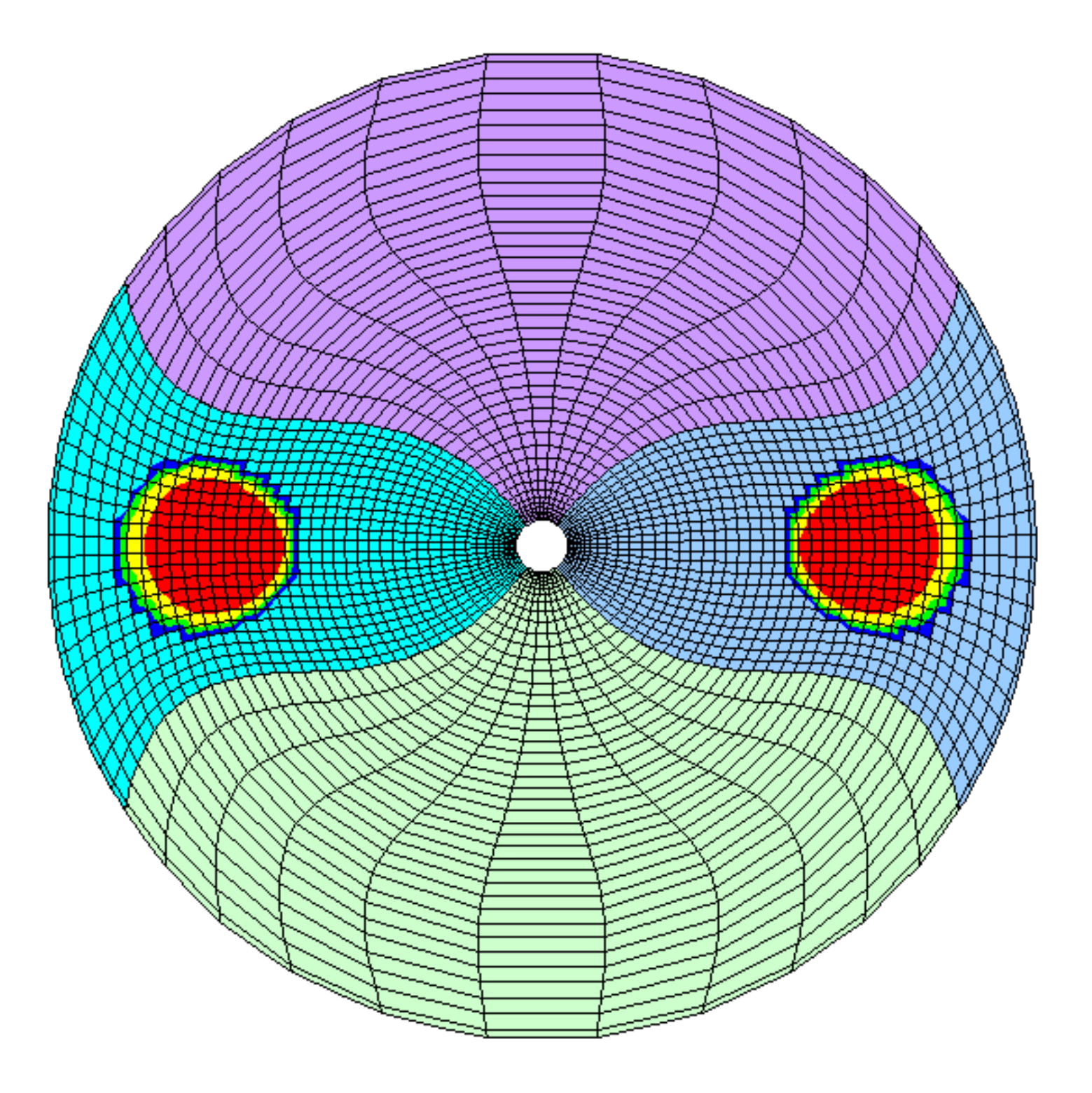}
}
\centerline{
\includegraphics[width=0.4\textwidth]{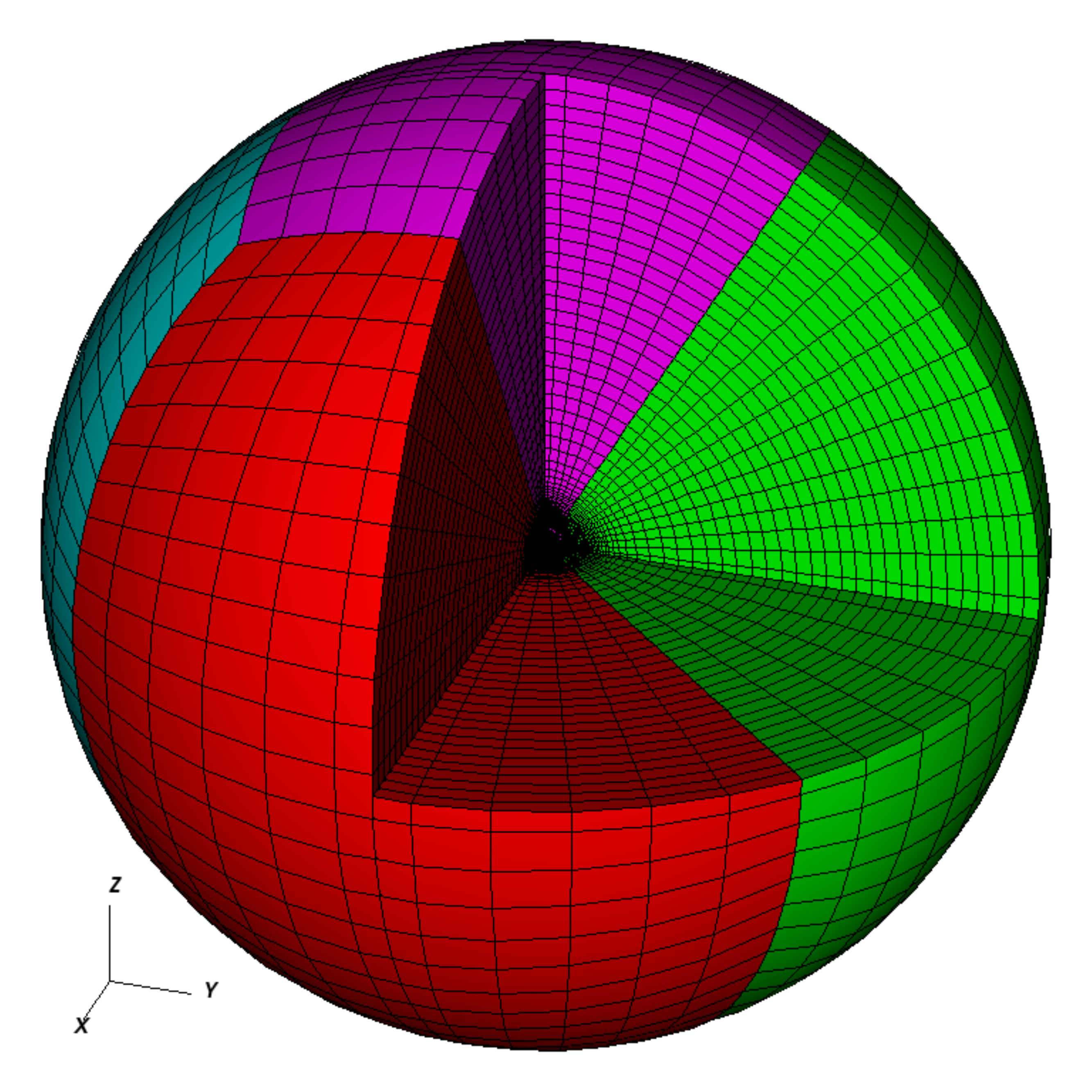}
\includegraphics[width=0.38\textwidth]{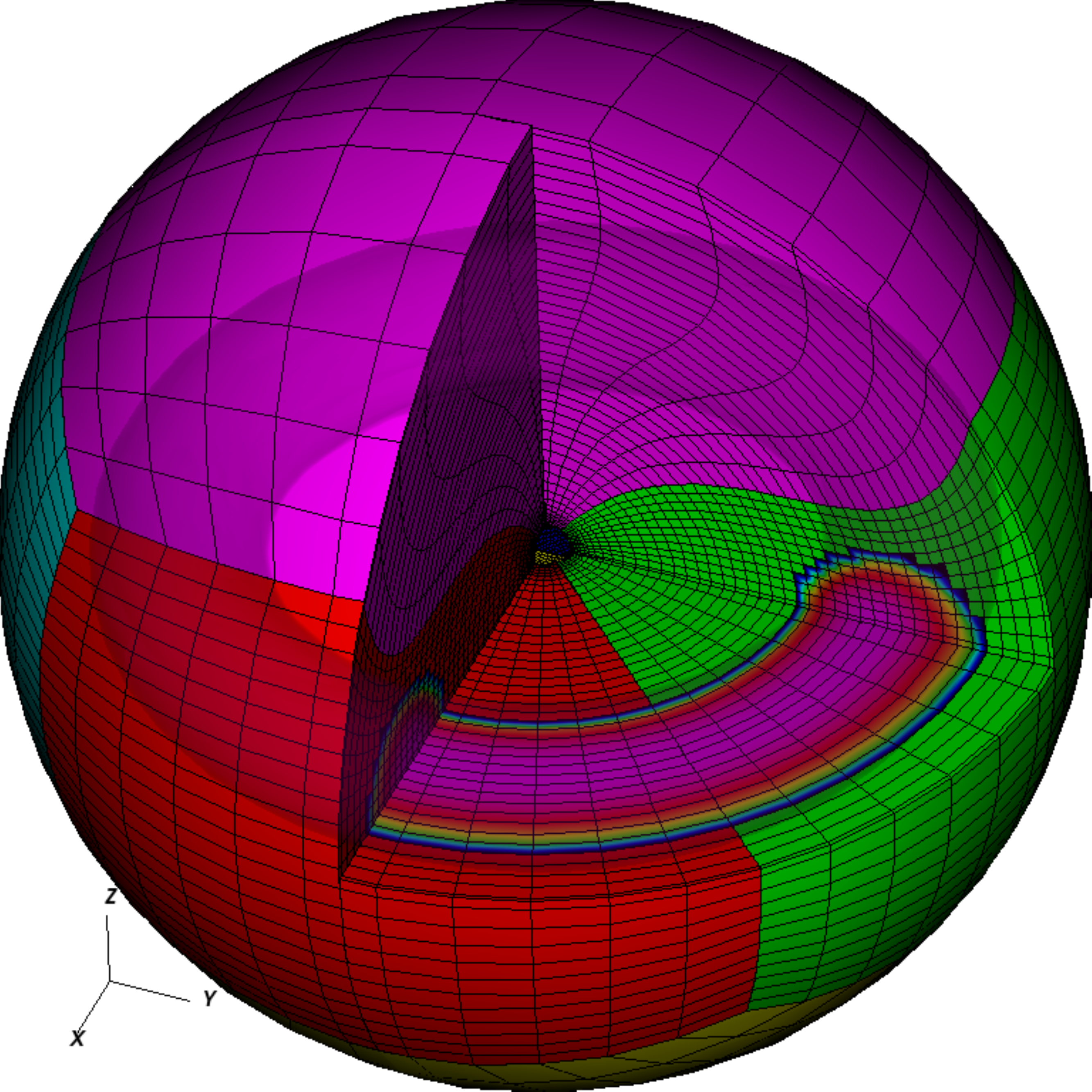}
}
\caption{Domain decomposition used in evolutions of accretion disks
  around black holes \cite{Korobkin:2010qh}, see the text for more
  details. Courtesy: Oleg Korobkin.}
\label{fig:oleg}
\end{figure}}

\subsection{Finite difference multi-block orbiting binary black holes simulations}

In~\cite{Pazos:2009vb} orbiting binary black hole simulations using a high order finite difference multi-domain approach were presented. The basic layout of the full domain used is shown in Figures~\ref{fig:domain} and \ref{fig:domainb}, where the centers of the excised black hole spheres are located along the $x$ axis at $x=\pm a$. The computational domain is based on two types of building blocks, shown in Figure~\ref{fig:jugg-in}, which are essentially variations of cubed spheres. In the simulations of \cite{Pazos:2009vb} the optimized $D_{8-4}$ difference operator described in Section~\ref{sec:sbp} was used, the patches were communicated using the penalty technique (see Section~\ref{sec:num_boundary}), and the resulting equations were evolved in time using an embedded fourth-fifth order Runge--Kutta stepper (see Section~\ref{sec:time_integration}).
The formulations of the equations used, boundary conditions and dual frame technique are exactly those of the spectral evolutions discussed above, only the numerical and domain decomposition approaches differ, which follow building work from~\cite{Lehner:2005bz, Diener:2005tn, Schnetter:2006pg, Dorband:2006gg, Pazos:2006kz}. Strong and weak scaling is observed up to at least several thousand processors.

\epubtkImage{}{
\begin{figure}[htbp]
\centerline{
\includegraphics[width=0.5\textwidth]{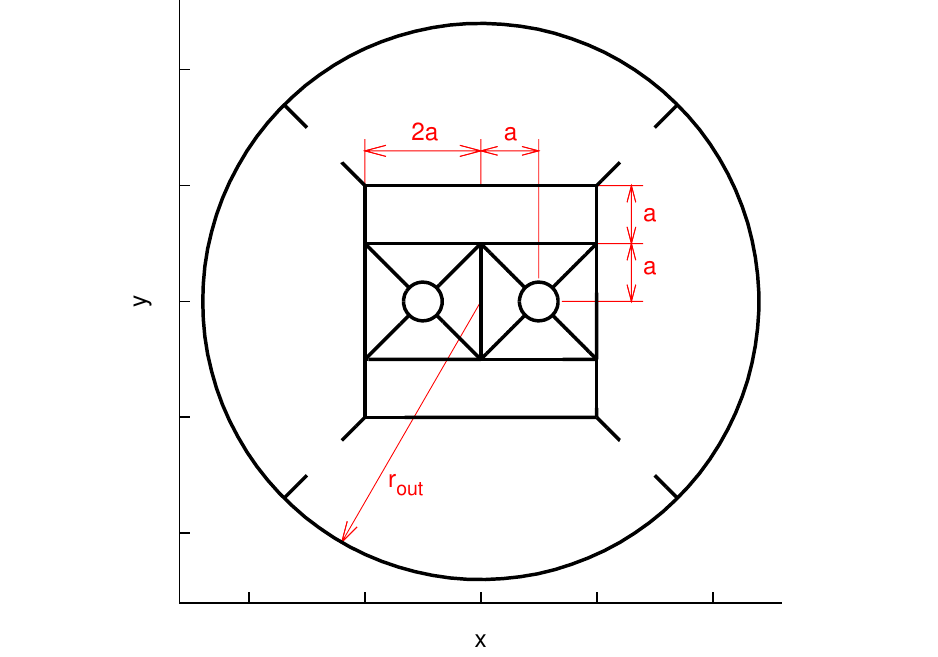}
\includegraphics[width=0.5\textwidth]{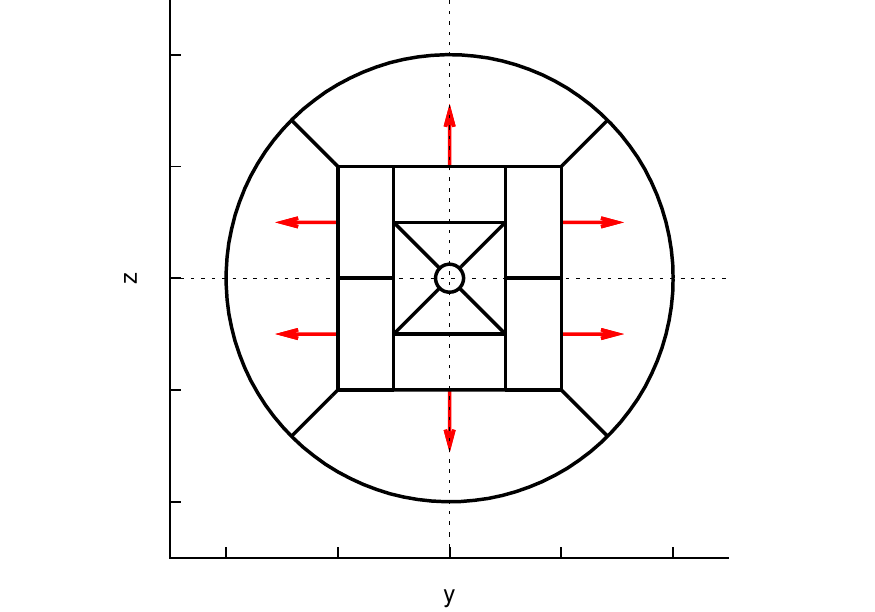}
}
\caption{Equatorial cut of the computational domain used in
  multi-block simulations of orbiting binary black holes
  (left). Schematic figure showing the direction considered as radial
  (red arrows) for the cuboidal blocks (right).}
\label{fig:domain}
\end{figure}}

\epubtkImage{}{
\begin{figure}[htbp]
\centerline{
\includegraphics[width=0.5\textwidth]{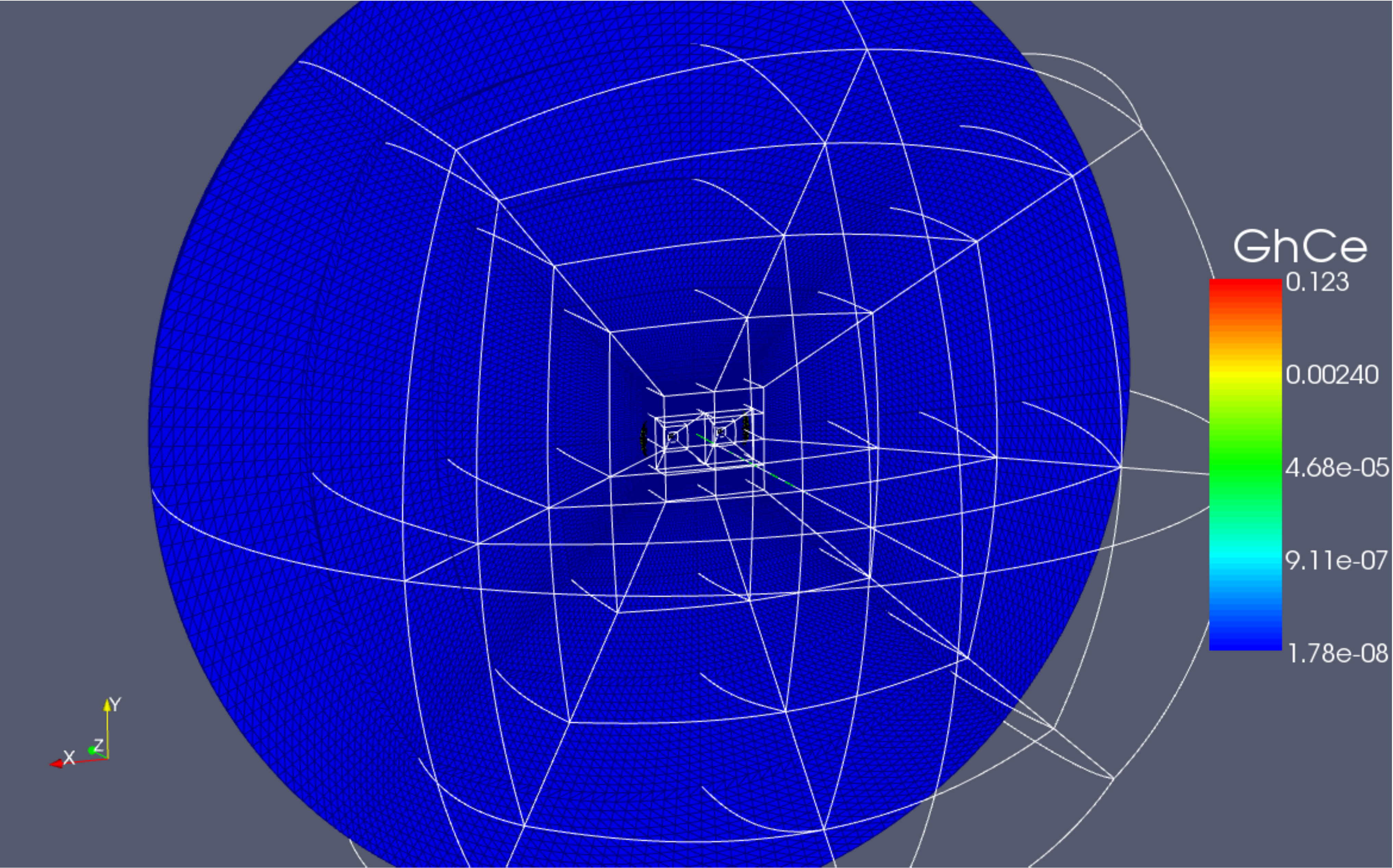}
\includegraphics[width=0.5\textwidth]{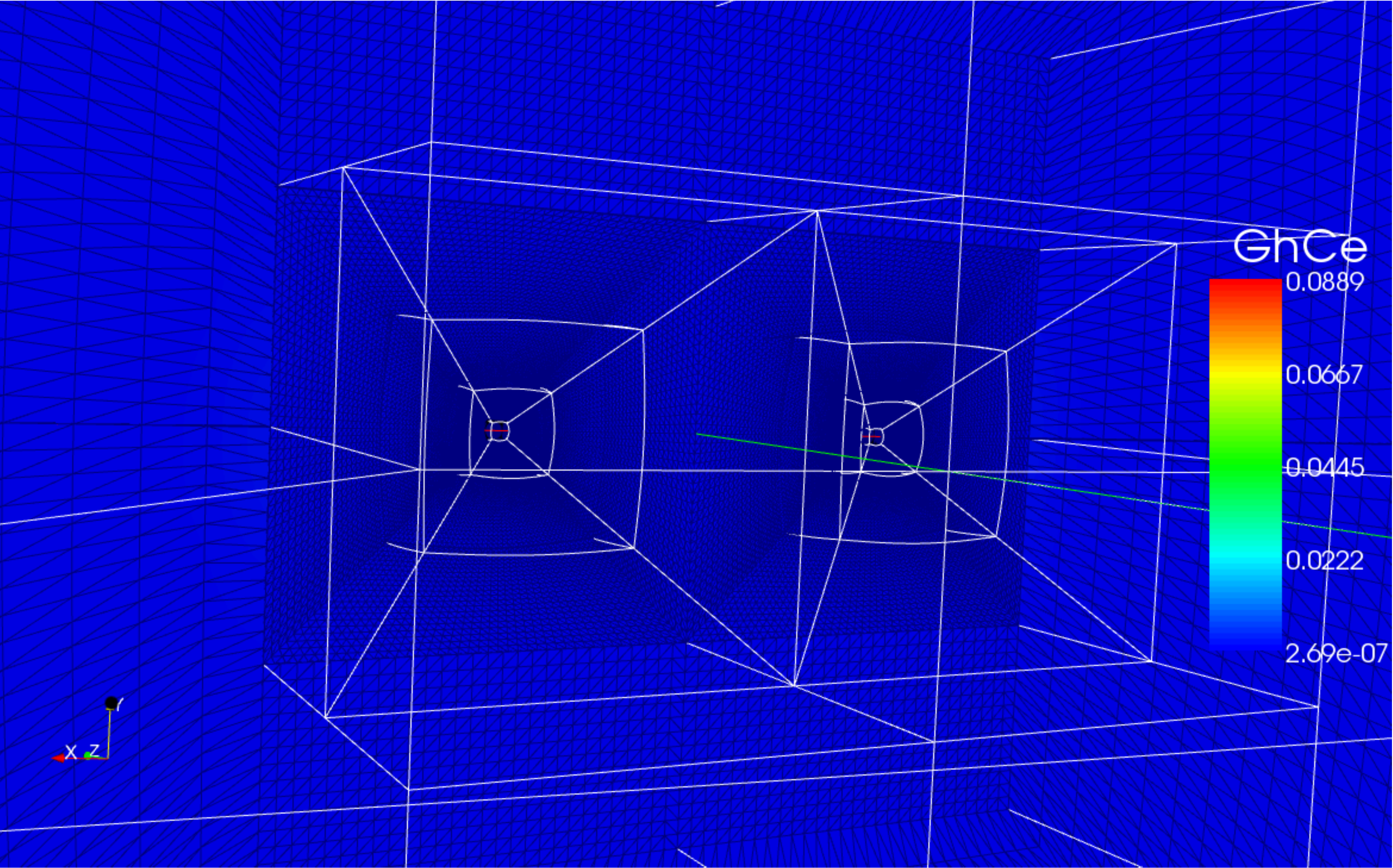}
}
\caption{Multi-block domain decomposition for a binary black hole simulation.}
\label{fig:domainb}
\end{figure}}

\epubtkImage{}{
\begin{figure}[htbp]
\centerline{
\includegraphics[width=0.45\textwidth]{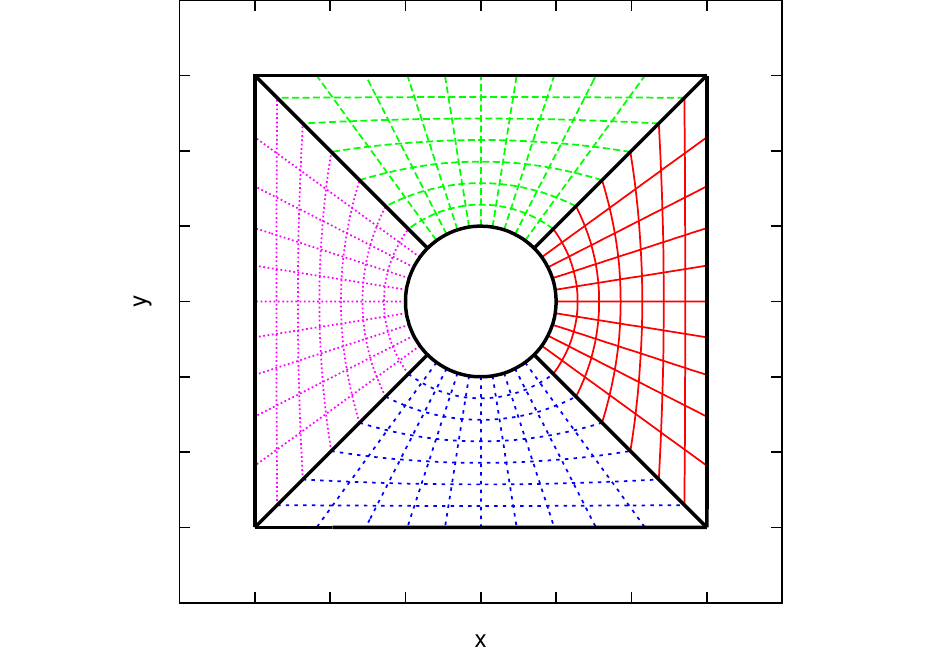}
\includegraphics[width=0.45\textwidth]{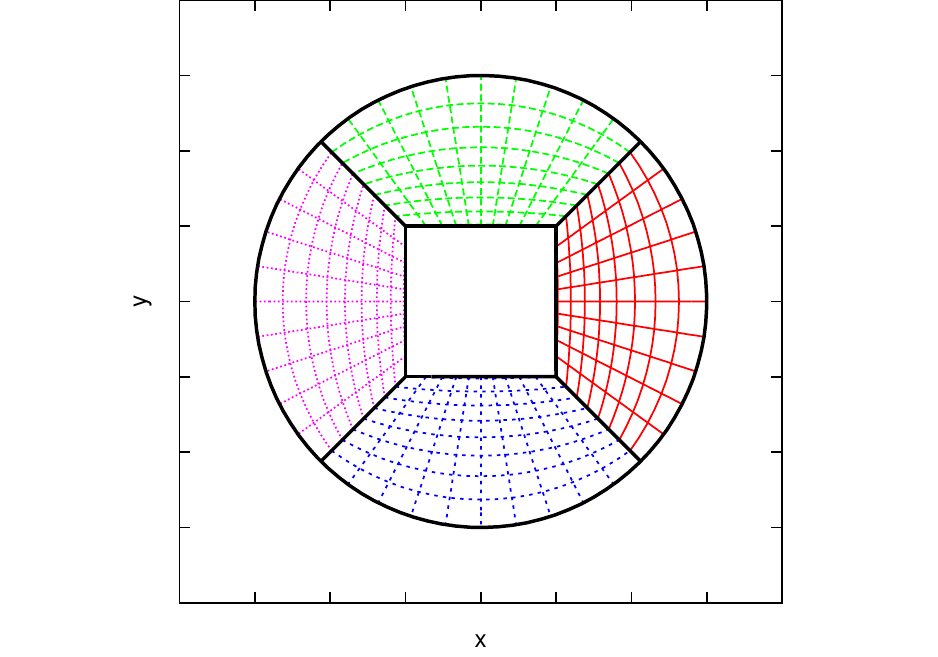}
}
\caption{Equatorial cross-section of varations of cubed sphere patches.}
\label{fig:jugg-in}
\end{figure}}

\clearpage
%===================================================================
\section{Acknowledgements}
\label{section:acknowledgements}
%===================================================================

Our views on the topics of this review have been shaped through
discussions, interactions and collaborations, through the years and/or
at the time of the writing of this review, with many colleagues, to
whom we are greatly indebted. Among many others: James Bardeen, Horst
Beyer, Luisa Buchman, David Brown, Gioel Calabrese, Peter Diener, Nils
Dorband, Scott Field, Helmut Friedrich, Carsten Gundlach, Ian Hawke,
Jan Hesthaven, Frank Herrmann, Larry Kidder, Oleg Korobkin, Heinz-Otto
Kreiss, Luis Lehner, Lee Lindblom, Ken Mattsson, Vincent Moncrief,
Gabriel Nagy, Dar\'{\i}o N\'u\~nez, Pelle Olsson, Enrique Pazos,
Harald Pfeiffer, Denis Pollney, Jorge Pullin, Christian Reisswig,
Oscar Reula, Oliver Rinne, Milton Ruiz, Mark Scheel, Erik Schnetter, Magnus Svard,
Bela Szil{\'{a}}gyi, Eitan Tadmor, Nicholas Taylor, Saul Teukolsky, Jonathan
Thornburg, James York, Jeffrey Winicour, Burkhard Zink, Anil Zengino\u{g}lu.

We especially thank Stephen Lau, Oscar Reula and Eitan Tadmor for going through a previous version of this manuscript in great detail and providing us with very helpful  feedback and Liam MacTurk for helpful edits.

OS thanks the Perimeter Institute for Theoretical Physics, where part
of this work has been done, for its hospitality. MT thanks the
Universidad Michoacana de San Nicol\'as de Hidalgo and Tryst DC
Coffeehouse Bar and Lounge, where parts of this work were done, for
their hospitality. This work has been partially supported by CONACyT
Grant 61173, Grant CIC 4.19 to the Universidad Michoacana de San
Nicol\'as de Hidalgo and by the National Science Foundation Grants
PHY0801213 and PHY0908457 to the University of Maryland.

\newpage

%===================================================================
\bibliography{refs}
%===================================================================

\end{document}